%% file: main.tex
\pdfoutput=1

\documentclass[%
    paper=A4,               %
    twoside=true,           %
    openright,              %
    parskip=full,           %
    chapterprefix=true,     %
    11pt,                   %
    headings=normal,        %
    bibliography=totoc,     %
    listof=totoc,           %
    titlepage=on,           %
    captions=tableabove,    %
    draft=false,            %
]{scrreprt}%

\input{my-thesis-setup}

\usepackage{soul,color}
\usepackage{textcomp}

\usepackage{subcaption}
\usepackage{enumitem}
\usepackage{rotating}
\usepackage{multirow}
\usepackage{makecell}

\usepackage[bottom]{footmisc}

\usepackage[final]{changes}

\usepackage{tabularx}
\newcommand{\tabincell}[2]{\begin{tabular}{@{}#1@{}}#2\end{tabular}}

\usepackage{array}
\newcolumntype{L}[1]{>{\raggedright\let\newline\\\arraybackslash\hspace{0pt}}m{#1}}
\newcolumntype{C}[1]{>{\centering\let\newline\\\arraybackslash\hspace{0pt}}m{#1}}
\newcolumntype{R}[1]{>{\raggedleft\let\newline\\\arraybackslash\hspace{0pt}}m{#1}}

\definecolor{maptrix-correct}{HTML}{0B467A}
\definecolor{maptrix-almost}{HTML}{6298C4}
\definecolor{maptrix-false}{HTML}{FB9E59}
\definecolor{maptrix-too-difficult}{HTML}{D0D0D0}

\definecolor{fifth}{HTML}{C7E9C0}
\definecolor{fourth}{HTML}{A1D99B}
\definecolor{third}{HTML}{74C476}
\definecolor{second}{HTML}{31A354}
\definecolor{first}{HTML}{006D2C}

\definecolor{second_fourth}{HTML}{BAE4B3}
\definecolor{third_first}{HTML}{238B45}

\definecolor{second_of_two}{HTML}{64BB63}
\definecolor{first_of_two}{HTML}{0C5C21}

\definecolor{first_conf}{HTML}{08519C}
\definecolor{last_conf}{HTML}{C6DBEF}

\definecolor{noDizzy}{HTML}{FDD0A2}
\definecolor{dizzy}{HTML}{D94801}

\setlist{nolistsep,topsep=-10pt}
\begin{document}

\renewcaptionname{english}{\figurename}{Fig.}
\renewcaptionname{english}{\tablename}{Tab.}

\pagenumbering{roman}			%
\pagestyle{empty}				%
\input{content/0-before-main-body/titlepages}		%
\cleardoublepage

\pagestyle{plain}				%
\input{content/0-before-main-body/0-copyright}

\input{content/0-before-main-body/1-declaration}

\input{content/0-before-main-body/2-abstract}		%

\input{content/0-before-main-body/acknowledgement} %
\cleardoublepage
\setcounter{tocdepth}{2}		%
\tableofcontents				%
\cleardoublepage

\pagenumbering{arabic}			%
\setcounter{page}{1}			%
\pagestyle{maincontentstyle} 	%

\input{content/1-introduction/0-index}

\input{content/3-interviews/0-index}
\input{content/2-related-work/0-index}

\input{content/4-maptrix/0-index}

\input{content/5-maps-and-globes-in-vr/0-index}

\input{content/6-od-flow-maps-in-vr/0-index}
\input{content/7-conclusion/conclusion}

{%
\setstretch{1.1}
\renewcommand{\bibfont}{\normalfont\small}
\setlength{\biblabelsep}{\labelsep}
\setlength{\bibitemsep}{0.5\baselineskip plus 0.5\baselineskip}
\printbibliography[title={References}]
} 

\cleardoublepage

\listoffigures
\cleardoublepage

\listoftables
\cleardoublepage

\end{document}

%% file: my-thesis-setup.tex
\PassOptionsToPackage{utf8}{inputenc}
\usepackage{inputenc}

\newcommand{\thesisTitle}{Visualising Geographically-Embedded Origin-Destination Flows:}
\newcommand{\thesisSubTitle}{in 2D and immersive environments}
\newcommand{\thesisName}{Yalong Yang}
\newcommand{\thesisSubject}{Doctor of Philosophy}
\newcommand{\thesisDate}{November, 2018}

\newcommand{\thesisFirstReviewer}{Prof. Gennady Andrienko}
\newcommand{\thesisFirstReviewerUniversity}{}
\newcommand{\thesisFirstReviewerDepartment}{Fraunhofer IAIS and giCentre, City University London}

\newcommand{\thesisSecondReviewer}{Prof. Jo Wood}
\newcommand{\thesisSecondReviewerUniversity}{\protect{City University London}}
\newcommand{\thesisSecondReviewerDepartment}{giCentre}

\newcommand{\thesisFirstSupervisor}{A/Prof. Tim Dwyer}
\newcommand{\thesisSecondSupervisor}{Prof. Kim Marriott}
\newcommand{\thesisThirdSupervisor}{A/Prof. Bernhard Jenny}
\newcommand{\thesisForthSupervisor}{Dr. Sarah Goodwin}
\newcommand{\thesisFifthSupervisor}{Dr. Haohui Chen}

\newcommand{\thesisUniversity}{\protect{Monash University}}
\newcommand{\thesisUniversityDepartment}{Caulfield School of Information Technology}
\newcommand{\thesisUniversityInstitute}{Faculty of Information Technology}
\newcommand{\thesisUniversityGroup}{Immersive Analytics Lab}
\newcommand{\thesisUniversityCity}{Melbourne}
\newcommand{\thesisUniversityStreetAddress}{900 Dandenong Rd, Caulfield East}
\newcommand{\thesisUniversityPostalCode}{3145}

\usepackage[english]{babel} %
\PassOptionsToPackage{%
    figuresep=colon,%
    sansserif=false,%
    hangfigurecaption=false,%
    hangsection=true,%
    hangsubsection=true,%
    colorize=full,%
    colortheme=bluemagenta,%
    bibsys=biber,%
    bibfile=bib-refs,%
    bibstyle=numeric,%
    wrapfooter=false,%
}{cleanthesis}
\usepackage{cleanthesis}

\hypersetup{%
    pdftitle={\thesisTitle},    %
    pdfsubject={\thesisSubject},%
    pdfauthor={\thesisName},    %
    plainpages=false,           %
    colorlinks=false,           %
    pdfborder={0 0 0},          %
    breaklinks=true,            %
    bookmarksnumbered=true,     %
    bookmarksopen=true          %
}

%% file: content/0-before-main-body/titlepages.tex
\begin{titlepage}
	\pdfbookmark[0]{Cover}{Cover}
	\flushright
	\hfill
	\vfill
	{\LARGE \color{ctcolortitle} \textbf{\thesisTitle}\\
		\LARGE \color{black} \thesisSubTitle
	\par}
	\rule[5pt]{\textwidth}{.4pt} \par
	{\Large\thesisName}
	\vfill
	\textit{\large\thesisDate} \\
\end{titlepage}

\begin{titlepage}
	\pdfbookmark[0]{Titlepage}{Titlepage}
	\tgherosfont
	\centering

 	\includegraphics[width=10cm]{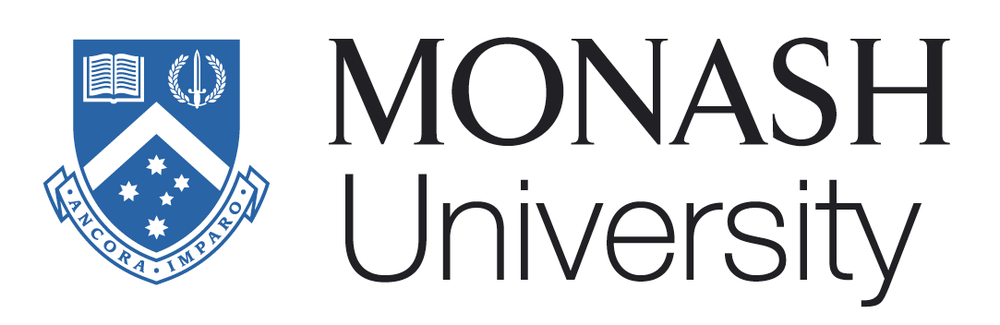} \\[2mm]

	{
		\LARGE \color{ctcolortitle}\textbf{\thesisTitle} \\
		\LARGE \color{black} \thesisSubTitle
		\\[10mm]
	}
	{\Large \thesisName} \\
	{Bachelor and Master of Engineering}

	\textsf{
		A thesis submitted for the degree of Doctor of Philosophy at \\
		\thesisUniversity\ in 2018 \\
		\thesisUniversityGroup, \thesisUniversityInstitute
	}

	\vfill
	\begin{minipage}[t]{.29\textwidth}
		\raggedleft
		\textit{Examiners:}
	\end{minipage}
	\hspace*{15pt}
	\begin{minipage}[t]{.65\textwidth} 
		{\thesisFirstReviewer} \vspace{-1mm}\\
	  	{\thesisFirstReviewerDepartment} \\
		{\thesisFirstReviewerUniversity} \vspace{-5mm}
	\end{minipage} \\[5mm]
	\begin{minipage}[t]{.29\textwidth}
		\raggedleft
		\textit{}
	\end{minipage}
	\hspace*{15pt}
	\begin{minipage}[t]{.65\textwidth}
		{\thesisSecondReviewer} \vspace{-1mm}\\
	  	{\thesisSecondReviewerDepartment}, {\thesisSecondReviewerUniversity}
	\end{minipage} \\[10mm]

	\begin{minipage}[t]{.29\textwidth}
		\raggedleft
		\textit{Main supervisor}
	\end{minipage}
	\hspace*{15pt}
	\begin{minipage}[t]{.65\textwidth}
		\thesisFirstSupervisor\
	\end{minipage} \\[0mm]

	\begin{minipage}[t]{.29\textwidth}
		\raggedleft
		\textit{Secondary supervisor}
	\end{minipage}
	\hspace*{15pt}
	\begin{minipage}[t]{.65\textwidth}
		\thesisSecondSupervisor\
	\end{minipage} \\[0mm]

	\begin{minipage}[t]{.29\textwidth}
		\raggedleft
		\textit{Associated supervisors}
	\end{minipage}
	\hspace*{15pt}
	\begin{minipage}[t]{.65\textwidth}
		\thesisThirdSupervisor \\
		\thesisForthSupervisor \\
		\thesisFifthSupervisor
	\end{minipage} \\[10mm]

	\thesisDate \\

\end{titlepage}

\hfill
\vfill
{
	\small
	\textbf{\thesisName} \\
	\textit{\thesisTitle\ \\ \thesisSubTitle} \\
	\thesisSubject, \thesisDate \\
	Examiners: \thesisFirstReviewer\ and \thesisSecondReviewer \\
	Supervisors: 
	\thesisFirstSupervisor, \thesisSecondSupervisor, \thesisThirdSupervisor, \thesisForthSupervisor\ and \thesisFifthSupervisor \\[1.5em]
	\textbf{\thesisUniversity} \\
	\textit{\thesisUniversityGroup} \\
	\thesisUniversityDepartment \\
	\thesisUniversityInstitute \\
	\thesisUniversityStreetAddress, \thesisUniversityCity, \thesisUniversityPostalCode\, Victoria, Australia \\ [1.5em]
	The style of this thesis is derived from \textit{Clean Thesis} (\url{http://cleanthesis.der-ric.de/}).
}

%% file: content/0-before-main-body/0-copyright.tex
\pdfbookmark[0]{Copyright}{Copyright}
\chapter*{Copyright notice}
\label{sec:copyright}
\vspace*{-10mm}

\textcopyright Copyright by\\
Yalong Yang (2018)

%% file: content/0-before-main-body/1-declaration.tex
\pdfbookmark[0]{Declaration}{Declaration}
\chapter*{Declaration}
\label{sec:declaration}
\vspace*{-15mm}

I hereby declare that this thesis contains no material which has been accepted for the award of any other degree or diploma at any university or equivalent institution and that, 
information derived from the published and unpublished work of others has been acknowledged in the text and a list of references is given.

This thesis includes two original papers published in peer reviewed journals and one accepted journal paper. The core theme of the thesis is visualising geographically-embedded origin-destination flows. The ideas, development and writing up of all the papers in the thesis were the principal responsibility of myself, the student, working within the Immersive Analytics Lab, Caulfield School of Information Technology, Monash University under the supervision of A/Prof. Tim Dwyer, Prof. Kim Marriott, A/Prof. Bernhard Jenny, Dr. Sarah Goodwin and Dr. Haohui Chen.

The inclusion of co-authors reflects the fact that the work came from active collaboration between researchers and acknowledges input into team-based research.

The following paper is accepted to present at InfoVis 2018 conference in Oct. 2018 and will be published in Jan. 2019:\\
Y. Yang, T. Dwyer, B. Jenny, K. Marriott, M. Cordeil and H. Chen. Origin-Destination Flow Maps in Immersive Environments. \textit{IEEE Transactions on Visualization and Computer Graphics (Proceedings of InfoVis 2018)}, 25(1):to appear, Jan. 2019. 

The following papers have previously been published: \\
Y. Yang, B. Jenny, T. Dwyer, K. Marriott, H. Chen, and M. Cordeil. Maps and Globes in Virtual Reality. \textit{Computer Graphics Forum (Proceedings of EuroVis 2018)}, 37(3):427-438, Jun. 2018.

Y. Yang, T. Dwyer, S. Goodwin, and K. Marriott, Many-to-Many Geographically-Embedded Flow Visualisation: An Evaluation,  \textit{IEEE Transactions on Visualization and Computer Graphics (Proceedings of InfoVis 2016)}, 23(1): 411–420, Jan. 2017.

Student signature:  			\hfill Date: \qquad\qquad\qquad\qquad\qquad%

%% file: content/0-before-main-body/2-abstract.tex
\pdfbookmark[0]{Abstract}{Abstract}
\chapter*{Abstract}
\label{sec:abstract}
\vspace*{-15mm}

This thesis develops and evaluates effective techniques for visualisation of flows (e.g. of people, trade, knowledge) between places on geographic maps.  This geographically-embedded flow data contains information about geographic locations, and flows from origin locations to destination locations.  We call this OD flow data for short.
Massive quantities of such data is collected by governments, industries and research organisations. The vast amount of data alone does not have an instant positive value to society, but the potential knowledge extracted from these data can lead to improvements to conventional procedures in many fields. Visualisation techniques have been used to support the analysis of OD flow data. However, due to its large scale and embedded geographic context, visualising such data is a challenging problem.

\vspace{-0.7em}
We explore the design space of OD flow visualisation in both 2D and immersive environments in this thesis. We do so by creating novel OD flow visualisations in both environments, and then conducting controlled user studies to evaluate different designs.

\vspace{-0.7em}
In 2D display space, we proposed a hybrid visualisation we call MapTrix. MapTrix overcomes the clutter associated with a traditional flow map while providing geographic embedding that is missing in standard OD matrix representations. We compared MapTrix with two state-of-the-art techniques (OD Maps and bundled flow maps) in two controlled user studies.  

\vspace{-0.7em}
In immersive environments, virtual objects can be rendered realistically and placed in the space around the user. The additional display space dimension offers the potential to reduce clutter when visualising OD flow data with flow maps, but also natural interactions afforded by motion tracked gesture interaction technologies offers new ways to explore such OD flow data. To systematically explore the design space of flow maps in immersive environments, we first investigate different ways to render geographic maps in immersive environments. We compared maps and globes as well as two novel representations in a controlled user study. We then look at different flow encodings on a flat map as well as flows on flat maps, globes and a novel design we call MapsLink. We conducted three controlled user studies to evaluate the effectiveness of these techniques.

\vspace{-0.7em}
Our exploration and evaluation extend the understanding of how to visualise geographically-embedded OD data in both 2D and immersive environments by providing guidance on design choices.

%% file: content/0-before-main-body/acknowledgement.tex
\pdfbookmark[0]{Acknowledgement}{Acknowledgement}
\chapter*{Acknowledgement}
\label{sec:acknowledgement}
\vspace*{-15mm}

First of all, I would like to thank Associate Professor Tim Dwyer and Professor Kim Marriott for being \textbf{\emph{awesome}} main and secondary supervisors. The uninterrupted and fruitful weekly meetings in the 3.5 years are the most powerful fuel to my research. Dr. Sarah Goodwin has been my supervisor twice, i.e. at my first 1.5 years and last half year. I would like to thank her for her thoughtful ideas, positive comments and how she helped me to structure and articulate part of my research. I would also like to thank Associate Professor Bernhard Jenny for bringing his expertise of maps into my research. I am also grateful to Dr Haohui Chen for his inputs from industrial and application perspectives. Collaboration with Dr Maxime Cordeil is also gratefully acknowledged. It was a privilege to work with all of them.

\vspace{-0.7em}
Thanks to Dr Aidan Slingsby and other members of the giCentre, City University London for discussion on OD Maps. Thanks to Prof. Tamara Munzner, Prof. Bettina Speckmann, Prof. Alan Borning, Martin von Wyss, Prof. Patrick Kennelly, Dr Amy Griffin, Craig Molyneux and industry professionals for providing insightful feedback for our immersive prototypes in various demonstrations.

\vspace{-0.7em}
I also would like to thank the interviewees (Edgar Scrase, Yaofu Huang, Jiafen Zheng, Dr Martin Tomko, Dr Ross Gawler and Dr Maxime Cordeil) for their engagement of discussing their experience of analysing geographically-embedded flow data. I would also like to thank all of our user study participants for their time and feedback. The valuable comments from the reviewers of our submissions are also highly appreciated.

\vspace{-0.7em}
I acknowledge Data61, CSIRO (formerly NICTA) and Caulfield School of Information Technology, Monash University for funding this PhD research. Thanks to Monash for giving me a desk (it finally becomes two because of all my equipment). Thanks to Immersive Analytics Lab for providing access to all of the latest AR/VR equipment and a large room to conduct user studies. I am also thankful to the supportive administration team at Caulfield School of Information Technology, with special thanks to Allison Mitchell, Julie Holden, Diana Sussman and Sidalavy Chaing.

\vspace{-0.7em}
Finally, I would like to thank all my friends and family, especially for my parents supporting my overseas study. A huge thank you goes to Xutong Hou for her love and being around me for the last few years and patiently listening to my research stuff. 

%% file: content/1-introduction/0-index.tex
\chapter{Introduction}
\label{chapter:intro}

\cleanchapterquote{The greatest value of a picture is when it forces us to notice what we never expected to see.}{J.W. Tukey}{Exploratory Data Analysis, 1977}

In many applications it is important to understand flows of some kind of commodities between different geographic locations. This thesis explores the design space of a challenging problem: visualising geographically-embedded origin-destination (OD) flows. It does so by creating novel OD flow visualisations in both 2D and immersive environments, and then conducting controlled user studies to evaluate different techniques.

\input{content/1-introduction/motivation}

\input{content/1-introduction/2D}

\input{content/1-introduction/immersive}

\input{content/1-introduction/aims}

\input{content/1-introduction/methodology}

\input{content/1-introduction/contribution}

\input{content/1-introduction/structure}

%% file: content/1-introduction/motivation.tex
\section{Research Context}
\label{sec:intro:motivation}
Many activities involve interactions between different geographic locations. For example, population migration~\citep{Tobler:1987kn}, commuting behaviour~\citep{Chiricota:2008ut}, animal movement~\citep{Gilbert:2005bd, Slingsby:2017vc} and commodity trading~\citep{Henderson:2011km}; or even intangible things, like disease~\citep{Guo:2007gi} and knowledge~\citep{Paci:2008iq}. Such activities have generated and are generating huge amounts of geographically-embedded flow data.  It is important for analysts (e.g., geographers, urban planners, health care departments, financial analysts study- ing trade) to be able to understand this data in order to gain insights and discover patterns so as to support better decision making.

Geographically-embedded flow data contains information about geographic locations and the directed links connecting them. 
It can be further broken down by the flow continuity into two types:
\begin{itemize}
	\item Trajectories, where the continuous physical routes between origins and destinations are available.
	\item OD flow data, where only the origin-destination pairs are available.
\end{itemize}
A trajectory, in some cases, can be treated as connected OD flow segments by simplifying the physical routes with intermediate stops between the origin and the destination. In this thesis, we restrict the research scope to OD flow data. Visualising trajectories is discussed as future work in Section~\ref{sec:conclusion:future}.

According to Munzner's data typology~\citep[Chap. 2]{Munzner:2014wj}, OD flow data can be classified as the combination of \emph{spatial} data and \emph{network} data:
\begin{itemize}
	\item Origins and destinations are nodes with geographic information.
	\item Flows are links (usually directed) with associated quantitative magnitudes.
\end{itemize}
For example, in the case of global international migration, nodes (origins and destinations) are countries and a link (flow) is the migration from one country to another country.
Timestamps are recorded in some data collection processes, which means OD flow data can also have \emph{time-varying} semantics. 

Small OD flow data may contain only a dozen locations and a few dozen flows, like interstate migration within Australia.  For  such data,  it is still feasible to look   at individual records in a spreadsheet, however, geographic patterns are difficult to perceive, for example, whether people tend to move to the east coast. For dense OD flow data, it is impossible to analyse such data in a spreadsheet. Examples of such data include interstate migration within United States (around 2,500 flows), or public transport transaction records in Shanghai for a working day (up to one million flows). \added{The patterns of OD flow data may also change substantially in different spatial resolutions. For example, people in different suburbs of Victoria, Australia may travel more often to nearby suburbs than distant ones which follows Tobler's first law of geography~\citep{Tobler1970} --- \textit{``everything is related to everything else, but near things are more related than distant things.''} However, this may not be the case for international migration patterns between countries.} Such large-scale and geographic (or spatial) context make understanding OD flow data a challenging problem.

%% file: content/1-introduction/2D.tex
\section{The Role of Visualisation}
\label{sec:intro:vis}
In analytic work flows, if people are clear about their questions, they can develop mathematical formalisation of these questions, and then use optimisation, machine learning or other mathematical techniques to solve such well-specified problems. However, in many circumstances where people are not clear about the targets, the problems cannot be easily formalised, or there are too many possible hypotheses.

\emph{Visual Analytics} is a new research field aiming to combine the strengths of human and electronic data processing with data visualisation to provide effective understanding, reasoning and decision making on the basis of data sets~\citep{Andrienko:2010co,Keim:2008gg}. 
It is an attractive way to support the analysis of OD flow data because of its large-scale and geographically-embedded nature (see Section~\ref{sec:intro:motivation}).

Firstly, data visualisation is an extremely effective way to understand large scale data. A single data visualisation can contain a huge number of graphic elements to present data. Within a graphic element, there are many visual channels (e.g. position, colour, shape, size) which can be used to encode different properties of a data item. In addition, visualisation takes advantage of the powerful human vision system. Much of the visual processing is pre-attentive and occurs in parallel~\citep{WARE2013139}. 

Secondly, visualisation techniques are central to analysing spatial data~\citep{Haslett:1992ew}. Spatial data analysis involves investigating the relationships between spatial locations and the interpretation of the data. Graphic representations for spatial information have been used for more than 2,000 years~---~i.e. maps and globes~\citep{Hruby:2018fn}. They facilitate a spatial understanding of things, concepts, conditions, processes, or events in the human world~\citep{Harley:1987}. For example, \deleted{a popular story in the visualisation literature describes Dr. John Snow’s use of visualisation to discover a link between cholera and the London water supply. By plotting the position of homes of victims he discovered that they were clustered around a water pump~\citep{Dwyer:2004wx,Gilbert:1958cl}.} \added{one domain expert we interviewed told us that governments and organisations in Europe are analysing the movements of refugees and migrants to better estimate the trend of arrivals so as to be prepared (see Section~\ref{sec:interviews:un} for more details and more analysis use cases are discussed in Chapter~\ref{chapter:interviews}).}

\begin{figure}[b!]
    \captionsetup[subfigure]{justification=centering}
    \centering
    \begin{subfigure}{0.45\textwidth}
        \includegraphics[width=\textwidth]{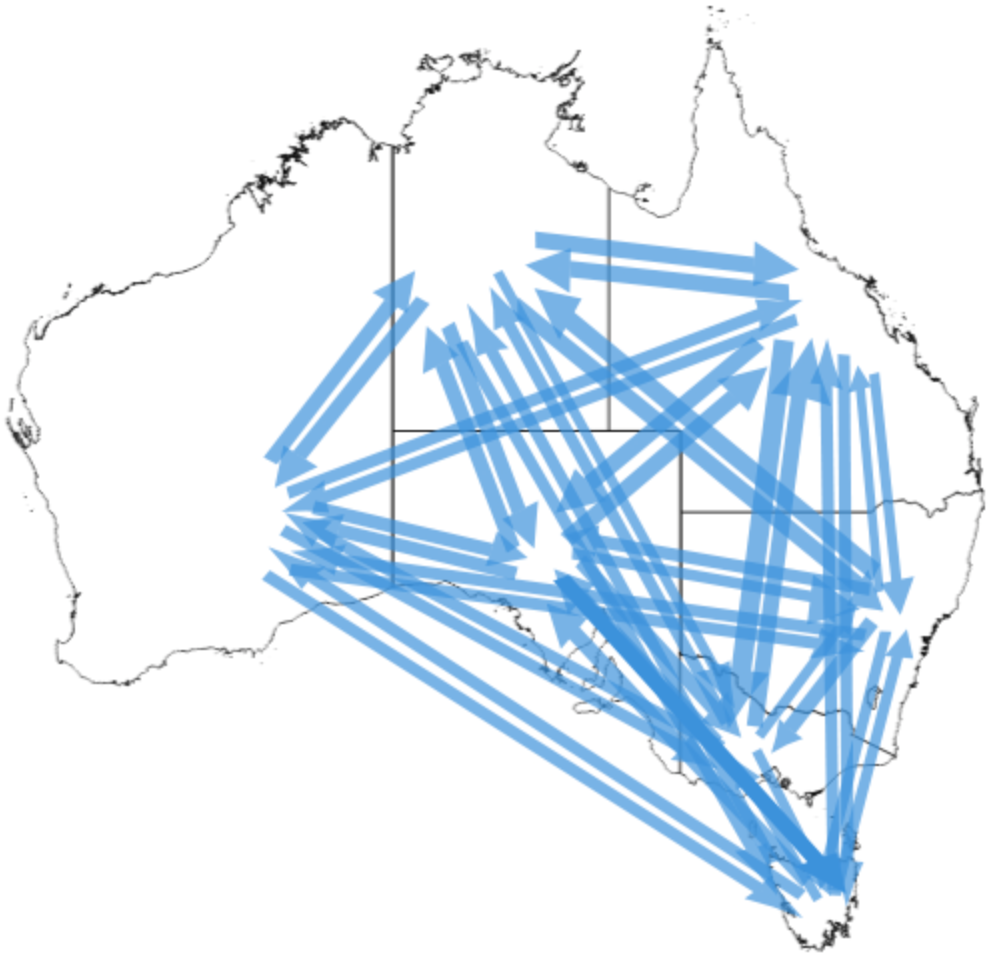}
        \caption{Flow Map.}
    \end{subfigure}
    \begin{subfigure}{0.45\textwidth}
        \includegraphics[width=\textwidth]{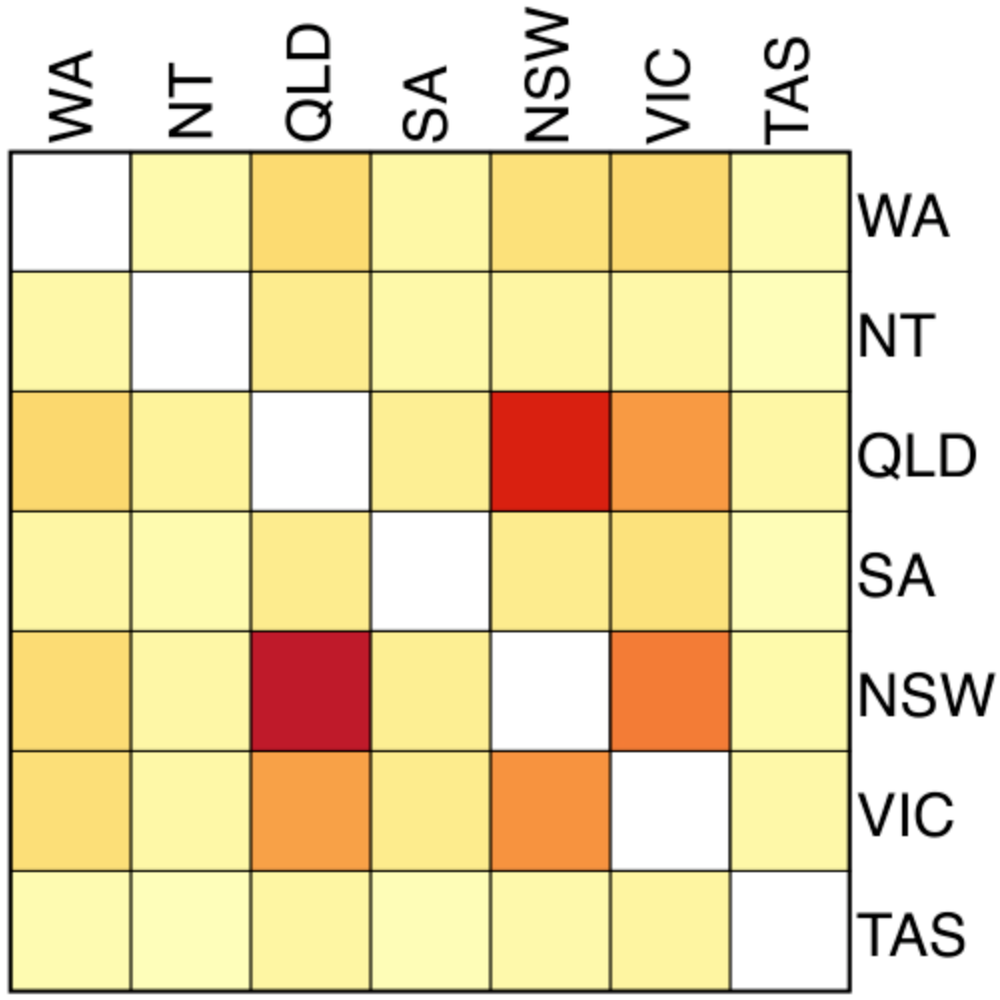}
        \caption{OD Matrix.}
    \end{subfigure}
    \caption{Two main approaches to visualise OD flow data.}
    \label{fig:intro:vis}
\end{figure}

Most research into OD flow data visualization has been conducted in the 2D visualisation design space. In particular, there are two main approaches to visually presenting OD flow data~---~\emph{flow maps} and \emph{OD matrices}:
\begin{itemize}
	\item Flow maps (see Fig.~\ref{fig:intro:vis}(a)) present origins and destinations on a map connected by lines or arrows. Thickness is commonly used to encode the magnitude of the flow.
	\item OD matrices~\citep{Voorhees:2013cx}(see Fig.~\ref{fig:intro:vis}(b)) use a row $r$ for each origin, a column $c$ for each destination, and a cell $(r,c)$ showing the flow from origin to destination. 
\end{itemize}
Whilst flow maps are intuitive and are well suited to show a small number of flows, they quickly become cluttered and difficult to read when the number of commodity sources increases. OD matrices scale better than flow maps, however the geographical embedding of the sources and destinations is missing. We need a new 2D OD flow visualisation that combines the benefits of both.

%% file: content/1-introduction/immersive.tex
\section{2D, 3D and Immersive Environments}
\label{sec:intro:immersive}
2D display media have existed for a long time and are familiar to most people, for example, paper and 2D screens. Due to this popularity and accessibility, most data visualisation research has focused on 2D visualisations. The 2D design space is much better explored than the 3D design space. However, with new emerging immersive technologies 3D visualisation is becoming more popular. 

\begin{figure}[b!]
    \captionsetup[subfigure]{justification=centering}
    \centering
    \begin{subfigure}{0.35\textwidth}
        \includegraphics[height=5.8cm]{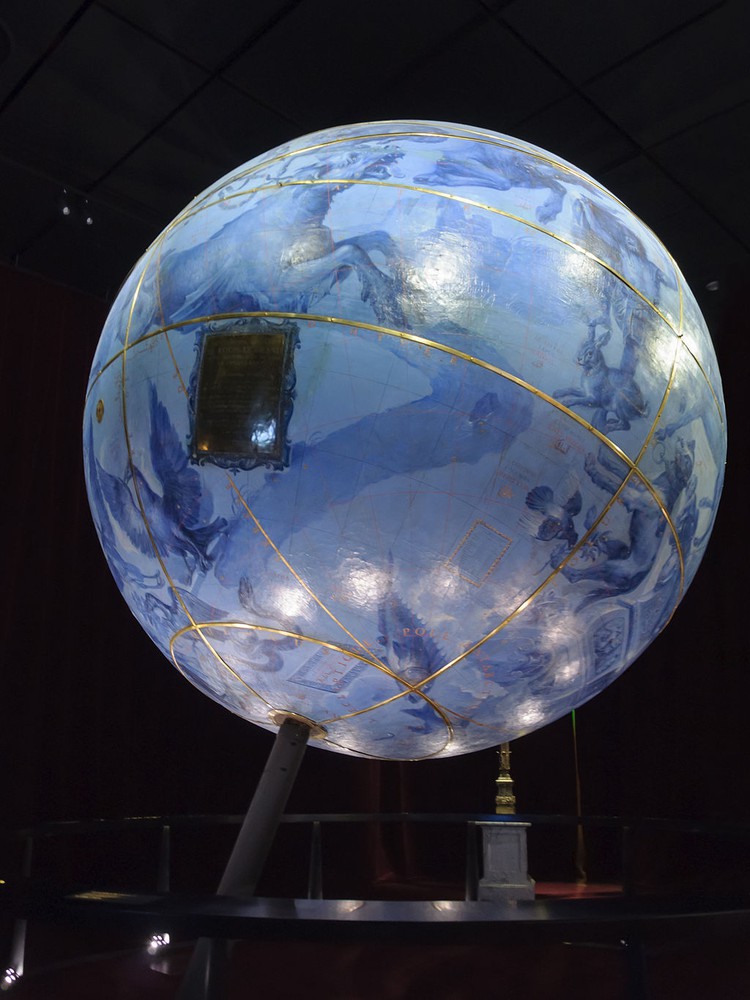}
        \caption{}
    \end{subfigure}
    \centering
    \begin{subfigure}{0.55\textwidth}
        \includegraphics[height=5.8cm]{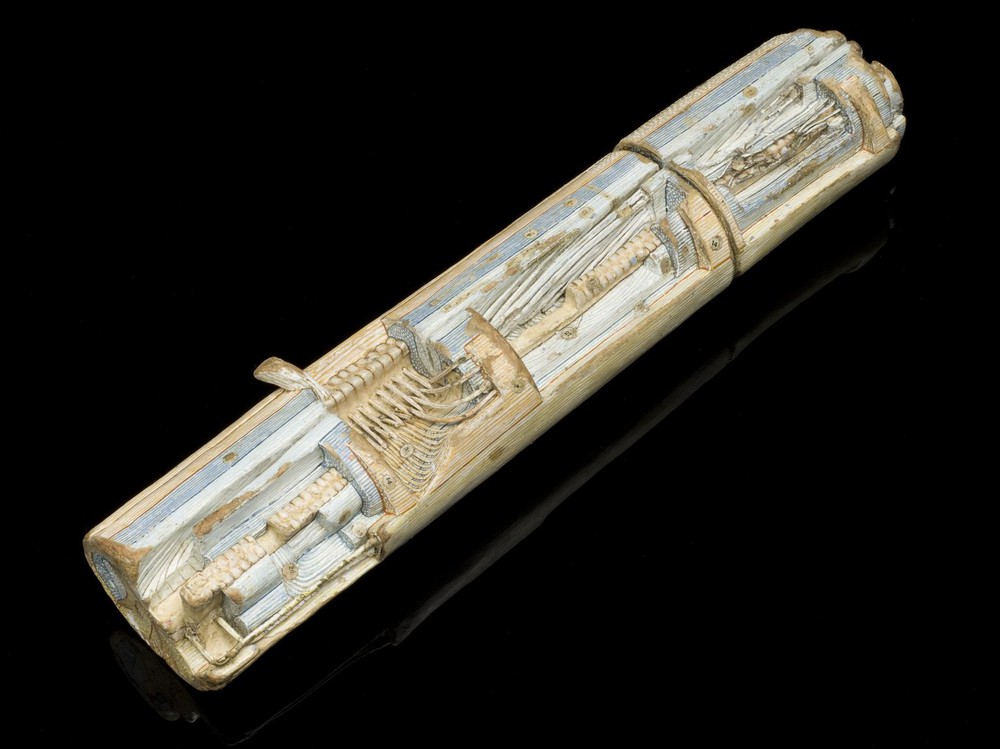}
        \caption{}
    \end{subfigure}
    \caption{Example 3D tangible physical objects. (a) globes created by Coronelli for Louis XIV of France, 1683. Image taken by Myrabella; (b) anatomical model of the spine by Dr Auzoux, French, 1901. © The Board of Trustees of the Science Museum, under CC BY-NC-SA 4.0 Licence.}
    \label{fig:intro:physical}
\end{figure}

\begin{figure}[t!]
    \captionsetup[subfigure]{justification=centering}
    \centering
    \begin{subfigure}{0.30\textwidth}
        \includegraphics[height=4cm]{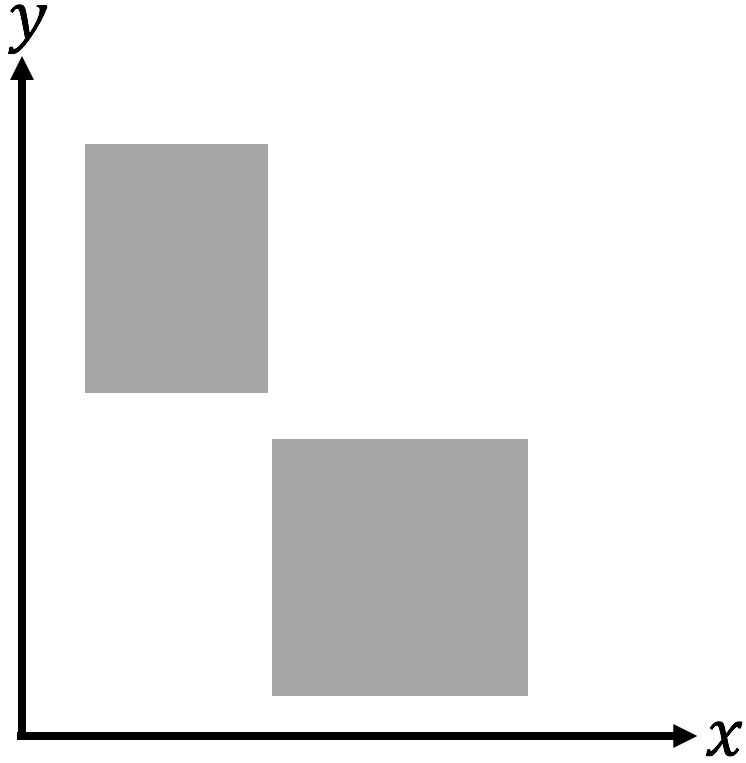}
        \caption{}
    \end{subfigure}
    \centering
    \begin{subfigure}{0.30\textwidth}
        \includegraphics[height=4cm]{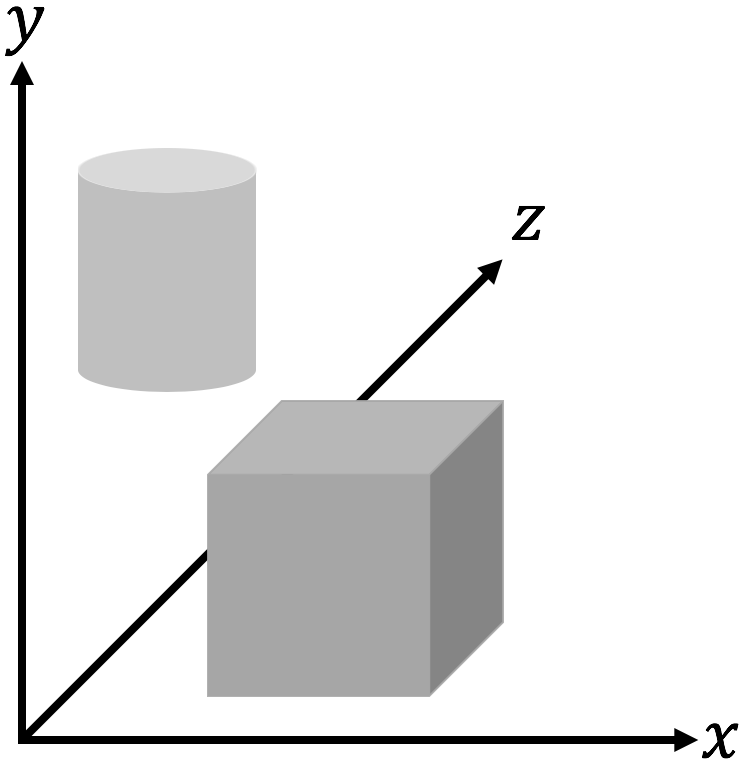}
        \caption{}
    \end{subfigure}
    \caption{3D objects in 2D medium from different view points. (a) from the front view, as a part of the three-view drawing. (b) from a perspective view point.}
    \label{fig:intro:3d}
\end{figure}

In early times, the most familiar 3D representations to people were usually tangible physical objects, for example, globes, sculptures, architectural or medical models etc (examples demonstrated in Fig.~\ref{fig:intro:physical}). However, physical objects are expensive to produce, bulky to carry and do not easily scale. Therefore, methods were developed to present 3D objects in a 2D medium. For example, using map projections to present all or part of the surface of a round body, especially the Earth, on a plane~\citep[Chap. 1]{Snyder:1987tk}; using three-view drawing (two side views and a top view) to describe 3D objects on a plane with views from orthographic projections~\citep{idesawa1973system} (see Fig.~\ref{fig:intro:3d}(a)). 

Another approach is to project and render the 3D objects more realistically with depth cues like occlusion, perspective and parallax motion (see Fig.~\ref{fig:intro:3d}(b)). With the rapid development of computer graphics technology, significantly driven by the game industry, such 3D rendered images and scenes with depth cues have become low-cost and more available~\citep{Wood:2005fl}. An obvious example of using the depth cues was discovered during my personal journey in Europe after the EuroVis 2018 conference. When I was visiting the art galleries in Italy, I realised that Renaissance artists had explored the use of colours to simulate lighting and shadows as well as developed a theory of linear perspective, thus adding to the viewer’s perceptions of depth in their paintings.

One of the most obvious benefits of using 3D visualisation with depth cues is the additional dimension in the display space. Mapping physical data  with  3D spatial information to the 3D display  space  is  straightforward. Many such visualisations have been developed, for example, medical and biological visualisations~\citep{Behrendt:2018ic,WeiChen:2009fo,Klein:2018hz}, meteorological visualisations~\citep{Kern:2018im,Rautenhaus:2017ic}, engineering simulation visualisations~\citep{Dutta:2016im,JankunKelly:2006dj,Li:2007bx}, terrain visualisations~\citep{Jenny:2013fo,wood:2009land,Wood:2005fl} etc. There is also evidence of 3D rendering with depth cues outperforming three-view drawing in shape understanding tasks~\citep{StJohn:2016kd}. 

However, there are concerns about using 3D visualisation with depth cues for abstract data~\citep[Chap. 6]{Munzner:2014wj}. The main arguments against its use are:
\begin{itemize}
    \item Occlusion hides information.
    \item A perspective view causes distortion. 
\end{itemize}

\begin{figure}[t!]
    \captionsetup[subfigure]{justification=centering}
    \centering
    \begin{subfigure}{0.30\textwidth}
        \includegraphics[height=5cm]{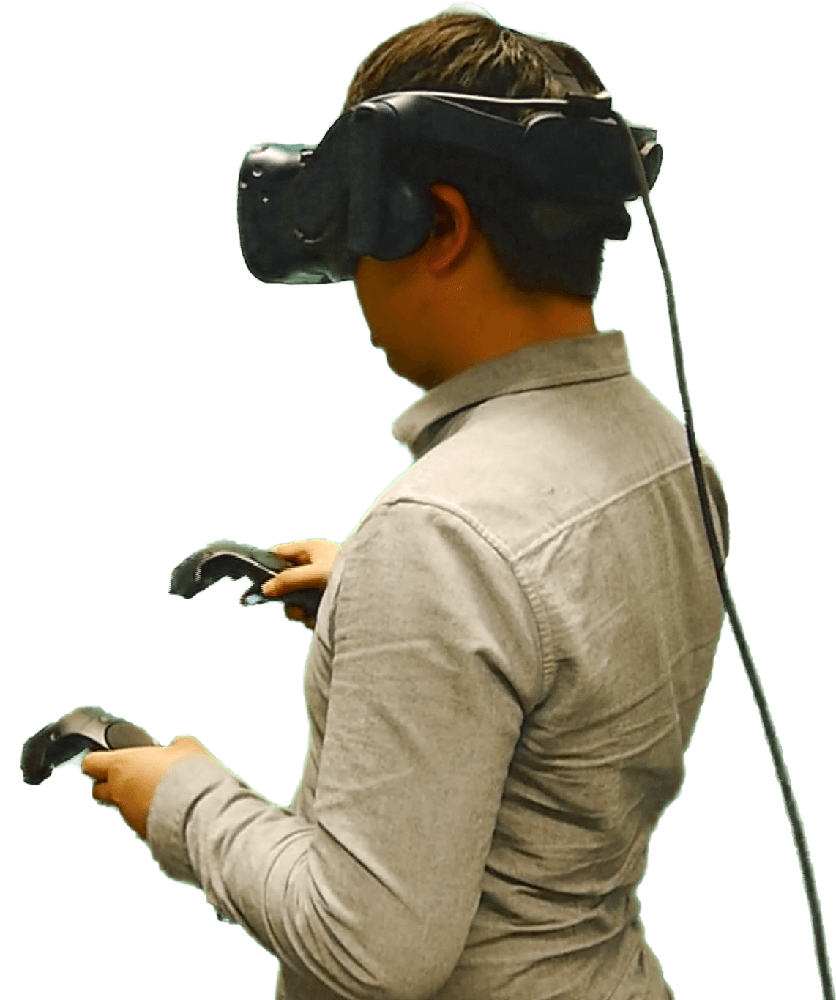}
        \caption{}
    \end{subfigure}
    \centering
    \begin{subfigure}{0.58\textwidth}
        \includegraphics[height=5cm]{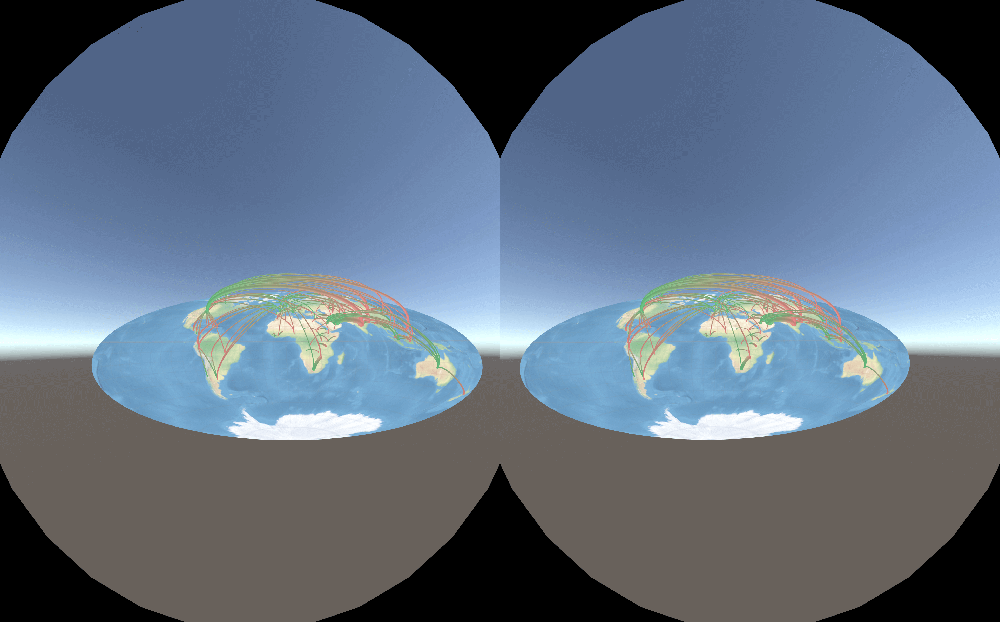}
        \caption{}
    \end{subfigure}
    \caption{A demonstration of using a VR headset: (a) a viewer wearing a space-tracked headset and holding two space-tracked controllers; (b) rendered images for different eyes.}
    \label{fig:intro:vr}
\end{figure}

Immersive visualisations have potential to alleviate these identified problems. Immersive environments provide realistic interaction and perception through technologies such as streaming stereo imagery per eye, and space-tracking (a demonstration of using a VR headset is shown in Fig.~\ref{fig:intro:vr}). 
These environments can overlay a flat image against any surface or hang virtual objects in the space around the user. The space-tracking functionality makes displays respond immediately to user's physical motion of walking and head movement. Such motion parallax is the same as in the physical environment which allows viewers to build a better understanding of the relative 3D position between objects. We discuss the detailed potential benefits in Section~\ref{sec:related:immersive}. 

In the long term, immersive displays may replace traditional flat panel displays in many situations. When displays are no longer intrinsically flat, we have to understand the best way to use the space around the user for data visualisation. In this thesis, therefore we explore OD flow data visualisations within an immersive 3D environment.

%% file: content/1-introduction/aims.tex
\section{Research Aim and Questions}
\label{sec:intro:aims}
Our fundamental aim is to better support analysis of flow data using data visualization:\\
\textbf{How can we visualise OD flow data to better support analysis tasks?} \\
\added{When an analyst explores data, either in pursuit of answers to specific questions that are already formulated, or to begin an investigation of new data without preconceived specific questions, that ``high-level'' analysis can be viewed as a series of more atomic and easily understood analysis tasks in order to better understand the requirements of a visualisation system.  To understand the analysis tasks conducted in real-world application scenarios, we interviewed domain experts about their existing workflows (see Chapter~\ref{chapter:interviews}). We formalised the work flows into some sequences of primitive analysis tasks in Section~\ref{sec:interviews:overview} and Section~\ref{sec:interviews:summary}.} 

Most visualisations for OD flow data have been created in the 2D space, so we first explore the 2D design space:
\begin{itemize}[leftmargin=1em]
	\item \textbf{RQ-1}: How can we visualise \emph{OD flow data} in \emph{2D} to better support analysis tasks?
\end{itemize}
\hangindent=2em \qquad As discussed in Section~\ref{sec:intro:vis}, there are two main approaches of visualising OD flow data. Each design has its relative merits and disadvantages. A key question is are there any other better ways to visualise such data which combine the merits of both.

\vspace{-1em}
Immersive environments demonstrate potential benefits for visualising OD flow data (see Section~\ref{sec:intro:immersive} and Section~\ref{sec:related:immersive}). However, immersive visualisation research is still at an early exploratory stage.
To maintain the feasibility of this thesis, we restrict our exploration of OD flow data visualisation in immersive environments to flow maps approaches. 
We focus on:
\begin{itemize}[leftmargin=1em]
	\item \textbf{RQ-2}: How can we visualise \emph{OD flow maps} in \emph{immersive environments} to better support analysis tasks ?
\end{itemize}
Visualisation of OD flow maps in immersive environments has not been explored before. In order to explore the design space in a systematic way, we first separate the design space into two orthogonal components following D\"ubel \emph{et al.}~\citep{Dubel:2015el}: the representation of OD flows and the representation of the geographic reference space. We look at the geographic reference space first:

\begin{itemize}[leftmargin=1em]
	\item \textbf{RQ-2.1}: How can we present the \emph{geographic reference space} in \emph{immersive environments} to better support analysis tasks?
\end{itemize}
\hangindent=2em \qquad Maps and globes have been used to present the geographic reference space for more than 2,000 years~\citep{Hruby:2018fn}, Determining the best way in immersive environments or whether other representations may be better was unexplored prior to the work presented in this thesis.

\vspace{-1em}
Then, we explore the representation of OD flows in immersive environments:
\begin{itemize}[leftmargin=1em]
	\item \textbf{RQ-2.2}: How can we visualise \emph{OD flows} in \emph{immersive environments} to better support analysis tasks?
\end{itemize}
\hangindent=2em \qquad OD flows were usually visualised as straight or curved lines in 2D visual representations. We want to explore, in immersive environments, whether such 2D representations are the best way, or whether some variant (like 3D lifted tubes) that makes use of a third dimension may be better.

\vspace{-1em}
Finally, we combine the representation of the geographic reference space (\textbf{RQ-2.1}) and the representation of OD flows (\textbf{RQ-2.2}) to investigate~\textbf{RQ-2}.

%% file: content/1-introduction/methodology.tex
\section{Research Methodology}
\label{sec:intro:methodology}
The research project explored how to visualise OD flow data in two ways:
\begin{itemize}
	\item Extending the design space by creating novel visualisations.
	\item Evaluating design choices within the design space.
\end{itemize}
This research is based on the \emph{Design Science Research Methodology}~\citep{Peffers:2008hl} which has three key steps:
\begin{enumerate}[leftmargin=1.5em]
	\item \textbf{Identify Problems \& Motivations}

	This was achieved by: (a) conducting semi-structured interviews with domain experts~\citep[Chap. 2]{Wilson:2014fs}, the interviews have been transcribed and analysed under the \emph{what-why-how} framework by Munzner~\citep{Munzner:2014wj}; (b) reviewing literature of existing visualisations.
	
	\item \textbf{Design \& Development}

	After identifying the pros and cons of existing visualisations, we attempted to create novel visualisations or improve existing ones to remove identified problems. Firstly, brainstorming meetings were held to propose design alternatives. Followed by an iterative development process of: (a) creating prototypes; (b) discussing pros, cons and possible improvements of prototypes. 
	
	\item \textbf{Evaluation}

	Both quantitative and qualitative methods were used for evaluation. Controlled user studies were designed and conducted to test finalised designs against other alternatives. Users' performance including accuracy and time were recorded for quantitative analysis; meanwhile, users were asked to rank all tested designs in a subjective manner, and to provide qualitative feedback.
\end{enumerate}

\section{Research Ethics} 
This research project includes a few user studies involving recruitment and observation of human participants. Thus, ethics approvals were required. The details are presented in this section.

For the expert interviews outlined in Chapter~\ref{chapter:interviews}, potential interviewees from different organisations were recruited in the field. Before starting the interview, they were provided an option to choose whether they wanted to be identified by name. This project has been approved by the Monash University Human Research Ethics Committee (MUHREC) with project number 273.

For the on-line user studies outlined in Chapter~\ref{chapter:evaluating-2d-od-flow-maps}, participants  were  recruited from a Monash university-wide bulletin and a range of mail lists including Microsoft Research (USA), HafenCity University (Germany) and two international map visualisation lists of GeoVis and CogVis. Participation was anonymous. This project has been approved by the MUHREC with project number CF15/3181 - 2015001358.

For the user studies conducted in the laboratory room of the Immersive Analytics Lab, outlined in Chapter~\ref{chapter:maps-globes-vr} and Chapter~\ref{chapter:flow-maps-vr}, participants were recruited from Monash University using permitted university mail lists. Participation was again anonymous. This project has been approved by MUHREC with project number 10217.

%% file: content/1-introduction/contribution.tex
\section{Contribution}
\label{sec:intro:contribution}
\begin{figure}[b!]
\centering
    \includegraphics[width=\columnwidth]{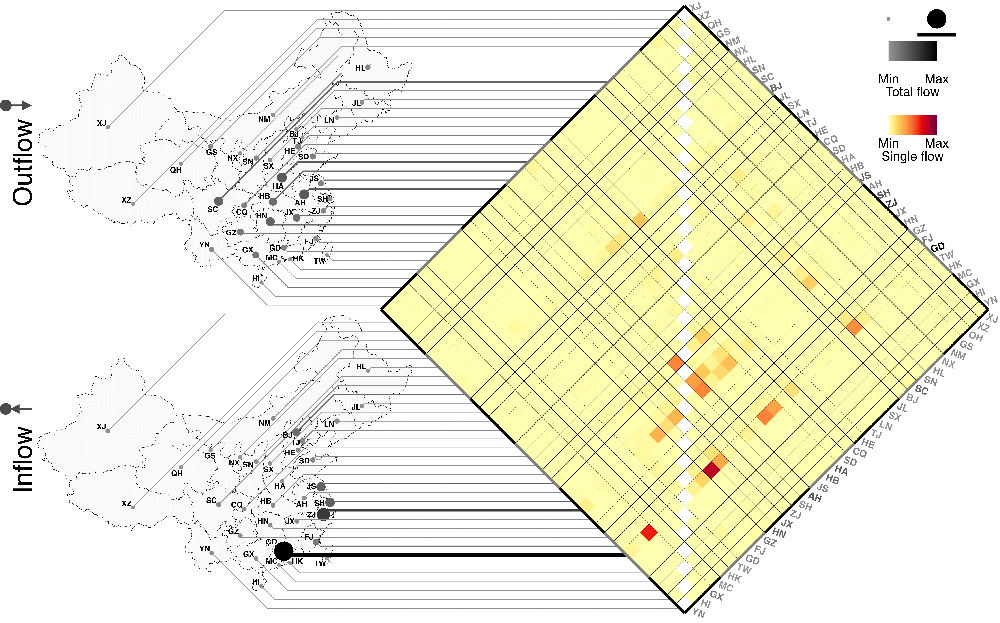}
    \caption{MapTrix showing inter-province migrations in China.}
    \label{fig:intro:maptrix} 
\end{figure}

The design space of visualising OD flow data is huge with many open questions and challenges. The contribution of our research is in the exploration of the design space in both 2D and immersive environments. In more detail:

\begin{itemize}[leftmargin=1em]
	\item %
	We proposed a new hybrid visualisation, \emph{MapTrix}, for showing OD flows that combines the OD matrix and flow map representations, preserving the benefits of both. An example of our novel MapTrix design is shown in Fig.~\ref{fig:intro:maptrix}. To evaluate the effectiveness of MapTrix, we conducted the first two quantitative user studies to evaluate different visual representations for dense many-to-many flow data. We compared MapTrix with two alternative state-of-the-art visualisation methods: a flow map using an edge bundling technique from Pupyrev \textit{et al.}~\citep{Pupyrev:2011bt} and the OD Maps by Wood \textit{et al.}~\citep{Wood:2010be}.
\end{itemize}

This contribution is in response to \textbf{RQ-1} which focuses on visualising OD flow data in 2D space, and is detailed in Chapter~\ref{chapter:od-flow-maps-2d} and Chapter~\ref{chapter:evaluating-2d-od-flow-maps}. This work was presented at the IEEE Conference on Information Visualization 2016 (InfoVis 2016) with a Best Paper Honourable Mention award. It has been published as~\citep{Yang:2017cy}:

\textbf{Y. Yang}, T. Dwyer, S. Goodwin, and K. Marriott, Many-to-Many Geographically-Embedded Flow Visualisation: An Evaluation, \textit{IEEE Transactions on Visualization and Computer Graphics}, 23(1): 411–~420, Jan. 2017.

\begin{itemize}[leftmargin=1em]
	\item %
	We explored different ways to render world-wide geographic maps in immersive environments, including two well-known visual representations: maps and globes; and two novel ones that are only possible in immersive environments. To evaluate their effectiveness, we conducted a controlled study with HMD in VR investigating user preferences and the efficacy (accuracy and time) of these four different interactive visual representations.

\end{itemize}

This contribution is in response to \textbf{RQ-2.1}. The details are presented in Chapter~\ref{chapter:maps-globes-vr}. This work was presented at the Eurographics Conference on Visualization 2018 (EuroVis 2018). It has been published as~\citep{Yang:2018mg}:

\textbf{Y. Yang}, B. Jenny, T. Dwyer, K. Marriott, H. Chen, and M. Cordeil. Maps and Globes in Virtual Reality. \textit{Computer Graphics Forum}, 37(3):427-438, Jun. 2018.

\begin{itemize}[leftmargin=1em]
	\item %
	We explored different OD flow maps in immersive environments differing in the representation of OD flows. We compared 2D flow link representations with straight and curved flow lines, and 3D flow link as tubes with constant height, height varying with quantity and distance between start and end points in a controlled user study in VR.
\end{itemize}

\begin{itemize}[leftmargin=1em]
	\item %
	We explored different OD flow maps in immersive environments primarily varying in the representation of the geographic reference space. We compared OD flows on flat maps, 3D globes and a novel interactive design we call \emph{MapsLink}, involving a pair of linked flat maps in a controlled user study in VR.
\end{itemize}

These two contributions are in response to \textbf{RQ-2.2} and \textbf{RQ-2} respectively, and are detailed in Chapter~\ref{chapter:flow-maps-vr}. This work was accepted and will be presented at IEEE Conference on Information Visualization 2018 (InfoVis 2018). It will be published as~\citep{Yang:2019bp}:

\textbf{Y. Yang}, T. Dwyer, B. Jenny, K. Marriott, M. Cordeil, and H. Chen. Origin-Destination Flow Maps in Immersive Environments. \textit{IEEE Transactions on Visualization and Computer Graphics}, 25(1):to appear, Jan. 2019.

%% file: content/1-introduction/structure.tex
\section{Thesis Structure}
\label{sec:intro:structure}
This thesis is structured in the following way:\\
\textbf{Chapter~\ref{chapter:interviews}} describes five expert interviews we conducted. The motivation for these interviews was to understand the role of OD flow data in real-world application scenarios and the existing work flows using it. The interviews provide the motivating user cases for this thesis.

\textbf{Chapter~\ref{chapter:related}} investigates the academic literature relating to data visualisation, OD flow visualisations in 2D and 3D environments, and immersive analytics.

\textbf{Chapter~\ref{chapter:od-flow-maps-2d}} describes two state-of-the-art 2D flow visualisations: bundled flow map and OD Maps, and our customised implementation of them. The iterative design processes of our novel visualisation, \emph{MapTrix}, is introduced. Our algorithm for leader line placement used in MapTrix is also described.

\textbf{Chapter~\ref{chapter:evaluating-2d-od-flow-maps}} describes the two user studies we conducted to compare state-of-the-art techniques with MapTrix. The analysis results of the collected quantitative and qualitative data are discussed.

\textbf{Chapter~\ref{chapter:maps-globes-vr}} explores different ways to render world-wide geographic maps in immersive environments. The design and implementation of four interactive visualisations of the earth’s geography are introduced. A user study comparing these visualisations for three analysis tasks and discuss the results.

\textbf{Chapter~\ref{chapter:flow-maps-vr}} reports on the findings of three studies exploring different flow map designs in immersive environments. The first experiment focused on different 2D and 3D encodings for flows on a flat map. The second and third experiment we compared flat maps, 3D globes and a novel interactive design we call \emph{MapsLink}, involving a pair of linked flat maps. Collected quantitative and qualitative data is analysed and reported.

\textbf{Chapter~\ref{chapter:conclusion}} presents discussions raised from this topic, overall contributions, conclusions and ideas for future works.

%% file: content/3-interviews/0-index.tex
\chapter{Motivating Use Cases: Expert Interviews}
\label{chapter:interviews}

\input{content/3-interviews/1-introduction}

\input{content/3-interviews/overview}

\input{content/3-interviews/un}

\input{content/3-interviews/urban}

\input{content/3-interviews/indoor}

\input{content/3-interviews/energy}

\input{content/3-interviews/airline}

\input{content/3-interviews/summary}

\input{content/3-interviews/conclusion}

%% file: content/3-interviews/1-introduction.tex
In the last chapter we introduced the context of visualising geographically-embedded flow data. In this chapter, we present our analysis of five expert interviews, each from a different application domain. Such analysis is crucial to understanding the real-world scenarios of analysing geographically-embedded flow data. We designed and conducted semi-structured interviews with experts in diverse domain areas. These experts were chosen because of their experiences of having analysed and/or visualised geographically-embedded flow data during their professional careers.

We begin by introducing the methodologies we used to conduct and analyse the interviews. We then give a brief overview of the results of our analysis, followed by the analysis of each individual interview in detail. Finally, we summarise the analysis results and suggest three different types of flows that should be considered when designing geographically-embedded flow visualisations.

\section{Introduction}
\label{sec:interviews:intro}
\subsection{Interview Preparation}
In order to tailor the interview questions to meet the specific needs and expertise of the interviewee and to ensure the interview time was kept to a minimum, prior to each interview each interviewee conducted a preliminary on-line survey. This survey contained three specific parts:
\begin{itemize}
	\item Consent form and whether they agreed to being identified by name;
	\item Personal information:
	\begin{itemize}
		\item Name;
		\item Degree;
		\item Proficiency self-rating in 
			\begin{itemize}
				\item Programming
				\item Statistics and Mathematics
				\item GIS (Geographic Information Systems)
				\item Visualisation and Designs
			\end{itemize}
	\end{itemize}
	\item Related projects using geographically-embedded flow data
	\begin{itemize}
		\item Overview;
		\item Related publications or visual examples (if available).
	\end{itemize}
\end{itemize}

Using this information a semi-structured interview was prepared for each candidate with a focus on understanding the real-world requirements for analysing and/or visualising geographically-embedded flow data, which was also termed spatial movement data, spatial flow data, trajectory or origin-destination data depending on the interviewee and/or project. The basic structure of the interview consisted of the following questions:
\begin{itemize}
	\item Can you give a brief overview of the related project(s)?
	\item Can you explain the data in more details?
	\item What are/were the initial motivations for the related project(s)?
	\item What tasks are involved in the analysis?
	\item Did you use visualisations/visual analytic tools in your analysis? If so, can you give us some examples?
	\item Do you have any suggestions on how to improve these visualisations/visual analytic tools?
\end{itemize}

\subsection{Method and Analysis}
\label{sec:interviews:method}
Interviews were conducted via VoIP (Voice over IP) software, like Skype, or face-to-face when possible. All interviews were audio-recorded and later transcribed. A pilot interview was also carried out internally in order to test the structure, questions and timing.

\begin{figure}[b!]
    \captionsetup[subfigure]{justification=centering}
    \centering
    \begin{subfigure}{\textwidth}
        \includegraphics[width=0.98\textwidth]{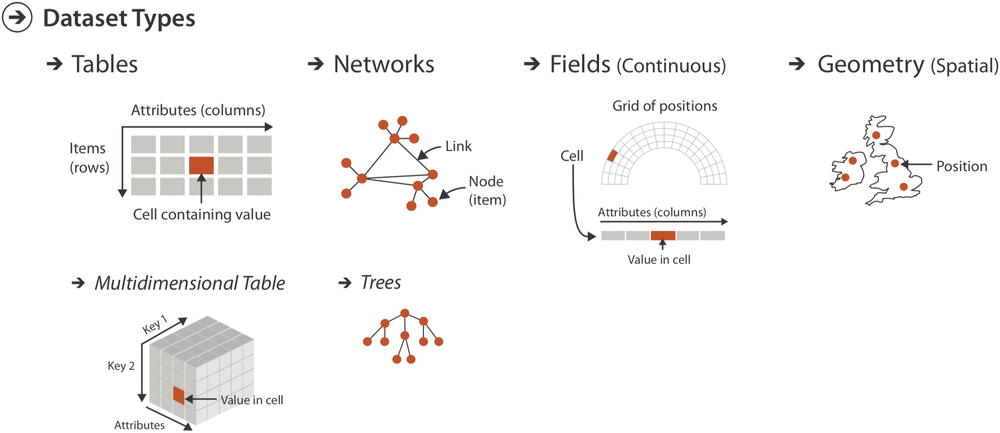}
        \caption{}
    \end{subfigure}
    \centering
    \begin{subfigure}{\textwidth}
        \includegraphics[width=0.98\textwidth]{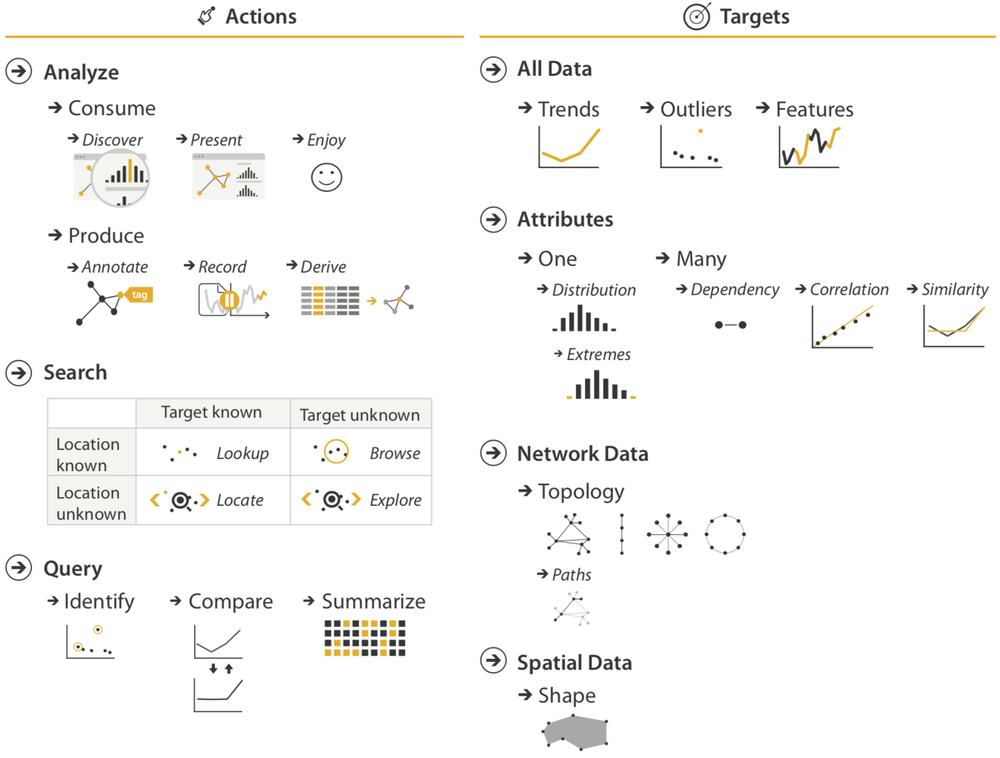}
        \caption{}
    \end{subfigure}
    \caption{Key elements in Munzner's \emph{what-why} framework. (a) data abstraction: four basic dataset types; (b) task abstraction: actions and target of tasks. Figures derived from~\citep[Chap. 2, 3]{Munzner:2014wj} under CC BY-4.0 Licence.}
    \label{fig:interviews:intro-what-why}
\end{figure}

We chose to interview experts from distinct disciplines to give us an in-depth yet diverse understanding of the problem domain. Whilst the data described in each interview is similar in structure (containing geographical locations and flows between them), the data is analysed and visualised in each domain for very different reasons. Each interview transcription therefore contained very detailed descriptions of each distinct project. After transcribing the interviews an initial qualitative coding exercise revealed some difficulties in this method of analysis due to diversity of the projects described. Therefore, the interview data was instead analysed using the 
\emph{what-why-how} framework by Munzner~\citep{Munzner:2014wj} with a particular focus on: 
\begin{itemize}
	\item \textbf{What: Data Abstraction}. Datasets are classified into five different types: tables, networks, fields, geometry or spatial (see Fig.~\ref{fig:interviews:intro-what-why}(a))~\citep[Chap. 2]{Munzner:2014wj}.
	\item \textbf{Why: Task Abstraction}. Tasks are analysed with two parts: \emph{actions} and \emph{targets} (see Fig.~\ref{fig:interviews:intro-what-why}(b))~\citep[Chap. 3]{Munzner:2014wj}. 
	\begin{itemize}
	 	\item For \emph{actions}, there are three levels of actions. The highest-level actions are to consume or produce information. The middle level actions are different cases of \emph{finding} required information. The bottom level is to query \emph{targets} at different scopes.
	 	\item For \emph{targets}, there are four kinds of abstract targets: all data, attributes, network data and spatial data.
	 \end{itemize}   
\end{itemize}

A visualisation task can be described as \emph{actions} with a \emph{target}.  
Compared to other analysis methods, this framework guides the translation of domain-specific problems into abstract visualisation tasks using a multi-level typology of tasks. This typology provides abstract and flexible descriptions of tasks which allows useful comparison to be made between different application domains~\citep{Brehmer:2013fq}.
We use the framework to impose a structure and allow us to think systematically about the similarities in analysis tasks and design choices, within the large and complex design space for visualising geographically-embedded trajectory and OD data~\citep{Munzner:2014wj}.

Prior to the analysis, we would like to point out some limitations to our methodology.  The topics and tasks described by the interviewee might be different from those undertaken by other analysts in a similar position in the same domain.  Also, due to our desire to gain a breadth in understanding, we had to spend time during each interview gaining knowledge of the problem domain. As each interview was limited to one hour we focussed on understanding the high-level tasks, domain and data characteristics. This did not allow time to explore more detailed aspects of the data such as spatial scale, spatial resolution, amount, quality, uncertainty or time dimension. In our approach, we tried to gain as much understanding as possible of the domain, the project and an overview of the types of analytical tasks carried out. There is a possibility that we have missed tasks through our method of query and analysis. Further interviews and continued discussions could reveal further details and important tasks in each domain.

%% file: content/3-interviews/overview.tex
\section{Overview}
\label{sec:interviews:overview}
\begin{figure}[b!]
    \centering
    \includegraphics[width=\textwidth]{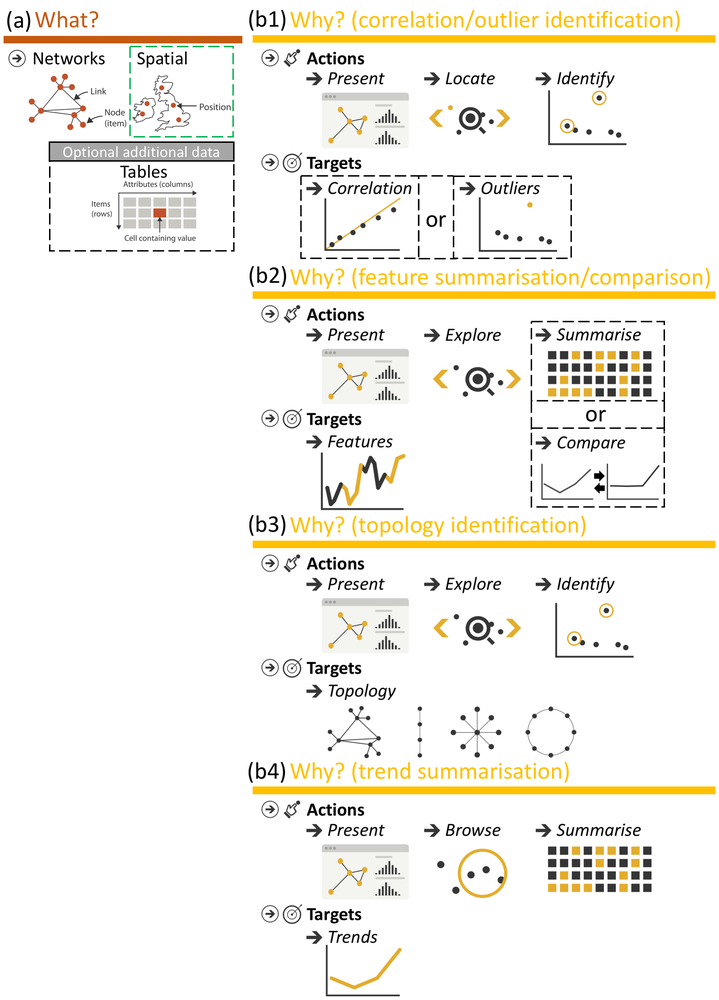}
    \caption{
    The overview of \emph{what-why} analysis for different interviews.
    (a) data abstraction; (b1,b2,b3,b4) task abstraction. Dashed box means optional in some analysis tasks. Icons are derived from~\citep[Chap. 2, 3]{Munzner:2014wj} under CC BY-4.0 Licence.
    }
    \label{fig:interviews:all} 
\end{figure}

In this section, we briefly introduce the abstracted data and tasks the five projects referenced in our interviews. Fig.~\ref{fig:interviews:all} provides a graphic depiction of these abstractions using icons from Fig.~\ref{fig:interviews:intro-what-why}.

\subsection{What: Data Abstraction}
As discussed in Section~\ref{sec:intro:motivation}, geographically-embedded flow data can be considered as the combination of \emph{spatial} data and \emph{network} data (see Fig.~\ref{fig:interviews:all}(a)). In all projects we interviewed, these two types of data are collected. \added{The spatial data in Munzner’s framework refers to data which is intrinsically spatial, including items like points, or one-dimensional lines or curves, or 2D surfaces or regions, or 3D volumes~\citep[Chap. 2]{Munzner:2014wj} from any sort of physical space, including not-only geographic data, but also medical imaging, engineering plans, and so on. Here we use the term spatial for consistency, but we only consider georeferenced data (i.e. longitude and latitude) for ODs and/or trajectories.}

In the case of energy flow analysis, although the interviewee showed a great interest in analysing geographic patterns, he stated that, in current work flows, analysts are more concerned about the network perspective of power grids and hardly investigate geographic patterns. Thus, we highlighted this circumstance using a dashed box in Fig.~\ref{fig:interviews:all}(a). 

Apart from the two main types of data, additional tabular data were also occasionally used in some analysis (this is shown as a grey dashed box in Fig.~\ref{fig:interviews:all}(a)). For example, analysts are investigating the correlation between weather and refugee arrivals (see Section~\ref{sec:interview:un:why}); demography data can be used to plan future energy grid systems (see Section~\ref{sec:interviews:energy:why}) etc.

\subsection{Why: Task Abstraction}
As discussed in Section~\ref{sec:interviews:method}, a visualisation task can be described as the combination of actions and a target. 

We first of all look at the actions. For the highest level of action~---~\textbf{Analyse}: in the interviews, both two main goals have been identified: analysts \textbf{consume} information to get better insights into the data as well as using other analytic tools to \textbf{produce} new information. 

\vspace{-1.5em}
\hangindent=2em \qquad \textbf{Consume} tasks mentioned in the interviews mainly related to \emph{presenting}. This is possibly because the interviewee mainly talked about their experience of decision-making, planning and forecasting instead of generating new hypotheses.

\vspace{-1.5em}
\hangindent=2em \qquad \textbf{Produce} tasks mentioned in the interviews mainly relate to \emph{deriving} new data using an analytical tool, e.g. constraints generation, modifying network structure etc.

For the middle level action~---~\textbf{Search}: different specific action was taken based on whether the identity and the location of the target is known or not.

For the lowest level action~---~\textbf{Query}: \emph{identify}, \emph{summarise} and \emph{compare} are commonly mentioned in the interviews with different targets.

We then explain the related targets discussed in the interviews. A target is some aspect of the data that is of interest to the viewer. \textbf{Trends} are high-level patterns in the data. \textbf{Outliers} are elements do not fit well with overall data distribution. \textbf{Features} are task dependent, can be some particular structures of interests. \textbf{Topology} describes the interconnections in network data.

In the following sections, we discuss each individual interview in terms of problem context, interviewee background, data and task abstraction. We also present visualisation examples for each domain when available.

%% file: content/3-interviews/un.tex
\section{Refugee and Migrants Movement}
\label{sec:interviews:un}
At times of violent conflict in the world, hundreds of thousands of innocent lives can be affected. In such disastrous situations, many organisations around the world work hard to ensure people's safety.
United Nations High Commissioner for Refugees (UNHCR) and International Organization for Migration (IOM) are two of these organisations. UNHCR and IOM both collect and analyse refugee and migrants data and produce situation reports to support evidence-based decision-making and advocacy on refugee and migrants issues~\citep{iom:reports,unhcr:2017reports}.

In November 2016, we conducted an interview with Edgar Scrase about his experience of analysing refugee and migrants movement. 
Edgar is an experienced Information Manager and Spatial Software Developer. He has a Masters Degree in Geographical Information Science and has been working for over 16 years in all areas of information collection, management, analysis and visualisation as well as software development. In particular, since 2010 Edgar has worked in humanitarian organisations specalising in work related to refugee and migrants movement. Specifically, 
Edgar worked as an Information Management Officer in UNHCR for approximately five years and as an Information Manager/Consultant in IOM for two years.

\subsection{What: Data Abstraction}
Data collection and quality assurance is particularly difficult in this domain. Data collection is a manual process and there is often missing, incomplete or inaccurate data recorded. Collection is particularly difficult when it involves multiple countries and different languages. These countries may have different standards for data collection. Data from some countries may not be detailed enough and can cause extra difficulties for data analysis; As stated by the interviewee, ``~\textit{UNHCR and IOM conduct regular discussions with governments to attempt to better harmonise this information.}''

One type of the most important data related to refugee and migrants movement analysis is, what is known as, \emph{arrival data}. The data is usually collected by local governments or agencies about individuals with the following attributes:
\begin{itemize}
	\item Country of Origin;
    \item Country of Arrival;
	\item Intended Destination;
	\item Sex;
	\item Age;
	\item Whether travelling with companion(s), and if so with whom. In the case of children, whether or not they are unaccompanied or have been separated from their parents.
\end{itemize}

In some cases, refugees and migrants are relocated to another country after their arrival. Such individual records are collected with:

\begin{itemize}
    \item Country of Arrival;
    \item Country of Relocation.
\end{itemize}

The analysis of these data is usually conducted with an aggregated number of people instead of individual records. There are many different ways to aggregate records (e.g. by OD pair, by time period or the combination of them). The analysts are  also interested in estimating the arrivals in the future. 
One way to achieve short term projections is to investigate the correlation between the refugee and migrants locations (e.g. geographic information) and other factors, for example, the weather (see Section~\ref{sec:interview:un:why} for more details).

These datasets can be described as a combination of \emph{Spatial} and \emph{Network} types, as they contain geographic locations and the links between them. All records also contain timestamps meaning the data also has \emph{time-varying} semantics. 

\subsection{Why: Task Abstraction}
\label{sec:interview:un:why}
The main motivation for this work is to improve the alignment of the scarce resources. This aim can be broken down into two distinct analytical tasks:
\begin{itemize}
	\item \emph{Estimate the arrival}. Better estimation can help governments and organisations prepare. There are different scenarios where visualisation can facilitate this task:
	\begin{itemize}
		\item \textbf{Trends Summarisation}. Analysts investigate if arrivals in different locations change across time (see Fig.~\ref{fig:interviews:all}(b4));

		\item \textbf{Correlation Identification}. Analysts tried to find out possible correlated factors with the arrivals (see Fig.~\ref{fig:interviews:all}(b1)). For example, they found out weather in some cases affected the arrivals, bad weather usually results in fewer arrivals. Probably because bad weather makes the journey more difficult.
	\end{itemize}

	\item Decide the priorities of relief operations. Due to limited resources, the most urgent cases, like unaccompanied children, should be prioritised~---~ these unusual cases are often found through \textbf{Outlier Identification} (see Fig.~\ref{fig:interviews:all}(b1)).
\end{itemize}

Besides the resources alignment motivation, there is also a focus on the identification of smuggling. First, analytical modelling and calculations to
\textbf{derive} 
the data are needed for this purpose. Then visual analytical tools can help identify possible smuggling routes via \textbf{Outlier Identification} (see Fig.~\ref{fig:interviews:all}(b1)). 

\begin{figure}[b!]
\centering
    \includegraphics[width=\columnwidth]{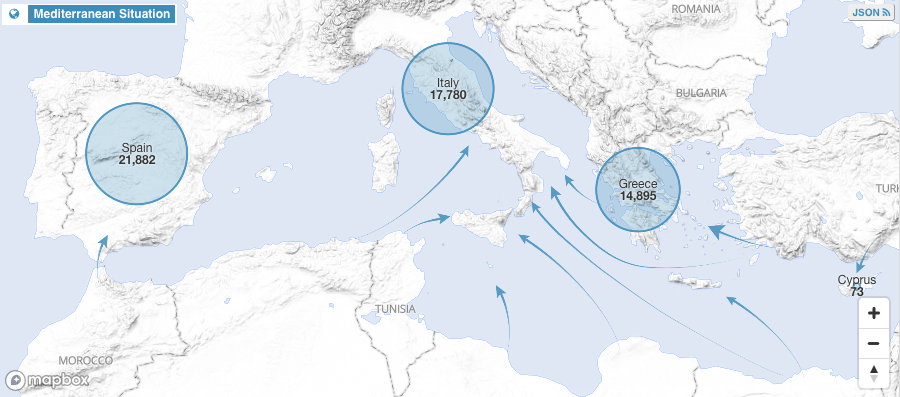}
    \caption{Total arrivals to Europe via the Mediterranean, 2017. Screen capture from~\url{https://data2.unhcr.org/en/situations/mediterranean}.}
    \label{fig:interviews:un:example} 
\end{figure}

In addition to analytical tasks, UNHCR and IOM also create high-level overview visualisations to communicate results to governments or the general public (see Fig.~\ref{fig:interviews:un:example}).

\begin{figure}[b!]
\centering
    \includegraphics[width=0.7\columnwidth]{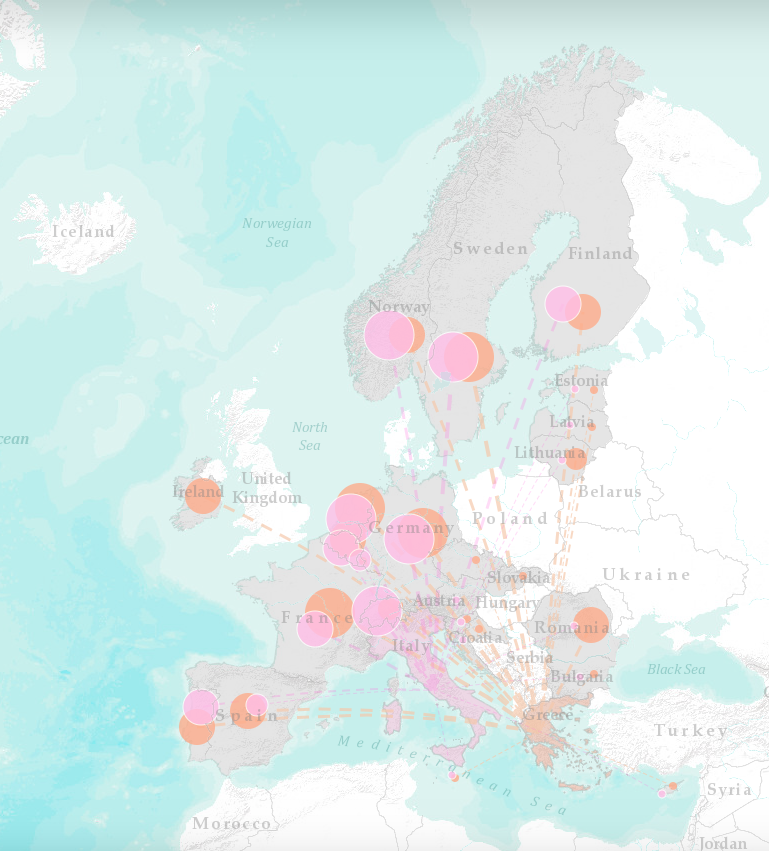}
    \caption{Relocation of refugees from Italy and Greece, updates as of 30 April 2018. Screen capture from~\url{http://migration.iom.int/europe/}.}
    \label{fig:interviews:un:relocate} 
\end{figure}

\subsection{Visualisation Examples}
Two examples of the types of visualisation representations created by humanitarian organisations to present refugee and migrants movement are shown in Fig.~\ref{fig:interviews:un:example} and Fig.~\ref{fig:interviews:un:relocate}. We have also identified a number of specific requirements for refugee and migrants movement  visualisation:
\begin{itemize}
    \item Tools need to be intuitive and easy to use, as many involved employees and volunteers do not have a technical backgrounds. Due to the nature of the work, time scales are particularly tight.
    \item Users need to be able to select and filter aggregated flows to drill down into the detail, for example to identify differences between gender or age groups.
\end{itemize}

%% file: content/3-interviews/urban.tex
\section{Urban Planning: a Bike Sharing Schema}
\label{sec:interviews:urban}
Urban planning focuses on the design of the urban environment, including land, environmental resources and public infrastructures. With the rapidly increased population and size of cities, many challenges emerge and affect people’s daily life, such as air pollution, traffic congestion and unaffordable house prices. On the other  hand, vast amounts of data can be collected from urban environments and activities. Such data can be used to improve urban planning practice~\citep{Zheng:2014ee}. In this section, we describe a bike sharing schema planning project with the analysis of various data sets.

In September 2016, Yaofu Huang was interviewed about his experience of working on a bike sharing schema planning project for the city of Xiamen (population of approximately 3.5 million). Subsequently in December 2016, Jiafen Zheng gave a detailed oral presentation of the same project organised by Prof. Xun Li, the Dean of Urbanization Institute of Sun Yat-sen University (UI-SYSU). The analysis described in this section was conducted with the information from both the interview and the presentation.

Yaofu Huang and Jiafen Zheng are both urban planners from UI-SYSU, and were involved in the bike sharing schema planning project. This project involved analysing the existing bike sharing schema's data and planning for an extension of the schema to encourage more bike sharing members and ultimately reduce car congestion in the city. A highlight is that the project proposed the first ``SkyCycle'' in China, and this dedicated elevated cycle path stretching a total length of 7.6 km, was built in Xiamen and opened to the public in January 2017. Part of their analysis and findings were published in~\citep{zjf2018}.

\subsection{What: Data Abstraction}
\label{sec:interview:urban:what}
A variety of data was used for the planning, A bike sharing schema in Xiamen has been in operation since July 2014 and bike sharing usage data had been collected since then. This is one of the main datasets in the analysis. The \textbf{Previous Bike Sharing Data} contains two different components:
\vspace{-0.5em}
\begin{itemize}
	\item \emph{Docking Stations}, with the following information for each docking station: \emph{geographic location}, \emph{capacity} and \emph{number of docked bikes} in real-time.
	\item \emph{Journey records}, with the following information for each journey: \emph{origin docking station}, \emph{start time}, \emph{destination docking station} and \emph{end time}.
\end{itemize}
For additional context of the city and how the population moves around the city, the following data sets (see Tab.~\ref{tab:interview:urban:data}) were also used.

\begin{table}[t!]
\begin{tabular}{|l|p{11.3cm}|}
\hline
\textbf{Dataset} & \textbf{Description of Geographical Data} \\ \hline
\makecell[tl]{\textbf{Road}\\ \textbf{Network}} & \emph{Roads}, route data for seven different types of roads in Xiamen. \\ \hline

\makecell[l]{\textbf{Public} \\ \textbf{Transport}} & 
    \begin{tabular}[c]{@{}p{11.3cm}@{}}
        \emph{Bus Stations}, with the geographic location for each bus station. \\ 
        \emph{Bus Routes} with each route containing a list of bus stations. \\ 
        \emph{Travel Card Journey} data with both touch on and off bus station locations for the Bus Rapid Transit (BRT) and only touch on locations for the standard bus system. \\ \\ The analysts interpolated destination locations where possible using historical user data, i.e. when a travel card is touched on at station A in the morning, and at station B in the afternoon every weekday, it is likely that the person with this card is travelling between station A and B as their daily commute.
    \end{tabular} \\ \hline

\makecell[tl]{\textbf{Taxi} \\ \textbf{Journeys}} & GPS routes for journeys containing \emph{Origins}, \emph{Destinations} and \emph{Trajectories}. \\ \hline

\makecell[l]{\textbf{Region} \\ \textbf{Category}} & 
    \begin{tabular}[c]{@{}p{11.3cm}@{}}
        This data was derived from the population and economic census data which contain information about residential space and industrial/office space in Xiamen. Regions in Xiamen have been divided into: \emph{Living regions}, \emph{Working regions} and \emph{Mixed regions}. 
    \end{tabular} \\ \hline

\makecell[tl]{\textbf{Points of} \\ \textbf{Interests}} & Data containing geographic locations of important locations including: schools, hospitals, parks, shopping malls, industrial parks etc. \\ \hline
\end{tabular}
\vspace{0.5em}
    \caption{Datasets used in bike sharing schema planning.}
    \label{tab:interview:urban:data}
\end{table}

\subsection{Why: Task Abstraction}
The overall aim of this project was to encourage people to use bike sharing instead of cars. In the context of bike sharing schema infrastructure planning, this aim can be achieved by expanding the catchment area (building new docking stations), adding more bikes to the schema (building new and increasing current docking station capacity), and facilitating safe cycling in the city by extending existing bike lanes or constructing new ones.

A key question for the planner is how to choose the locations for these new stations and lanes to best serve people's travel goals and also fit within the constraints of the built environment. To better answer this question, the data in Section~\ref{sec:interview:urban:what} was analysed.

Firstly, the previous bike sharing data was analysed, and it was evident that the majority of previous bike sharing was used for short distance travel. To explore this in detail, short distance travel routes in the form of OD data and trajectories were extracted from previous bike sharing data, public transport data, and taxi data. To plan for new docking stations, three types of candidate locations were identified through \textbf{Outlier Identification} (see Fig.~\ref{fig:interviews:all}(b1)) relating to popular origins and destinations, dense intersections of trajectories and Points of Interests (POIs). To plan for new and extended bike lanes, they first analysed the existing bike lane network, and identified the missing short connections between docking stations and other facilities, e.g. POIs or public transit stations (using \textbf{Outlier and Topology Identification}, see Fig.~\ref{fig:interviews:all}(b1) and (b3). Those short connections became candidates for new bike lanes.

\begin{figure}[b!]
\centering
    \includegraphics[width=0.6\columnwidth]{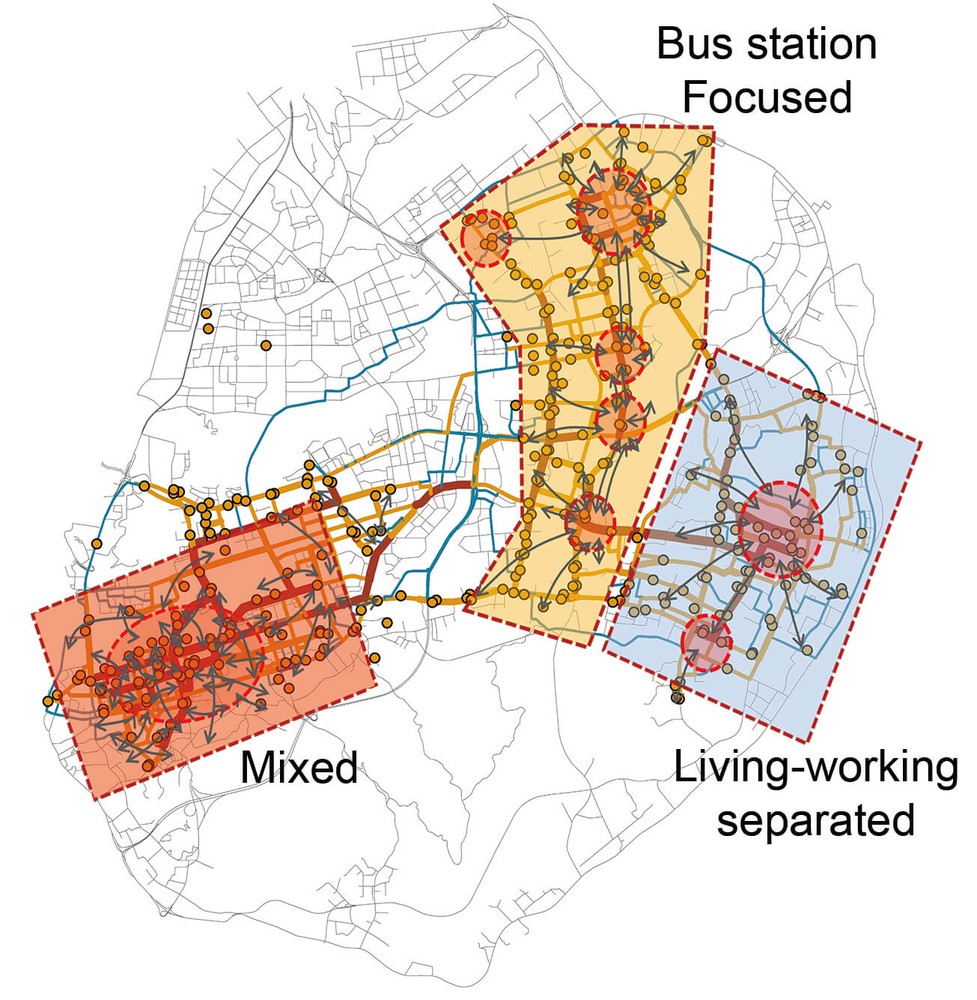}
    \caption{Different urban OD types in a part of Xiamen, China. Figure provided by UI-SYSU.}
    \label{fig:interviews:urban-0}
\end{figure}

Apart from the main aim, the interviewees described some interesting findings they discovered in the analysis, which conform to existing literature on bike sharing schema use in other countries, such as in London, UK~\citep{Beecham:2014he}:
\begin{itemize}
	\item When analysing the correlation between bike usage and bike lanes (see Fig.~\ref{fig:interviews:all}(b1)~---~\textbf{Correlation Identification}), they found out that whilst many of the existing bike lanes were on main roads (likely due to space), these lanes were hardly used by the bike sharing riders. They stated one possible reason could be the safety consideration of riding a bike on busy roads.

	\item When analysing the topology of bicycle usage network (see see Fig.~\ref{fig:interviews:all}(b3)~---~\textbf{Topology Identification}), they identified different patterns of bicycle use in different regions (see Fig.~\ref{fig:interviews:urban-0}):
	\begin{itemize}
		\item Mixed; there is no clear pattern of using bicycles in such region.
		\item Bus station focused; there is a strong demand for using bicycles around bus stations in such regions.
		\item Living-working separated; the area for living and working are well separated in such regions, and bicycles are heavily used as a commuting tool between living and working areas.
	\end{itemize}
\end{itemize}

\begin{figure}[b!]
    \captionsetup[subfigure]{justification=centering} 
    \centering
    \begin{subfigure}{0.4\textwidth}
        \includegraphics[height=6cm]{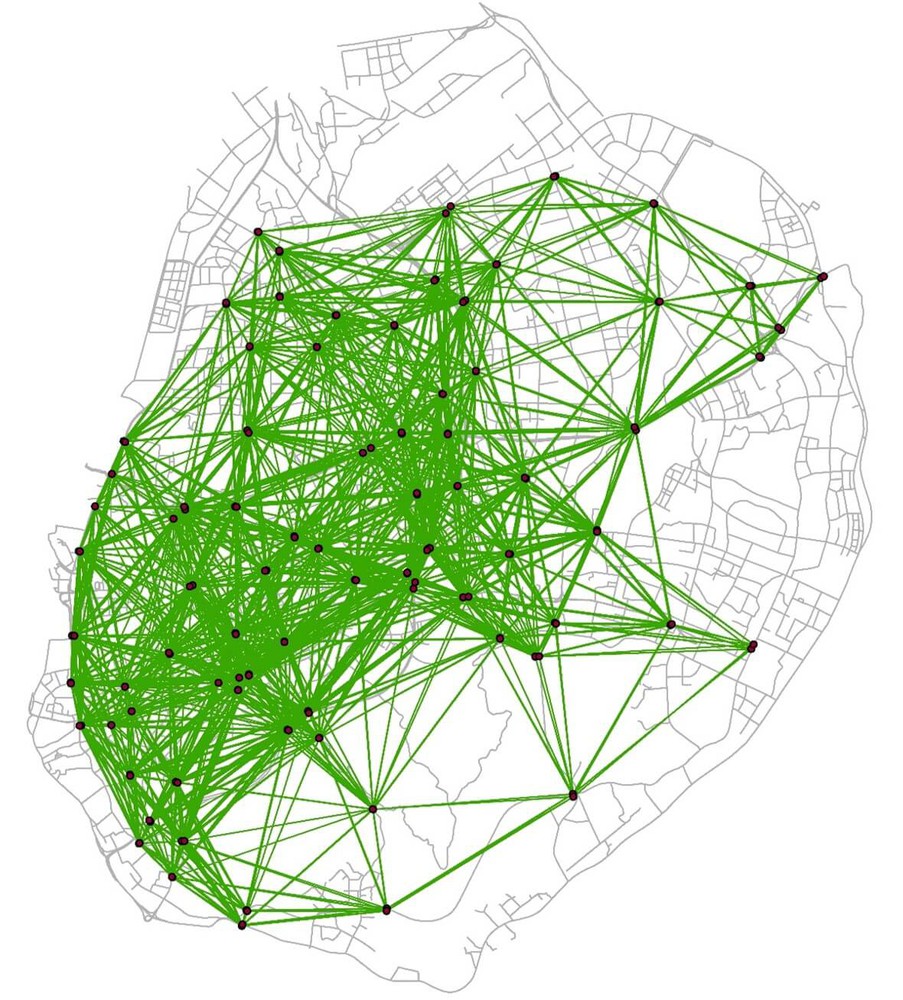}
        \caption{}
    \end{subfigure}
    \centering
    \begin{subfigure}{0.58\textwidth}
        \includegraphics[height=6cm]{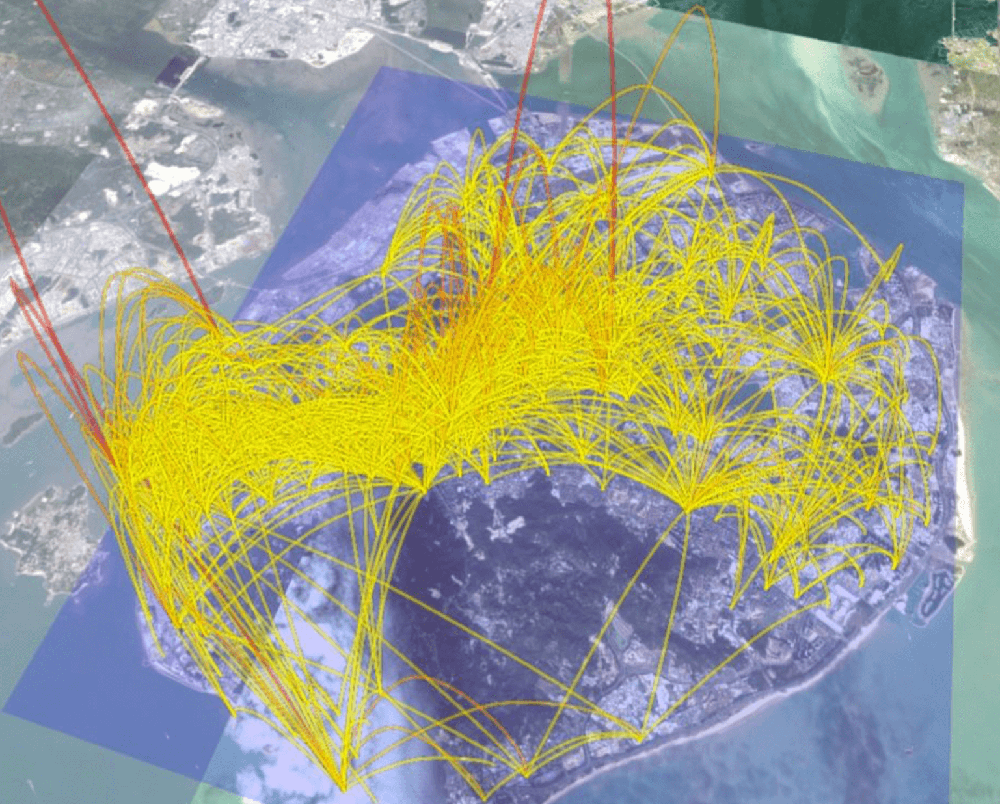}
        \caption{}
    \end{subfigure}
    \caption{Using flow maps to show part of short-distance OD flows. (a) in 2D space; (b) in 3D space, elevating the 2D straight lines. Figures provided by UI-SYSU.}
    \label{fig:interviews:urban-all}
\end{figure}

\subsection{Visualisation Examples}
Flow maps were used as the main visual analytic tool in the analysis of OD flow data in this project. The analysts initially used a 2D flow map (see Fig.~\ref{fig:interviews:urban-all}(a)) but found it difficult to explore due to overlapping and clutter. So they tried to elevate the flows to the third dimension and carefully choose a view point to present the 3D flow map in perspective view (see Fig.~\ref{fig:interviews:urban-all}(b)). They told us that their clients (Xiamen Municipal Planning Committee) preferred the 3D version for aesthetic reasons as well as giving them a better sense of the overall geographical patterns. They also admitted there was a large potential for improving the design of the 3D flow map, for example improving the colour encodings. However, due to the limited time, they have not explored all the options of the design space in their project.

%% file: content/3-interviews/indoor.tex
\section{Flu Forecasting}
\label{sec:interviews:indoor}
Influenza is a relatively common infection, particularly during the winter season and in colder climates, that not only affects people’s daily lives but the local economy and health services. Influenza epidemics vary substantially in size, location, timing and duration from year to year, making it a challenge to deliver timely and proportionate responses~\citep{MOSS:2017il}. Understanding, and predicting epidemics is important to assess the risk and effectively manage the public health resources.

Martin Tomko was interviewed about his experience in spatial analytics in May 2018. Martin Tomko is a Lecturer at the Department of Infrastructure Engineering, The University of Melbourne. Martin received his PhD in GIScience in 2007 from The University of Melbourne, and has continued to work in the geospatial analytics field in Academia, both at the University of Zurich and The University of Melbourne. Martin specialises in computational approaches to spatial communication problems. In the interview, Martin discussed three distinct projects:
\begin{itemize}
	\item Indoor customer movement analysis;
	\item Global postal data analysis; and
	\item Mapping urban mobility for flu forecasting.
\end{itemize}
For reasons of brevity we have chosen to focus on a single representative project, flu forecasting. Despite research by computational epidemiologists focusing on flu forecasting ~\citep{MOSS:2017il,Moss:2018ib,Moss:2016ka}, the focus is mainly on the timing and intensity of the epidemic, treating a large region or metropolitan area (like Greater Melbourne) as a single target. For this project~\footnote{\url{https://networkedsociety.unimelb.edu.au/research/projects/active/urban-flu-forecasting}}, Martin is working with computational epidemiologists, introducing spatial tracking data into the prediction model aiming at fine-grained spatial resolution, to allow more targeted predictions for resources. At the time of the interview, the project was at an early stage, and the team was focused on data exploration.

\subsection{What: Data Abstraction}
There were two types of data set in this project: flu notifications and urban mobility. Flu notification data contains weekly (time) updates of positively tested cases of flu by postcode (geographical location). Urban mobility data was acquired from two different sources: (a) Australian Bureau of Statistics (ABS) journey to work data; (b) Satellite Navigation (satnav) tracking data of around 120,000 - 150,000 people across Australia. All data contains spatial information, while the urban mobility data also form networks. Records are investigated in an aggregated manner, usually by time and/or spatial regions. A time-varying semantic is also particularly important in this project.

\subsection{Why: Task Abstraction}
The main goal of this project is improving flu forecasting modelling by integrating spatial flow data. As the project was at an early stage at the time of interview, the tasks abstracted describe the data analysis and expected tasks in later stages.

\begin{figure}[b!]
\centering
    \includegraphics[width=0.45\columnwidth]{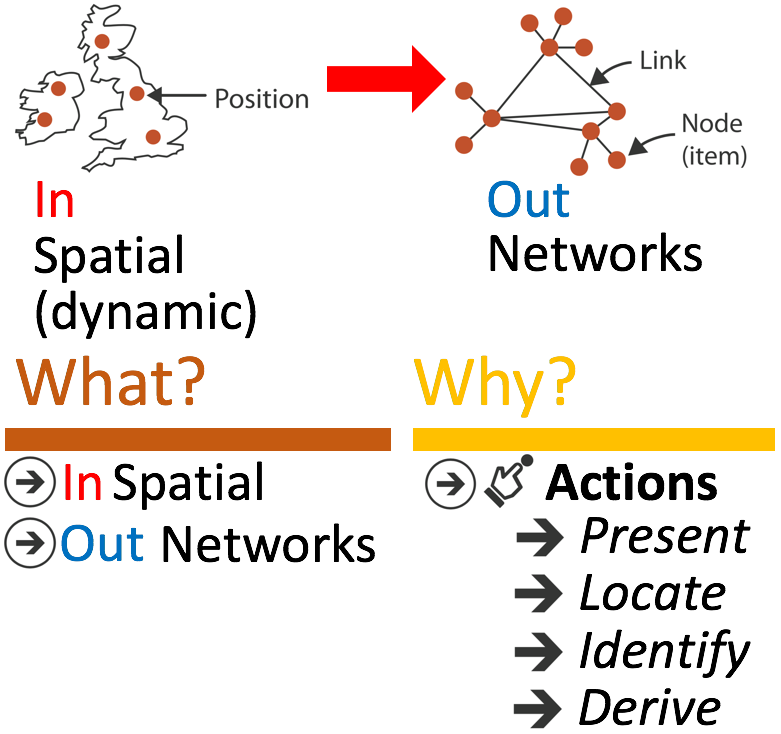}
    \caption{Visual derivation of flu spread network.}
    \label{fig:interviews:flu:derive} 
\end{figure}

One of the tasks was to investigate ``flu spread'' in both a spatial and a temporal manner. The time-varying semantic is essential for this task along with the geographical locations of flu reports. By first presenting the change of location of flu reports visually and then connecting these locations heuristically, a network of flu spread structures can be approximated (see Fig.~\ref{fig:interviews:flu:derive}). Subsequently the topology can be analysed based on the dervied networks using the geographical information (see Fig.~\ref{fig:interviews:all}(b3)~---~\textbf{Topology Identification}). The interviewee and his collaborators identified some interesting findings regarding flu spread in Melbourne: ``~\textit{the CBD (Central Business District) acts as a hub, clearly, but it does not have contagion to and from the local population of the CBD, but from the people that work there. This (phenomenon) is shown very nicely in the data.}''

To introduce spatial flow data, they first investigated the spatial correlation between urban mobility and the flu spread, which involves visually comparing two geographic networks and identifying correlation between them (see Fig.~\ref{fig:interviews:all}(b1)~---~\textbf{Correlation Identification}). They now plan to integrate urban mobility mathematically into the existing statistical prediction model.

The interviewee noted that only limited visualisations have been created for this project, and whilst they store the OD data as a matrix they have presented most of their findings as bar and line charts to show changes over time.  The interviewee also identified that scalability was a potential issue when visualising the urban mobility data, as visual clutter can make it difficult to read useful information.

%% file: content/3-interviews/energy.tex
\section{Energy Network Flows}
\label{sec:interviews:energy}
Power transmission systems are some of the most complex networks. Put simply, a power system has a network of transmission lines connecting power stations and customers. At present electricity is not economically storable. To optimise a power system, the demand from customers needs to be met at every moment~\citep{Bouts:2015fr}. Moreover, due to embedded micro generation in the form of solar panel and/or wind turbine installations on customer sites in recent years, maintaining the power system in an effective way has become an even more challenging problem.

In October 2016, Ross Gawler was interviewed about his experience with power transmission system analysis, network design and planning. Ross Gawler obtained his PhD of Electrical Engineering at Monash University in 1977. Since then, he has worked in the energy sector for nearly 40 years in different roles (Engineering Manager, Director and Principal Consultant) in different companies/organisations (including the State Electricity Commission of Victoria, the Essential Services Commission and Jacobs). Since 2014 he has worked as a Senior Research Fellow in the Monash Energy Materials and Systems Institute (MEMSI) focusing on transmission planning research.

\subsection{What: Data Abstraction}
There are two high-level components in a power system: \emph{elements} and \emph{transmission lines}. \emph{Elements} in a power system is a collective name for transformers, voltage regulators, generators, etc. For the \emph{elements}, different information are captured for different types of devices. Voltage and temperature are commonly measured in real time. Their geographic locations are also recorded. \emph{Transmission lines} are the links connecting \emph{elements}. Along the \emph{transmission lines}, there are two types of flows: \emph{active} and \emph{reactive} power flows. Current is also measured on the \emph{transmission lines}. For emergency alert and analysis, additional data sets are used, like weather, to help the reasoning process.

\emph{Elements} and \emph{transmission lines} belong to the combination of \emph{Spatial} and \emph{Network} types. The data also contains timestamps, which means it has \emph{time-varying} semantics.

\subsection{Why: Task Abstraction}
\label{sec:interviews:energy:why}
The interviewee identified two main problems for energy flow analysis described during the interview: (a) the operating problem, that is instantaneously adjusting elements to increase the capacity of a portion of the power system; (b) the planning problem, which focuses on how to economically relieve constraints where they may occur in the future.

\begin{figure}[b!]
    \captionsetup[subfigure]{justification=centering}
    \centering
    \begin{subfigure}{0.49\textwidth}
        \centering
        \includegraphics[height=5cm]{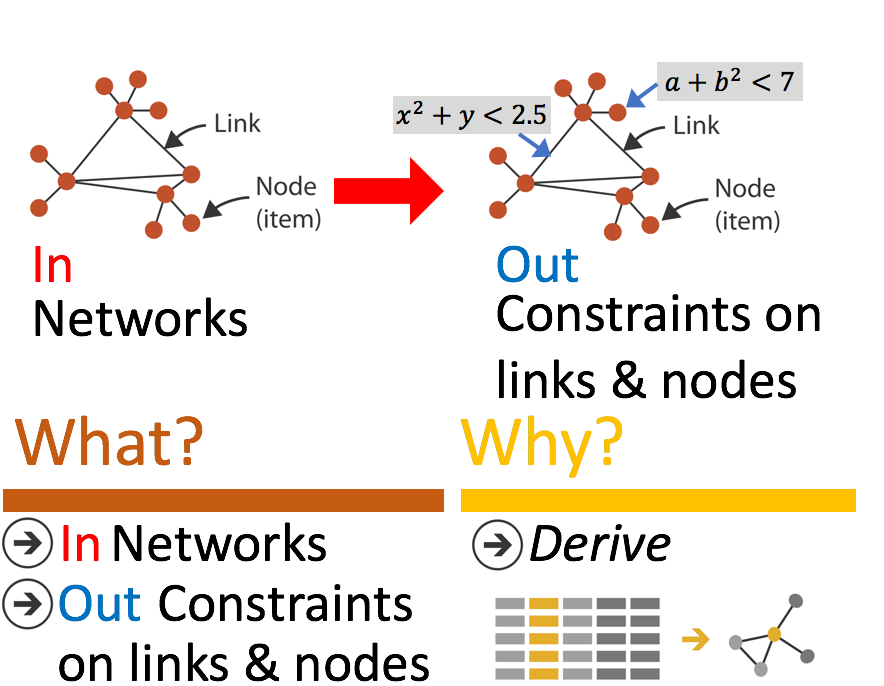}
        \caption{}
    \end{subfigure}
    \centering
    \begin{subfigure}{0.49\textwidth}
        \centering
        \includegraphics[height=5cm]{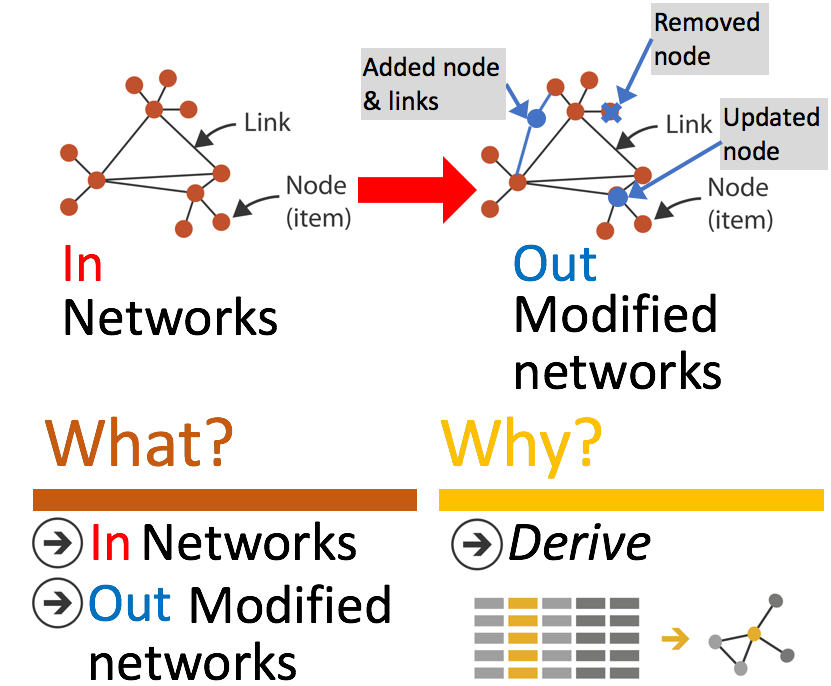}
        \caption{}
    \end{subfigure}
    \caption{Extra data derivation in energy flow analysis: (a) modelling constraints from the \emph{elements} and \emph{transmission lines} data. (b) modifying the network structure to improve capacity.}
    \label{fig:interviews:energy:derive}
\end{figure}

For both problems, constraints were modelled first based on the \emph{elements} and \emph{transmission lines} (see Fig.~\ref{fig:interviews:energy:derive}(a)). There are different types of constraints:
\begin{itemize}
	\item Thermal constraint; Current in the transmission exceeds the rating of a particular element, which can generate too much heat, exceeding the limit and can melt the element, or make the transmission line sag down to an unsafe level.
	\item Voltage collapse constraint; Voltages can become outside normal operating limits, and damage elements. The fact that customers are able to generate energy at their premises makes the situation more complex.
	\item Other constraint; There are other transient and dynamic oscillatory type behaviours that can affect the power system. 
\end{itemize}
According to Ross, there are more than 30,000 constraints in the system he previously worked with, among them, around 300 constraints have higher priorities.

For operating problems, operators monitor the constraints in real time, and identify cases when some constraints enter an insecure margin (see Fig.~\ref{fig:interviews:all}(b1)~---~\textbf{Outlier Identification}). 
To relieve the constraints, operators first propose several possible solutions (see Fig.~\ref{fig:interviews:energy:derive}(b)), summarise and compare the advantages and disadvantages of each, then finally choose the most economical proposal (see Fig.~\ref{fig:interviews:all}(b2)~---~\textbf{Feature Summarisation and Comparison}).

For planning problems, the work flow is the same as demonstrated in Fig.~\ref{fig:interviews:all}(b1)~---~\textbf{Outlier Identification}. Experts analyse the past records to identify where constraints occur and how frequently they occur.
They then modify the model (Fig.~\ref{fig:interviews:energy:derive}(b)) to improve the compatibility for future years. This process usually combines analysing the trends of additional \emph{tabular} data, like local demography (see Fig.~\ref{fig:interviews:all}(b4)~---~\textbf{Trends Summarisation}). For extreme system collapse cases (e.g. tornado, storm), analysis will be conducted afterwards with external \emph{tabular} data, like weather, to understand the potential reasons (see Fig.~\ref{fig:interviews:all}(b1)~---~\textbf{Correlation Identification}).

\begin{figure}[b!]
\centering
    \includegraphics[width=0.65\columnwidth]{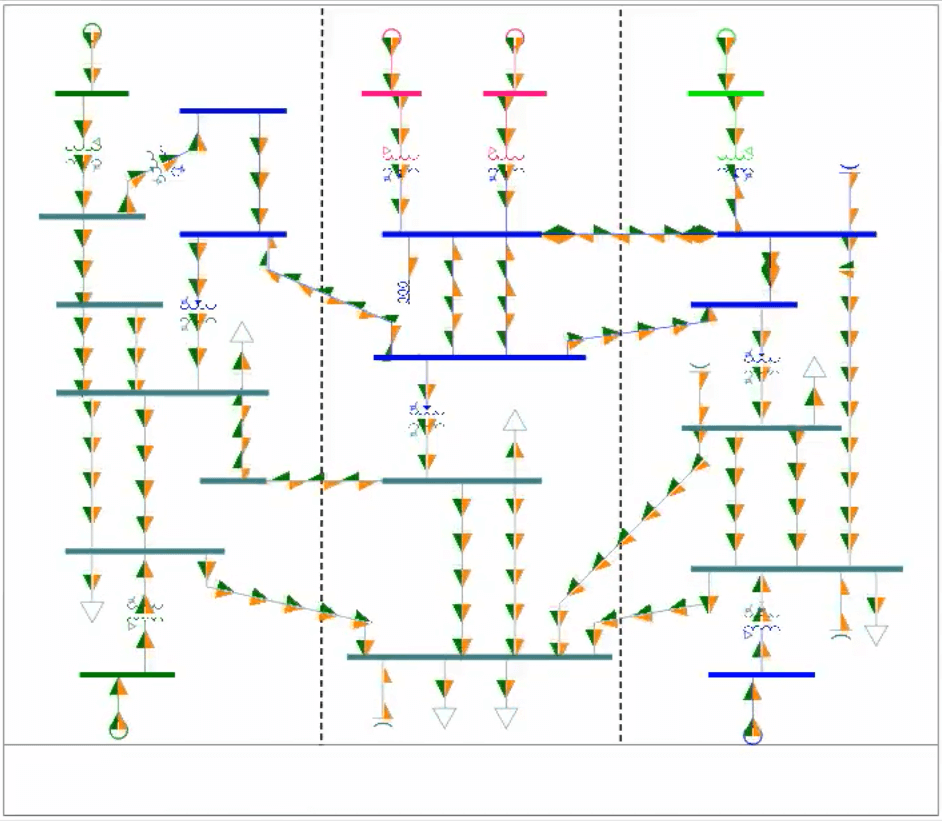}
    \caption{A screen capture of PSS®E, a transmission planning and analysis software.}
    \label{fig:interviews:energy:example} 
\end{figure}

\subsection{Visualisation Example}
As discussed in Section~\ref{sec:interviews:overview}, in the current work flows of analysing energy flows, domain experts mainly focus on the network perspective of the data (see Fig.~\ref{fig:interviews:energy:example}). Geographic data was noted as being important in emergency case reasoning and future model planning. However, the interviewee stated that there is a great potential in using geographic data in other processes as well. Whilst geographical locations are not explicitly used in the first instance, it is also noted that the network visualisations are usually spatial in their layout and therefore often form abstract representations of the geography. The interviewee was enthusiastic for the future of visualisation in the energy sector, concluding the interview with: ``~\textit{visualization of those (energy flow) patterns will become useful both for system operations purposes, and also for trying to develop more stochastic planning models}.''

%% file: content/3-interviews/airline.tex
\section{Aircraft Trajectories}
\label{sec:interviews:airline}
Aircraft trajectory data is used in both Air Traffic Control (ATC) and Management (ATM). ATC looks at real-time aircraft trajectories to maintain traffic fluidity and security, while ATM analyses off line large amounts of past aircraft trajectory data to understand past situations and improve future procedures~\citep{Cordeil:2016ed}. In the interview, in addition to these two activities, we also discussed the problem of redesigning airline routes or adding new airways.

In July 2018, Maxime Cordeil was interviewed about his experience of aircraft trajectories analysis. Maxime Cordeil obtained his PhD in computer science and human computer interaction in 2013. Part of his PhD research project was about studying user interfaces used in air traffic control and management. After his PhD, he worked as a software engineer in Steria Group designing and developing new user interfaces for air traffic controllers. Since 2015 he has worked as a Research Fellow focusing in Immersive Analytics within Monash Computer Human Interaction and Creativity (CHIC) discipline research group. One of his research interests is using immersive technologies for air traffic control and management.

\subsection{What: Data Abstraction}
Information about each aircraft is collected during each flight, including:
\begin{itemize}
	\item Unique identifier.
	\item 3D Position; i.e. longitude, latitude and altitude.
	\item Speed and direction.
	\item Weather, e.g. temperature, wind speed etc.
	\item Timestamps.
\end{itemize}
As the take off and landing of a aircraft make great noise, such parts of airways have to avoid living areas. As a result, demography data is usually used to inform redesign of existing airways or design of new airways.

\subsection{Why: Task Abstraction}
In ATC, the pilots in the aircraft and the air traffic controllers in the control tower are actively communicating with each other. Air traffic controllers give instructions to pilots and pilots report back their situation. Some emergency cases were discussed in the interview:
\begin{itemize}
	\item Conflict between two aircraft; i.e. if two aircraft retain their direction of flying, accidents can occur. Automatic systems are commonly used to give warnings of such situation to both pilots and air traffic controllers.
	\item Accidents happen on the runway; air traffic controller will ask pilots to take an alternative runway for take off or landing.
	\item Weather; for landing in bad weather, pilots could be instructed to put on hold, and in some cases, to land at another nearby airport.
\end{itemize}
These actions involve identifying outliers in both trajectories and other \emph{tabular} data sets, like weather, like weather (see Fig.~\ref{fig:interviews:all}(b1)~---~\textbf{Outlier Identification}). In some cases, the journeys of other aircraft may be affected, in which case urgent response proposals need to be derived. Air traffic controllers need to summarise and compare different proposals to chose the most suitable one (see Fig.~\ref{fig:interviews:all}(b2)~---~\textbf{Features Summarisation and Comparison}).

In ATM, the main goal is to get a better understanding of unusual historical events and to avoid similar issues occurring in the future. The outliers of either aircraft or route trajectories are identified. Then the main task is to find the correlation between these outliers and other information, such as the aircraft condition or bad weather, to infer possible reasons (Fig.~\ref{fig:interviews:all}(b1)~---~\textbf{Correlation Identification}). According to the interviewee, analysts mainly concentrate on individual trajectories, however they also show interests in the overview of a large number of trajectories, for example, investigating the topology structure of trajectories (see Fig.~\ref{fig:interviews:all}(b3)~---~\textbf{Topology Identification}). 

In cases of redesigning existing airways or creating new airways, several proposals are usually derived. Features of these proposals, usually combined with external information like demography data, are summarised and compared (see Fig.~\ref{fig:interviews:all}(b2)~---~\textbf{Features Summarisation and Comparison}) to select the most appropriate proposals given all the context.

\subsection{Visualisation Examples}
\begin{figure}[t!]
    \captionsetup[subfigure]{justification=centering}
    \centering
    \begin{subfigure}{0.49\textwidth}
        \includegraphics[height=5.8cm]{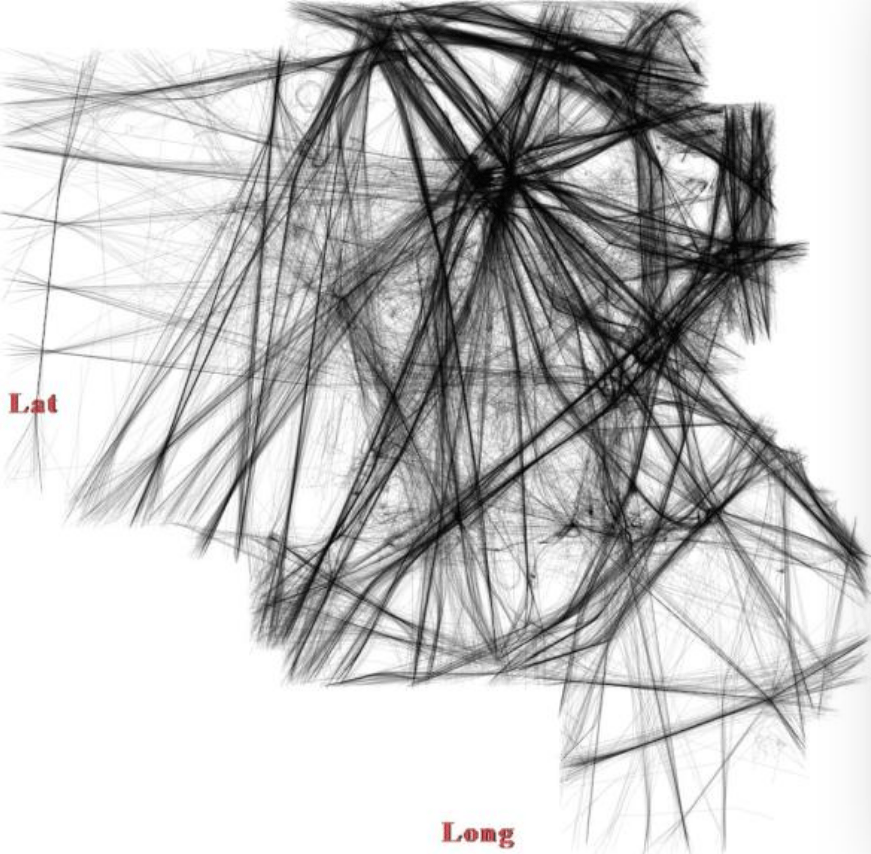}
        \caption{}
    \end{subfigure}
    \centering
    \begin{subfigure}{0.49\textwidth}
        \includegraphics[height=5.8cm]{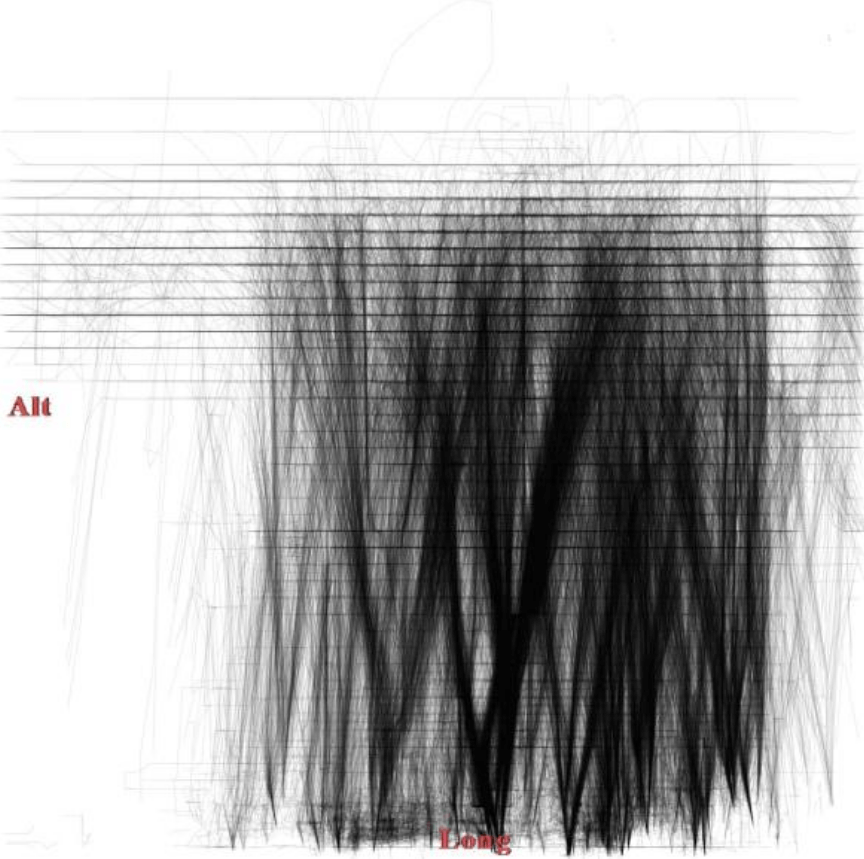}
        \caption{}
    \end{subfigure}
    \caption{A demonstration of aircraft trajectory visualisation: (a) top view (longitude-latitude view); (b) vertical view (longitude-altitude view). Figure from~\citep{Cordeil:2013wu}.}
    \label{fig:interviews:air}
\end{figure}

Visualisations are widely used in aircraft trajectory analysis. An aircraft trajectory is basically a list of 3D positions and it is commonly visualised as a 3D line connecting them. Fig.~\ref{fig:interviews:air} demonstrates that these 3D trajectories can be presented as two views: top view (longitude-latitude view) and vertical view (longitude-altitude view). Cordeil \textit{et al.} also explored the use of animation between vertical and top view to help the user understand the traffic, find pattern and find landing points~\citep{Cordeil:2013wu}.

Due to the complexity of the large number of trajectories, interactions are essential for the analysis. Interactive tools on traditional 2D screens are available to support iterative exploration of the trajectories, e.g. FromDaDY~\citep{Hurter:2009hy}. Immersive technologies, such as Virtual Reality (VR), have been preliminarily explored for analysing aircraft trajectories by the interviewee and his collaborators~\citep{Cordeil:2016ed}. The interviewee commented in the interview ``~\textit{the nature of this data (trajectories) is 3D, 3D mapping of these trajectories is straightforward.}'' The interviewee also discussed the difference between the interactions on 2D screen and in immersive environments ``~\textit{When you're manipulating a 3D view on the screen, you are just moving the mouse play around with rotation. But if you're in VR, it's like as if the object is here, so you can grab the object.}'' The interviewee also pointed out that using immersive environments in aircraft trajectory analysis is still at an early exploratory stage: the industry is conducting small scale experiments, but immersive solutions have not yet been widely adopted in practice. The interviewee is currently working with his collaborators to bring more interactions in immersive environments to improve aircraft trajectory analysis~\citep{Hurter:2019hz}.

%% file: content/3-interviews/summary.tex
\section{Summary}
\label{sec:interviews:summary}

In this section we aim to summarise the common analytical tasks identified in each of the five domains. In the five interviews we identified six different \textbf{Consume} tasks as well as a number of \textbf{Produce} tasks. These \textbf{Produce} tasks usually involve additional modelling processes (e.g. \emph{deriving} constraints in energy network flow analysis) which are outside the scope of this thesis. For completeness we described all tasks in the individual interviews in detail, however, as our focus is on how to visually represent geographically-embedded flow data for analytical tasks, we concentrate on the \textbf{Consume} tasks in this summary. The six \textbf{Consume} tasks identified in the interviews are presented in Tab.~\ref{tab:interview:tasks}.

\begin{table}[t!]
\small
\begin{tabular}{|l|c|l|c|c|l|c|}
\hline
\textbf{Consume}                                                     & \multicolumn{6}{c|}{Present}                                                                                                                                                                       \\ \hline
\textbf{Search}                                                      & \multicolumn{2}{c|}{Locate}                                     & \multicolumn{3}{c|}{Explore}                                                                & Browse                             \\ \hline
\textbf{Query}                                                       & \multicolumn{2}{c|}{Identify}                                   & Summarise                    & Compare                      & \multicolumn{1}{c|}{Identify} & Summarise                          \\ \hline
\textbf{Targets}                                                     & Outliers                                                                        & \multicolumn{1}{c|}{Correlation}                    & Features                                                                 & Features                                                                 & \multicolumn{1}{c|}{\begin{tabular}[c]{@{}c@{}}Topology\\ (space)\end{tabular}} & \begin{tabular}[c]{@{}c@{}}Trends\\ (time)\end{tabular}                       \\ \hline
\begin{tabular}[c]{@{}l@{}}\textbf{Graphic}\\ \textbf{Depiction}\end{tabular} 
	& \multicolumn{2}{c|}{Fig.~\ref{fig:interviews:all} (b1)}
    & \multicolumn{2}{c|}{Fig.~\ref{fig:interviews:all} (b2)} 
    	& 
    		\begin{tabular}[c]{@{}c@{}}
    			Fig.~\ref{fig:interviews:all} \\ (b3)
    		\end{tabular} 
    	& \multicolumn{1}{c|}{
    		\begin{tabular}[c]{@{}c@{}}
    			Fig.~\ref{fig:interviews:all} \\ (b4)
    		\end{tabular}} \\ \hline
\textbf{Domains}                                                  & 
	\multicolumn{1}{l|}{
		\begin{tabular}[c]{@{}l@{}} 
			Refugee\\ Urban\\ Flu  \\ Energy \\ Aircraft
		\end{tabular}
	} & 

	\multicolumn{1}{l|}{
		\begin{tabular}[c]{@{}l@{}} 
			Refugee\\ Urban \\ Energy \\ Aircraft
		\end{tabular}
	} & 

	\multicolumn{1}{l|}{
		\begin{tabular}[c]{@{}l@{}} 
			Energy \\ Aircraft
		\end{tabular}
	} & 

	\multicolumn{1}{l|}{
		\begin{tabular}[c]{@{}l@{}} 
			Energy \\ Aircraft
		\end{tabular}
	} & 

	\multicolumn{1}{l|}{
		\begin{tabular}[c]{@{}l@{}} 
			Urban\\ Flu \\ Aircraft
		\end{tabular}
	} & 

	\multicolumn{1}{l|}{
		\begin{tabular}[c]{@{}l@{}} 
			Refugee\\ Energy
		\end{tabular}
	} \\ \hline
\end{tabular}
\vspace{0.5em}
	\caption{Tasks mentioned in each of the interviews.}
	\label{tab:interview:tasks}
\end{table}

It is evident, from Tab.~\ref{tab:interview:tasks} that there are quite a few commonalities between the tasks identified in the five interviews. In particular nearly all analysts were interested in \emph{Locate > Identify > Correlation or Outliers} tasks as well as \emph{Explore > Identify > Topology} tasks. This task taxonomy has given us a method to demonstrate the common themes between our interviews, however, when we look into the individual tasks from the different domains, we notice that the taxonomy may over simplify some of these tasks. We recognise that there is a difference in target when performing analytical tasks relating to the geographically-embedded flow data. 

The geovisualization and cartography task taxonomy research~\citep{Andrienko:2006ek,Roth:2013ja} help us to explore this in greater detail. These taxonomies distinguish tasks into two levels according to the level of data analysis: \emph{elementary} and \emph{synoptic}. \emph{Elementary tasks} refer to individual elements; \emph{synoptic tasks}~\citep{Andrienko:2006ek} (refered to as ``general'' in~\citep{Roth:2013ja}) involve the whole reference set or its subsets.

For the \emph{Locate > Identify > Correlation or Outliers} tasks, our domain experts are either interested in:
\vspace{-0.5em}
\begin{itemize}
 	\item a flow between two locations; e.g. energy flow between two elements or the frequency of bicycle usage between two docking stations. Or
 	\item the total collection of flows into or out of a location; e.g. migrations into one country or total number of bikes coming out of a bicycle docking station. 
\end{itemize} 
\vspace{-0.5em} 
These tasks can be considered as \emph{elementary tasks}.

In addition, the domain experts were also often interested in the spatial patterns (\emph{Explore > Identify > Topology}), e.g. the spatial movement of flu spread. This often involves the mental calculation of multiple flows into and out of multiple locations to see the regional pattern, or topology of the data. These tasks can be considered as \emph{synoptic tasks}.

\begin{table}[b!]
\begin{tabular}{|l|l|l|}
\hline
\multirow{9}{*}{SF} & Refugee & \textbf{Identify [outliers]} possible smuggling routes. \\ \cline{2-3} 
    & \multirow{3}{*}{Urban}  & \textbf{Identify [outliers]}  dense intersections of short distance travels. \\ \cline{3-3} 
                    &                                                                        & \textbf{Identify [outliers]} missing connections between docking stations.              \\ \cline{3-3} 
                    &                                                                        & \textbf{Identify correlation} of bicycle usage and bike lanes.              \\ \cline{2-3} 
                    & Flu                                                                    & \textbf{Identify correlation} of urban mobility and the flu spread.         \\ \cline{2-3} 
                    & \multirow{2}{*}{Energy}                                                & \textbf{Identify [outliers]} insecure transmission lines.                               \\ \cline{3-3} 
                    &                                                                        & \textbf{Summarise and compare features} of newly proposed network.           \\ \cline{2-3} 
                    & \multirow{2}{*}{\begin{tabular}[c]{@{}l@{}}Aircraft\end{tabular}} & \textbf{Identify correlation} of trajectories and emergency cases.                   \\ \cline{3-3} 
                    &                                                                        & \textbf{Summarise and compare features} of newly proposed network.           \\ \hline
\multirow{9}{*}{TF} & \multirow{2}{*}{Refugee}                                               & \textbf{Summarise trends} of arrivals of a locations across time.                         \\ \cline{3-3} 
                    &                                                                        & \textbf{Identify correlation} of arrivals of a locations and other factors. \\ \cline{2-3} 
                    & Urban                                                                  & \textbf{Identify [outliers]} popular origins and destinations.                          \\ \cline{2-3} 
                    & Flu                                                                    & \textbf{Identify topology} of flu spreading, and found the CBD as a hub.     \\ \cline{2-3} 
                    & \multirow{4}{*}{Energy}                                                & \textbf{Identify [outliers]} insecure elements.                                         \\ \cline{3-3} 
                    &                                                                        & \textbf{Summarise and compare features} of newly proposed network.           \\ \cline{3-3} 
                    &                                                                        & \textbf{Summarise trends} of demography across time.                                      \\ \cline{3-3} 
                    &                                                                        & \textbf{Identify correlation} of collapse cases and other factors.           \\ \cline{2-3} 
                    & \begin{tabular}[c]{@{}l@{}}Aircraft\end{tabular}                  & \textbf{Identify topology} of large amount of trajectories.                  \\ \hline
\multirow{5}{*}{RF} & Urban                                                                  & \textbf{Identify [topology]}: different patterns of using bicycles in regions.        \\ \cline{2-3} 
                    & \multirow{2}{*}{Flu}                                                   & \textbf{Identify topology} of flu spreading.                                 \\ \cline{3-3} 
                    &                                                                        & \textbf{Identify correlation} of urban mobility and the flu spread.         \\ \cline{2-3} 
                    & Energy                                                                 & \textbf{Summarise and compare features} of newly proposed network.           \\ \cline{2-3} 
                    & \begin{tabular}[c]{@{}l@{}}Aircraft\end{tabular}                  & \textbf{Identify topology} of large amount of trajectories.                  \\ \hline
\end{tabular}
\vspace{0.5em}
	\caption{Tasks in the interviews categorised based on the targets of single flow (SF), total flow (TF) and regional flow (RF), showing the tasks from Tab.~\ref{tab:interview:tasks} in bold.}
	\label{tab:interview:targets}
\end{table}

\vspace{-1em}
We therefore identify that there are three types of targets for analytical tasks relating to geographically-embedded flow data. These distinct differences in targets from the literature are fundamental in not only understanding the different tasks used in these domains, but also for describing the design space for geographically-embedded flow visualisation. To fill this gap, we propose three different targets for analysing geographically-embedded flow data: single flow, total flow and regional flow:
\begin{itemize}
	\item Single flow (SF) is the aggregated number of records (or magnitude) between two locations.

	\item Total flow (TF) is the aggregated number of records (or magnitude) coming in/out of a specific location.

	\item Regional flow (RF) is the aggregated number of records (or magnitude) between or within sets of locations. It can be considered as single flow in different geographic resolutions, i.e. considering multiple locations as a single region. 

\end{itemize}
We categorise the tasks mentioned in interviews into these three types of targets and found these three types of targets were mentioned in almost all of the five interviews (see Tab.~\ref{tab:interview:targets}). One thing worth noting is that in Munzner's framework, the \textbf{Features} are described as ``\textit{any particular structures of interest}''. In terms of the interviews, we see that the \textbf{Features} can refer to all three types of targets (see Tab.~\ref{tab:interview:targets}).
This analysis help us gain a clearer understanding of the common tasks that analysts in different domains perform with geographically-embedded flow data.
\vspace{-1em}

%% file: content/3-interviews/conclusion.tex
\section{Conclusion}
\label{sec:interviews:conclusion}
This chapter analysed five expert interviews to better understand the real-world practice of analysing geographically-embedded flow data. Data and tasks discussed in the interviews were abstracted using Munzner’s what-why-how framework~\citep{Munzner:2014wj}. This was useful to structure our analysis of the interviews and demonstrate common tasks described by our interviewees, but we also recognised that a more detailed classification of the targets of potential analytical tasks is required to better describe the role of flow map visualisation in the domain. Thus, we propose three different types of targets for analysing geographically-embedded flow data: \emph{single flow}, \emph{total flow} and \emph{regional flow}. We continue to explore the identified tasks and targets in this thesis, in particular, in Chapter~\ref{chapter:evaluating-2d-od-flow-maps}, where we systematically evaluated state-of-the-art visualisation techniques using specific user tasks designed for these three types of targets.

%% file: content/2-related-work/0-index.tex
\chapter{Related Work}
\label{chapter:related}
In the last chapter we showed that geographically-embedded flow data is widely collected and analysed in many real-world applications, and visualisations are commonly used to support these analyses.

This chapter surveys previous work in visualising geographically-embedded OD flow data. We start by discussing general concepts and definitions in data visualisations. We then focus on examining existing visualisation techniques for visualising OD flow data in 2D and 3D display environments to gather research findings as well as to identify gaps in the existing research. Immersive technologies demonstrate a great potential for improving OD flow visualisation, however the design space of OD flow visualisation in such immersive environments has not been well explored. Therefore, in the last section, we introduce an emerging research area called \emph{Immersive Analytics}.

\input{content/2-related-work/visual-analytics}
\input{content/2-related-work/2D}

\input{content/2-related-work/3D}
\input{content/2-related-work/immersive}

\input{content/2-related-work/conclusion}

%% file: content/2-related-work/visual-analytics.tex
\section{Data Visualisation}
\label{sec:related:va}
Two general aims for using visualisations are commonly cited. These are briefly explained here:

\vspace{-1em}
\hangindent=2em \qquad \textbf{Exploration and discovery.} The core part of data analysis is exploring the data and discovering patterns and trends~\citep{larose2014discovering}, both to confirm expected hypotheses, and to discover unexpected insights~\citep{Munzner:2014wj}. Data exploration is complementary to computational analysis and provides a more detailed view of the data than just a few statistical features. A combination of different visualisations are usually used in this process.  

\vspace{-1em}
\hangindent=2em \qquad \textbf{Presentation and communication.} There are generally two types of audiences for this aim: 
\begin{itemize}[leftmargin=4em]
	\item \textbf{Analysts}: These can be colleagues working on the same project or researchers in the same domain. They usually have a good understanding of the problem context and possibly are well-trained to interpret visualisations in the domain, e.g. different statistic charts, heatmaps. Selected visualisations used in exploration and discovery could be directly presented to this audience, sometimes with highlighted findings.
	
	\item \textbf{Other stakeholders}: These can be managers, policy makers, the general public, etc. They usually do not know the details of the analysis and only have a high-level overview of the problem context. They may also not have enough experience in data analysis or may not be familiar with the interpretation of domain specific visualisations. In such situations, visualisations need to be easy to understand to communicate a specific message or narrative. For this purpose, charts and graphs used need to be easy to understand.
\end{itemize}

Visualisation research is conducted to target these two general aims. We briefly introduce three related research directions here:

\vspace{-1em}
\hangindent=2em \qquad \textbf{Information Visualisation} uses interactive visual representations of data to amplify cognition. It is mainly concerned with abstract data that has no inherent spatial structure and transforms such data to screen space~\citep{Card:1999ut,Purchase:2008bz}, for example, a line chart for stock market prices, a node-link diagram for social network relationships, etc.

\vspace{-1em}
\hangindent=2em \qquad \textbf{Thematic Cartography and Geovisualisation} uses visual representations to display the spatial pattern of a theme or attribute~\citep{robinson1995elements,slocum2009thematic}. Maps are commonly used as a general-reference space and background for the visualisations. Geographic data alone is not considered abstract (in the sense identified above for Information Visualisation), but in thematic cartography, the geographic map is often overlaid with abstract attributes, e.g. population density, relationships between locations (e.g. migration flows). Thematic Cartography can be considered the intersection between Information Visualisation and cartography.

\vspace{-1em}
\hangindent=2em \qquad \textbf{Visual Analytics} combines automated analysis techniques with interactive visualizations to support effective understanding, reasoning and decision making for very large and complex data sets~\citep{Andrienko:2010co,Keim:2008gg}. Visual analytics emphasises analytical reasoning with visualisation techniques. An example is a visual analytic tool which combines visualisation of high-dimensional data with automatic algorithms for detecting clusters or other patterns to guide the user’s visual exploration.

\vspace{-1em}
These three research directions overlap with one another. The problem of visualising OD flow data is at the intersection of all three. It involves visualising geographic information as well as abstract values (e.g, relationships between locations, flow magnitude by area, topological flow hubs) to support analytic work flows. 

To better design an effective visualisation, designers have to consider three resource limitations: those of computers, of humans, and of displays~\citep[Chap. 1]{Munzner:2014wj}.

\vspace{-1em}
\hangindent=2em \qquad \textbf{Computers} are very effective at processing a sequence of commands. As a result, computer scientists are working in various areas aiming to transform domain specific problems to computer based models. However, as discussed in Section~\ref{sec:intro:vis}, many analysis problems are ill specified. Modelling such problems to the point that they can be solved without human inspection or intervention is difficult or impossible with current technologies. 
 
\vspace{-1em}
\hangindent=2em \qquad \textbf{Human} capability can be thought of a combination of perceptual, cognitive and motor systems~\citep[Chap. 2]{Card:1983ui}. The subtle cooperation between these systems allows humans to be adaptive to different task environments~\citep{simon1971human}. Humans are good at highly dynamic tasks, like concept formalisation, exploratory analysis and decision-making. However, humans are limited to their computational ability, cognitive capacity and memory capacity~\citep{miller1956magical}.

\vspace{-1em}
\hangindent=2em \qquad \textbf{Displays} are external devices that can augment the internal capacity of cognition and memory of human capability. However, displays are fixed in size and limited in resolution. Reasoning about information presented on a computer display is not a ``natural'' activity. It requires humans to have experience with and capability to use metaphors that have become conventional, for example, the standard windows, icons, menus, pointer (WIMP) based user interface. 

A visualisation system uses \emph{displays} to achieve a complementary combination of \emph{computer} and \emph{human} capabilities. In the following two sections, we are going to describe and discuss OD flow visualisation techniques in different display spaces: 2D and 3D respectively.

%% file: content/2-related-work/2D.tex
\section{2D OD Flow Visualisations}
\label{sec:related:2d}
The presentation of multiple flows on a map is a classic problem in cartography and geographic visualisation. A significant amount of research has been conducted to explore this problem in 2D display space, however, as the design space is huge, there is still room for further exploration. Here, we discuss three visualisation approaches: two broad approaches as \emph{flow map approaches} and \emph{matrix-based approaches}, as well as \emph{hybrid approaches} which aim to combine these two broad approaches.

\subsection{Flow Map Approaches}
\begin{figure}[t!]
    \captionsetup[subfigure]{justification=centering}
    \centering
    \begin{subfigure}{0.95\textwidth}
        \includegraphics[width=\textwidth]{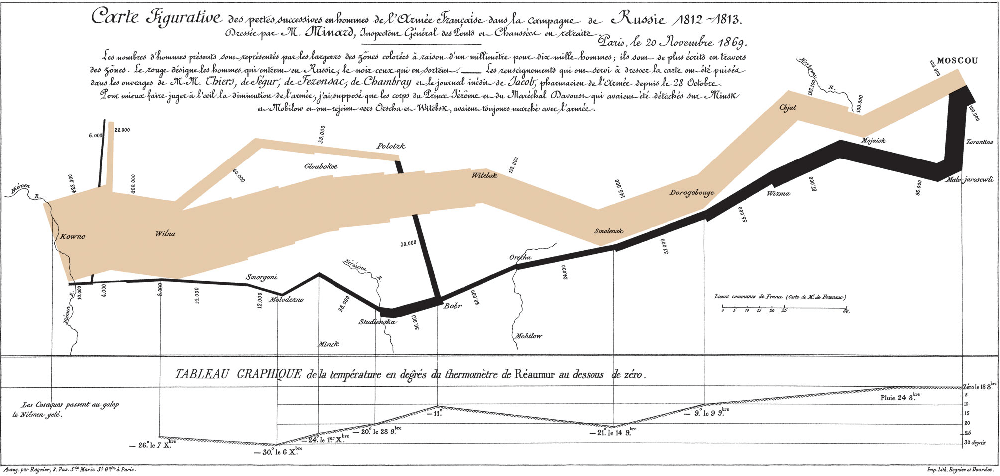}
        \caption{Napoleon's disastrous Russian campaign of 1812.}
    \end{subfigure}
    \begin{subfigure}{0.95\textwidth}
        \includegraphics[width=\textwidth]{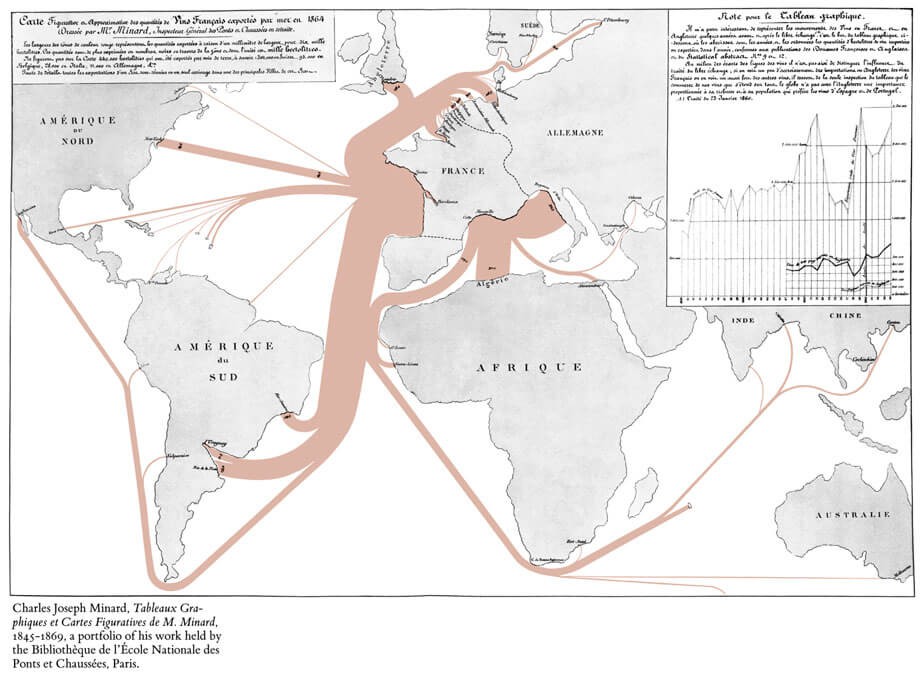}
        \caption{French wine exports of 1864.}
    \end{subfigure}
    \caption{Flow maps mannually designed by Charles Joseph Minard.}
    \label{fig:related:charles}
\end{figure}

The earliest known flow map was created by Henry Drury Harness in 1837 to show rail usage~\citep{Robinson:1955hz}. Shortly after, Charles Joseph Minard popularised their use with sophisticated depictions of emigration and trade~\citep{Robinson:1967cj}, examples demonstrated in Fig.~\ref{fig:related:charles}. Such well-designed, manually-prepared flow maps are mainly used for presentation and communication purpose. Although they are presenting small scale data (relatively few origins and destinations and few flows between them), the design and manual fine-tuning of such pictures takes a long time, for example, distorting some parts of the map in Fig.~\ref{fig:related:charles}(b) to make space for flow lines.

Early digital cartographers extended and improved automatic techniques to map quantitative origin-destination flow links using straight lines with width proportionately varying with quantitative attributes~\citep{chicagoAreaTransportStudy,kadmon1971komplot,kern1969Mapit,Tobler:1981kw,Tobler:1987kn,wittick1976}. Unfortunately visual clutter and line crossings are inevitable even in small datasets using these methods.

Research has been conducted in various aspects to improve flow map designs. In this section, we classify such research into three categories: \emph{layout}, \emph{interaction} and \emph{other visual encodings}.

\textbf{Layout}\\
How to best layout nodes and links is one of the core topics in graph visualisation research\footnote{\url{http://www.graphdrawing.org/}}. While flow maps share some similarities with graph visualisations, e.g. nodes (locations) and links (flows) are the main visual elements, flow map design has more constraints. The most obvious one is that nodes in flow maps are geographic locations and thus cannot be moved freely. This is unlike graph visualisations in which nodes usually can be placed at any position to best show the graph structure (e.g. to avoid link crossings and keep the nodes well-spaced). The main challenge of flow map layout is to route the links (flows) to improve the readability. 

\begin{figure}[b!]
\centering
    \includegraphics[width=0.98\columnwidth]{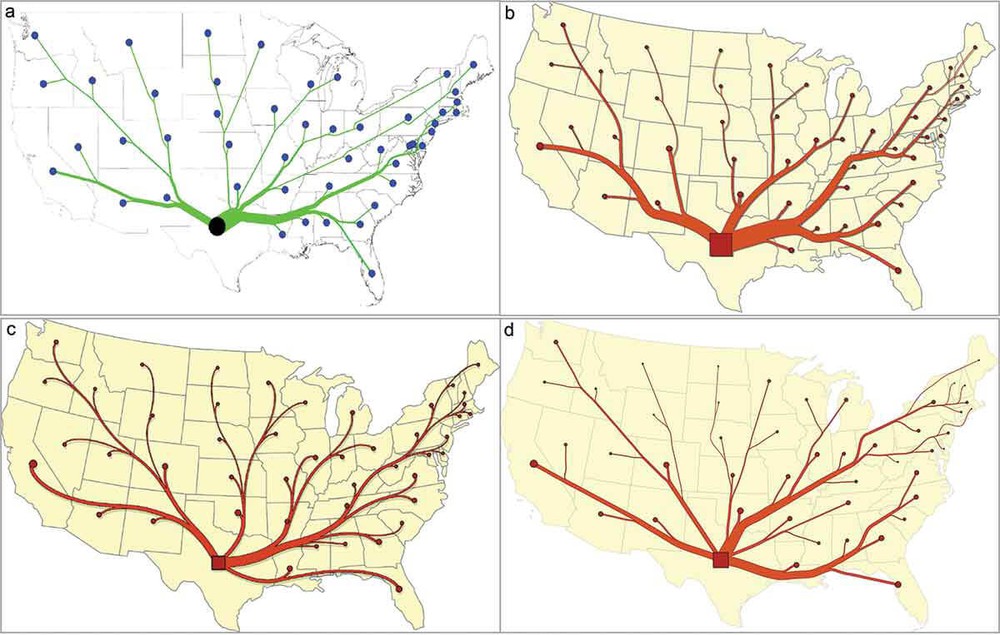}
    \caption{Different one-to-many flow map layouts: (a) Force-directed layout by Debiasi \textit{et al.}~\citep{Debiasi:2014tq}; (b) Spiral tree by Buchin \textit{et al.}~\citep{Verbeek:2011fk}; (c) Stub bundling by Nocaj and Brandes~\citep{Nocaj:2013gd}; and (d) Triangulation, network, shortest path, and smooth (TNSS) by Sun~\citep{Sun:2018db}. Figure derived from Figure 10 in~\citep{Sun:2018db}, © 2018 Taylor \& Francis.}
    \label{fig:related:2Ds}
\end{figure}

Visual clutter is a major concern for maps even with only a few dozen flows, let alone hundreds or even thousands of flows. For one-to-many flow maps, i.e. one origin connecting many destinations, or the reverse, can be visulised as tree-like styles rooted at one location. Such techniques have been in use for a long time and several algorithms for their automatic generation have been proposed~\citep{Debiasi:2014tq,Nocaj:2013gd,Phan:2005cn,Sun:2018db,Verbeek:2011fk} (see Fig.~\ref{fig:related:2Ds}). While these techniques offer aesthetically appealing results for one-to-many flow maps, it is unclear how to best adapt them to many-to-many flows. Another problem with these techniques is that it obscures individual flow lines where they share common end points. Merged lines could be interpreted incorrectly (e.g. as actual geographic features such as highways or as physical flows joining together).

\begin{figure}[b!]
\centering
    \includegraphics[width=0.98\columnwidth]{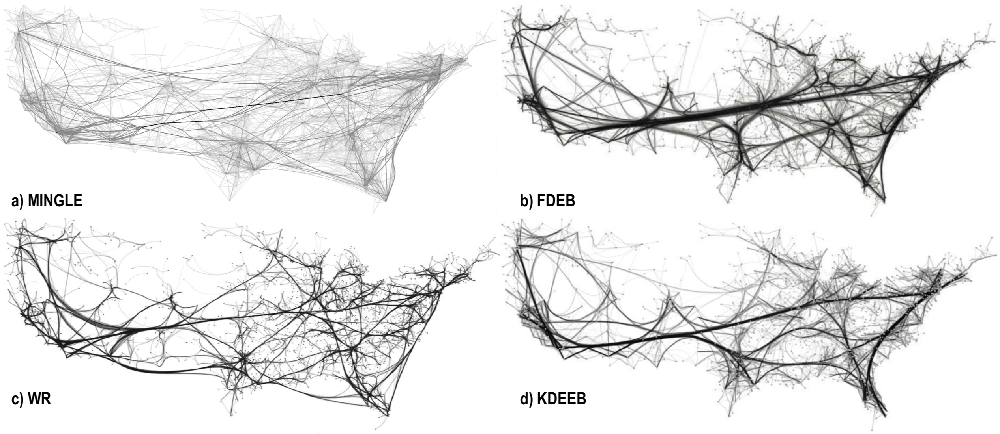}
    \caption{Different many-to-many flow map layouts showing US migrations with 1715 locations and 9780 flows: (a) by Gansner \textit{et al.}~\citep{Gansner:2011iya}; (b) by Holten \textit{et al.}~\citep{Holten:2009kga}; (c) by Lambert \textit{et al.}~\citep{Lambert:2011fi}; and (d) by Hurter~\citep{Hurter:2012bm}. Figure derived from Figure 4 in~\citep{Lhuillier:2017gm}, © 2017 John Wiley and Sons.}
    \label{fig:related:bundling}
\end{figure}

\vspace{-0.5em}
One solution to the clutter produced by many-to-many flow maps, is to ``bundle'' the flows together into ``highways'' between highly connected parts of the map. Although these edge-bundling techniques were originally investigated for general graph (network) visualisation~\citep{Cui:2008hx,Ersoy:2011ht,Gansner:2011iya,Holten:2009kga,Hurter:2012bm,Hurter:2018kk,Lambert:2011fi,Lhuillier:2017dy,Peysakhovich:2015ch,Selassie:2011hw}, they can be applied to visualise many-to-many OD flow data as flow maps. Some examples demonstrated in Fig.~\ref{fig:related:bundling}; more discussion and examples are available in~\citep{Lhuillier:2017gm}). 
These techniques provide an overview for many-to-many flow maps, however, it is difficult to distinguish or follow a single OD flow. For example, when a flow line enters a bundle with multiple outgoing paths, the precise connectivity is ambiguous. For example, so-called confluent bundling techniques can merge flows without ambiguity but can be challenging to interpret~\citep{Bach:2016br}.
There are a couple of recent edge bundling methods that are able to neatly offset individual flows within bundles using a “metro-map” style visual metaphor~\citep{Bouts:2015ip,Pupyrev:2011bt,Pupyrev:2016cd} (see Section~\ref{sec:maptrix:flow-map}). These techniques make individual flows more readable, but our user study suggests they do not scale well (see Section~\ref{sec:maptrix:first-study}).

\vspace{-1em}
An alternative to layout the flows is to ``fan-out'', or maximise the angular separation between flows incident to origins or destinations~\citep{Riche:2012ht}. Such a technique was recently found to offer improved readability in a controlled study~\citep{Jenny:2017ci}. We tested such a technique in our user studies (see Section~\ref{sec:vr-flow-maps:study-01}).

\textbf{Interaction}\\
Interaction is another way to overcome visual clutter in flow maps. Tobler~\citep{Tobler:1987kn} identifies filtering methods and guidelines for simplifying flow data to increase the readability of maps, such as sub-setting (to only show flows of a selected area), thresholding (to only display the largest flows), or merging (to aggregate flows with spatially close origins and destinations). Van den Elzen and Van Wijk~\citep{vandenElzen:2014em} recently introduced a system providing interactive filtering and aggregation that restricts the set of origins and destinations to a manageable amount. Obviously, in printed or public displays such interaction is unavailable, yet even with interaction each individual view should, ideally, be as informative and unambiguous  as possible  with respect to the underlying data~\citep{Nguyen:2013et}. Thus, designs for OD flow maps that are as readable as possible on static displays are still desirable.  Interactions built on top of such visualisation can then extend the scalability of such ``base-line'' views.

\textbf{Other Visual Encodings}\\
Other visual encodings have been explored as well. \added{For example, direction information is important in some uses of flow maps. Fekete \textit{et al.} used asymmetric curves in Treemaps to indicate directed links between nodes~\citep{Fekete:2003wya}, for example the curve of an arrow was more pronounced towards the end of the link, giving it the appearance of a ``ballistic'' trajectory falling more steeply towards the target.} Holten \textit{et al.}~\citep{Holten:2011fp,Holten:2009eq} conducted a series of user studies to compare different directed-edge representations in respect to graph visualisation. They found tapered and animated compressed lines were overall the best techniques. Jenny \textit{et al.}~\citep{Jenny:2017ci} tested the similar styles of link rendering in flow maps and found arrows indicatating direction are more effectively than tapered line widths. They discussed that the different results could be due to long paths in geographic contexts where tapered lines perform poorly with~\citep{Netzel:2014ca}. Koylu \textit{et al.} compared five flow line symbolizations regarding both presenting the direction and the magnitude. They recommend that design choices should be guided by task taxonomy. Their results provide some support for using monotone arrowhead and fading arrowhead in 2D flow maps.

\textbf{Summary}\\
Design strategies of 2D flow maps have long been discussed in popular visual design and cartography texts, e.g.~\citep{bertin:1967,Dent:2008vd,slocum2009thematic}. Additionally, with the some of the research discussed in this section, Jenny \textit{et al.}~\citep{Jenny:2017ci,Stephen:2017jb} compiled the following design principles to declutter 2D flow maps: curving flows, minimising overlap among flows~\citep{Purchase:1996} and between flows and nodes~\citep{Wong:2003es}, avoiding acute angle crossings~\citep{Huang:2014dg}, radially distributing flows~\citep{Huang:2007ir}, and stacking small flows on top of large flows~\citep{Dent:2008vd}. 

\subsection{Matrix-Based Approaches}
\label{sec:related:od-map}
Adjacency matrix representation of flow networks are called OD matrices. These present flow using a table where rows and columns represent origin and destination locations. Each cell indicates the quantity of movement from one location to another. The original OD matrix dates from 1955~\citep{Voorhees:2013cx}. More generally, adjacency matrices have long been useful for presenting a network of relationships in a compact and structured format where reordering of rows or columns can reveal patterns~\citep{bertin:1967}. A user study by Ghoniem \textit{et al.}~\citep{Ghoniem:2004jk} found that adjacency matrices perform better than node-link diagrams for quickly reading adjacencies. Whilst the original OD matrices were purely numerical, colour shading using a heatmap approach~\citep{Wilkinson:2009et} is often used to encode the size of the flow.

One major drawback of the classic OD matrix is that it lacks a mapping from OD locations to geographical positions. The identification of geographically related rows, columns or cells can be difficult and so spatial patterns in the dataset can be hard to determine~\citep{Wood:2010be}. Marble \textit{et al.}~\citep{marble1997recent} attempt to preserve the spatial properties of the OD locations by reordering columns and rows by approximate spatial position but only limited spatial information is retained due to dimension reduction. 

\begin{figure}[b!]
\centering
    \includegraphics[width=0.98\columnwidth]{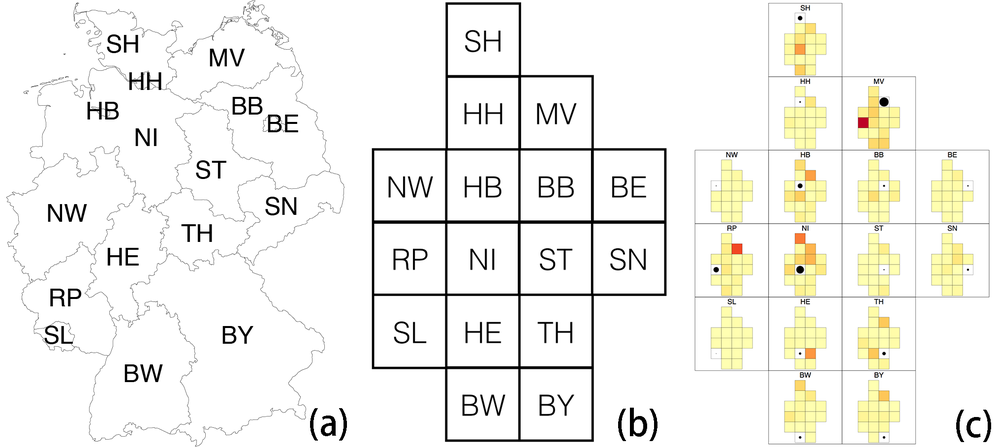}
    \caption{Demonstrating the OD Maps design for Germany showing abbreviations for all 16 states (a) standard map with administrative boundaries; (b) empty OD Maps layout; and (c) nested OD Maps coloured by example flow data.}
    \label{fig:related-ODmapLayout} 
\end{figure}

OD Maps~\citep{Wood:2010be} attempt to overcome this limitation through a nested small-multiples design (see Fig.~\ref{fig:related-ODmapLayout}). They provide schematic geographical information by dividing the canvas into a regular grid based on the actual geographical locations on the map using a spatial treemap structure~\citep{Wood:2008gm}. A second level of spatial treemaps is embedded within the first to present the OD information using colour shading. To aid readability, some cells of the grid may be left blank to indicate the outline shape of the country, e.g. Fig.~\ref{fig:related-ODmapLayout}(b). Like the OD matrix, spatial locations in OD Maps are presented as squares. As all locations have similarly sized cells, this allows data for small, highly populated areas to be seen at the same level of detail as more sparsely populated larger regions, e.g. in Fig.~\ref{fig:related-ODmapLayout} compare Berlin (BE) to Brandenburg (BB). \added{A key step in creating OD Maps is to transform the geographic map to a tile map (also known as grid map), McNeill and Hale developed a system to generate multiple tile maps for the specific geographic map and rank them according to a adjustable cost function~\citep{McNeill:2017gp}.}

Whilst spatial treemaps are currently being tested for their performance in a number of tasks~\citep{Ali:2013tg}, OD Maps had not yet been evaluated in a quantitative user-study prior to the work presented in this thesis. Less formal studies show they are useful for presenting geographical commodity flows to data experts~\citep{Kelly:2013wm,Wood:2011ez}, but OD Maps have not yet been tested on a wider audience or compared with other visualisations prior to the work presented in this thesis.

Flowstrates~\citep{Boyandin:2011fa} connects a temporal heatmap with two maps presenting the geographical locations of origin and destination and shows how flow changes over time. The resulting visual representation is superficially similar to the MapTrix visualisation presented in Section~\ref{sec:maptrix:maptrix}. However, while Flowstrates does present OD data it is not designed to present a complete OD matrix as we do in MapTrix. In Flowstrates each row corresponds to a single flow as it uses columns for the temporal scale. In contrast, a single MapTrix cell corresponds to a single flow. Thus to show all flows between $M$ sources and $N$ destinations Flowstrates requires $M \times N$ leader lines but MapTrix only $M + N$ leader lines. Furthermore, it is possible to avoid leader line crossings with MapTrix but not with Flowstrates for larger multi-way flows.

%% file: content/2-related-work/3D.tex
\section{In 3D Display Space}
\label{sec:related:3d}
D\"ubel \emph{et al.}~\cite{Dubel:2015el} categorise geovisualisations based on whether the reference space (i.\,e.\ the map or surface) is shown in 2D or 3D and whether the abstract attribute is shown in 2D or 3D. In the case of flow maps, this categorisation implies the design space has two orthogonal components: the representation of geographic region and the representation of flow. 

\begin{figure}[b!]
\centering
    \includegraphics[width=0.8\columnwidth]{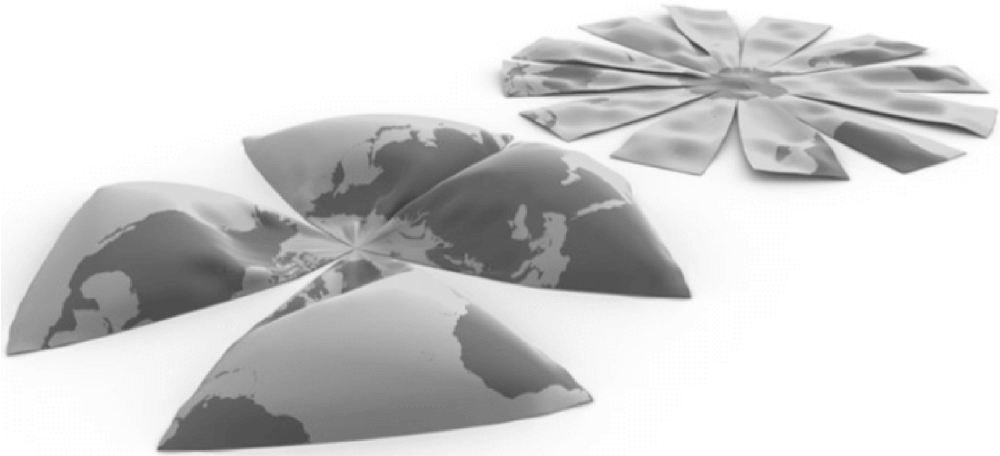}
    \caption{The earth is impossible to unfold onto a planar surface without distortion. Figure from Fig.~9.1 in~\citep{Jenny2017mmm}. © 2017 Springer Nature.}
    \label{fig:related-map-distortion} 
\end{figure}
\begin{figure}[b!]
\centering
    \includegraphics[width=0.8\columnwidth]{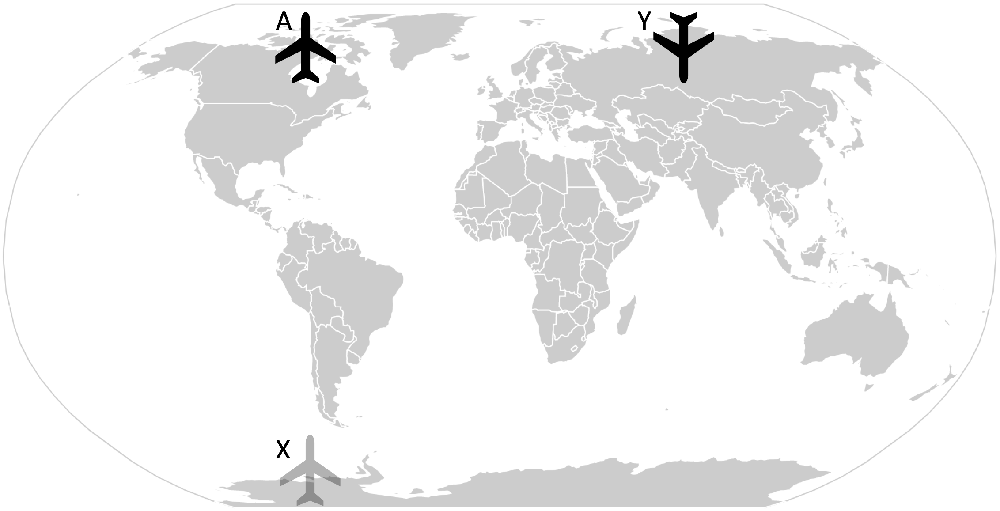}
    \caption{An example demonstrating possible misinterpretation of a trajectory that crosses the edge of a world map:  a plane at A following its flying direction. After crossing the edge, people may think it will reach X, while it actually will get to Y.}
    \label{fig:related-map-edge} 
\end{figure}

\subsection{Representation of Geographic Reference Space}
In 2D flow maps the geographic reference space is always presented as an underlying flat map. However, there are two well-known problems with using flat maps to represent geographic regions:
\begin{itemize}
	\item Distortion of map projections (see Fig.~\ref{fig:related-map-distortion}): a map projection transforms a sphere to a flat surface~\citep{Snyder:1987tk}. Hundreds of map projections have been devised~\citep{snyder1997flattening}, however, all map projections either distort angles or the relative size of areas. No projection can preserve the distances between all locations on a map and most only preserve distance along very few, carefully selected lines. Cartographers consider the choice of projection based on the area displayed and the purpose of the map (i.e. how important is it to accurately preserve distance, angle, shape etc.)~\citep{Jenny2017mmm,Savric:2016do}.

	\item Edges of world maps~\citep{Hennerdal:2015do,Ayala:2016eu,Hruby:2018fn}; the shape of the earth is close to a sphere which means the surface is continuous in any direction, however, the edges of flat world maps inevitably ``cut'' certain regions into parts. Moreover, such discontinuity can mislead viewers while (for example) observing a trajectory that crosses the edge (see Fig.~\ref{fig:related-map-edge}). \added{Please note that this issue primarily exists when presenting global geography. For visualising data on a relatively smaller geographic extent, for example Victoria, the edge of the map is less problematic.}
\end{itemize}

In 3D display space, it makes sense to investigate the 3D exocentric globe representation, where the user’s viewpoint is outside the globe. These virtual globes have become familiar to most VR users, e.g.~\citep{GoogleEarthVR}.
In addition to the familiar flat map and globe representation, Zhang \textit{et al.} also explored an egocentric globe representation with the viewpoint inside a large globe in VR~\citep{Zhang:2016,Zhang:2018jo}. 

A key question, however, is whether 3D exocentric globes are the best way to present a geographic reference space or whether maps or some other visualisation may be better. Surprisingly, given the fundamental importance of this question for the design of geovisualisation applications, to the best of our knowledge, prior to the work presented in this thesis, no user study has been conducted. This thesis fills the gap in Chapter~\ref{chapter:maps-globes-vr}.

\subsection{3D OD Flow Maps}
\begin{figure}[b!]
\centering
    \includegraphics[width=0.55\columnwidth]{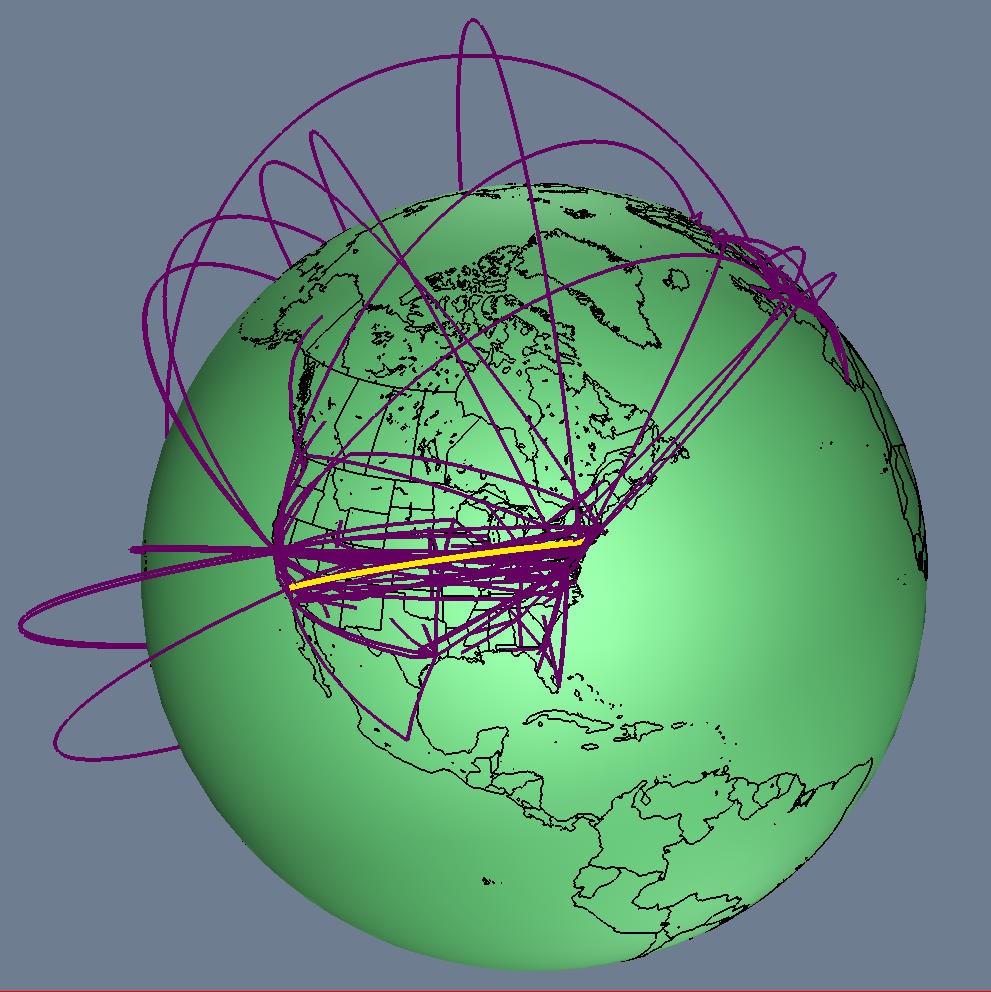}
    \caption{A 3D interactive map of the MBone tunnel structure by Munzner \textit{et al.}~\citep{Munzner:1996co}. © 1996 IEEE.}
    \label{fig:related-3d} 
\end{figure}
The design space of 3D OD flow visualisation has mainly focused on flow map approaches. Lifting origin-destination flows into the third dimension is not a new idea. Early examples of flow maps drawn with three-dimensional arcs elevated above 2D reference maps and 3D globes were introduced more than 20 years ago~\citep{Cox:1995fp,Cox:1996gv,Munzner:1996co} (an example is demonstrated in Fig.~\ref{fig:related-3d}). The explicit goal of these early maps was to increase readability by untangling flows.

The height of flows above the reference map or globe can vary with the total volume of flows~\citep{Eick96aspectsof}, the distance between endpoints~\citep{Cox:1995fp}, the inverse of the distance~\citep{Vrotsou:2017im}, time \cite{hagerstrand1970people} or any other attribute~\citep{Vrotsou:2017im}. 
Discussion of such 3D ``geovirtual'' environments typically focuses on interactive filtering to deal with clutter (e.g.~\citep{buschmann2012challenges}).
We are not aware of any study evaluating the effectiveness of these different 3D encodings prior to the work presented in this thesis. 

%% file: content/2-related-work/immersive.tex
\vspace{-0.5em}
\section{Immersive Analytics}
\label{sec:related:immersive}
Immersive Analytics is the use of engaging, embodied analysis tools to support data understanding and decision making~\citep{Chandler:2015eb,Marriott:2018ig}. With the arrival of commodity head-mounted displays (HMDs) for VR, e.g. HTC Vive, Oculus Rift, and AR, e.g. Microsoft Hololens, Meta2 and Magic Leap, there is growing interest in how to better design visualisations in immersive environments~\citep{Bach:2016dd,Chandler:2015eb,dwyer2016immersive,Marriott:2018ig}. While there is considerable caution about the use of 3D in abstract data visualisation, e.g.~\citep[Chap. 6]{Munzner:2014wj} (details in Section~\ref{sec:intro:immersive}), there is also a realisation that data visualisation in immersive environments will be increasingly common because of the growing flexibility that MR HMDs offer over traditional desktop environments and potential benefits:
\begin{itemize}[leftmargin=1em]
	\item Display data with an additional dimension.
\end{itemize}
\vspace{-0.5em}
\hangindent=2em \qquad There seems considerable potential to use the third dimension for visualisation of geographically-embedded data~\citep{Dubel:2015el,Wood:2005fl}. This is because the geographic reference space typically takes up two-dimensions. Adding a third dimension offers the possibility of an additional spatial encoding for data attributes. Indeed common geographic representations such as space-time cubes or prism maps routinely use a third dimension even though they are to be displayed on standard desktop displays.  Some Evaluations of such 3D geographic visualisation on flat displays has yielded positive results (e.g.~\citep{Kristensson:2009cj}) while some has not (e.g.~\citep{kaya20143d}). However, there has been little research into the effectiveness of such visualisation using modern head-tracked binocular HMDs.

\begin{itemize}[leftmargin=1em]
\item Exploit spatial awareness and visual working memory.
\end{itemize}
\vspace{-0.5em}
\hangindent=2em \qquad The construction of immersive environments enables us to exploit our spatial awareness and visual working memory~\citep{Wood:2005fl}. An immersive mediums bring a much larger display space than the traditional 2D medium. Although only part of the space is available in the Field of View (FOV) at one time, our spatial awareness and visual working memory can possibly help us navigate in such environments, e.g. people can remember the existence and spatial position of an object for some time after they face away from it.

\begin{figure}[b!]
    \centering
    \begin{subfigure}{\textwidth}
         \centering
        \includegraphics[width=0.85\textwidth]{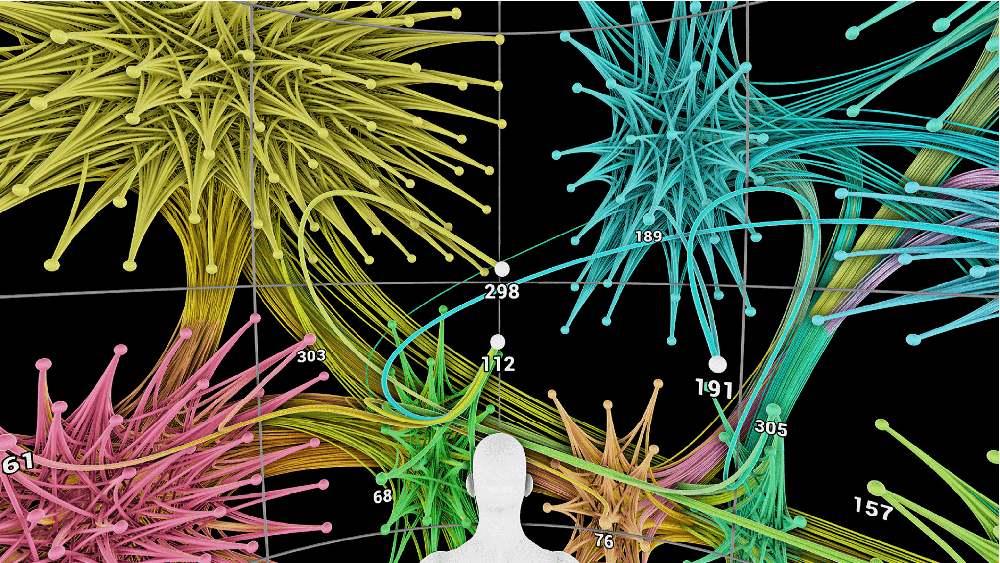}
        \caption{A graph visualisation in an immersive environment. Figure from Fig. 1 in~\citep{Kwon:2016go}, © IEEE.}
    \end{subfigure}
    \centering
    \begin{subfigure}{\textwidth}
         \centering
        \includegraphics[width=0.85\textwidth]{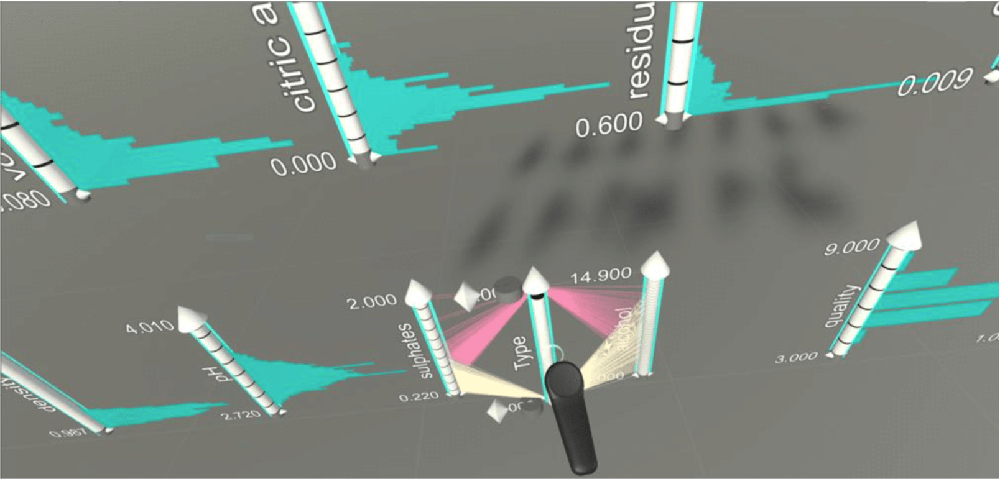}
        \caption{ImAxes: a multivariate data visualisation in immersive environments. Figure from Fig. 12 in~\citep{Cordeil:2017hy}, © ACM.}
    \end{subfigure}
    \caption{Examples of immersive visualisations.}
    \label{fig:related:ia}
\end{figure}

\begin{itemize}[leftmargin=1em]
\item Interact with 3D visualisations more intuitively. 
\end{itemize}
\vspace{-0.5em}
\hangindent=2em \qquad It has been identified that manipulating (e.g. moving and rotating) 3D visualisations in 2D display space can make the viewer easily lose context~\citep{Cordeil:2013up}. Thus, the viewer has to invest extra effort to build a mental model of the 3D visualisation when it is portrayed via a flat screen. Taking advantage of space-tracking in immersive environments, the motion parallax in the immersive environment is the same as in the physical environment. As a result, manipulation of 3D objects can be well simulated in immersive environments to lower the mental load of the viewer~\citep{Cordeil:2017dz}. Meanwhile, the tracked head movement provides an intuitive way to change viewpoints, which helps the viewer to navigate in the immersive environments as well as facilitate the process of finding a good viewpoint to avoid occlusion.

\begin{itemize}[leftmargin=1em]
\item Facilitate collaborations. 
\end{itemize}
\vspace{-0.5em}
\hangindent=2em \qquad For collaborative scenarios, immersive environments allow two or more people to see and to interact with the visualisations in a large (virtual in VR or mixed-reality in AR) space, while still seeing each other directly (or through avatars) for communication~\citep{Cordeil:2016io}. And even when users are distributed in different physical locations~\citep{Beck:2013hp}.

\begin{itemize}[leftmargin=1em]
\item Enable Situated Analytics.
\end{itemize}
\vspace{-0.5em}
\hangindent=2em \qquad Immersive technologies, especially AR, are suitable for applications that require interactive analytical reasoning embedded in the physical environment~\citep{elsayed2016situated} such as in the field, surgery or factory floor.

\vspace{-0.5em}
With these potential benefits, visualisations in immersive environments have been explored in regard to 3D graph layout e.g.~\citep{Cordeil:2017dz,YiJhengHuang:2017iga,Kwon:2016go,Ware:2005boa} and multivariate data visualisation e.g.~\citep{Butscher:2018bw,Cordeil:2017hy,wagner2018immersive,WagnerFilho:2018is} (see Fig.~\ref{fig:related:ia}). Experiments showed a much greater benefit for 3D viewing in immersive environments than using traditional desktops for some visual analytic tasks~\citep{Kwon:2016go,Ware:2005boa,WagnerFilho:2018is}.
However, to the best of our knowledge, there is no previous research exploring the design space of visualising OD flow data in immersive environments.

%% file: content/2-related-work/conclusion.tex
\section{Conclusion}
\label{sec:related:conclusion}
The design space of 2D OD flow visualisations has been explored for flow map approaches and matrix-based approaches. However, each type of approach has its limitations. It makes sense to think about designing a new method to obtain the benefits of both.  Furthermore, we are not aware of any research evaluating the effectiveness of these different 2D OD flow visualisations prior to the work presented in this thesis. In Chapter~\ref{chapter:od-flow-maps-2d} and Chapter~\ref{chapter:evaluating-2d-od-flow-maps}, we will present our novel design \emph{MapTrix} and describe our evaluation comparing \emph{MapTrix} to two other state-of-the-art 2D OD flow visualisations.

Some aspects of the design space of 3D OD flow maps have been explored. However, these studies were conducted around 20 years ago with limited computation and display resources and in many cases only using 2D screens without head tracking or other parallaxes. Immersive Analytics demonstrates strong potential for improving some visual representations and interactions. It is necessary to revisit these early 3D OD flow maps in immersive environments, to see whether these 3D visualisations  can gain extra benefits from new display and interaction technologies than was observed in these early studies.

%% file: content/4-maptrix/0-index.tex
\chapter{2D OD flow visualisations: designs}
\label{chapter:od-flow-maps-2d}

\begin{figure}[ht!]
    \captionsetup[subfigure]{justification=centering}
    \centering
    \begin{subfigure}{0.28\textwidth}
    	\label{subfig:arrow-au}
        \includegraphics[width=\textwidth]{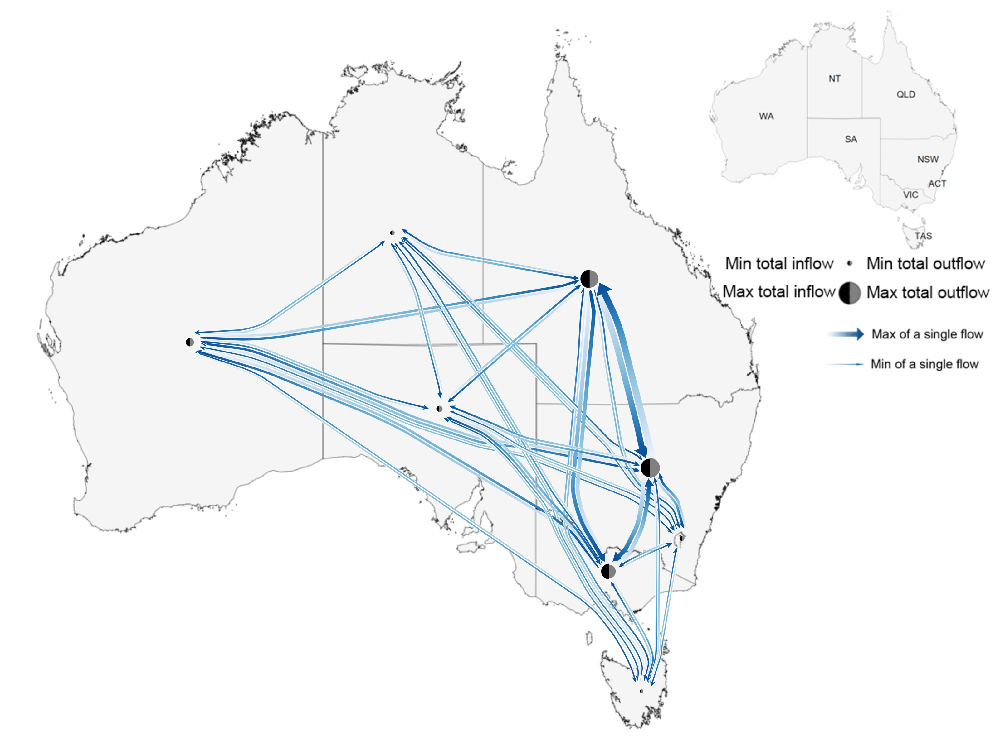}
        \caption{Bundled Flow Map}
    \end{subfigure}
    \begin{subfigure}{0.33\textwidth}
    	\label{subfig:maptrix-od-map-au}
        \includegraphics[width=\textwidth]{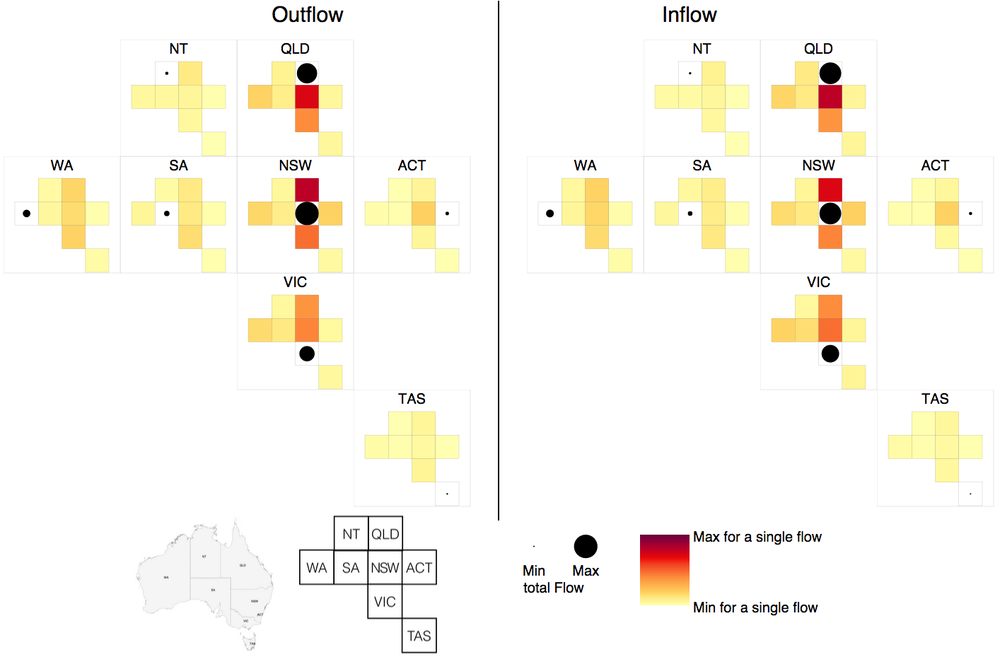}
        \caption{OD Maps}
    \end{subfigure}
    \begin{subfigure}{0.36\textwidth}
        \includegraphics[width=\textwidth]{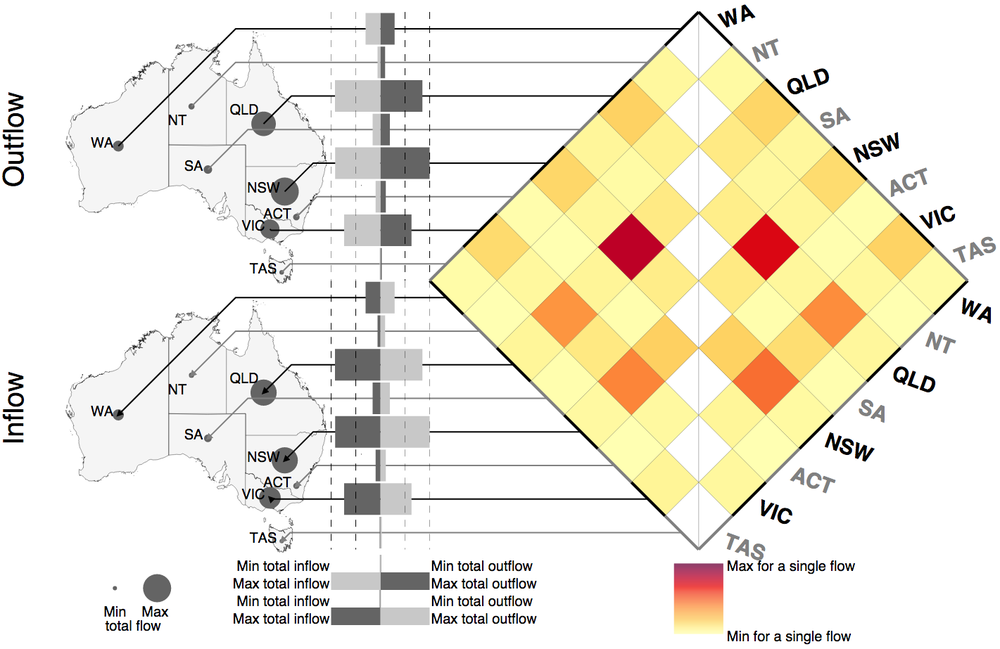}
        \caption{MapTrix}
        \label{subfig:maptrix-au}
    \end{subfigure}
    \caption{The three 2D visualisation methods discussed in this chapter.}
    \label{fig:maptrix-teaser}
\end{figure}
In the last chapter, more specifically Section~\ref{sec:related:2d}, we introduced two main approaches of 2D OD flow visualisations: flow map approaches and matrix-based approaches. 

In this chapter, we first introduce \emph{bundled flow map} (see Fig.~\ref{fig:maptrix-teaser}(a)) and \emph{OD Maps} (see Fig.~\ref{fig:maptrix-teaser}(b)) as the representatives of flow map approaches and matrix-based approaches respectively. We then demonstrate our iterative design process to create a novel visualisation MapTrix (see Fig.~\ref{fig:maptrix-teaser}(c)) including an improved algorithm for leader line placement. MapTrix aims to combine the benefits of both flow map approaches and matrix-based approaches.

The work presented in Chapter~\ref{chapter:od-flow-maps-2d} and Chapter~\ref{chapter:evaluating-2d-od-flow-maps} is a collaborative effort with my supervisors: Tim Dwyer, Sarah Goodwin and Kim Marriott. This work was accepted and presented at IEEE Conference on Information Visualization (InfoVis) 2016 and received a Best Paper Honourable Mention award. In this work, I collaboratively formulated the hypothesis and research questions, was solely responsible for the implementation and user studies, did the data analysis, and collaboratively wrote the research paper.

\input{content/4-maptrix/1-edge-bundling}
\input{content/4-maptrix/1-od-map} 
\input{content/4-maptrix/1-maptrix} 
\input{content/4-maptrix/1-conclusion}

\chapter{2D OD flow visualisations: an evaluation}
\label{chapter:evaluating-2d-od-flow-maps}
In the last chapter we discussed two state-of-the-art OD flow visualisations as well as proposing our novel hybrid visualisation: MapTrix.

In this chapter, we report, to the best of our knowledge, the first quantitative user studies to compare different (static) visual representations of dense many-to-many OD flows. These user studies enable us to systematically compare the effectiveness of different 2D OD flow visualisations.

\input{content/4-maptrix/2-introduction}

\input{content/4-maptrix/2-first-study}
\input{content/4-maptrix/2-redesign}
\input{content/4-maptrix/2-second-study}

\input{content/4-maptrix/3-interaction}

\input{content/4-maptrix/4-conclusion}

%% file: content/4-maptrix/1-edge-bundling.tex
\section{Bundled flow map}
\label{sec:maptrix:flow-map}

\begin{figure}[t!]
\includegraphics[width=0.26\columnwidth]{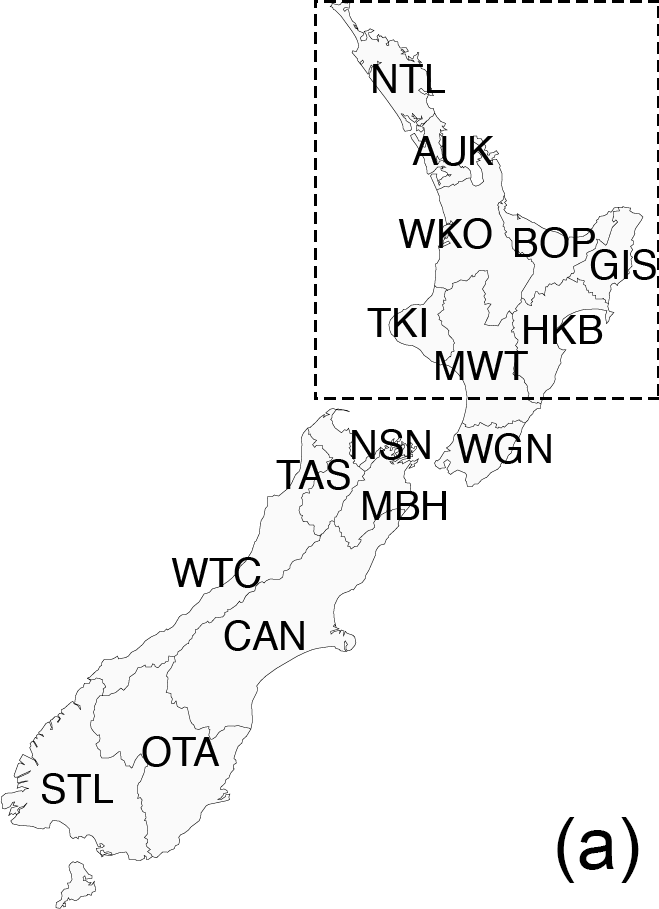}
\includegraphics[width=0.36\columnwidth]{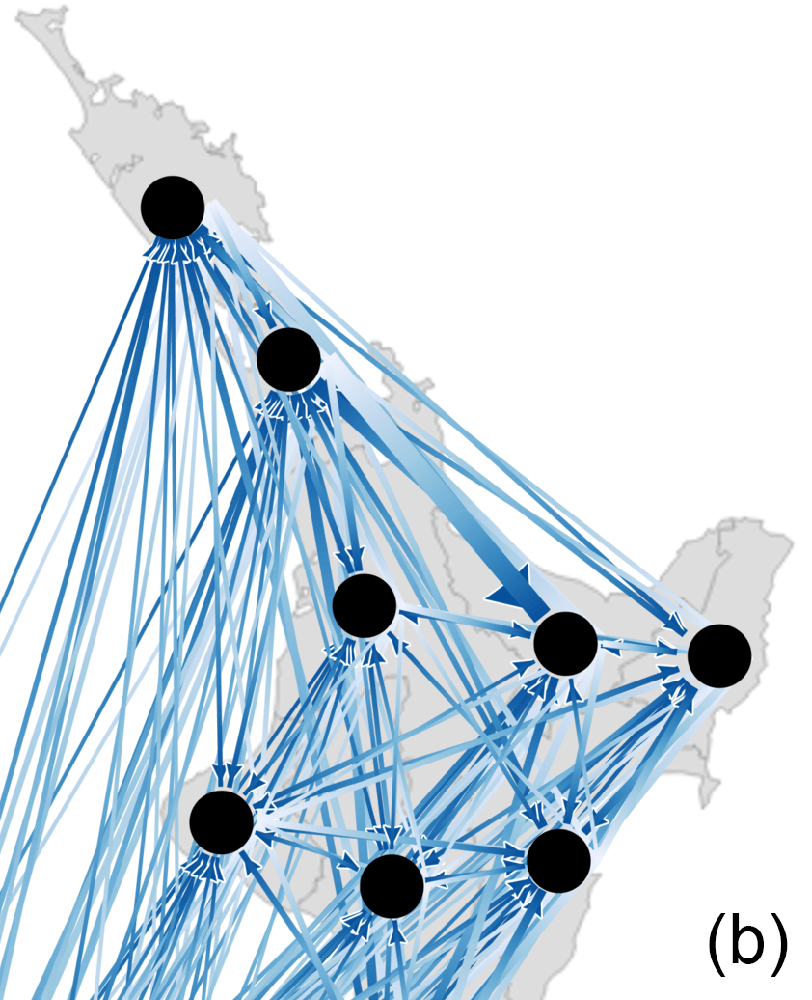}
\includegraphics[width=0.36\columnwidth]{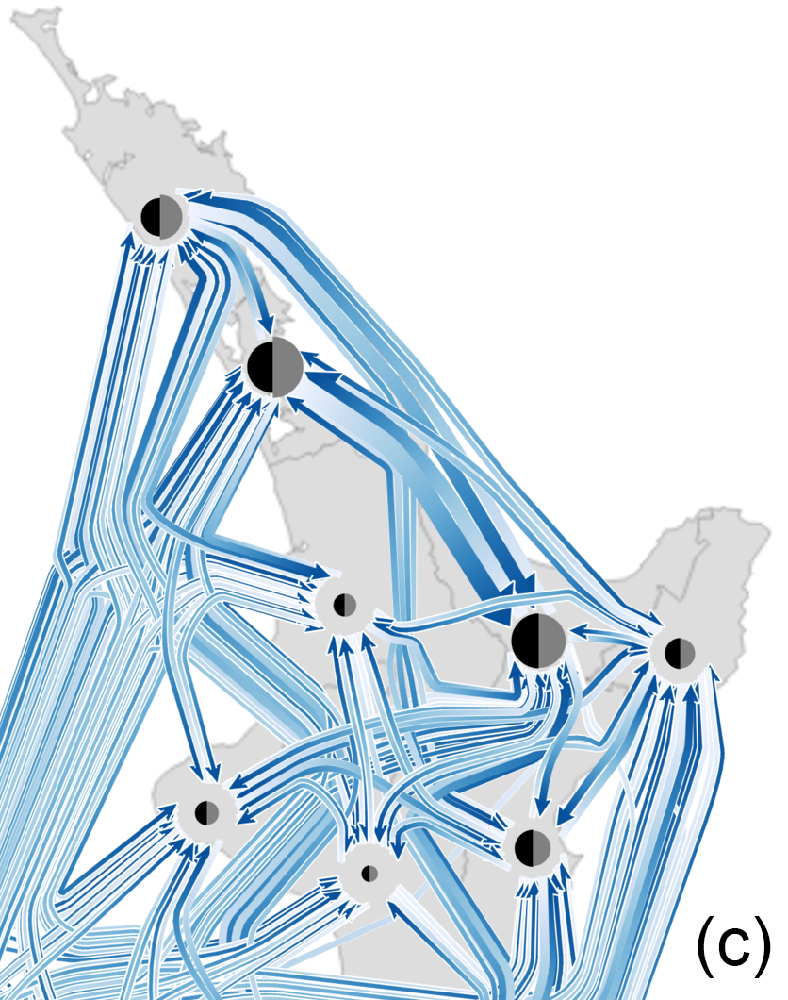}
\caption{Flow map design decisions. Arrow thickness indicates magnitude of flow. Arrow head and colour gradient shows direction. (a) Zoom into a sub region of NZ to have a detailed view; (b) Straight lines with full black circles for locations; and (c) Bundled lines with half circles for locations and magnitude of total in (black) and out (grey) flow.} 
 	\label{fig:maptrix-bundling} 
\end{figure}

Flow maps present origins and destinations on a map connected by lines or arrows. Whilst flow maps are intuitive and are well suited to show the flows from a single source, they quickly become cluttered and difficult to read when the number of commodity sources increases.

In order to study denser flow maps, we searched for a flow map design solution which could minimise data occlusion by reducing overlap without removing individual flows such as through flow aggregation. There are a couple of recent edge bundling methods that are able to neatly offset individual edges within bundles~\citep{Bouts:2015ip,Pupyrev:2011bt}. We adopt the method by Pupyrev \textit{et al.} \citep{Pupyrev:2011bt} which groups edges on shared paths that are centred between obstacles. It then neatly offsets the curves so that all are visible and uses a heuristic to minimise crossings as lines join and leave the bundles. To demonstrate the reduction of line overlap Fig.~\ref{fig:maptrix-bundling} shows straight and bundled arrows.

We investigated the use of colour together with arrow size to indicate magnitude of flow. We found that due to line occlusion around arrowheads the flow direction was often difficult to determine. Following Holten \textit{et al.}\citep{Holten:2011fp}, we therefore decided to encode line direction using colour gradient. The darker section of the line shows inflow direction, while the lighter section depicts outflow as shown in Fig. \ref{fig:maptrix-bundling}(c). The continuous blues colour scheme from colorbrewer is used \citep{Harrower:2003jm}. The key aspect of using a continuous gradient from source to target is that the directionality of the line can be understood at any part of the line, so the reader does not need to follow the line to find an arrow head. Note that, since we use line width to encode flow magnitude, the tapered line representation advanced by Holten \textit{et al.} would not work in this situation. Furthermore, in the context of OD flow maps a study by Jenny \textit{et al.}\citep{Jenny:2017ci} revealed difficulties interpreting tapered connections in geographic context.

To embed the information of total in/out flows we use proportional circle sizes. We therefore replaced the solid black circle for each location as shown in Fig.~\ref{fig:maptrix-bundling}(b) with two half circles as shown in Fig.~\ref{fig:maptrix-bundling}(c). The left half circle in black indicates total inflow, while the right half circle in grey shows total outflow.

%% file: content/4-maptrix/1-od-map.tex
\section{OD Maps}
\label{sec:maptrix:od-map}

\begin{figure}[b!]
\centering
	\vspace{-1em}
    \includegraphics[width=0.9\columnwidth]{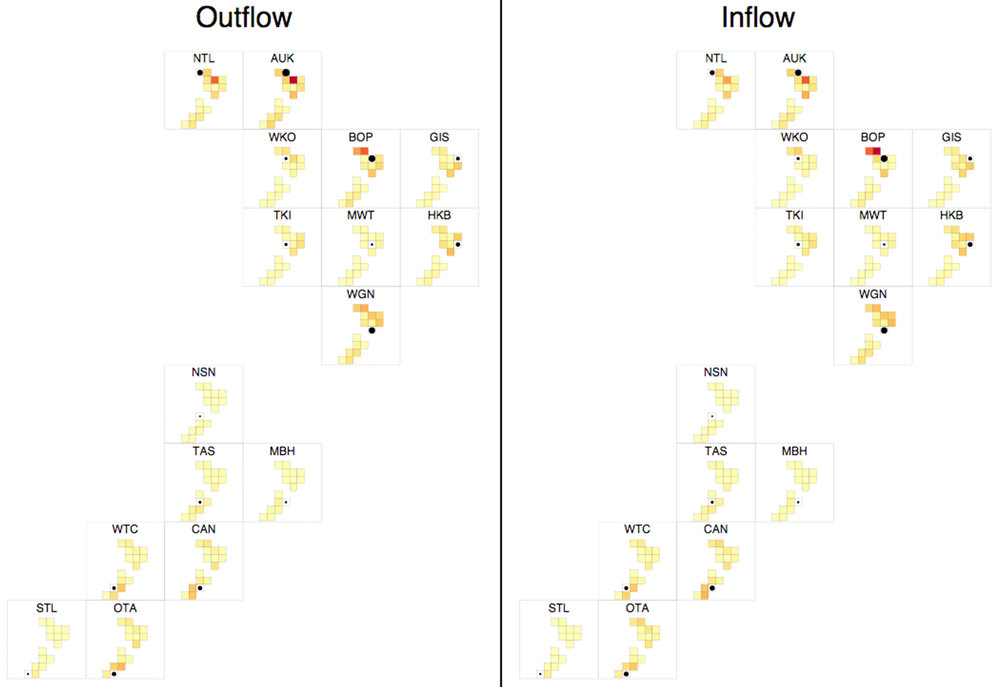}
    \caption{OD map and DO map showing interstate migrations within New Zealand.}
    \label{fig:maptrix-odmap} 
\end{figure}

In Section~\ref{sec:related:od-map}, we discussed OD Maps in detail as it is one of the most important matrix-based approaches to present OD flow data. OD Maps have a better scalability than flow map approaches, however, geographic information in OD Maps has been distorted. In general, we would expect OD Maps to best suit countries with similar width and height such as Ireland, Germany or Australia, where less blank space may be introduced in the process of reorganising the layout.  However, they may be less suitable for countries with elongated proportions such as Japan or New Zealand (see Fig.~\ref{fig:maptrix-odmap}) where \replaced{more blank space is needed to preserve their shapes.}{the distortion of map location to grid location may cause cognitive difficulties.}

OD Maps preserve the geographical aspects of OD matrices without including lines or arrows and introducing occlusion. Having discussed OD Maps implementation with the authors~\citep{Kelly:2013wm,Wood:2010be,Wood:2008gm} we manually created grid layouts for the necessary countries to ensure the grid structure was as intuitive and as similar to the country shape as possible, as shown for Germany in Fig.~\ref{fig:related-ODmapLayout} and New Zealand in Fig.~\ref{fig:maptrix-odmap}. We used the same colour scheme as shown in the MapTrix matrix for the flow data and slightly modified the Wood \textit{et al.} OD Maps design~\citep{Wood:2010be} to include a proportional circle at the associated origin or destination cell of the small multiple to show the total in/out flow for each location. We also show both the OD Maps for outflows and the reverse `DO' map for inflows, to allow for two way comparison. Fig.~\ref{fig:maptrix-odmap} and Fig.~\ref{fig:maptrix-teaser}(b) show the dual OD/DO Map visualisation for Australia and New Zealand.

%% file: content/4-maptrix/1-maptrix.tex
\section{MapTrix}
\label{sec:maptrix:maptrix}
Flow map approaches and matrix-based approaches each have their merits and shortcomings. A representation that can combine the advantages of both should be considered. An example of such a successful hybrid technique for visualisation of regular (non-geographic) networks is NodeTrix which combines node-link diagrams and adjacency matrices to present network/graph data~\citep{Henry:2007er}. Our novel flow visualisation, \textit{MapTrix}, is intended to show quantitative multi-source flow data together with its associated geographical information. It has three main components: an origin map, a destination map, and an OD matrix with a single line connecting each origin and destination to the corresponding matrix row or column.

\subsection{Design}
\begin{figure}[t!]
\centering
    \includegraphics[width=0.5\columnwidth]{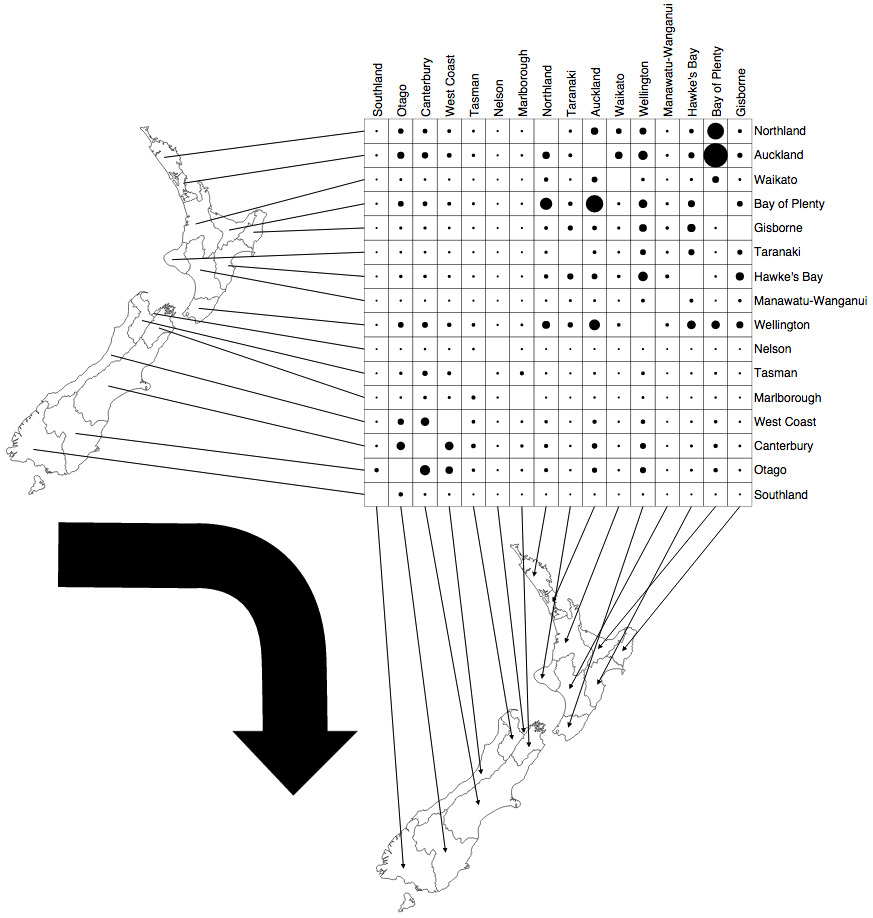}
    \caption{First attemp to link two maps to an OD matrix, rows and columns of matrix are ordered differently.}
    \label{fig:maptrix-first-try} 
\end{figure}

\begin{figure}[t!]
    \centering 
    \begin{subfigure}{0.495\textwidth}
        \includegraphics[width=\textwidth]{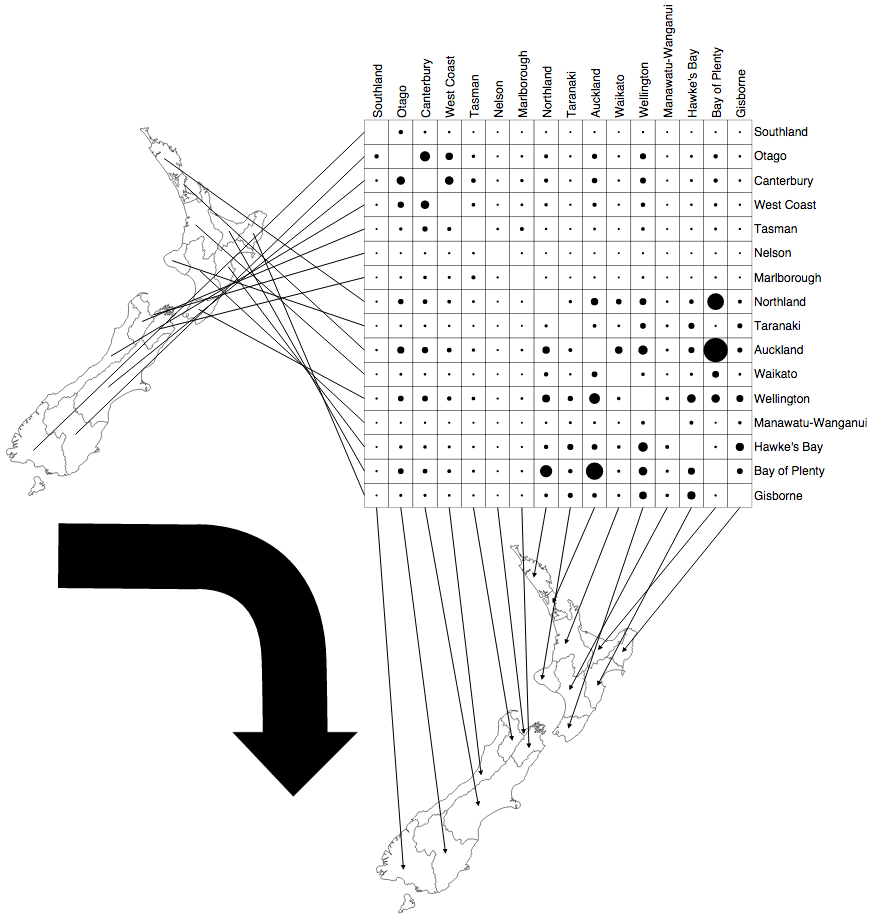}
        \caption{Rows follow the order of columns.}
    \end{subfigure}
    \begin{subfigure}{0.495\textwidth}
        \includegraphics[width=\textwidth]{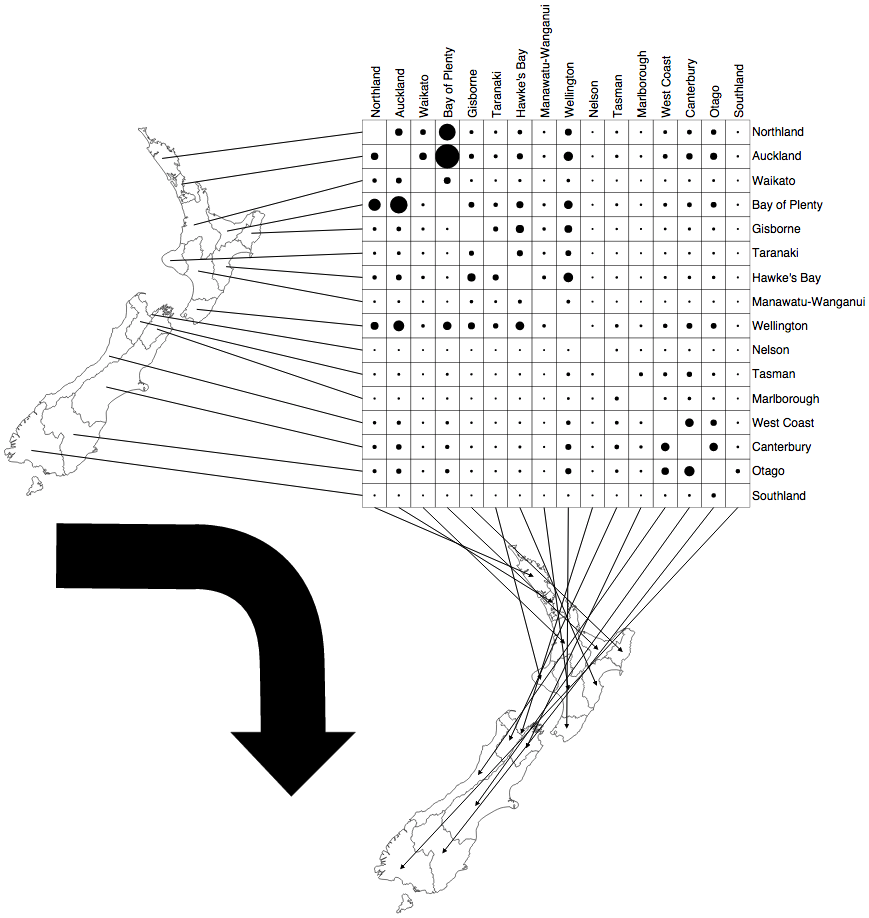}
        \caption{Columns follow the order of rows.}
    \end{subfigure}
    \caption{Attempt to keep the same order for rows and columns.}
    \label{fig:maptrix-order-try}
\end{figure}

Our first attempt to connect the OD matrix to the two maps ordered rows and columns by their map locations’ $y-$ and $x-$coordinate, respectively and used straight line leaders connected map locations to their corresponding matrix row or column (see Fig.~\ref{fig:maptrix-first-try}). Crossings are inevitable in this design, furthermore, the different orders for rows and columns were reported to be confusing by several pilot users. Then, we tested keeping the same order for rows and columns with different ordering strategies. In Fig.~\ref{fig:maptrix-order-try}, we can see that the matrix become easy to read, while more crossings occur among the straight lines.

\begin{figure}[t!]
    \centering
    \begin{subfigure}{0.495\textwidth}
        \includegraphics[width=\textwidth]{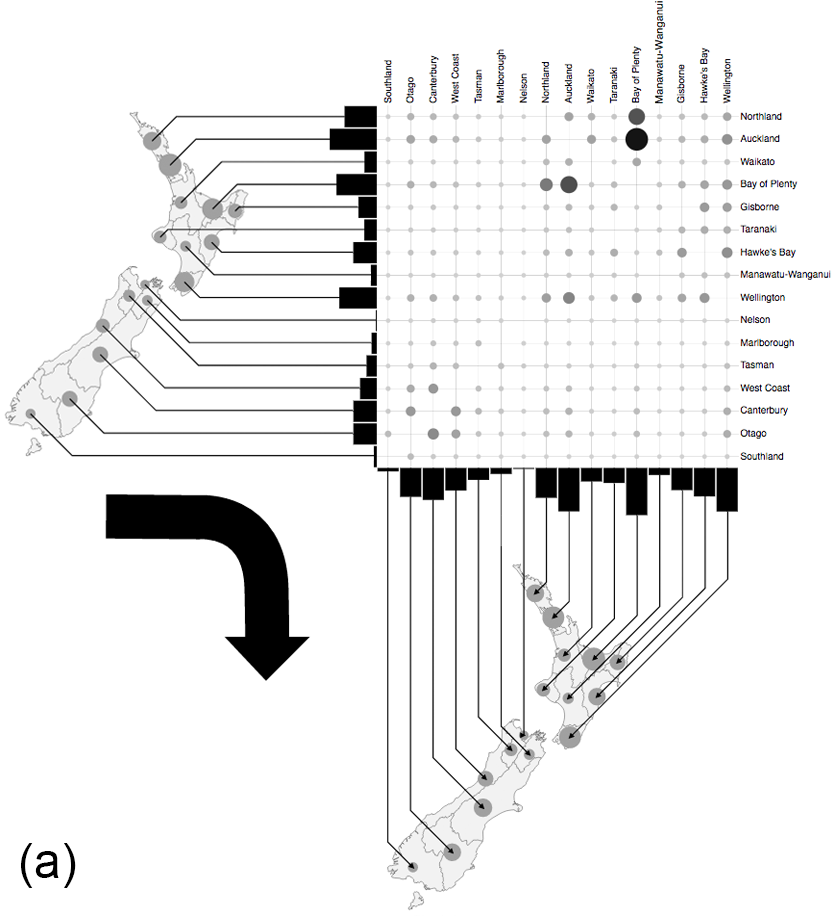}
    \end{subfigure}
    \begin{subfigure}{0.495\textwidth}
        \includegraphics[width=\textwidth]{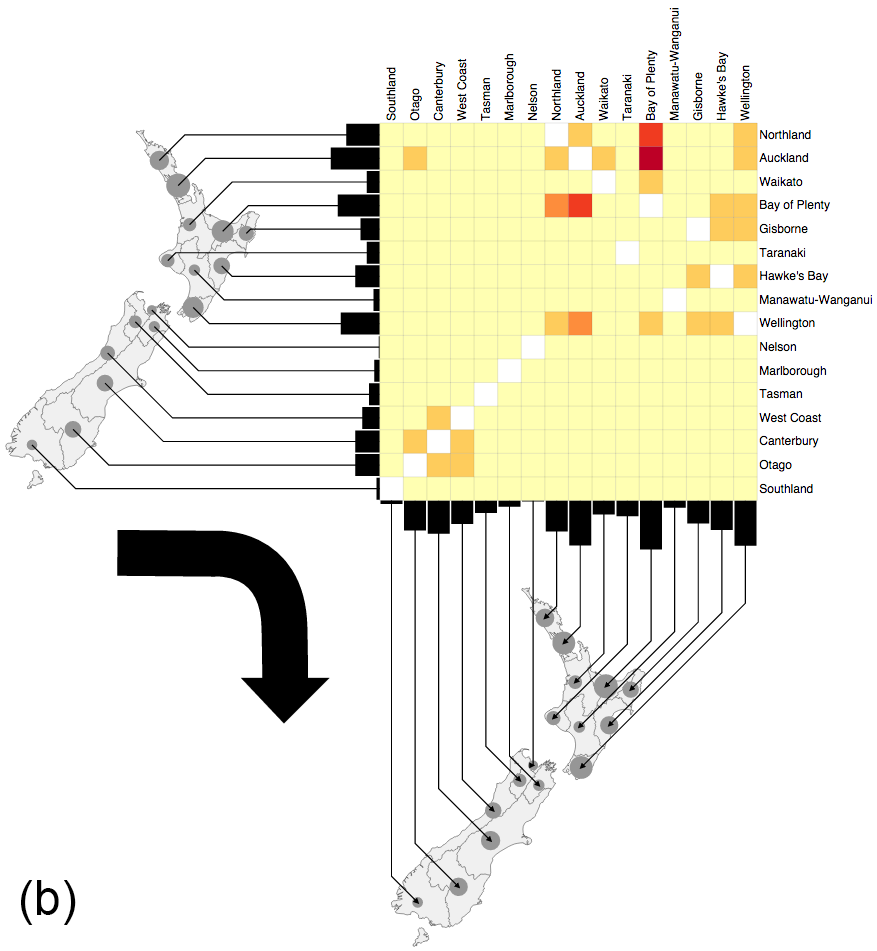}
    \end{subfigure}
    \caption{One-sided boundary labelling model to keep crossing-free. (a) using proportional circles for flow magnitude; (b) using continuous color scale for flow magnitude.}
    \label{fig:maptrix-boundary-labelling}
\end{figure}

Our next attempt led to the design shown in Fig.~\ref{fig:maptrix-boundary-labelling}(a) which ensures that the connection between maps and matrix was \textit{clear, easy to track} and \textit{unambiguous}. To achieve this we solved a so-called \textit{boundary labelling} problem which finds an ordering for the matrix rows and columns that permits leaders to connect map locations without crossings. There are various models for aesthetic boundary labelling for different situations~\citep{Bekos:2009id,Bekos:2007hn,Bekos:2010tp}. Our design uses a \textit{one-sided boundary labelling model} to generate crossing-free connections with a horizontal and a diagonal segment between points in the figure and labels at one side of the figure. We introduce a novel leader adjustment algorithm (next subsection) to more evenly space the leader lines. We also try to use colour shading (“YlOrRd” continuous scale from colorbrewer~\citep{Harrower:2003jm}) to show magnitude of flow between states. Geographical locations’ total in/out flows are indicated by proportional-sized circles in the map, Fig.~\ref{fig:maptrix-boundary-labelling}(b). Choropleth maps were also investigated, but as the scale of the total and single flows could be very different, multiple colour schemes would be needed. In addition to the proportional circles, bar charts were added to help the reader to follow the line (e.g. from large circle to large bar) between map and matrix and to emphasise total in and out flows.

\begin{figure}[t!]
\centering
    \includegraphics[width=\textwidth]{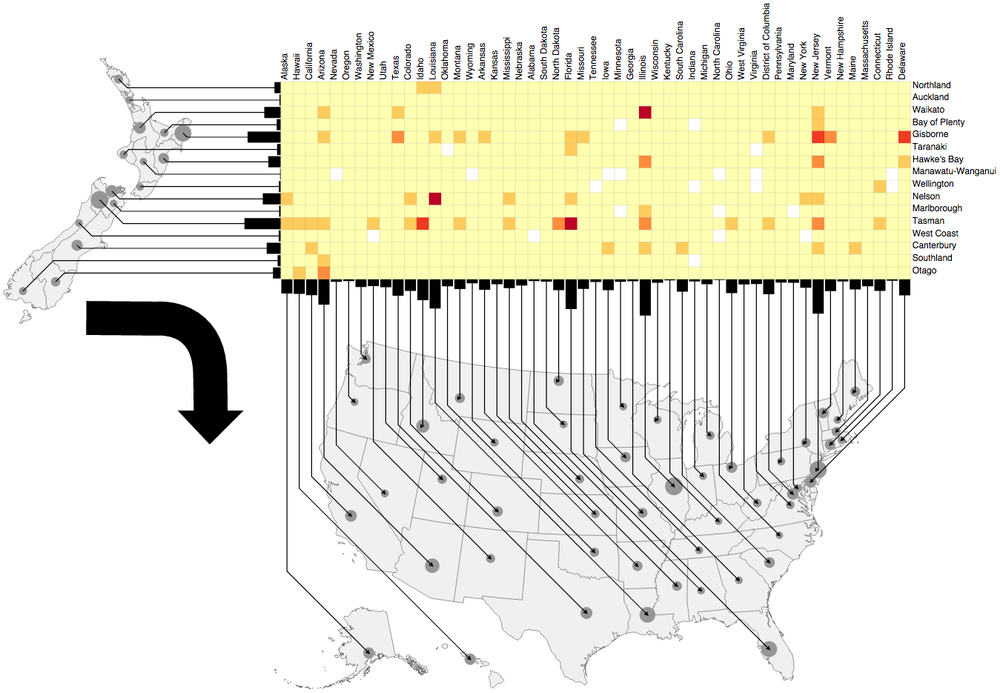}
    \caption{OD flows between two different countries (NZ to USA).}
    \label{fig:maptrix-different-countries} 
\end{figure}

The design is also well suited to showing flow between different countries, as shown in Fig.~\ref{fig:maptrix-different-countries}. However, for showing flow within a single country the asymmetrical ordering of rows and columns in the OD matrix can be confusing. A consistent ordering for the rows and columns is critical for revealing patterns~\citep{Guo:2007gi}.

\begin{figure}[t!]
\centering
    \includegraphics[width=\textwidth]{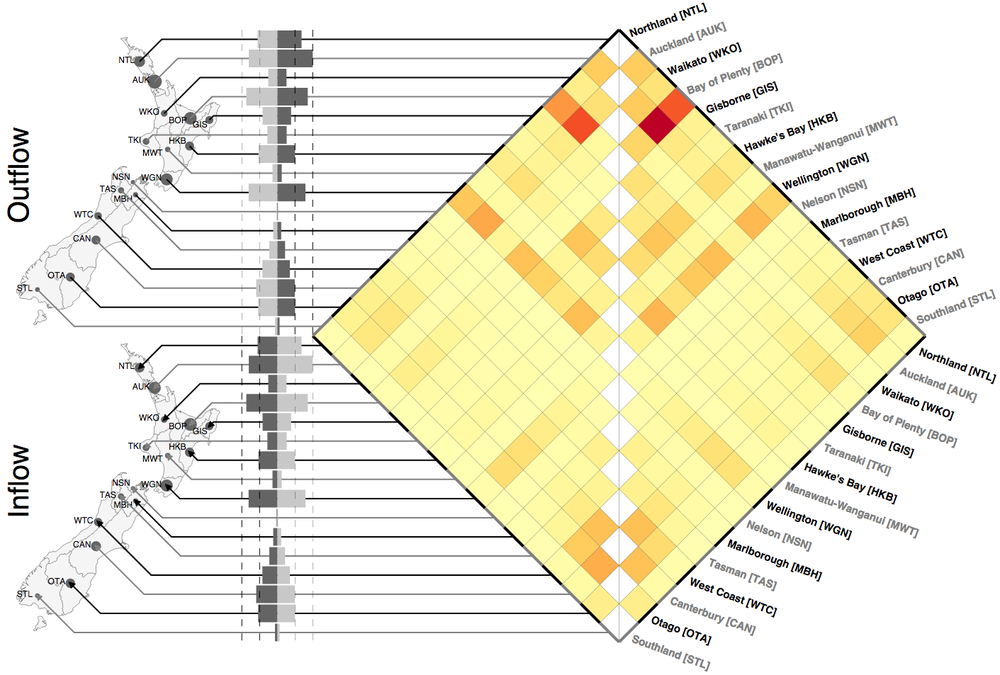}
    \caption{Final design for crossing-free leader lines connecting maps and OD matrix with identical column and row ordering.}
    \label{fig:maptrix-final} 
\end{figure}

\begin{figure}[t!]
\centering
    \includegraphics[width=0.95\textwidth]{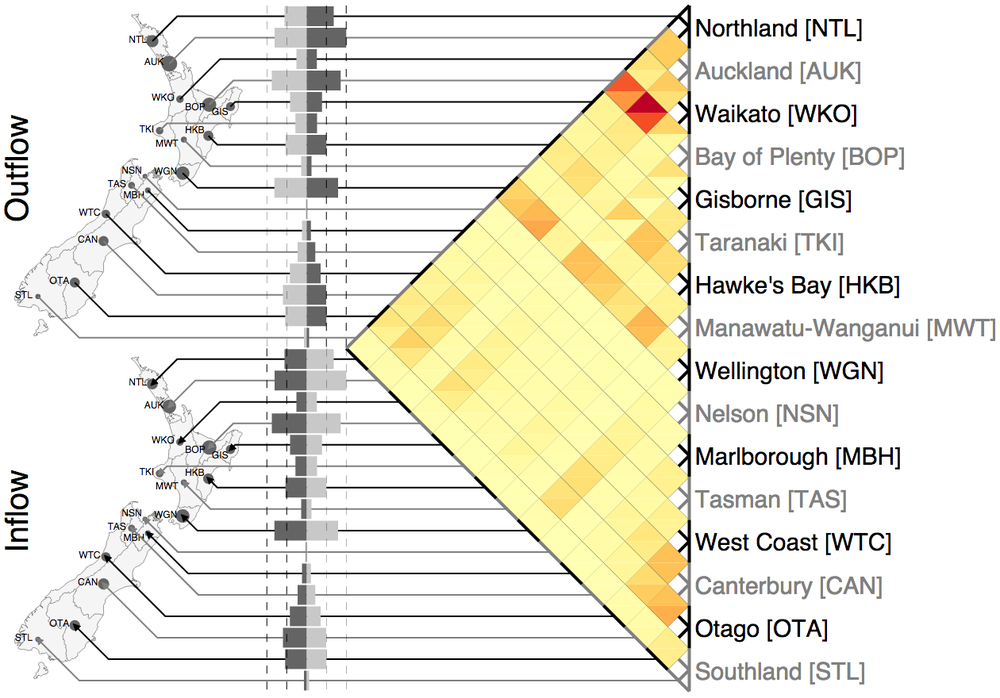}
    \caption{Design variant of MapTrix with half matrix.}
    \label{fig:maptrix-half} 
\end{figure}

Our final design~---~shown in Fig.~\ref{fig:maptrix-final}~---~permits the same ordering to be used for rows and columns by rotating the OD matrix. The destination map is placed under the origin map and the OD matrix is rotated to allow both symmetric ordering and crossing-free leader lines to the maps. An additional advantage of the rotated matrix is that the labels are easier to read. To utilise the additional space and aid leader line connection, the bar charts showing total in/out flow are centred on the leader-lines. We also add total inflow and outflow to both bar charts, differentiated by colour, to allow net flow to be easily determined. Instead of the large arrow, we use the darker bar charts to indicate direction of flow.

In a further adaptation of the design we explored whether we could compress a full OD matrix into a half OD matrix. We investigated this by splitting the matrix cell in half and presenting the flow from OD flow in the upper triangle and the reverse (DO flow) in the lower triangle, as shown in Fig.~\ref{fig:maptrix-half}. Such design (a) saves screen space, (b) makes it easier to compare two reverse flows for an associated location (c) and allows labels to be larger with duplication in rows and columns. However, this means that the space for a single flow reduces to half and the patterns revealed in the full OD matrix may be difficult to locate in the half matrix.

\subsection{Algorithm for Leader Line Placement}
When connecting sites in the maps with rows and columns in the OD matrix we would like:
\begin{enumerate}
	\item connection lines to be crossing-free;
	\item adjacent connection lines to be clearly separated;
	\item clear separation between lines and map locations (\textit{sites}) to avoid ambiguity.
\end{enumerate}

\begin{figure}[b!]
    \captionsetup[subfigure]{justification=centering}
    \centering
    \begin{subfigure}{0.3\textwidth}
        \includegraphics[width=\textwidth]{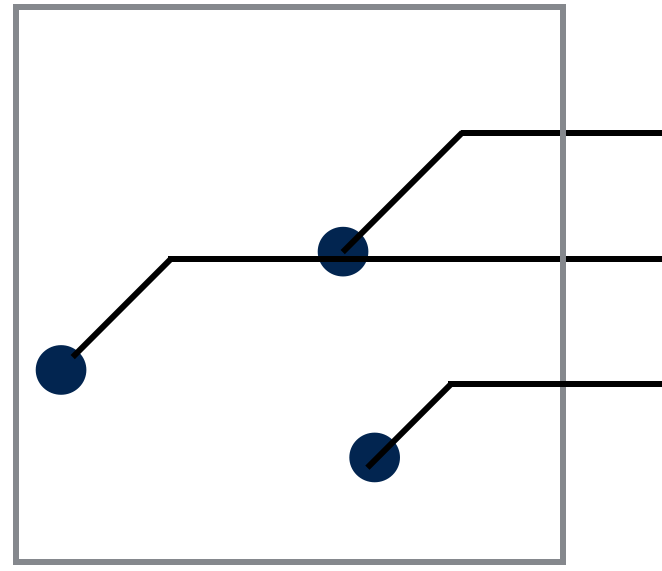}
        \caption{Line through site}
    \end{subfigure}
    \begin{subfigure}{0.3\textwidth}
        \includegraphics[width=\textwidth]{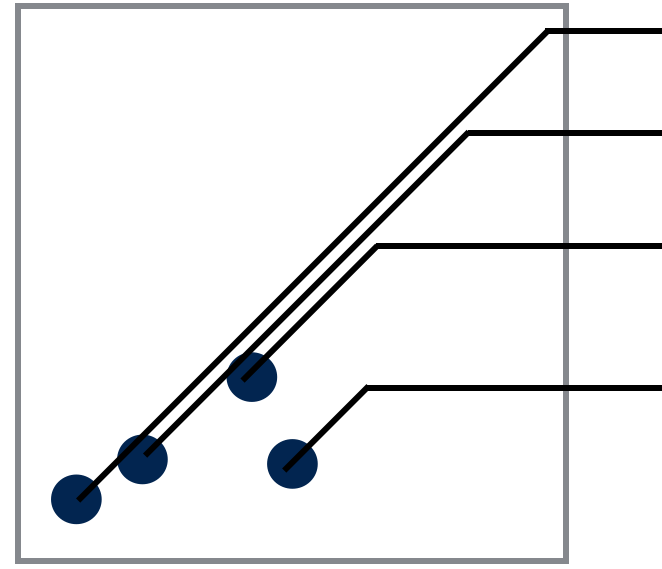}
        \caption{Lines too close}
    \end{subfigure}
    \begin{subfigure}{0.3\textwidth}
        \includegraphics[width=\textwidth]{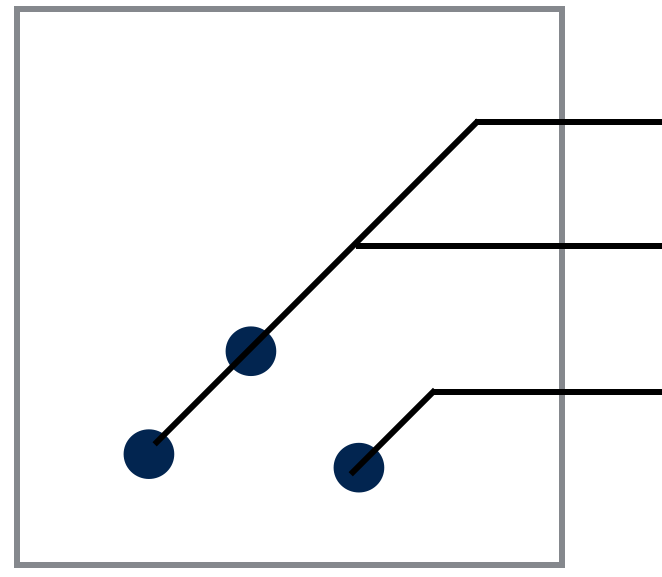}
        \caption{Line overlap}
    \end{subfigure}
    \caption{Issues of one-sided boundary labelling model.}
    \label{fig:maptrix-algorithm-bad}
\end{figure}

The starting point for our algorithm is the one-sided boundary labelling method of Bekos \textit{et al}. which orders and spaces labels evenly at one side of the figure. The model by Bekos \textit{et al}. produces crossing-free, minimal length leaders, each with a diagonal segment of uniform gradient. However, while Bekos \textit{et al}. can ensure no crossings, their method cannot ensure adequate separation between leaders and connection sites or other leaders which may lead to serious ambiguity (see Fig.~\ref{fig:maptrix-algorithm-bad}).

Fortunately, with the MapTrix visualisation we are showing flows between areal regions within which there is typically some freedom to move the connection site of the leader. This means that in a second stage of the layout we can fine-tune the connection site placement so as to increase the separation between leader lines. We use a quadratic program to do this.

We associate penalties with close leader-line segments and displacement of connection points from their initial position. We define hard linear constraints to preserve the ordering of leaders and keep the connection points inside their state boundaries.

Input to the Bekos \textit{et al.} one-sided boundary labelling is a connection site for each map location, typically at the centre of a region. The output is a label ordering permitting crossing-free connection to map locations. That is, for $n$ sites $c_{xi}$, $c_{yi}$, $1 \le i \le n$ we have an ordering such that the leaders for sites $i$ and $i + 1$ are adjacent and crossing free. There are two types of leaders: those with diagonals pointing upward from the sites and those with downward diagonals.

\begin{figure}
    \captionsetup[subfigure]{justification=centering}
    \centering
    \begin{subfigure}{0.34\textwidth}
        \includegraphics[height=3cm]{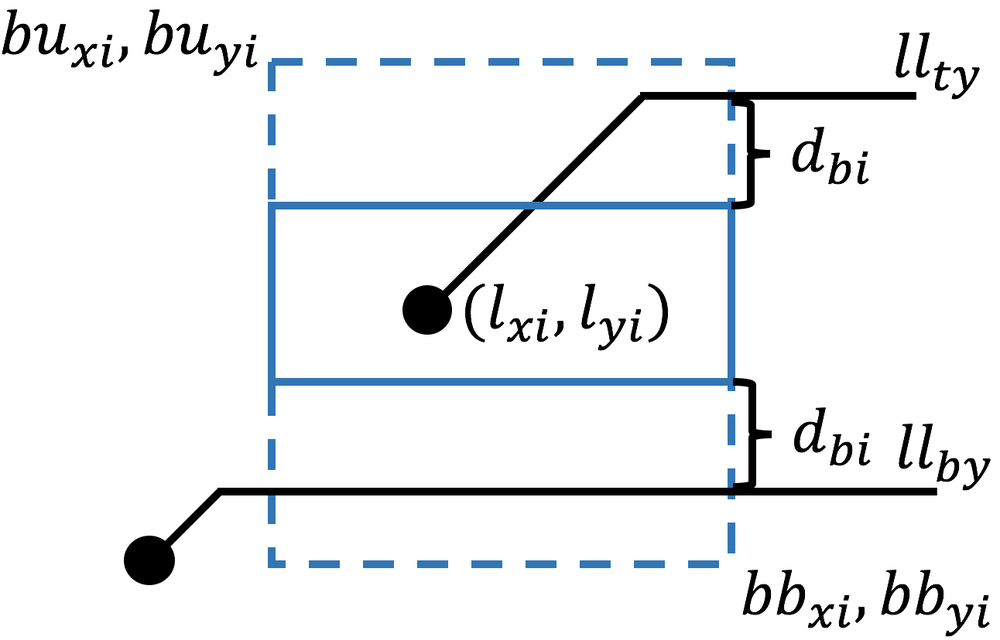}
        \caption{Bounding Box\\Constraint}
    \end{subfigure}
    \begin{subfigure}{0.3\textwidth}
        \includegraphics[height=3cm]{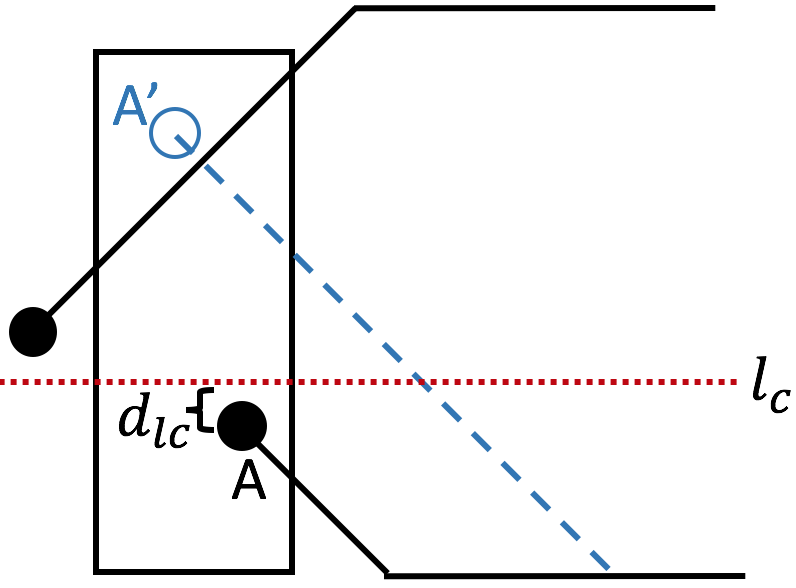}
        \caption{Center Distance Constraint}
    \end{subfigure}
    \begin{subfigure}{0.3\textwidth}
        \includegraphics[height=3cm]{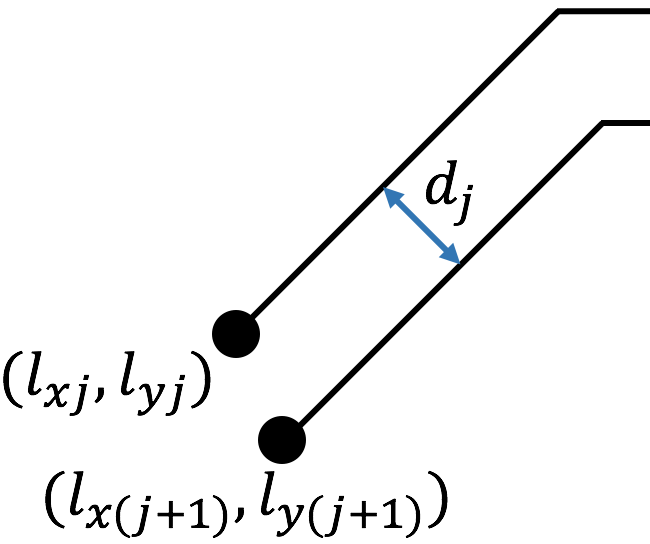}
        \caption{Line Distance Constraint}
    \end{subfigure}
    \caption{Linear constraints applied on top of Bekos \textit{et al.} one-sided boundary labelling model.}
    \label{fig:maptrix-algorithm-constraints}
\end{figure}

The quadratic program to reposition connection sites to achieve good leader separation is as follows.
Let $l_{xi}, l_{yi}$ be variables for leader connection coordinates.
The first set of goal terms penalise displacement of connection sites from their initial position:
$$
  \mathit{PCentre} = \sum_{i=1}^n (l_{xi} - c_{xi})^{2} + (l_{yi} - c_{yi})^{2}
$$
Inside each state boundary we find a rectangle in which the connection site can be safely positioned. Ideally, in order to maximise freedom in placing the connection site, this should be a rectangle with maximal width and height centered around the initial site position. We use a simple heuristic to find such a rectangle. We start from the initial position of each location and grow a rectangle within the state boundary using binary search, alternating between growing the width and height.  This gives us, for each region $i$ the rectangle with upper-left corner $bu_{xi}, bu_{yi}$ and bottom-right corner $bb_{xi}, bb_{yi}$. However, this rectangle may permit the connection site to cross another leader line and introduce a crossing (\emph{Line through site}) as shown in Fig.~\ref{fig:maptrix-algorithm-bad}(a). Thus, we prune the rectangle to ensure the site remains a minimum distance from all other leader lines ($d_{bi}$). Constraints keep the leader connections inside the (pruned) rectangle boundaries (see Fig.~\ref{fig:maptrix-algorithm-constraints}(a)):
  \begin{equation}
    \label{eq:maptrix-algorithm-within}
    \begin{array}{rcl}
      l_{xi} > bu_{xi}, l_{xi} < bb_{xi}\quad \wedge \quad  l_{yi} > ll_{by} + d_{bi}, l_{yi} < ll_{ty} - d_{bi}
    \end{array}
  \end{equation}
The second part of the quadratic model aims to increase the separation between leader lines without introducing crossings. The Bekos \emph{et al.} Algorithm produces alternating bands of leader lines with upward or downward bends. Furthermore, there is horizontal separation between each pair of bands. To ensure that we do not introduce overlap between leaders in two adjacent bands we simply add a separating line  between the adjacent bands (red dashed line $l_c$ in Fig.~\ref{fig:maptrix-algorithm-constraints}(b)) and add a constraint to ensure sites in each band maintain at least a certain distance above or below this boundary line ($d_{lc}$). 

We now consider the case of adjacent leaders in the same band. The distance between adjacent, similarly oriented leader line diagonals (in Fig.~\ref{fig:maptrix-algorithm-constraints}(c)) is given by 
\begin{equation}
  \label{eq:maptrix-algorithm-distance}
  d_j = \frac{(k l_{x_{(j+1)}} - l_{y_{(j+1)}} - k l_{xj} + l_{yj})} {\sqrt{k^2+1}}
\end{equation}
where $1 \le j < n$ and $k$ is the gradient of the leader diagonals.  Since $k$ is constant the relationship is linear.  We introduce another variable to our quadratic program for each $d_j$ and the above relation between $d_j$ is added as a hard constraint.
The constraint:
\begin{equation}
\label{eq:maptrix-algorithm-ordering}
d_j > 0
\end{equation}
preserves the ordering of parallel leader lines ensuring they remain crossing free.  
A final set of penalty terms encourages equal separation between  adjacent leader lines:
$$
  \mathit{PSep} = \sum_{j=1}^{n-1} (d_j - D)^2
$$
\noindent where $D$ is the maximum initial separation between adjacent leader diagonals output by the Bekos \emph{et al.}\ algorithm. 

The full quadratic goal is $\mathit{PCentre} + w(\mathit{PSep})$ where the weight $w \ge 0$ can be varied to trade-off displacement of connection sites and equal separation of leader diagonals.   To obtain connection sites with good separation we minimise this goal subject to the linear constraints of Equ. \ref{eq:maptrix-algorithm-within}, \ref{eq:maptrix-algorithm-distance} and \ref{eq:maptrix-algorithm-ordering}.  Since the number of variables and constraints is linear in the number of input regions, solving this quadratic program with a standard solver is very fast.  Placement of hundreds of connection sites takes a fraction of a second on a standard computer.

%% file: content/4-maptrix/1-conclusion.tex
\section{Conclusion}
\label{sec:maptrix:conclusion-1}
In this chapter, we discussed two state-of-the-art techniques for visualising many-to-many OD flows: bundled flow map and OD Maps. We have also proposed a new method, \emph{MapTrix}, for visualising such data by connecting an OD matrix with origin and destination maps. Compared to these two techniques, MapTrix can scale to large data while still preserving undistorted geographic context within its visual representation. We have provided a detailed description of the alternatives produced in the iterative designing process of MapTrix. We have also given an algorithm for computing an arrangement with crossing-free leader lines.

%% file: content/4-maptrix/2-introduction.tex
\section{Introduction}
\label{sec:maptrix:introduction-2}
Given the practical importance of understanding many-to-many commodity flows, it is surprising that—--to the best of our knowledge—--there have been no user studies comparing different visualisations for showing flow prior to this work presented in this chapter. The most relevant study is by Ghoniem \textit{et al.} \citep{Ghoniem:2004jk} which demonstrates that a matrix representation of a general network performs better than a node-link diagram for large or dense datasets. However, their evaluation does not consider the specific application of commodity flows and the need to embed this in a geographic context.

We conducted two quantitative user studies. The first investigated user preferences and task performance for three very different visualisations: bundled flow map, OD Maps and MapTrix. Example stimuli from the study are shown in Fig.~\ref{fig:maptrix-teaser}. 62 participants completed our on-line questionnaire. We found that MapTrix was the preferred representation while MapTrix and OD Maps had very similar task performance which was much better than with the bundled flow map.

In our second user study we compared MapTrix and OD Maps on larger dense data sets with up to 51 sources and destinations. At this scale it was infeasible to use the bundled flow map representation. 49 participants completed this on-line questionnaire. Again, we found that task performance was very similar. For both representations it was very difficult to compare aggregated flows between or within regions comprising several sources or destinations.

The results of our studies provide strong guidance on how interaction could be used to improve task performance with the different representations. We discuss this more fully in Section~\ref{sec:maptrix:interaction}.

This chapter is structured as follows: Section~\ref{sec:maptrix:first-study} describes our first user study and discusses the results; Section~\ref{sec:maptrix:redesign} demonstrates the redesign of visualisations according to the results and feedback from the first study; Section~\ref{sec:maptrix:second-study} describes our second user study and discusses the results; Section~\ref{sec:maptrix:interaction} demonstrates preliminary interactions for \textit{MapTrix} according to the results and feedback from the studies; at last, in Section~\ref{sec:maptrix:conclusion}, we summarise this chapter.

%% file: content/4-maptrix/2-first-study.tex
\section{First user study}
\label{sec:maptrix:first-study}

In order to test how the visualisations perform for different numbers of locations we investigated their use for different countries. We decided to use real rather than fictional countries and locations to implicitly emphasize the use of such visualisations for common commodity flows such as population migration. This also allowed us to explore the possible impact of prior knowledge of geography on performance.

\begin{table}[b!]
\scriptsize
\setlength{\abovecaptionskip}{.1cm}
\setlength{\belowcaptionskip}{-0.8cm}
  \centering
\begin{tabular}{ | C{0.075\textwidth} | L{0.05\textwidth} | L{0.44\textwidth} | L{0.34\textwidth} | }
    \hline
    Group & Abbr & Description & Example \\
    \hline
    \multirow{2}{*}{\tabincell{c}{Total\\Flow}} 
    	& TFI & \textit{\underline{\textbf{I}}dentify} two total in/out flows for two named locations and compare their magnitude. & Comparing the two locations QLD and TAS, which has the greater total inflow? \\ \cline{2-4}
 		& TFS & \textit{\underline{\textbf{S}}earch} for the largest/smallest total in/out flow. & Which state has the largest total outflow? \\ 
 		\hline
 	\multirow{3}{*}[-5pt]{\tabincell{c}{Single\\Flow}} 
 		& SFI & \textit{\underline{\textbf{I}}dentify} two single flows between named locations and compare their magnitude. & For the two flows from WA to ACT and TAS to SA, which is greater? \\ \cline{2-4}
 		& SFSo & \textit{\underline{\textbf{S}}earch} for the greatest single in/out flow for \textit{\underline{\textbf{o}}ne} named location. & ACT receives the largest single flow from which state? \\ \cline{2-4}
 		& SFSm & \textit{\underline{\textbf{S}}earch} for the largest single flow across all (\textit{\underline{\textbf{m}}any}) locations. & Which is the largest single flow? \\ 
 		\hline
 	\tabincell{c}{Regional\\Flow} & RF & if the flow is predominantly within the \textit{\underline{\textbf{r}}egions} or among the \textit{\underline{\textbf{r}}egions}. & Using the regions A and B defined in the above map, is the flow predominantly within A or B? \\
 	\hline
\end{tabular}
  \vspace{2.8em}
  \caption{Task description, abbreviations and examples questions from AU}~\label{tab:maptrxi-tasks}
\end{table} 

\subsection{Tasks}
We identified a variety of tasks that commodity flow visualisations should support by reviewing the geographical visualisation literature~\citep{Andrienko:2006up,Ali:2013tg, Tobon:2005}. We propose three different targets for analysing geographically-embedded flow data: \textit{single flow (SF)}, \textit{total flow (TF)} and \textit{regional flow (RF)} (more discussion available in Section~\ref{sec:interviews:summary}). For single and total flows we are mainly interested in flows from a given target location(s), or identifying which location(s) corresponds to a given characteristic. These tend to be lookup or comparison tasks which may refer to identifying and comparing total flow (TF) values of 1, 2 or many locations, or single flows (SF) between 2 or many locations. A further important task involves determine the geographical or regional distribution of the flow (RF). This involves identifying if flow is predominantly within a certain area on the map or between two different areas. We designed our questions of the study into the following six task categories: \textit{TFI, TFS, SFI, SFSo, SFSm and RF}. These are defined along with examples of exact questions in Tab.~\ref{tab:maptrxi-tasks}. \added{Although these tasks were identified before we conducted the expert interviews (see Chapter~\ref{chapter:interviews}), we found that these different task types can be well-linked to real-world practice of analysing geographically-embedded flow data (see Tab.~\ref{tab:interview:targets}).}

\subsection{Countries and Datasets}
To represent actual commodity flow data, we created synthetic datasets based on real internal population migration data. The first country we chose was Australia (AU) as it has a large spacious country shape with relatively few federal states (and territories). With 8 states there are only $8 \times 8$ between state flows to present. The original dataset for AU is based on 2013-14 internal migration for AU\footnote{http://stat.abs.gov.au//Index.aspx?QueryId=1233}. 

To investigate larger number of flows, Germany (DE) was chosen as a comparison as it again has a large and spacious shape but double the number of federal states and therefore $16 \times 16$ individual flows. For DE between state migration was not openly available so we allocated data from USA internal migration data from 2009-10\footnote{https://www.census.gov/hhes/migration/data/acs/state-to-state.html}. 

A third country, New Zealand (NZ), was chosen to allow us to investigate the effect of country shape. It has the same number of national states as DE but is more elongated. The original NZ dataset is based on regional migration from 2001--06\footnote{http://www.stats.govt.nz/browse\_for\_stats/population/Migration}. 

During our pilot sessions we also investigated countries with larger number of locations, including the United States of America (US) with $51 \times 51$ flows. This number of flows was found to be too confusing and difficult for users, particularly for the bundled flow map design. We therefore removed US from the first study (but used it in the second study Section~\ref{sec:maptrix:second-study}). In order to train the participants we introduce the problem and explain the visualisations using the United Kingdom. It has a distinguishable country shape and only four national states (in this case countries) so therefore only $4 \times 4$ flows. 

\subsection{Procedure}
\label{sec:maptrix:first-study:procedure}
The structure of the study was slightly amended following the pilot study as the study took too long with all three countries. As all tasks were shown to be important to the analysis we chose to split the countries so each participant was asked questions about only one pair of countries (AU-NZ; AU-DE; NZ-DE). The choice of country pair was counterbalanced. After receiving information about the study through the explanatory statement and agreeing to the consent form the study took the following structure: 
\begin{enumerate}

\item Background knowledge: participants were asked about their prior experience using maps -- rarely use, navigation only or often use maps to read statistical information -- and their knowledge of the administrative structure of their pair of countries; 

\item Training: participants were given an overview of the problem and explanation of each of the visualisation methods. Upon finishing the training for each method the participant was showed two sample questions with the answers and explanation. They were then asked to answer another two questions to verify that they understood the method. The training order was counterbalanced.

\item 
Tasks: participants were asked to answer $36 = 6 \times 3 \times 2$ questions: one for each kind of task for each of the three visualisation methods and each of the two countries.  Question order was randomised.

\item
Ranking and Feedback: participants were asked to rank the three visualisation methods in terms of visual design and in terms of effectiveness of reading information for each of the two countries that had been shown. They also had the opportunity to comment on the strengths and weaknesses of each visualisation method.
\end{enumerate}

\subsection{Participants}
\label{sec:maptrix:first-study:participants}
To attract a range of skill-levels amongst participants the study was advertised at Monash University (Australia) using a university-wide bulletin and through email lists at Microsoft Research (USA), HafenCity University (Germany) and two international map visualisation lists of GeoVis and CogVis. Three \$50 gift cards were offered as an incentive, where participants could optionally provide their contact details and be placed in the prize draw. 

In total we had 62 complete responses, with an equal split of country pairs -- 20 AU-DE, 21 AU-NZ and 21 NZ-DE. Of these 2 participants were excluded from the final analysis due to the exceedingly quick completion time of 5m and an average task time of 8s. Upon analysis we also trimmed 1\% of response times -- those over 300s/5m -- this removed large outliers ranging from 305s to 3352s. On average the 60 participants spent 39s per task and the entire online study took an average of 51m:52s to complete.

\subsection{Statistical Analysis Methods}
We consider response time and accuracy for each question. We investigate the effect of the three conditions of visualisation (\emph{Vis}) (these are abbreviated to \emph{BD} for Bundled Flow Map, \emph{OD} for OD Maps, \emph{MT} for MapTrix in this section), country and task, and to what degree these conditions differ significantly. 

In our analysis we treat all conditions as being independent. Although question order was randomised, we validated task independence by plotting results against question order. No clear pattern was  evident. 

To compare error rates between different conditions we use standard non-parametric statistics~\cite{Andy:2012ds}: For multiple (more than 2) conditions, we use Friedman's ANOVA to check for significance and apply Post hoc tests with Bonferroni correction to compare groups while for two conditions, we use the Wilcoxon signed-rank test. Both tests require the same participants in all conditions so when comparing across countries (3 conditions) we could not directly use Friedman's ANOVA as participants only completed the study for two countries.  Instead we split the results into 3 groups, one for  each pair of countries and used a Wilcoxin signed-rank test for each group.

To compare response time we consider only times for correct and almost correct responses. To test for significance we use a multilevel model for analysing mixed design experiments~\cite{Andy:2012ds}.  Here we breakdown the analysis by each condition and their interactions (\emph{Vis/Country}, \emph{Vis/Task}, \emph{Country/Task} and \emph{Vis/Country/Task}).

For the user preference results we again use Friedman's ANOVA and Post hoc tests to test for significance. 

\subsection{Results}\vspace{-1em}
\subsubsection{Error Rate}
\begin{figure}[b!]
\centering
    \includegraphics[width=\linewidth]{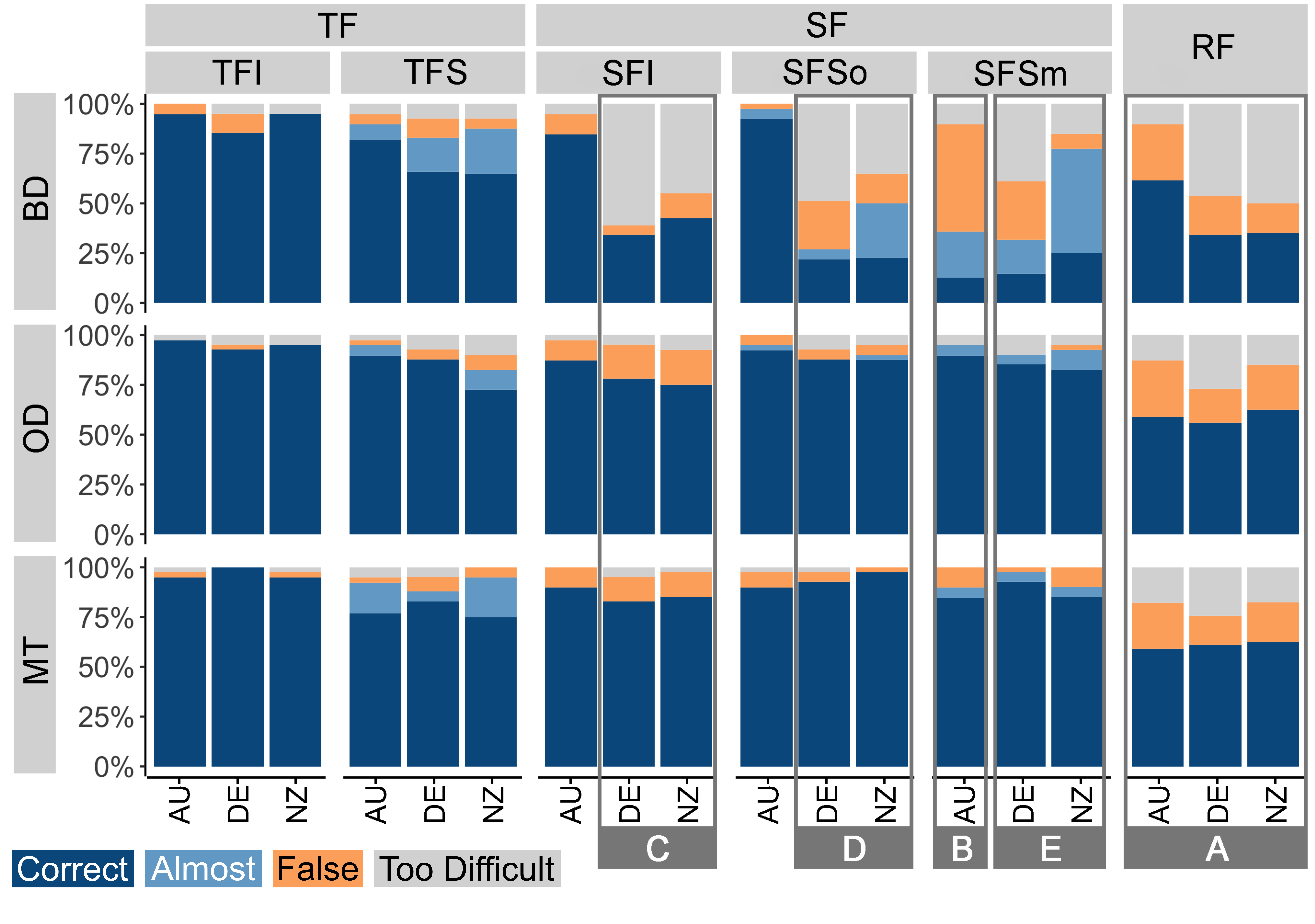}
    \caption{\label{fig:maptrix:error-1}First study accuracy. Highlights A-E are statistically significant as described in the text.}
\end{figure}

\vspace{-2em}
Responses were in four categories of accuracy: \setlength{\fboxsep}{1.5pt}\colorbox{maptrix-correct}{\textcolor{white}{Correct}}, \setlength{\fboxsep}{1.5pt}\colorbox{maptrix-almost}{\textcolor{white}{Almost}}, \setlength{\fboxsep}{1.5pt}\colorbox{maptrix-false}{\textcolor{black}{False}}, \setlength{\fboxsep}{1.5pt}\colorbox{maptrix-too-difficult}{\textcolor{black}{Too Difficult}}, Fig.~\ref{fig:maptrix:error-1}. 
We see notable differences in the performance of BD compared to OD and MT, in particular for the SF tasks for the two larger datasets (DE and NZ). All vis methods perform well in the TF tasks, especially TFI. The RF task also shows a far lower accuracy across all vis methods (A in Fig.~\ref{fig:maptrix:error-1}).

Our smallest dataset (AU) consistently out-performs DE and NZ in almost all tasks for all vis methods.  There is one notable exception (see highlight B: Fig.~\ref{fig:maptrix:error-1}): 
BD performs far worse for SFSm with 13\% correct + 23\% almost correct, compared to 90\%+5\% for OD and 85\%+5\% for MT. Statistical significance is shown between BD:OD and between BD:MT (both $p< .0001$). No statistical significance is evident between OD:MT. 

The other two countries DE and NZ have the same number of flows. There are some similarities and notable differences when comparing the two sets of results. Most notably, BD is less accurate for SF tasks, see C, D, E in Fig.~\ref{fig:maptrix:error-1}. 
For all SF tasks using DE and NZ, Wilcoxon signed-rank tests show statistical significance between BD:OD (SFI: $p= .0012$, both SFSo and SFSm $p<.0001$), and between BD:MT ($p$ values: SFI $< .0001$, SFSo $< .0001$ and SFSm $< .0001$).

For SFS(o/m), compared to BD not only does response rate improve using OD and MT (all $p$ for OD:BD \& MT:BD in SFS(o/m) $<.0001$), but the ability to differentiate the dominant answer (i.e. \emph{correct} rather than \emph{almost correct}) is far higher, see D and E in Fig.~\ref{fig:maptrix:error-1}. 

For TF tasks we observe more similarity between methods. All vis perform well particularly for TFI, with BD performing slightly worse for DE. For TFS we see some differences between vis methods with OD and MT performing better than BD, but no statistical significance is found. 

For RF tasks, not only do we see a difference in performance between all tasks for all vis, but BD performed notably worse than OD and MT (See Fig.~\ref{fig:maptrix:error-1} A).
Friedman's ANOVA (details Section~\ref{sec:maptrix:first-study:participants}) for DE reveals a statistical significance between BD:OD ($p< .0001$) and between BD:MT ($p< .0001$), the same for NZ; between BD:OD ($p=.0108$) and between BD:MT ($p = .0108$). Similar percentages are reported for both DE and NZ, with 34 and 35\% for BD and between 56 and 63 for OD, and 61 and 63\% for MT.  No statistical significance is again found between OD and MT.

\vspace{-1.5em}
\subsubsection{Response Time}\vspace{-2em}
We extract the results of all \emph{Correct} and \emph{Almost Correct} responses (1689 timed responses) from all 2160 responses and plot these for all conditions, as shown in Fig.~\ref{fig:maptrix:time-1}.  
These box plots, together with multilevel model analysis method, reveal: 

\begin{figure}[b!]
\centering
    \includegraphics[width=\textwidth]{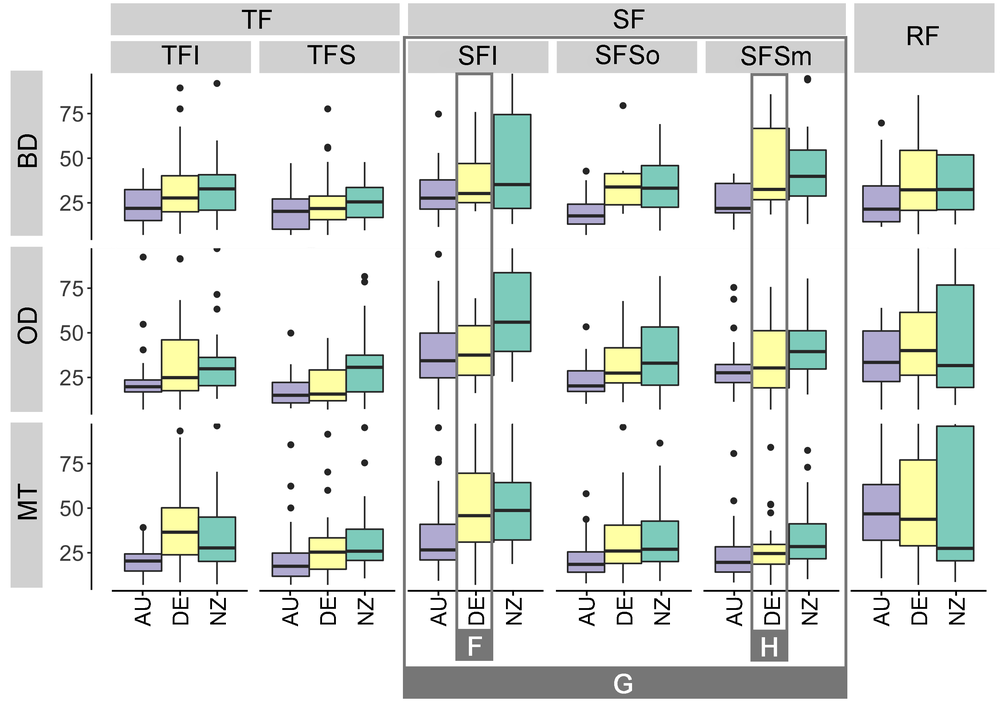}
    \caption{\label{fig:maptrix:time-1}First study response times in seconds. Highlights F-H are statistically significant.}
\end{figure}

For DE, Task SFI takes increasingly longer from BD, OD and MT (i.e. MT $>$ OD $>$ BD -- see F in Fig.~\ref{fig:maptrix:time-1}). This is shown to be statistical significant ($p=0.0087$);

For DE, SFSm the trend is the opposite (i.e. MT $<$ OD $<$ BD) -- see G in Fig.~\ref{fig:maptrix:time-1}. Again, there is a statistical significance ($p=0.0485$);

Although their accuracy is higher, OD and MT took notably more time on RF than BD. MT longer than OD. Correct responses have a wider range for NZ. No statistical significance is found.

Finally, as the size of the dataset increases from AU to DE/NZ we see increasing response time for all tasks, especially for SF tasks (multilevel comparison: DE $>$ AU \& NZ $>$ AU, $p=0.0003$) (see Fig.~\ref{fig:maptrix:time-1} H). NZ often takes longer than DE.

\begin{figure}[b!]
	\captionsetup[subfigure]{justification=centering}
    \centering 
    \begin{subfigure}{0.35\textwidth}
        \includegraphics[width=\textwidth]{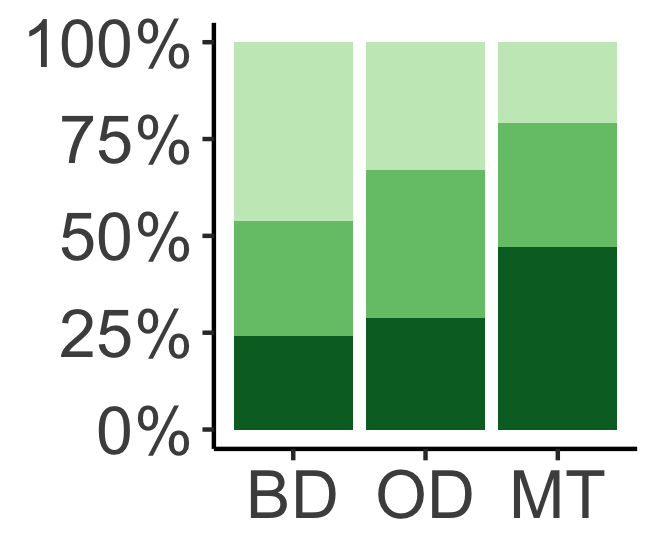}
        \caption{Visual Design Ranking}
    \end{subfigure}
    \begin{subfigure}{0.35\textwidth}
        \includegraphics[width=\textwidth]{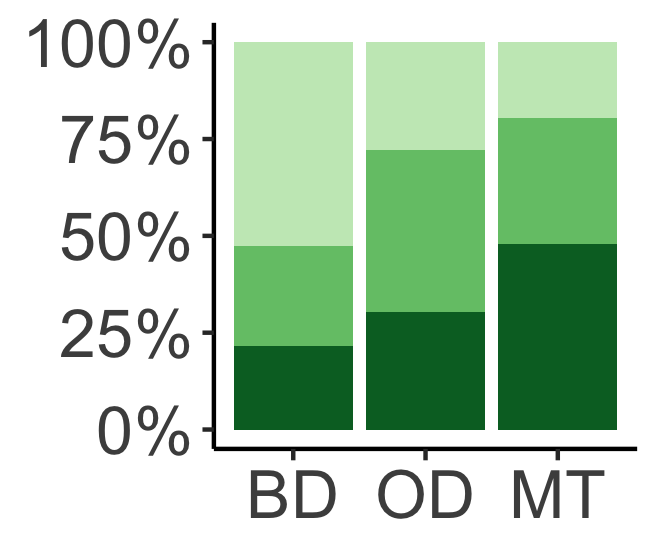}
        \caption{Readability Ranking}
    \end{subfigure}
    \caption{User preference ranking (\setlength{\fboxsep}{1.5pt}\colorbox{third_first}{\textcolor{white}{$1^{st}$}}, \setlength{\fboxsep}{1pt}\colorbox{third}{\textcolor{black}{$2^{nd}$}} and \setlength{\fboxsep}{1pt}\colorbox{second_fourth}{$3^{rd}$}).}
    \label{fig:maptrix-preference-1}
\end{figure} 

\vspace{-1.5em}
\subsubsection{User Preference \& Feedback}\vspace{-2em}
Participant ranking of visual design for each of the three methods---by percentage of respondents---is shown here by colour (see Fig.~\ref{fig:maptrix-preference-1}).  The strongest preference is for MT, with almost 50\% of respondents voting MT first place. The other two methods each received approximately 25\% of votes.  

These differences have statistical significance $\mathrm{OD}>\mathrm{BD}$ with $p=0.0035$, $\mathrm{MT}>\mathrm{OD}$ and $\mathrm{MT}>\mathrm{BD}$ both $p<0.0001$.
Participant ranking of readability is similar and, again, statistically significant with $\mathrm{MT}>\mathrm{OD}>\mathrm{BD}$, all $p<0.0001$.

The final section of the study allowed participants to give feedback on the pros and cons of each design. Qualitative analysis of these comments reveal (overall):
\begin{itemize}
	\item \textbf{BD} was intuitive and familiar: \textit{``it is good for anyone with geographic knowledge and spatial cognition''}. But arrows overlap, arrows are too long, the visualisation does not help when there are many locations and was hard for the RF task: \textit{``Too many locations means many arrows, they occlude, it's hard to see which is which''} and \textit{``Hard to follow arrows over long distances or through intersections... impossible to answer the between or within regions questions''}.
	\item \textbf{OD} was easy to comprehend and participants often liked the geographical layout. Others found it good for comparison and easy to read the flows. Some also commented on the novelty: \textit{``It is creative and clear''}. Yet, it was also seen as the most unfamiliar and sometimes difficult to comprehend: \textit{``Arrows are missing, I was confused to identify inflow and outflow''}. The small square sizes were also frustrating:  \textit{``the visualisation (grids) can become rather small and more difficult to interpret''}.   
	\item \textbf{MT} was visually attractive, easy for larger flows and intuitive: \textit{``clear, it is easy to quantify the flows''}. Yet, some found it confusing or unfamiliar.  A few commented that it looked complex: \textit{``It may look complicated but it is the best visualization for information extraction''}. Others found it difficult to follow the lines or read the labels in the matrix, especially with more locations: \textit{``When dataset is large, it becomes difficult to follow the flow''}. Some also commented on there being too much information and there being redundant visual elements (e.g. bars and circles).
\end{itemize}

\vspace{-1.5em}
\subsubsection{Summary}\vspace{-2em}
These results reveal that:
\begin{itemize}
\item AU is the fastest and best performing of all countries. All vis methods are suitable for such small datasets, with the exception of BD for the SFSm task;
\item Error rate worsens with scaling data from AU to NZ/DE, especially for BD for all the SF tasks, where OD and MT out-perform BD with statistical significance; 
\item SFI takes the longest of the SF tasks and on average has the highest error rate;
\item The RF task takes the longest and has the highest error rate compared to all other tasks. All vis methods performed poorly;
\item OD and MT show no significant differences in performance across all conditions;
\item Participants prefer MT for design and readability of information.
\end{itemize}

A central design goal of OD and MT is to overcome the problem of occlusion of flows as data increases. 
For the larger datasets (DE and NZ) both OD and MT were significantly better than BD, but there is no significant difference between the two for any condition. User ranking indicates a preference for MT. The fact that BD performs worse is unsurprising given the known problem of overlapping flows; however, the remarkably similar performance of OD and MT is unexpected. We now examine to what extent these methods scale and whether the similarities in task performance continue with increasing scale.

%% file: content/4-maptrix/2-redesign.tex
\section{Redesign}
\label{sec:maptrix:redesign}

Feedback and suggestions from the first study and modifications to make it more capable with large dataset, led to some design improvements for both methods explained below.

\begin{figure}
\centering
  \includegraphics[width=\textwidth]{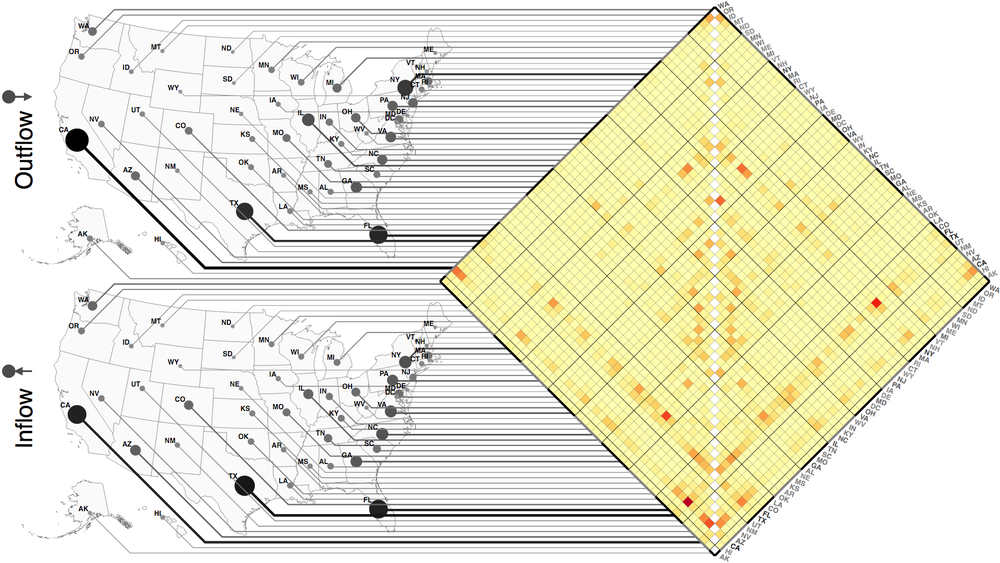}
  \caption{Redesign of MapTrix.}\label{fig:maptrix:maptrix-redesign}
\end{figure}

MT, shown in Fig.~\ref{fig:maptrix:maptrix-redesign}:
\begin{itemize}
    \item Removed TF barcharts to allow more lines, improve tracking and reduce redundancy of information;
    \item Scaled lines thickness and grey shading of line and label proportional to TF circle size and aid line tracking;    
   \item Added separation lines within matrix every 5 rows / columns to aid user tracking;
   \item Minimised overlapping of circles and labels in maps;
  \item Removed full names in matrix. All labels refer to abbreviations;
   \item Removed the arrows in the destination maps to give more space for labels and circles; instead, we  used a destination icon next to the map label. 
\end{itemize}

\begin{figure}[b!]
\centering
  \includegraphics[width=\textwidth]{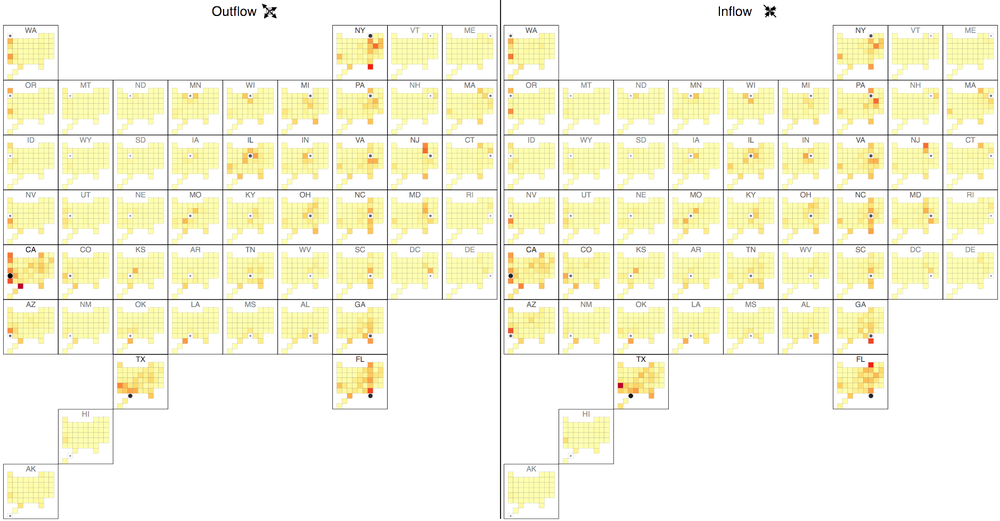}
  \caption{Redesign of OD Maps.}\label{fig:maptrix:od-map-redesign}
\end{figure}

OD, shown in Fig.~\ref{fig:maptrix:od-map-redesign}: 
\begin{itemize}
   \item Removed white space to increase grid square size;
   \item Moved and enlarged legend to improve lookups and to allow more space for grid;
   \item Extra care with grid layout to ensure neighbouring regions were adjacent and limited white space -- the downside being that the country is more abstract;%
  \item Increased text size and added label shading to relate to TF and match to the proportional labels in MT;
   \item Added destination icons to indicate direction to match new icons in MT, with multiple arrows in/out compared to only one for MT.
\end{itemize}
\added{It should be noted that there is emerging research focusing on optimising map layout~\citep{McNeill:2017gp,Meulemans:2017fo}. These guidelines were not available at the time we made the redesign choices.}

%% file: content/4-maptrix/2-second-study.tex
\section{Second user study: scalability}
\label{sec:maptrix:second-study}
In this section we concentrate on the scalability of MT and OD. Our second study followed the same structure and participant recruitment method (see Section~\ref{sec:maptrix:first-study:procedure} and ~\ref{sec:maptrix:first-study:participants}) as the first, but the countries investigated were amended together with improvements made to the tasks and visual designs.

\subsection{Data}
To investigate larger countries than NZ and DE ($16 \times 16$ flows) we wanted to use \emph{The United States of America (US)} with $51 \times 51$ flows as our previous pilot revealed that participants got frustrated with BD for US, but less so with OD or MT. We also chose \emph{China (CN)} with $34 \times 34$ flows as it is almost half way between the two.  For CN the original data set is available for the internal migration from 2005-10~\footnote{http://www.stats.gov.cn/tjsj/pcsj/rkpc/6rp/indexch.htm}. For US we use 2009-10 internal migration data~\footnote{https://www.census.gov/hhes/migration/data/acs/state-to-state.html}. Again we randomised the locations of the data for each question.

\subsection{Tasks}
In the first study, the RF task was found to be extremely difficult. However, when considering flow in a geographical context it is important to be able to easily compare multiple groups of locations.  In the description of tasks below, we define a \emph{region} as being a collection of locations on the map that are geographically contiguous (adjacent).  
Due to the design choices of both OD and MT the marks corresponding to flows for such regions in the map may not be adjacent in the visualisation. 

For more detailed comparison, we divide the RF task from Study 1 into subtasks related to the adjacency of regions in the visualisation and whether the flow is occurring ``between'' regions or ``within'' a region.  The six subtasks are labelled: RFBb, RFBw, RFBn, RFWb, RFWw, RFWn.  These codes are explained as follows (examples are provided as supplementary material):

Assume two regions A \& B each consisting of multiple contiguous locations.
    Are the flows predominantly \\
\null\quad\textit{\underline{\textbf{B}}etween}: between A to B or B to A? \\
\null\quad\textit{\underline{\textbf{W}}ithin}: within A or B?\\
Different adjacency conditions for visuals of regions and locations: \\
\null\quad\textit{\underline{\textbf{b}}etween}: locations within region and regions are adjacent in vis; \\
\null\quad\textit{\underline{\textbf{w}}ithin}: only locations in each region are adjacent in vis; \\
\null\quad\textit{\underline{\textbf{n}}one}: both are not adjacent in vis.

For each question we manually identified appropriate regions for the task for each visualisation method to ensure comparability. These were combined with the same tasks as the first study. In total, participants were asked to answer $44 = 11 (Tasks) \times 2 (Vis) \times 2 (Country)$ questions.

\subsection{Pilot Test and Highlighting}
The first study indicated that the RF task was the most difficult and time consuming across all vis techniques. Our redesign of the RF question to investigate adjacency was intended to investigate this task in more detail; however, pilot testing revealed difficulties.

The RF tasks, although now possible to answer, still took considerable time and were a particular cause of frustration. One participant took over 1h:30m to complete the pilot, with the majority of this time spent manually connecting flows or identifying the squares for the regional tasks. 

To continue to investigate scalability and to allow us to determine whether one visualisation out-performs the other for the aggregation of flows we opted to aid the users in finding the right locations by highlighting them on the OD Maps or MapTrix.  Our assumption is that such simple highlighting is easily made available with interaction. We eventually implemented this, see Section~\ref{sec:maptrix:interaction}. Subsequent pilots revealed much more satisfied users and much faster completion time. 

To encourage participants to think over their answers, instead of showing ``Too difficult'' option straight away we revealed it after 1 minute for every question.

\subsection{Results}
\begin{figure}[b!]
\centering
    \includegraphics[width=\linewidth]{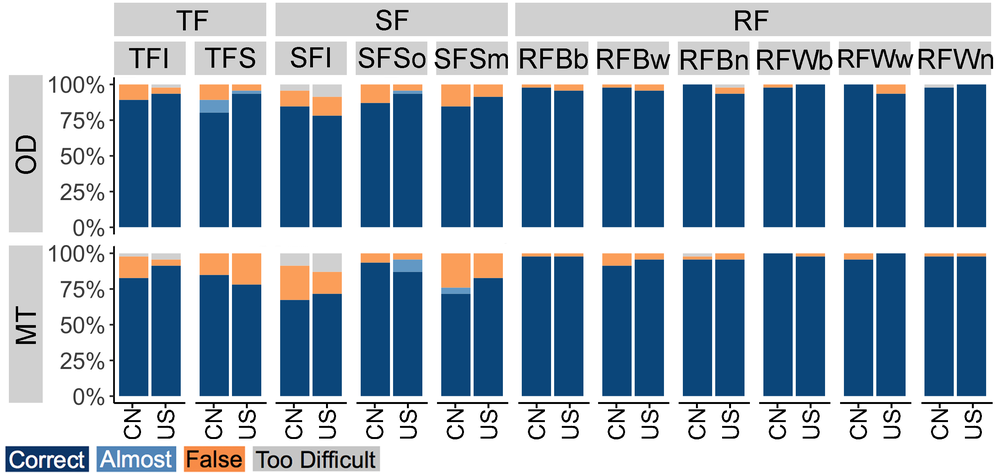}
    \caption{\label{fig:maptrix:error-2}Second study accuracy.}
\end{figure}

The study had 46 valid responses from an original 49 (3 with impossibly quick responses were excluded). On average, individual task completion time was 31.74s and the entire study took 45m:12s. We present the results for error rate and response time in Fig.~\ref{fig:maptrix:error-2} and Fig.~\ref{fig:maptrix:time-2}. For the response time analysis, we took the 1861 \emph{correct} and \emph{almost correct} responses from 2024 total responses.

\vspace{-1.5em}
\subsubsection{Error Rate}\vspace{-2em}
Fig.~\ref{fig:maptrix:error-2} shows remarkably similar results across all conditions. No differences are evident in the RF tasks, which all performed very well. Some differences are evident in Fig.~\ref{fig:maptrix:error-2} between the vis methods for the SF and TFS tasks, but these are not consistent between countries and these differences are not statistically significant. Considering the increase in data flows, it is surprising to see that the results often show an improvement for US over CN. Investigating whether task performance improved with country knowledge, we compare the results for those who claimed good knowledge of the states of US (12 participants) or CN (11) to those who claimed little to no knowledge of the states.  As expected identification tasks (TFI and SFI) increase in speed as well as accuracy for those with knowledge of US, however, only SFI shows an increase for CN. Perhaps the US map is more well known than participants realise, or perhaps it is better suited for these designs.  Feedback from one of the pilot participants suggested that the block shapes of US states helped identification. 

The differences for SF tasks show for CN, OD (82\%) outperformed MT (62\%) in SFI, with slight improvement for both for the US. For CN, OD (82\%) outperformed MT (65\%) in SFSm, with improvements for both for US. This is the only task where one method outperforms the other for both data sets. For CN, MT (91\%) slightly outperformed OD (85\%) in SFSo, but for US the results reduce for MT and increase for OD. 

The final notable difference is for TFS, where for US OD (98\%) outperformed MT (74\%), but for CN results for relatively similar for both vis methods.

\vspace{-1.5em}
\subsubsection{Response Time}
\begin{figure}[b!]
\centering
    \includegraphics[width=\textwidth]{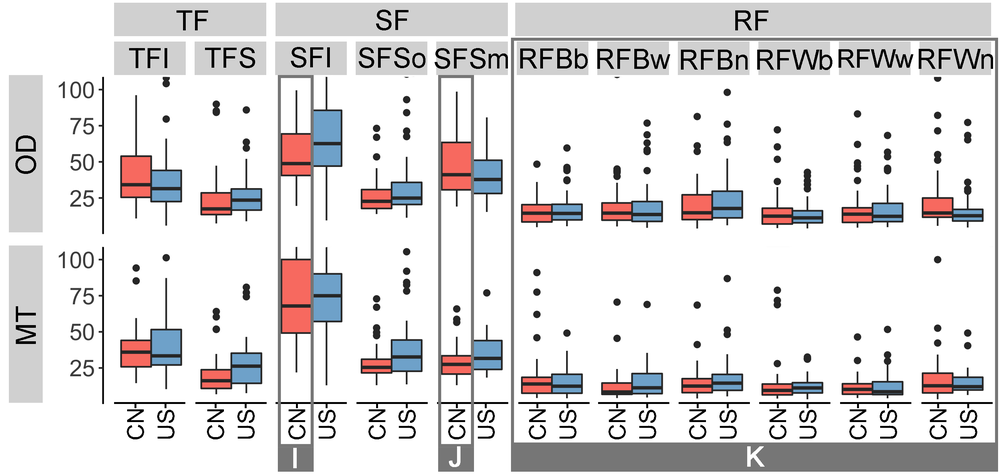}
    \caption{\label{fig:maptrix:time-2}Second study response times in seconds. Highlights I-K are statistically significant as described in the text.}
\end{figure}

\vspace{-2em}
In our response time analysis, Fig.~\ref{fig:maptrix:time-2} shows all conditions. Some notable and statistical significant differences are evident:
\begin{itemize}
  \item In CN for SFI, MT took longer than OD ($p=0.0373$), see I in Fig.~\ref{fig:maptrix:time-2}, reflecting the increased difficulty indicated through the higher error rate;
  \item In CN for SFSm, OD took longer than MT ($p<0.0001$), see J in Fig.~\ref{fig:maptrix:time-2}, despite a lower error rate;
   \item With OD, SFSm took longer than SFSo ($p < 0.0001$);
  \item In general, SFI took longer than all other tasks. This is statistically significant for OD in US and MT in both US and CN with ($p=0.0373$).
  \item RF is significantly quicker than the rest ($p=0.0173$). 
 \item RFB[bwn] take significantly longer than RFW[bwn] ($p=0.0404$), see K in Fig.~\ref{fig:maptrix:time-2}. 
 \item MT is quicker than OD for RF, but not significantly.
\end{itemize}

\vspace{-1.5em}
\subsubsection{User Preferences and Feedback}
\begin{figure}
	\captionsetup[subfigure]{justification=centering}
    \centering 
    \begin{subfigure}{0.3\textwidth}
        \includegraphics[width=\textwidth]{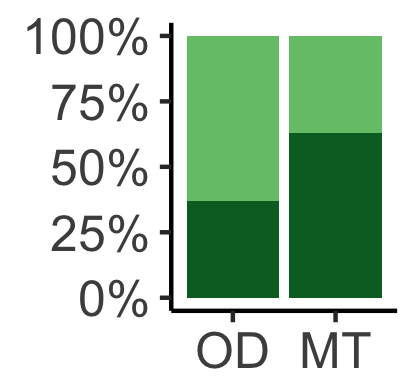}
        \caption{Visual Design Ranking}
    \end{subfigure}
    \begin{subfigure}{0.3\textwidth}
        \includegraphics[width=\textwidth]{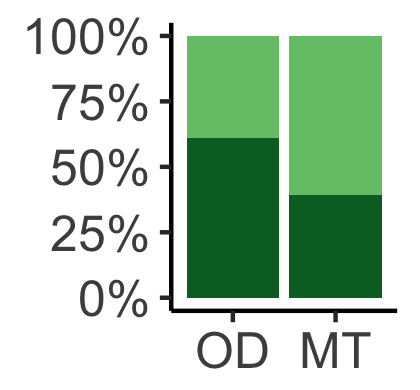}
        \caption{Readability Ranking}
    \end{subfigure}
    \caption{User preference ranking (\setlength{\fboxsep}{1.5pt}\colorbox{first_of_two}{\textcolor{white}{$1^{st}$}} and \setlength{\fboxsep}{1pt}\colorbox{second_of_two}{\textcolor{black}{$2^{nd}$}}).}
    \label{fig:maptrix-preference-2}
\end{figure}

\vspace{-2em}
Participant ranking for each of the two methods, by percentage of respondents is shown here by colour(see Fig.~\ref{fig:maptrix-preference-2}). The majority of participants ranked MT first for design (63\%).  Compared to the previous study, OD now replaces MT as first for readability (60.9\%). The difference in percentages are marginal and not statistically significant (visual design: $p=0.0641$; readability: $p=0.1228$).  Nevertheless, we investigated whether these rankings relate to participants' knowledge of the country or their map knowledge but found no difference between these groups and the overall ranking. Investigating whether previous experience of these designs affect these rankings, we see only slight differences. Removing the 6 participants who had participated in the first study from the results we see the rankings for readability of OD increase to 65\%, whilst design remains the same.

The qualitative analysis of the feedback quotes again reveals quite conflicting preferences:
\begin{itemize}
	\item \textbf{OD} was seen as easy to link locations with visualisations. Some participants found it easy to compare single flows: \emph{``OD is easy to find the flow from one location to another without losing your place''} and to find the location names.  Many found the visual elements (grids, cells, circles, labels) far too small. Some disliked the abstract geography: \emph{``losing some geographical reference make locations confusing, given prior map knowledge''}, whilst others recognised that although it can be difficult at first, you can learn the representation: \emph{``you would quickly learn their locations''}.
	\item \textbf{MT} was found to be familiar because it has real maps and related to the geographical locations. For some the matrix display was also familiar: \emph{``Closer to familiar matrix display. The way of connecting the maps on the left with the rows and columns of the matrix works well.''}. Many participants commented on the difficulty of finding locations, e.g. \emph{``It was too dense with the labels too small to identify the place.''}. Whilst some commented that it was difficult to trace the leader lines and there was a need for a marker: \emph{``Sometimes I had to use a ruler to find the intersection.''}
\end{itemize}

In general, many participants requested interaction, such as highlighting and selecting. Some noted that locations need to be easier to find in MT, i.e. through reordering the matrix or by allowing text based searching. A few participants also commented that the RF task would be near impossible without the highlighting.

%% file: content/4-maptrix/3-interaction.tex
\section{Interaction}
\label{sec:maptrix:interaction}
\begin{figure}[b!]
\centering
\includegraphics[width=\textwidth]{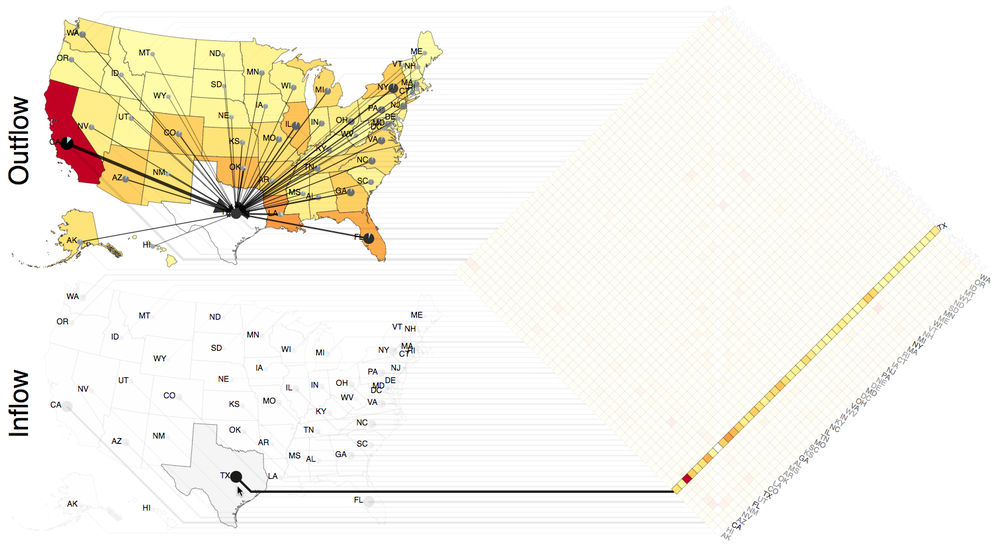} 
\caption{
\label{fig:maptrix:interaction-hovering}
One location in maps can be hovered to presented detailed information.
}
\end{figure}

\begin{figure}
\centering
\includegraphics[width=.945\textwidth]{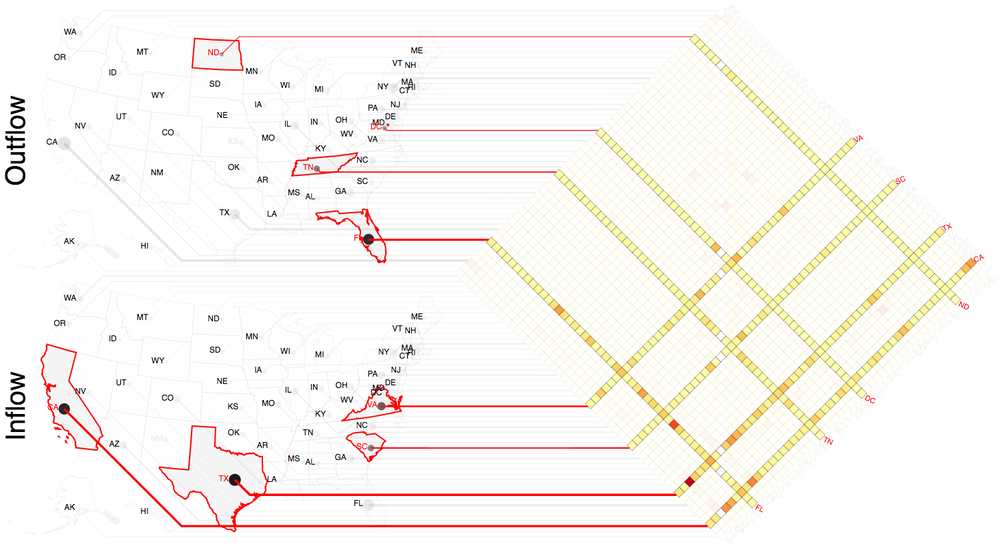} 
\caption{
\label{fig:maptrix:interaction-selection}
Locations in maps or cells in matrix can be clicked to persist highlighting of related elements. 
}
\end{figure}

\begin{figure}
\centering
\includegraphics[width=.945\textwidth]{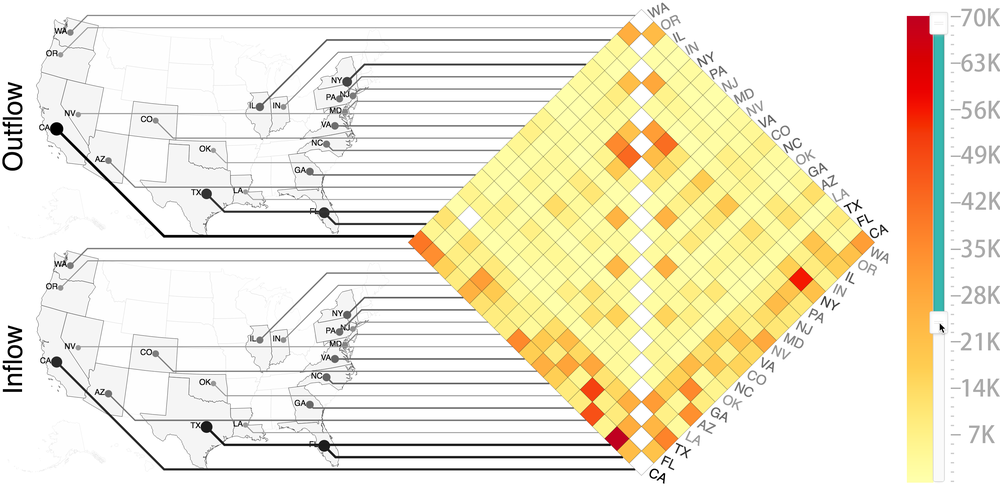} 
\caption{
\label{fig:maptrix:interaction-label}
The maptrix display can be limited to show only a certain range of values, this triggers a relayout of the matrix and leaders 
}
\end{figure}

\begin{sidewaysfigure}
\centering
\includegraphics[width=\textwidth]{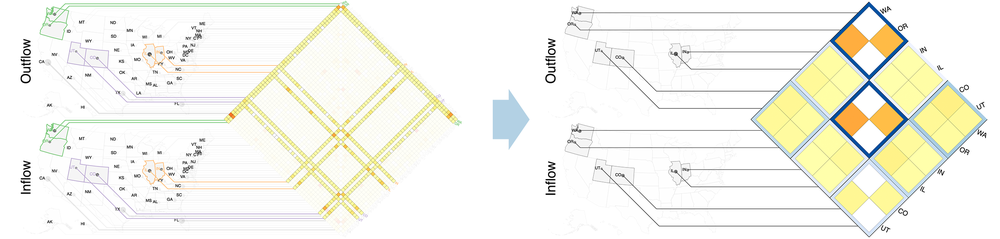} 
\caption{
\label{fig:maptrix:interaction-region}
A set of contiguous regions can be selected for comparison in a detailed MapTrix view, this also causes a relayout. 
}
\end{sidewaysfigure}

We learned from our second study that while MapTrix makes it possible to read a single flow value between a given source and destination in larger datasets it did become more difficult.  For comparing clusters of locations our pilots revealed that highlighting of paths (from origins via leaders and matrix cells to destinations) was essential.  We have constructed a prototype interactive system which allows users to interactively create these highlights through various selection mechanisms~\citep{yalong:2016}. 
In particular, the following interactions directly support the indicated tasks from Tab~\ref{tab:maptrxi-tasks}:
\begin{itemize}
	\item \textbf{SFSo}--Highlighting of the associated row/column on mouse-hover over a map region, cell, label or leader line, Fig.~\ref{fig:maptrix:interaction-hovering}.
	\item \textbf{TFI, SFI, SFSo}--Mouse-click makes such cell highlighting persist such that multiple flows can be compared simultaneously, Fig.~\ref{fig:maptrix:interaction-selection}.
	\item \textbf{TFS, SFSm}--The colour key beside the MapTrix is an interactive widget allowing for filtering the MapTrix to a particular range of flow values, Fig.~\ref{fig:maptrix:interaction-label}.
	\item \textbf{RF}--Aggregate selection for Regional Flow comparison tasks, Fig.~\ref{fig:maptrix:interaction-region}.
\end{itemize}

The last two interactions (demonstrated in Fig.~\ref{fig:maptrix:interaction-label} and Fig.~\ref{fig:maptrix:interaction-region}) both reduce the number of regions shown in the MapTrix and induce a re-layout of the MapTrix and leader lines.  Such re-layout is fast to compute; for the US with 51 locations it is in the order of a few milliseconds.    This dynamic rearrangement, together with the smooth transition animations we use, are demonstrated at~\cite{yalong:2016}.

\newpage

%% file: content/4-maptrix/4-conclusion.tex
\section{Conclusion}
\label{sec:maptrix:conclusion}
In this chapter, we discussed two user studies of visual representations of many-to-many flows. In our first study we compared MapTrix with a flow map with bundled edges and with OD Maps for different country maps: Australia (AU – 8 locations), New Zealand (NZ - 16 locations) and Germany (DE – 16 locations). All three visualisations performed well for the smallest data set (AU), but MapTrix and OD Maps were far better for DE and NZ. There was no statistically significant difference between MapTrix and OD Maps on data sets of this size. Surprisingly, we did not find that country shape affected performance: in particular we had expected this to affect OD Maps. In our second study we compared MapTrix and OD Maps on two larger data sets China (CN - 34 locations) and the United States (US - 51 locations). Both performed relatively well for all tasks and we did not find that one method outperformed the other even for individual tasks. We did find in the pilot that analysing flow between or within regions for data sets of this size was extremely difficult with both methods, though slightly easier with OD Maps. Thus, in the study we used highlighting to help with analysis of regional flow. In the first study, users ranked MapTrix highest in terms of design and readability while in the second study MapTrix is preferred for design but OD Maps for readability.

Our user studies concentrate on static visual representations of dense many-to-many flows. However, in our second user study, we did explore the usefulness of highlighting for analysis of regional flow. The results of our studies led us to implement several types of interaction, not only highlighting but also filtering and region zooming. 

Preliminary results from our study indicate that flow maps have a more serious scalability issue than OD Maps and MapTrix. However, flow maps have other advantages.  Some of our study participants commented that flow maps were more intuitive and, as we found from our interviews with experts in Chapter~\ref{chapter:interviews}, flow maps are widely used in real-world analysis by domain specialists. Hence, there is a strong motivation to investigate how we can improve the scalability of flow map approaches. One way to do this might be using emerging display and interaction technologies.

As discussed in Section~\ref{sec:related:immersive}, developing display and interaction technologies brings many benefits to visualising data in immersive environments. We are keen to see how we can improve flow maps in such environments. However, before we start exploring the design space of flow maps in immersive environments, we recognised that the representation of a geographic reference, i.e. the underlying map, had not been well explored in such environments previously. In the next chapter, we explore different ways to present geographic reference space in immersive environments.

%% file: content/5-maps-and-globes-in-vr/0-index.tex
\chapter[Maps and Globes in Immersive Environments]{Maps and Globes in \\ Immersive Environments}
\label{chapter:maps-globes-vr}

\begin{figure}[ht!]
    \captionsetup[subfigure]{justification=centering}
    \centering
    \begin{subfigure}{0.4\textwidth}
    	\label{subfig:vr-maps-globes:exo-globe-teaser}
        \includegraphics[width=\textwidth]{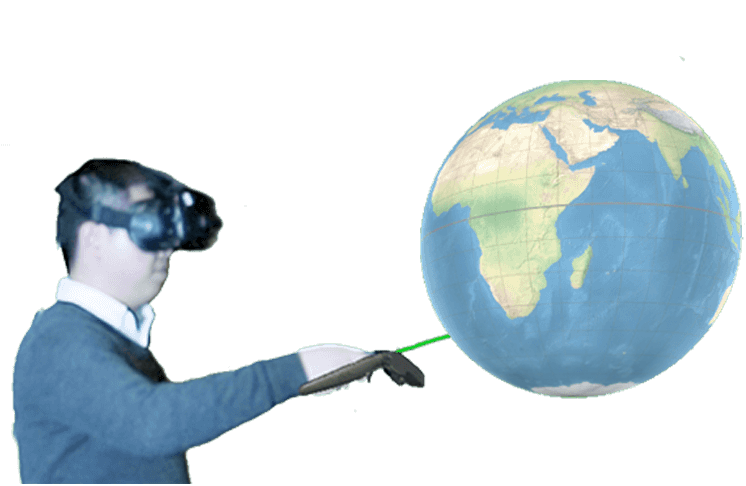}
        \caption{Exocentric globe}
    \end{subfigure}
    \begin{subfigure}{0.4\textwidth}
    	\label{subfig:vr-maps-globes:flat-map-teaser}
        \includegraphics[width=\textwidth]{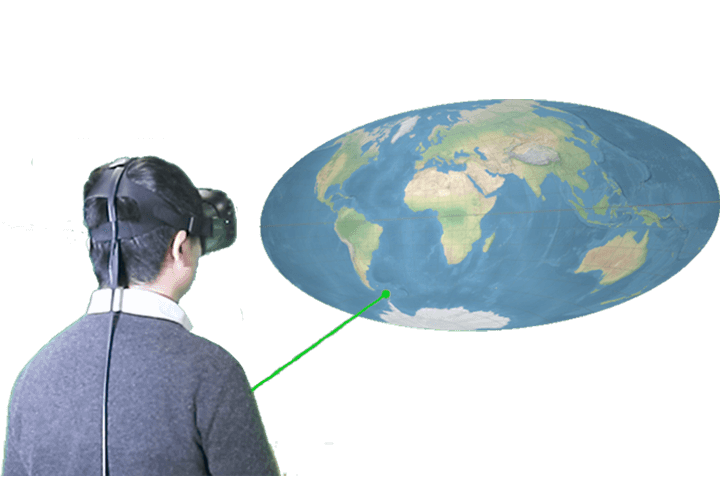}
        \caption{Flat map}
    \end{subfigure}

    \begin{subfigure}{0.4\textwidth}
    	\label{subfig:vr-maps-globes:ego-globe-teaser}
        \includegraphics[width=\textwidth]{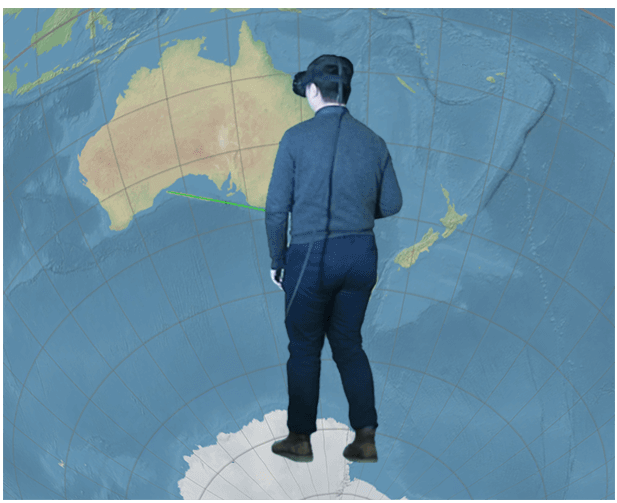}
        \caption{Egocentric globe}
    \end{subfigure}
    \begin{subfigure}{0.4\textwidth}
    	\label{subfig:vr-maps-globes:curved-map-teaser}
    	\vspace{2em}
        \includegraphics[width=\textwidth]{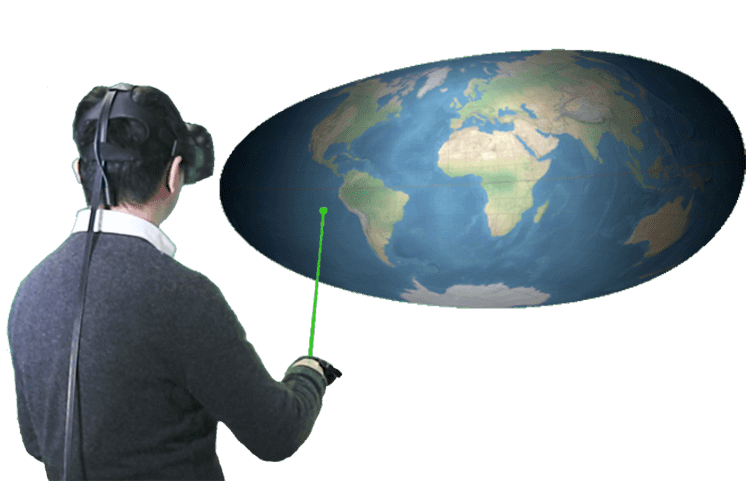}
        \caption{Curved map}
    \end{subfigure}

    \caption{Four interactive visualisations for geographic data in Virtual Reality (VR).}
    \label{fig:vr-maps-globes-teaser}
\end{figure}

In the previous two chapters, we explored the design space of visualising OD flow data in 2D. We also evaluated three different 2D OD flow visualisations with controlled user studies. Flow maps are intuitive and widely used but suffered from scalability issues. Therefore, we decide to look at how we can improve flows maps in immersive environments.

In this chapter, we take our first step into the exploration of flow map representation in immersive environments. We investigate different ways to render the underlying geographic reference space in such environments.

We begin by examining the history of using maps and globes and we reflect upon how both 2D and 3D representations have long-standing precedents. We then consider the design space for geographic representation in immersive environments and the opportunities these afford (such as reduced distortion of the map) as well as the problems introduced by faithful globe representation (such as only being able to see part of the globe from a single point of view).  Then we move to our realisation of several points from this design space in the form of four interactive systems using different representations of the world-wide geographic reference space in immersive environments. We then report our user study of evaluating these representations for three types of geographic tasks. Finally, we provide guidelines for presenting global geography in immersive environments.

The work presented in this chapter is a collaborative effort with my supervisors and collaborators: Bernhard Jenny, Tim Dwyer, Kim Marriott, Haohui Chen and Maxime Cordeil. This work was accepted and presented at Eurographics Conference on Visualization (EuroVis) 2018. In this work, I collaboratively formulated the hypothesis and research questions, was solely responsible for the implementation and user studies, did the data analysis, and collaboratively wrote the research paper.

\input{content/5-maps-and-globes-in-vr/1-introduction}

\input{content/5-maps-and-globes-in-vr/2-earth-vis} 
\input{content/5-maps-and-globes-in-vr/3-user-study}    
\input{content/5-maps-and-globes-in-vr/4-results}

\input{content/5-maps-and-globes-in-vr/5-discussion}

\input{content/5-maps-and-globes-in-vr/6-conclusion}

%% file: content/5-maps-and-globes-in-vr/1-introduction.tex
\section{Introduction}
\label{sec:vr-map-globe:intro}
\added{To investigate OD flow maps in immersive environments systematically, we follow~\citep{Dubel:2015el} to separate the design space of flow maps into: the geographic reference and the flows. In this chapter, we explore the design choices for the geographic reference.}

Maps and globes are widely used to visualise geographic data. They underpin how we understand the world and our place on it.  They are the foundation for geovisualisation and thematic cartography, in which qualitative and quantitative data with a spatial aspect is overlaid on its geographic location~\citep{slocum2009thematic}. Geovisualisation is widely used to understand both social and physical data, for instance  analysing census data~\citep{Martin:1989ii}, studying epidemiology~\citep{Moore:1999cc}, planning urban transportation policies~\citep{Arampatzis:2004hw}, and exploring animal migration patterns~\citep{Kolzsch:2013vq,Slingsby:2017vc}. 

Terrestrial maps and globes were used by the Ancient Greeks more than two thousand years ago. During the Renaissance matched pairs of celestial and terrestrial globes were the main tool for teaching geography and cosmology~\citep{dekker2007globes}, but in subsequent centuries the use of globes declined as they are more expensive to produce than maps, much bulkier to store, and do not scale. Additionally, from the Renaissance onwards many sophisticated map projections were invented~\citep{snyder1997flattening} that at least partly overcame the great disadvantage of a map: that it is not possible to draw the surface of the earth on a 2D surface without significant spatial distortion.

In the 21st century, however, the globe has made a remarkable comeback. Virtual globes have become familiar to most VR users, e.g.\ ~\citep{GoogleEarthVR}. With the arrival of commodity head-mounted displays (HMDs) for VR (e.g. HTC Vive) and also augmented- and mixed-reality (AR and MR, e.g. Microsoft Hololens, Meta2 and so on), we can expect to see more virtual globes being used in geographic visualisation applications. MR in particular, has great appeal in situated analytics scenarios~\citep{ElSayed:2015ji} where visualisations are made available in challenging situations (in the field, surgery, or factory floor), and also for collaborative visualisation scenarios, where two or more people wearing HMDs can each see and interact with the globe, while still seeing each other directly for communication~\citep{Cordeil:2016io}. 

A key question, however, is whether virtual globes are the best way to show global geographic data in immersive environments or whether maps or some other visualisations may be better. Surprisingly, given the fundamental importance of this question for the design of geovisualisation applications in VR and mixed-reality, it has not been formally tested previously. This chapter fills this gap by making two main contributions:

First, in Sec.~\ref{sec:vr-map-globe:vis} we present four different interactive visualisations of the earth's geography designed for use in VR for head mounted displays (HMD). The first two are well-known: the \textit{exocentric globe} (Fig.~\ref{fig:vr-maps-globes-teaser}(a)) and the \textit{flat map} (Fig.~\ref{fig:vr-maps-globes-teaser}(b)). The other two are more novel. The standard globe is an \emph{exocentric} visualisation such that the viewer stands outside the globe.  An alternative approach is to place the viewer inside the globe~\cite{Zhang:2016}. With our \emph{egocentric globe} (Fig.~\ref{fig:vr-maps-globes-teaser}(c)) the viewer sees a map projected onto a surrounding 360$^\circ$ sphere. One possible advantage of the egocentric sphere is that, if the viewpoint is close to the centre of the sphere, the inside surface of the globe is a constant distance away, reducing perceptual distortion. Our fourth visualisation is a novel VR visualisation we call the \emph{curved map} (Fig.~\ref{fig:vr-maps-globes-teaser}(d)). This is a map projected onto a section of a sphere. The viewer faces the concave side of the map, so again a possible advantage is that the distance from the viewpoint to the surface of the map is relatively constant resulting in reduced perceptual distortion. 
All four visualisations support the same basic interaction, detailed in Section \ref{sec:vr-map-globe:vis:interaction}.  The user can interactively move any geographic location to the centre of the view, an interaction we found to be essential to enable the tasks tested in our study. Users can also change their viewpoint through headtracked motion standard to modern VR.

Our second main contribution is a controlled study with HMD in VR investigating user preferences and the efficacy (accuracy and time) of these four different interactive visualisations (Sec.~\ref{sec:vr-map-globe:study}). We evaluate three fundamental spatial analysis tasks: distance and area comparison, as well as estimation of orientation between two locations.  We also analyze physical movement and user interaction with the visualisations.  

The ego- and exo-centric globes and curved maps are naturally 3D visualisations, while the flat map view is a 2D visualisation. 
Testing across different devices (e.g.\ flat map on regular screen versus 3D visualisations in VR) would introduce a large number of variables into the evaluation.  For example: resolution; head-tracked vs non-headtracked interaction; comfort of headset; and so on -- all of which are purely a function of the limitations of the current (rapidly developing) technology.  
Thus, we evaluate all four visualisations in VR such that these variables are eliminated.  Rather, we can focus on the geometry of the geographic surface over differences between devices. 

The results of our study (Sec.~\ref{sec:vr-map-globe:discussion}) show that the exocentric globe is the best choice for most of the tested tasks.  This surprised us, since in such a display only half of the geographic surface is visible to the user at any time.  Further, it has the most perceptual distortion (Sec.~\ref{sec:vr-map-globe:vis}), though less distortion due to projection than the flat and curved map. 
Though not as effective overall, for some tasks the flat map and curved map also perform well and were preferred by some participants.  
This result motivated a prototype implementation of a novel interaction for smoothly transitioning from exocentric globe to flat map and back (Sec.~\ref{sec:vr-map-globe:conclusion}).

%% file: content/5-maps-and-globes-in-vr/2-earth-vis.tex
\section{Showing the Earth in VR}
\label{sec:vr-map-globe:vis}
In this section we describe the four geospatial VR visualisations that we tested in the user study. We  compare the different visualisations in terms of the amount of distortion introduced by projecting the earth's surface onto the visualisation surface, the amount of perceptual distortion, and also the amount of the earth's surface that is visible to the viewer (Tab.~\ref{tab:vr-maps-globes:comparison}). 
Perceptual distortion arises because the viewer essentially sees the VR image as a 2D image projected onto each eye and the visual system must reconstruct the position of elements on the surface using depth cues such as linear perspective, texture gradient, etc.~\citep{Ware:2012ej}. The degree of perceptual distortion depends upon the relative position of the viewer and representation. It is also influenced by the depth cues provided by the VR environment. All visualisations were implemented in the Unity3D engine for the HTC Vive headset. The HTC Vive provides head-tracked stereoscopic VR and we used the Unity3D engine to provide linear perspective, texture gradient and shadows from a light source placed above and behind the viewer. 

\begin{table}[b!]
	\centering
	\begin{tabular}{l|l|l|l|}
\cline{2-4}
& \multicolumn{2}{l|}{Amount of distortion} & \multirow{2}{*}{\begin{tabular}[c]{@{}l@{}}Approximate area in field of view\end{tabular}} \\ \cline{2-3}
& Projection& Perceptual& \\ \hline
\multicolumn{1}{|l|}{Exocentric Globe} & None & High on edge & Hemisphere \\ \hline
\multicolumn{1}{|l|}{Flat Map}         & High & Medium       & Entire sphere \\ \hline
\multicolumn{1}{|l|}{Egocentric Globe} & None & Low          & Hemisphere \\ \hline
\multicolumn{1}{|l|}{Curved Map}       & High & Very low     & Entire sphere  \\ \hline
\end{tabular}
	\vspace{0.5em}
	\caption{Estimate of distortion and field of view of the four visualisations. Perceptual distortion and field of view are estimated for the initial position of the viewer.}
	\label{tab:vr-maps-globes:comparison}
\end{table}

\subsection{Exocentric Globe}
A direct three-dimensional rendering of the spherical model (Fig.~\ref{fig:vr-maps-globes-teaser}(a)). Our exocentric globe has an invariable radius of 0.4 metres. The globe is initially positioned 1 metre in front of  the user at head height.
Because of occlusion the user can see at most one hemisphere. There is no distortion due to projecting the earth's surface onto the surface of the globe, but there is areal and angular distortion along the edges of the visible hemisphere.

\subsection{Flat Map}
An elliptical projection is texture-mapped onto a quad measuring 1$\times$0.5 metres and is placed 1 metre from the user (Fig.~\ref{fig:vr-maps-globes-teaser}(b)). For vector data, we use our own partial port of the D$^3$ library for spherical rotation, cutting, clipping and resampling \cite{Bostock:2013ix}. For raster images we use a GPU rendering technique \cite{Jenny:2015hf}.

We chose the \emph{Hammer} map projection, which preserves the relative size of areas. 
To reduce the distortion of shapes and also for aesthetic reasons, we chose to use an equal-area projection with an elliptical boundary. We did not to use a projection that shows the poles as lines, when the map centre can be adjusted by the user as this is potentially confusing. 
We preferred the Hammer to the area-preserving Mollweide projection for computational efficiency. In comparison to other projections for world maps, the Hammer projection adds small distortion to distances in the central area of the map.

The great advantage of a map over a globe is that the entire surface of the earth is visible. While the Hammer map projection does not distort the relative size of areas, it introduces angular distortion, which increases away from the center of the map. There is some perceptual distortion because the distance between the viewer and the display surface varies; perceptual distortion increases as the viewer nears the map. Foreshortening distortion results when the map is viewed with an oblique angle.

\subsection{Egocentric Globe}
In a design (Fig.~\ref{fig:vr-maps-globes-teaser}(c)) following Zhang \textit{et al}.~\citep{Zhang:2016}, we initially set the radius to 3 meters, and positioned the user at the centre of the globe.  However, pilot participants reported that they felt their field of view was limited when they are at the centre and wished to move to the edge of the globe to enlarge their field of view. So we adjusted the position of the globe to allow the participants to stand at the edge of the globe (80\% of radius away from the centre). To ensure that participants cannot walk out of the sphere, we increased the radius to 8 meters, which is larger than the walkable space.

\begin{figure}[b!]
	\centering
	\includegraphics[width=\textwidth]{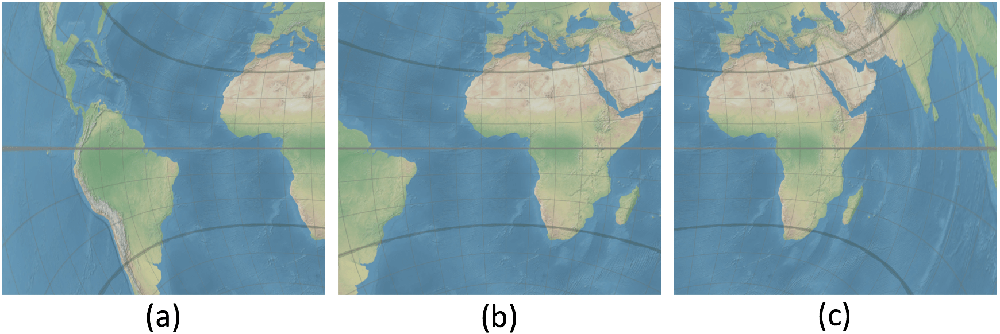}
	\caption{Distortion of an egocentric globe changes with user position: (b) view at the initial location; (a) view from left close to the sphere's hull; (c) close view from right. %
	}
	\label{fig:vr-maps-globes:lg-distortion}
\end{figure}

The motivation for this design is to create a visualisation for VR with a maximum immersive experience. Areas, distances and shapes are not distorted on the egocentric sphere model, as there is no map projection involved.  When the head is positioned at the center of the sphere, there is little perceptual distortion, but if the user moves closer to the sphere and views it under an oblique angle, there is considerable distortion (see Fig.~\ref{fig:vr-maps-globes:lg-distortion}). Slightly more than half of the sphere is visible to the user in the initial position.

In our pilot study, participants reported experiencing motion sickness in the egocentric globe. A stable external horizon reference is known to reduce motion sickness~\citep{Bos:2006vf}. Thus, we added two static rings (Fig.~\ref{fig:vr-maps-globes:lg-distortion}) to create a stable horizon. These rings align with two lines of constant latitude when the north pole is placed above the user. The rings remain in this static position when the user rotates the sphere. After adding this artificial horizon, participants reported that  dizziness was reduced and manageable.

\subsection{Curved Map}
We project an area-preserving map onto a spherical section (Fig.~\ref{fig:vr-maps-globes-teaser}(d)). The user stands at the centre of this sphere, which has a radius of 1 metre. The map covers a horizontal angle of 108$^\circ$ and a vertical angle of 54$^\circ$. Using a section of a sphere as a projective surface is not a new idea in VR, e.g., Kwon \textit{et al}.~\citep{Kwon:2016go} use spherical projection surfaces for graph analysis. However, these previous applications did not use geospatial data or geographic maps.

The motivation for the curved map is to:
\begin{itemize}
 	\item create a more immersive experience than with flat maps;
 	\item allow the user to view the entire map with minimum head or eye movements;
 	\item reduce perceptual distortion created by an oblique viewing angle.
 \end{itemize}

Our curved map is rendered with an additional ramp shader with a gradient texture to create the impression of a concave surface. As for the flat map, we use the Hammer projection when transforming the sphere to an initial flat map. We then apply a second transformation, which linearly maps the Hammer map onto the sphere. The resulting curved map does not preserve the relative size of areas. 

\subsection{Interactions}
\label{sec:vr-map-globe:vis:interaction}
\begin{figure}[t!]
	\centering
	\includegraphics[width=\textwidth]{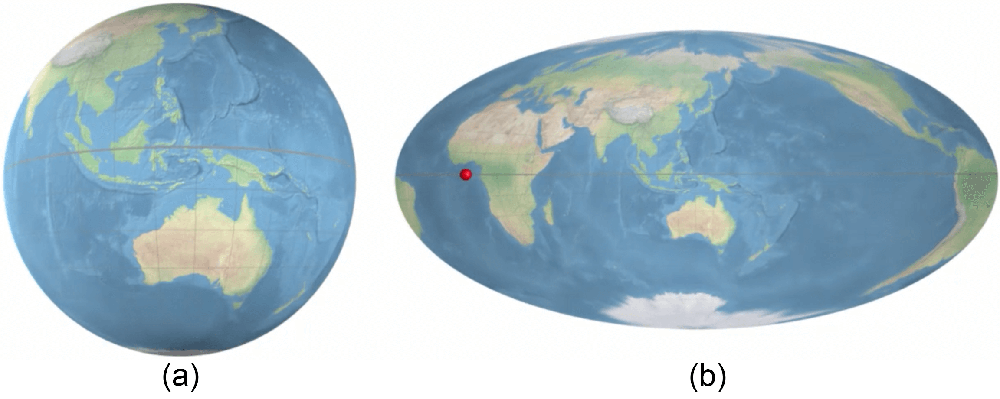}
	\caption{(a)Spherical rotation applied to change the central meridian of a globe; (b) project the rotated globe, i.e. (a), to a flat map (red dot presents 0$^\circ$ longitude and 0$^\circ$ latitude).
	}
	\label{fig:vr-maps-globes:geo-ratation}
\end{figure}

We provide similar interaction across the four visualisations.
First, the VR model is fixed in space allowing the viewer to approach and move around it. 
Second, we allowed the viewer to adjust the centre of the geographic area in the visualisation. They could pick any location and drag it to a new position using a standard VR controller. The geographic location smoothly follows the beam that is sent from the controller and intersects the map or globe surface. A spherical rotation is applied to the geometry model before the model is projected (for the flat and curved maps) or rendered (for the exocentric and egocentric globes)~\citep{Bostock:2013ix,Davies:2013ug} (see Fig.~\ref{fig:vr-maps-globes:geo-ratation}). With this interactive adjustment of the map centre, users can fine-tune the visualisations such that the area of interest is displayed at the centre of the map or globe.

Adjusting the geographic centre allows for:
\begin{itemize}
	\item bringing features of interest to the centre of the visualisation;
	\item reducing distortion in shape of features of interest;
	\item avoiding the path or area of interest being split into two by the edge of the maps or out of view with the globes.
\end{itemize}

A degree of zooming interaction is possible by the viewer physically moving closer to the surface of the map. For simplicity, the experiment was designed such that more extreme zooming was unnecessary. That is, all targets were sufficiently visible at the natural view position.  While other zoom interactions were not included in this study, it is worth noting that they would be required to support tasks where multiple view scales were necessary, e.g.\ moving from country to city scale.

%% file: content/5-maps-and-globes-in-vr/3-user-study.tex
\section{User Study}
\label{sec:vr-map-globe:study}
Following~\cite{Battersby:2009jo,Friedman:2000ux,Friedman:2006ux,Carbon:2010bo} we evaluated four visualisations with three tasks essential to geospatial data visualisation: distance comparison, area comparison, and direction estimation. These tasks relate to real-world analysis scenarios, for example: area comparison is used to analyse the global forest cover change \cite{Hansen:2013iy}, global air quality monitoring \cite{Chu:2003ca}; distance comparison and direction estimation are used to help analyse the global movement of ocean animals \cite{Block:2011be}, cargo ship movements\cite{Kaluza:2010hb}, and air transportation networks \cite{Guimera:2005dm}.
 
\subsection{Stimuli and tasks}
In order to rule out the influence of previous geographic knowledge, we avoided basing tasks on real-world geographic features, such as comparing the size of two existing countries, and used artificial distances and areas instead.

\vspace{-1.5em}
\subsubsection{Distance Comparison Task}\vspace{-2em}
Given two pairs of points, find the pair separated by the largest (spherical great circle) distance. 

Following Feiner \emph{et al.}~\cite{Feiner:1993ip}, we use leader lines to link the points with labels: ``A'', ``B'' for one pair and ``X'', ``Y''  for the other. 
Labels were kept horizontal and oriented towards the viewer via rotation in real time. 
Two factors that may affect user performance:
\emph{Variation} of the relative distance between each pair of points;
\emph{Geographic distance} between geographic midpoints of the two pairs.

The coefficient of variation (CV)~\cite{Brown:1998dj} was used to measure the variation between the distances of two pairs. We designed three groups of tasks with different difficulty levels:
\begin{itemize}
	\item \emph{easy}, with large variation in the pairwise distances and close geographic distance between the pairs;
	\item \emph{small variation} between the pairwise distances and close geographic distance between the pairs;
	\item \emph{far distance} with large variation in the pairwise distance and far geographic distance between the pairs.
\end{itemize}
In our pilot testing, far distance with small variation was found to be too difficult.

Initially, 20\% and 10\% CV were used as large and small variations, however, the resulting error rates were very low in our first pilot study. We therefore adjusted the CV values to 10\% and 5\%, respectively.
For the geographic distance, we chose 60$^\circ$ for the short distance, which approximately spans the size of a continent, and 120$^\circ$ was chosen for the long distance.
Two pairs of points were randomly generated. The distance between a pair of points is restricted to lie  in the range of $40^\circ$-$~60^\circ$.

\vspace{-1.5em}
\subsubsection{Area Comparison Task}\vspace{-2em}
Given two labeled polygons, identify the polygon with the larger area. Steradian (spherical area) was used to calculate the reference areas of polygons. The  leader line system was again used to link the centroids of the polygons with labels.

As with the distance comparison task, user performance may be affected by: \emph{variation} between the the areas of the polygons; \emph{geographic distance} between the centroids of polygons.

A convex polygon was created by linking eight random points generated with the same geographic distance (8$^\circ$) from a centroid. To ensure the generated polygons are in similar shape, the minimum central angle between two adjacent points was 30$^\circ$. To avoid the effect of color on area perception, the two polygons were placed on a similar background (either both in the sea or both on land).

\vspace{-1.5em}
\subsubsection{Direction Estimation Task}\vspace{-2em}
Given a short arrow, estimate whether or not the path continuation passes through a given location. 

The main factor that might affect user performance is
\emph{geographic distance} between the two locations.
Again, 60$^\circ$ and 120$^\circ$ were used to distinguish the geographic distance. Two groups with different difficulty levels were designed for direction estimation tasks.

\subsection{Measures: response time, accuracy and interactions}\
The response time was the interval between the initial rendering of the visualisation and the double-click of the trigger button on the controller. After double-click, the visualisation was hidden
and two answer options shown. As our tasks provide binary options we used the accuracy score from \cite{Willett:2015fv}: $(\frac{number\ of\ correct\ responses}{number\ of\ total\ responses} - 0.5) / 0.5$, where 1 indicates a perfect performance and 0 indicates a result equal to chance (i.e. randomly guessing).

We also measured the degree of interaction. The number of clicks has been widely used in conventional interaction evaluations, as well as in VR \cite{Kwon:2016go}. However, unlike those systems, the HTC Vive allows users to move in a larger open space, thus, not only the  number of interactions, but also the user's movements need to be considered. In our experiment, the positions and rotations of the user's head and the controller were recorded every 0.1s while the participants were viewing the visualisations. We used the aggregated changes in positions (in euclidean distance) and rotations (in degrees) between records to analyse user interactions.

\subsection{Experimental Setup}
\begin{figure}[t!]
	\centering
	\includegraphics[width=\textwidth]{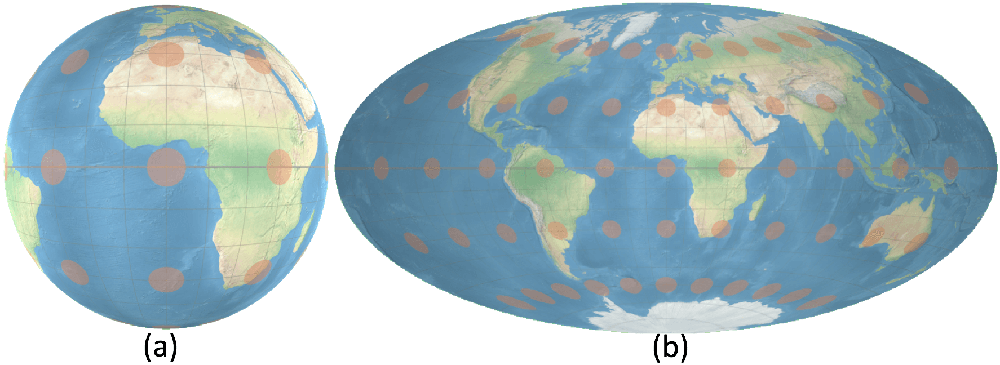}
	\caption{Tissot indicatrices on the Natural Earth raster map with a graticule, (a) on an exocentric globe, (b) on a flat map.}
	\label{fig:vr-map-globe:tissot}
\end{figure}

An HTC Vive with 110$^\circ$ field of view and 90Hz refresh rate~\citep{Dempsey:2016hn} was used as the VR headset in the experiment. The PC was equipped with an NVIDIA GeForce GTX 1080 graphics card and Intel i7-6700K 4.0GHz processor. Only one controller was needed in the experiment; a pointer from the controller (a shooting beam) was available at all times. We configured all interactions to work with the controller trigger. Participants could hold the trigger and manipulate the geographic centre.  The framerate was around 110FPS throughout the experiment, i.e.\ computation was faster than the display refresh rate. 
The Natural Earth raster map from \url{naturalearthdata.com} was used as the base texture for all tasks. A graticule was created for every 10$^\circ$ both in longitude and latitude. A thicker line was used for the equator. Tissot indicatrices (discussed below) shown in the training were rendered every 30$^\circ$ longitude and latitude (Fig.~\ref{fig:vr-map-globe:tissot}). The initial position of the globes and maps for each question was adjusted to ensure the user was looking at 0$^\circ$ longitude and 0$^\circ$ latitude.

\subsection{Participants}
We recruited 32 participants (11 female) from our university campus, all with normal or corrected-to-normal vision. Participants included university students and researchers. 
22 participants were within the age group 20--30, 7 participants were between 30--40, and 3 participants were over 40. VR experience varied:  25 participants had less than 5 hours of prior VR experience, 6 participants had 6--20 hours, and 1 participant had more than 20 hours. 

\subsection{Design}
The experiment was within-subjects: 32 participants $\times$ 4 visualisations $\times$ 3 tasks $\times$ 9 repetitions = 3,456 questions (108 questions per participant) with performance measures and lasted one hour on average. Each of the three difficulty conditions for distance and area comparison tasks was repeated 3 times. For direction estimation, we trialled 5 repetitions for the two difficulty conditions in the pilot study. Pilot participants reported the direction estimation for long distance was too difficult. We modified the repetition of the direction estimation to 3 repetitions for far distance and 6 for close distance. 
The mapping between tasks and techniques was counterbalanced across subjects by keeping the order of tasks constant (in the order of distance comparison, area comparison, and direction estimation) and using a Latin square design to balance the order of visualisations.

\subsection{Procedure}
Participants were first given a brief introduction about this project, the four types of visualisations, and the three types of tasks. Two types of training were included in this experiment: interaction training and task training.

\emph{Interaction training} was conducted when each visualisation was presented to the participants for the first time. The participants were introduced to the interactive visualisations and given sufficient time to familiarise themselves with the interaction. They were then asked to use the interaction to move Melbourne to the centre of their view and double-click on it. This activity familarised participants with the VR headset and controller, as well as each interactive visualisation.

\emph{Task training} was conducted when each condition (task $\times$ visualisation) was presented to the participants for the first time. Two sample tasks, different from the experimental tasks, were given to participants with unlimited time. After participants finished a training task, we highlighted the correct answer and presented additional geographic information related to the task when applicable: for distance comparison tasks, lines connecting each pair showing actual geographic distances were shown; for direction estimation tasks, actual geographic trajectories were shown. We reminded the participants to test their strategies both when they were doing the training tasks and when the correct answers were shown. For the two training direction estimation tasks, participants were presented with one ``hit''  and one ``miss'' condition. 

Participants were asked to finish the different types of tasks one by one. Within one task, they were presented the four visualisations one after the other. They were not explicitly informed about the area and angular distortion that can be caused by map projections, but during the training we displayed Tissot indicatrices (Fig.~\ref{fig:vr-map-globe:tissot}) to show scale variation and angular distortion on all four visualisations. A Tissot indicatrix appears as a circle on maps without angular distortion, and appears as an ellipse on maps with angular distortion. The size of Tissot indicatrices changes with area distortion, resulting in larger ellipses where area is inflated, and smaller ellipses where area is compressed \cite{Snyder:1987tk}. Tissot indicatrices were not shown during the experimental tasks.

A posthoc questionaire recorded feedback on:
\begin{itemize}
	\item preference ranking of visualisations in terms of visual design and ease of use for the experimental tasks;
	\item experience of motion sickness in the different visualisations;
	\item advantages and disadvantages for each visualisation;
	\item strategies for different types of tasks;
	\item background information about the participant.
\end{itemize}

%% file: content/5-maps-and-globes-in-vr/4-results.tex
\section{Results}
\label{sec:vr-map-globe:results}
Histograms and Q$-$Q plots revealed that the error rate distribution was not normal. As there were more than two conditions, we used the Friedman test to check for significance and applied the Wilcoxon-Nemenyi-McDonald-Thompson post-hoc test to conduct pairwise  comparisons~\citep{Hollander:1999ns}. 
To compare response time and user interactions we considered only times and interactions for correct responses. Histograms and Q$-$Q plots showed both distributions were approximately normal. Due to the unbalanced number of correct answers per participant, we chose linear mixed-effects (LME) ANOVA to check for significance and applied Tukey's HSD post-hoc tests to conduct pairwise comparisons~\citep{McCulloch:2013kl,Pinheiro:2000id}.
For user preferences and motion sickness ratings we again used the Friedman test and the Wilcoxon-Nemenyi-McDonald-Thompson post-hoc test to test for significance. 

\subsection{Distance Comparison}
\begin{figure}[b!]
	\centering
	\includegraphics[height=6cm]{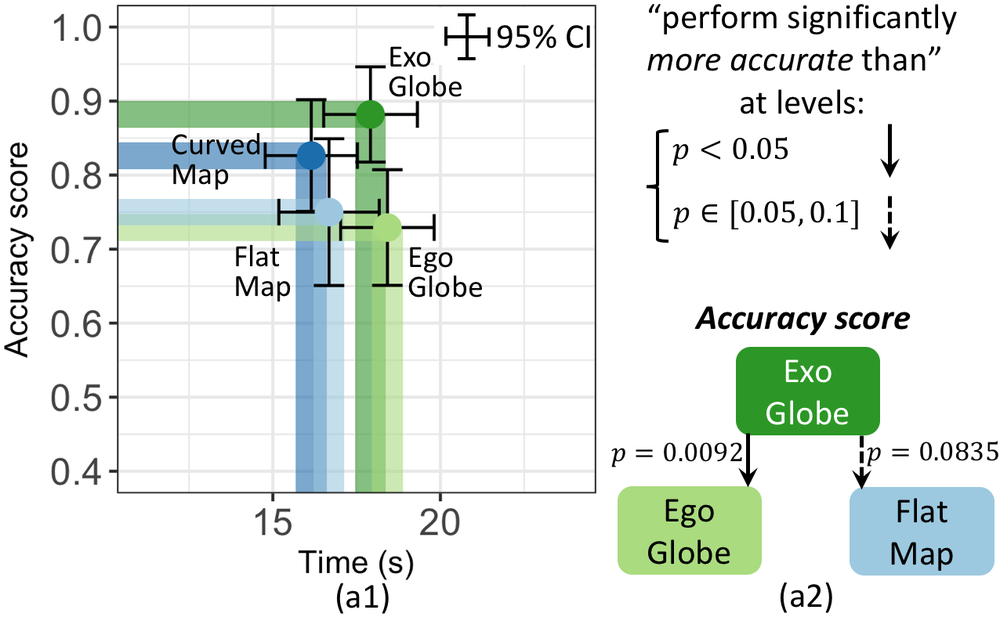}
	\caption{(a1) Average performance of \textbf{distance comparison} task per visualisation, with 95\% confidence interval, (a2) graphical depiction of results of pairwise post-hoc test.}
	\label{fig:vr-maps-globes:distance-vis}
\end{figure}

The Friedman test revealed a statistically significant effect of visualisations on accuracy ($\chi^2(3) = 11.453, p = .0095$). Fig.~\ref{fig:vr-maps-globes:distance-vis}(a1) shows the average accuracy score of exocentric globe (0.88) was higher than that of egocentric globe (0.73) and flat map (0.75). While curved map (0.83) outperformed egocentric globe and flat map, this was not found to be statistically significant. A post-hoc test showed statistical significances as per Fig.~\ref{fig:vr-maps-globes:distance-vis}(a2). 

The LME ANOVA analysis showed significant effect of visualisations on time ($\chi^2(3) = 6.837, p = .0773$). Fig.~\ref{fig:vr-maps-globes:distance-vis}(a1) shows avg.\ response times with curved map (16.1s) and flat map (16.7s) were less than the avg.\ response times for exocentric (17.9s) and egocentric globe (18.4s), however, the post-hoc test did not find statistical significance.

\begin{figure}[b!]
	\centering
	\includegraphics[height=18.5cm]{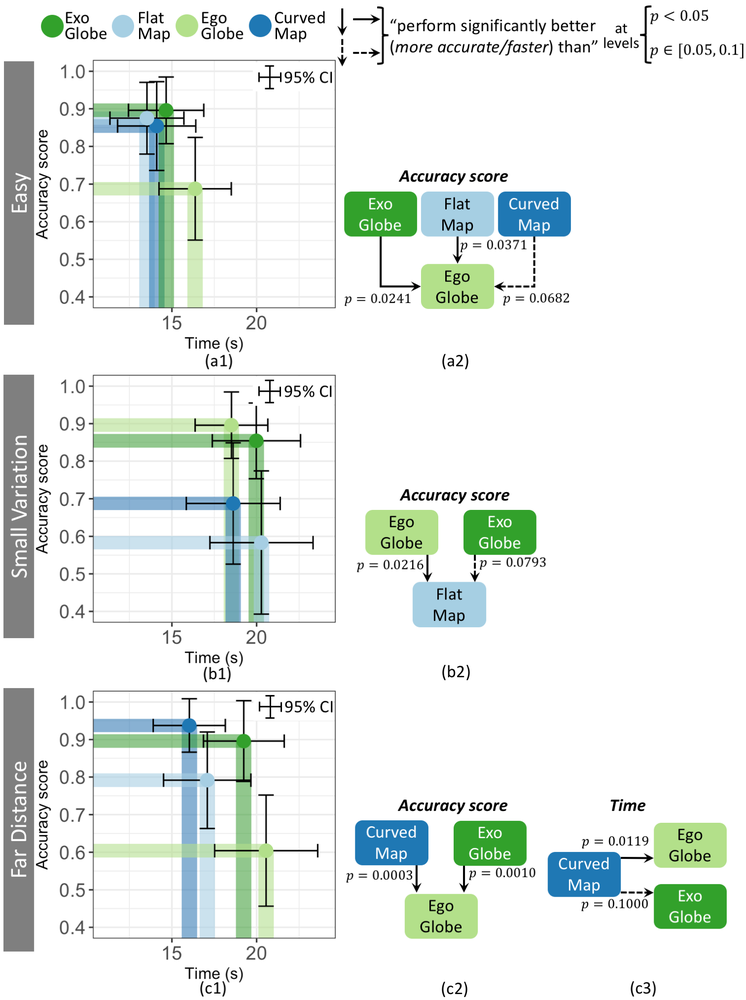}
	\caption{Break down of \textbf{distance comparison} task into different difficulty conditions. (a1,b1,c1) Average performance per visualisation with 95\% confidence interval, (a2,b2,c2,c3) graphical depiction of result of pairwise post-hoc test.}
	\label{fig:vr-maps-globes:distance-vis-break}
\end{figure}

By difficulty condition the Friedman test revealed significant effect for \textbf{avg.\ accuracy}:\\
\noindent\textbf{\textit{Easy}:}
$\chi^2(3)=10.711, p = .0134$. Fig.~\ref{fig:vr-maps-globes:distance-vis-break}(a1) shows avg.\ accuracy of egocentric globe (0.69) $<$ exocentric globe (0.90), flat map (0.88), and curved map (0.86). Significance in post-hoc testing detailed in Fig.~\ref{fig:vr-maps-globes:distance-vis-break}(a2).\\
\noindent\textbf{\emph{Small variation}:} 
$\chi^2(3)=10.938, p = .0120$. Fig.~\ref{fig:vr-maps-globes:distance-vis-break}(b1): avg.\ accuracy of flat map (0.58) $<$ egocentric globe (0.90) and exocentric globe (0.86), post-hoc significance Fig.~\ref{fig:vr-maps-globes:distance-vis-break}(b2)). Curved map accuracy (0.69) is similar to flat map.\\
\noindent\textbf{\textit{Far distance}:} 
$\chi^2(3)=20.131, p = .0001$. Fig.~\ref{fig:vr-maps-globes:distance-vis-break}(c1): avg.\ accuracy of egocentric globe (0.60) $<$ curved map (0.94) and exocentric globe (0.90). Post-hoc testing indicated statistical significance (see Fig.~\ref{fig:vr-maps-globes:distance-vis-break}(c2)). Avg.\ accuracy of flat map (0.79) is higher than that of the egocentric globe, however, the post-hoc test did not find statistical significance.

\noindent For \textbf{avg. response time}, LME ANOVA analysis revealed:\\
\noindent\textbf{\textit{Easy}:} 
no significance ($\chi^2(3) = 4.903, p = .179$). Although in Fig.~\ref{fig:vr-maps-globes:distance-vis-break}(a1), egocentric globe (16.3s) tended to be slower than others (flat map 13.5s, exocentric globe 14.7s and curved map 14.1s). \\
\noindent\textbf{\textit{Small variation}:} 
no significance ($\chi^2(3) = 1.147, p = .765$).\\
\noindent\textbf{\textit{Far distance}:}
a significant effect ($\chi^2(3) = 10.451, p = .0151$). Fig.~\ref{fig:vr-maps-globes:distance-vis-break}(c1) shows curved map (16.0s) was faster than either exocentric globe (19.2s) or egocentric globe (20.6s). Post-hoc: Fig.~\ref{fig:vr-maps-globes:distance-vis-break}(c2)). Flat map response time (17.1s) was similar to curved map.

\noindent\textbf{\textit{Participant strategies}:} From the questionnaires, we found two general strategies for distance comparison with all visualisations:\\
\noindent\textit{ - Using the graticule grid to calculate distance} -- two explicitly mentioned using Manhattan distance;\\
\noindent\textit{ - Moving each pair in turn to the centre of the map or globe} -- usually  more than once and alternating between them.

\noindent There were also specific strategies for different visualisations:\\
\noindent\textit{ - For flat and curved maps:} moving the two pairs so they are placed symmetrically around the centre. ``It seems they will have the same scale''. This strategy was not usually possible in the globes because of the more limited field of view.\\
\noindent\textit{ - For exocentric and egocentric globes:} some participants used proprioception to estimate the distance between the points in each pair. They used interaction to rotate the sphere to move the points past a reference point (such as the center of the globe) and the  effort required for each pair gave an estimate of the relative distance. \\
\noindent\textit{ - For the exocentric globe:} use the top of the visible hemisphere as a reference point, in turn placing one point of each pair at this position and memorising the position of the other point in the pair, then switching to the other pair. Often they ensured  the pairs were vertically aligned.

\subsection{Area Comparison}
\begin{figure}[b!]
	\centering
	\includegraphics[height=6cm]{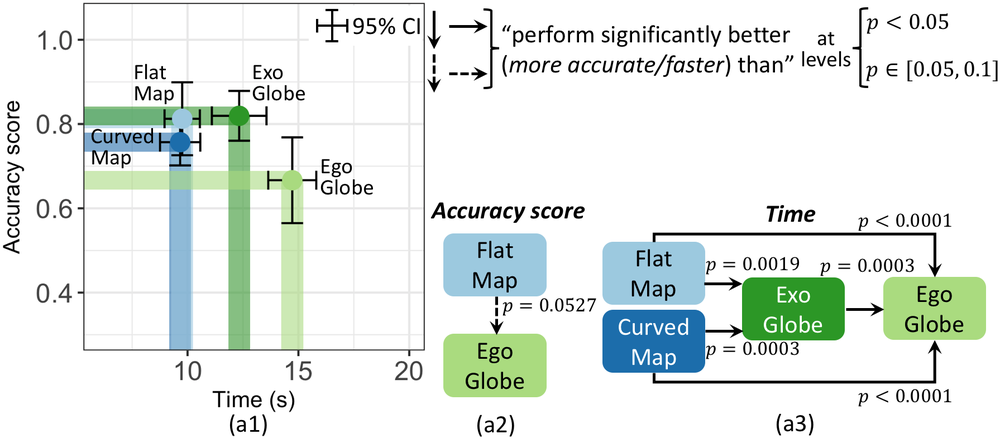}
	\caption{(a1) Average performance of \textbf{area comparison} task per visualisation, with 95\% confidence interval, (a2, a3) graphical depiction of results of pairwise post-hoc test.}
	\label{fig:vr-maps-globes:area-vis}
\end{figure}

The Friedman test revealed a statistically significant effect of visualisations on accuracy for the area comparison task ($\chi^2(3)=7.218, p = .0652$). Fig.~\ref{fig:vr-maps-globes:area-vis}(a1) shows the average accuracy score of egocentric globe (0.67) was lower than that of the others (exocentric globe 0.82, flat map 0.81, and curved map 0.76). The post-hoc test only shows a significant difference between the comparison of flat map and egocentric globe (see Fig.~\ref{fig:vr-maps-globes:area-vis}(a2)).

The LME ANOVA analysis revealed a statistically significant effect of visualisations on time ($\chi^2(3)=46.762, p < .0001$). Fig.~\ref{fig:vr-maps-globes:area-vis}(a1) shows the average response times with curved map (9.6s) and flat map (9.7s) were less than those of exocentric globe (12.3s) and egocentric globe (14.7s). The exocentric globe also performed significantly faster than egocentric globe. A post-hoc test showed statistical significances as per Fig.~\ref{fig:vr-maps-globes:area-vis}(a3). 

\begin{figure}[t!]
	\centering
	\includegraphics[height=18.5cm]{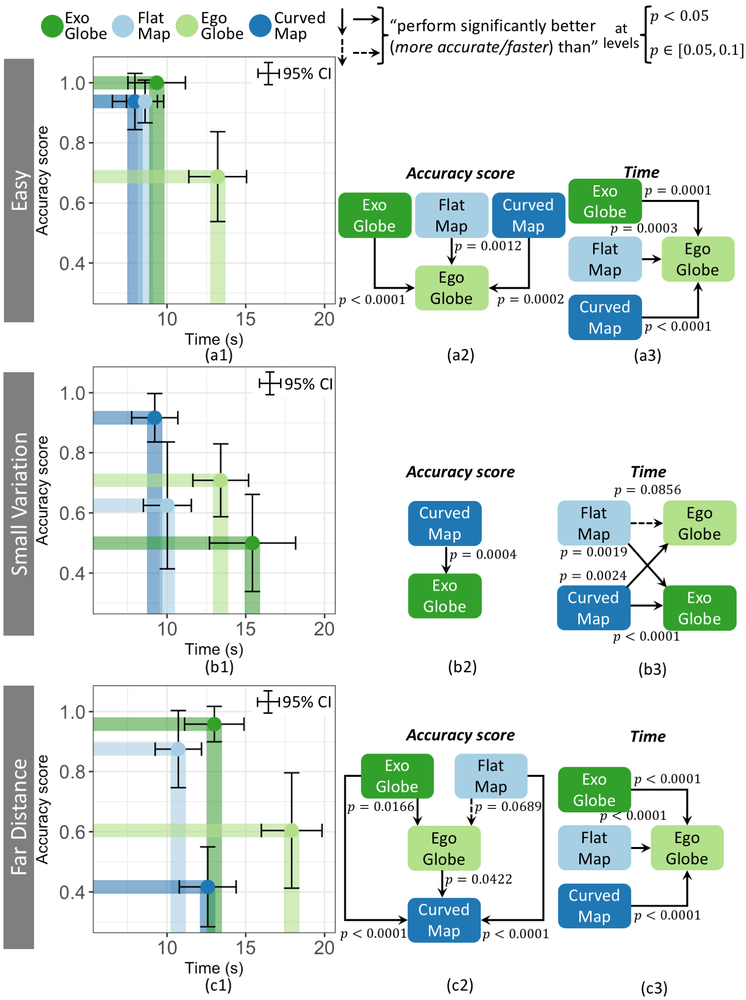}
	\caption{Break down of \textbf{area comparison} task into different difficulty conditions. (a1,b1,c1) Average performance per visualisation with 95\% confidence interval, (a2,a3,b2,b3,c2,c3) graphical depiction of results of pairwise post-hoc test.}
	\label{fig:vr-maps-globes:area-vis-break}
\end{figure}

By difficulty condition the Friedman test revealed significant effect for \textbf{avg. accuracy}:\\
\noindent\textbf{\textit{Easy}:} 
$\chi^2(3)=27.545, p < .0001$. Fig.~\ref{fig:vr-maps-globes:area-vis-break}(a1) shows the avg.\ accuracy score of egocentric globe (0.69) < the other visualisations (exocentric globe 1.00, flat map 0.94, and curved map 0.94). A post-hoc test showed statistical significances as per Fig.~\ref{fig:vr-maps-globes:area-vis-break}(a2). \\
\noindent\textbf{\textit{Small variation}:} 
$\chi^2(3)=15.451, p = .0014$. Fig.~\ref{fig:vr-maps-globes:area-vis-break}(b1) shows the avg.\ accuracy score of curved map (0.92) > the other visualisations (egocentric globe 0.71, flat map 0.63, and excocentric globe 0.50). The post-hoc test showed statistical significance for the difference between curved map and exocentric globe (see Fig.~\ref{fig:vr-maps-globes:area-vis-break}(b2)). \\
\noindent\textbf{\textit{Far distance}:} 
$\chi^2(3)=39.346, p < .0001$. Fig.~\ref{fig:vr-maps-globes:area-vis-break}(c1) shows the avg.\ accuracy scores with excocentric globe (0.96) and flat map (0.88) > egocentric globe (0.60) and curved map (0.42). The egocentric globe also performed more accurately than curved map. A post-hoc test showed statistical significances as per Fig.~\ref{fig:vr-maps-globes:area-vis-break}(c2).

\vspace{-1em}

\noindent LME ANOVA revealed significant effect for \textbf{avg. response time}:\\
\noindent\textbf{\textit{Easy}:}
$\chi^2(3)=31.269, p < .0001$. Fig.~\ref{fig:vr-maps-globes:area-vis-break}(a1) shows egocentric globe (13.2s) was slower than that of the other visualisations (curved map 7.9s, flat map 8.6s, and exocentric globe 9.3s). A post-hoc test showed statistical significances as per Fig.~\ref{fig:vr-maps-globes:area-vis-break}(a3). \\
\noindent\textbf{\textit{Small variation}:}
$\chi^2(3)=21.663, p < .0001$. Fig.~\ref{fig:vr-maps-globes:area-vis-break}(b1) shows curved map (9.2s) and flat map (10.0s) were faster than those of egocentric globe (13.4s) and exocentric globe (15.4s). A post-hoc test showed statistical significances as per Fig.~\ref{fig:vr-maps-globes:area-vis-break}(b3). \\
\noindent\textbf{\textit{Far distance}:}
$\chi^2(3)=36.511, p < .0001$. Fig.~\ref{fig:vr-maps-globes:area-vis-break}(c1) shows egocentric globe (17.91s) was slower than those of the other visualisations (flat map 10.7s, curved map 12.5s, and  exocentric globe 17.9s). A post-hoc test showed statistical significances as per Fig.~\ref{fig:vr-maps-globes:area-vis-break}(c3)).

\noindent\textbf{\textit{Participant strategies}:} 
From the questionnaire we found two general strategies for area comparison used for all visualisations \\
\noindent\textit{ - Using the graticule grid} to estimate the area. \\
\noindent\textit{ -  Moving each polygon in turn} to the centre of the map or globe, usually  more than once and alternating between them.

\noindent One specific strategy for flat and curved maps was identified:\\
\noindent\textit{ - For flat and curved maps}, moving the two polygons so they are placed symmetrically around the centre. This strategy was not usually possible in the exocentric globes and egocentric globes because of the more limited field of view.

\subsection{Direction Estimation}
\begin{figure}[t!]
	\centering
	\includegraphics[height=6cm]{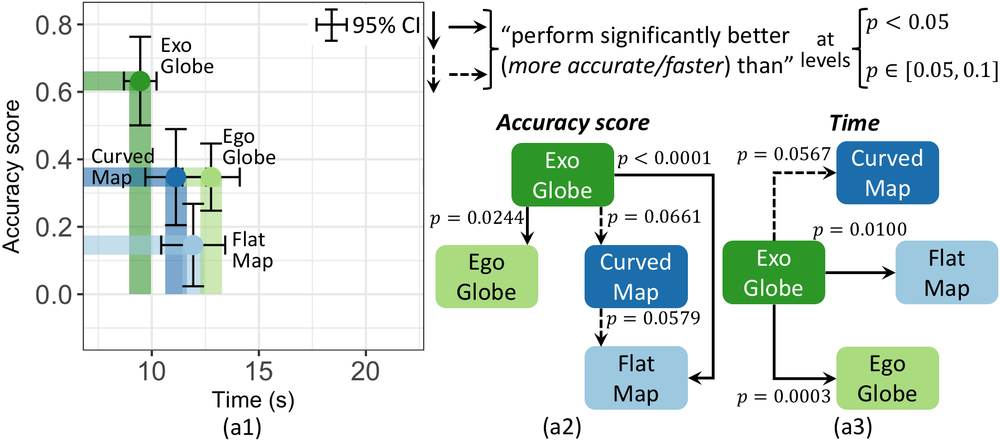}
	\caption{(a1) Average performance of \textbf{direction estimation} task per visualisation, with 95\% confidence interval, (a2, a3) graphical depiction of results of pairwise post-hoc test.}
	\label{fig:vr-maps-globes:nav-vis}
\end{figure}

The Friedman test revealed a statistically significant effect of visualisations on accuracy for the direction estimation task ($\chi^2(3)=24.937, p < .0001$). Fig.~\ref{fig:vr-maps-globes:nav-vis}(a1) shows the average accuracy score of exocentric globe (0.63) is higher than those of the other visualisations (curved map 0.35, egocentric globe 0.35, and flat map 0.15). The post-hoc test showed statistical significance for these differences. The curved map also had a statistically significantly higher accuracy score than the flat map (see Fig.~\ref{fig:vr-maps-globes:nav-vis}(a2)).

The LME ANOVA analysis revealed a statistically significant effect of visualisations on time ($\chi^2(3)=11.846, p = .0079$). In Fig.~\ref{fig:vr-maps-globes:nav-vis}(a1), we can see the avg.\ response time with exocentric globe (9.4s) was faster than curved map (11.1s), flat map (11.9s), and egocentric globe (12.7s). The post-hoc test showed statistical significance (see Fig.~\ref{fig:vr-maps-globes:nav-vis}(a3)). 

\begin{figure}[t!]
	\centering
	\includegraphics[height=12.5cm]{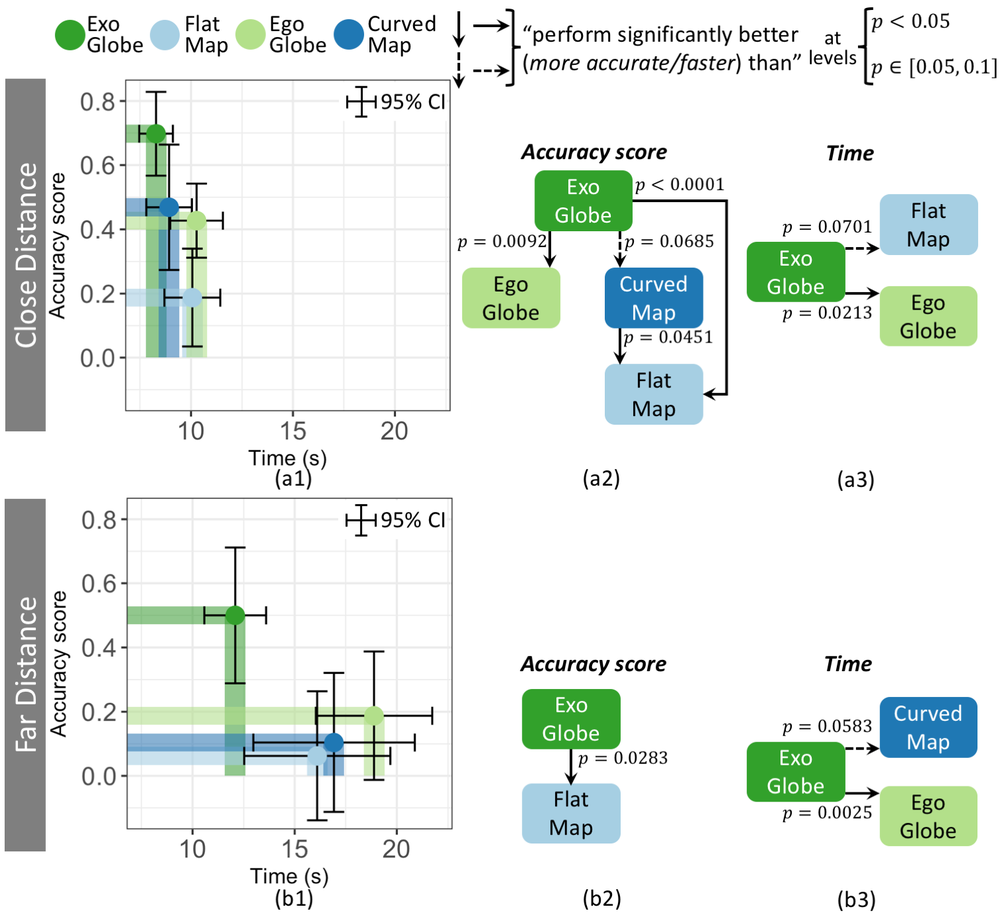}
	\caption{Break down of \textbf{direction estimation} task into different difficulty conditions. (a1,b1) Average performance per visualisation with 95\% confidence interval, (a2,a3,b2,b3) graphical depiction of results of pairwise post-hoc test.}
	\label{fig:vr-maps-globes:nav-vis-break}
\end{figure}

By difficulty condition the Friedman test revealed significant effect for \textbf{avg. accuracy}:\\
\noindent\textbf{\textit{Close distance}:} 
$\chi^2(3)=26.162, p < .0001$. Fig.~\ref{fig:vr-maps-globes:nav-vis-break}(a1) shows the avg.\ accuracy score of exocentric globe (0.70) > other visualisations (curved map 0.47, egocentric globe 0.43, and flat map 0.19). Curved map also had a significantly higher accuracy score than flat map. A post-hoc test showed statistical significances as per Fig.~\ref{fig:vr-maps-globes:nav-vis-break}(a2). \\ 
\noindent\textbf{\textit{Far distance}:}
$\chi^2(3)=8.496, p = .0368$. Fig.~\ref{fig:vr-maps-globes:nav-vis-break}(b1) shows the avg.\ accuracy score of the exocentric globe (0.50) > other visualisations (egocentric globe 0.19, curved map 0.10, and flat map 0.06). The post-hoc test only showed statistical significance between the exocentric globe and the flat map (see Fig.~\ref{fig:vr-maps-globes:nav-vis-break}(b2)).

\noindent LME ANOVA revealed significant effect for \textbf{avg. response time}:\\
\noindent\textbf{\textit{Close distance}:}
$\chi^2(3)=7.444, p = .0590$. Fig.~\ref{fig:vr-maps-globes:nav-vis-break}(a1) shows texocentric globe (8.2s) was significantly faster than those of flat map (10.0s) and egocentric globe (10.2s). The exocentric globe tended to outperform the curved map (8.9s), however, no statistical significance is evident from the post-hoc test. A post-hoc test showed statistical significances as per  Fig.~\ref{fig:vr-maps-globes:nav-vis-break}(a3). \\
\noindent\textbf{\textit{Far distance}:}
$\chi^2(3)=10.335, p = .0159$. Fig.~\ref{fig:vr-maps-globes:nav-vis-break}(b1) shows exocentric globe (12.0s) tended to be faster than flat map (16.1s), curved map (16.9s), and egocentric globe (18.8s). The post-hoc test showed the statistical significance between exocentric globe and curved map and egocentric globe (see Fig.~\ref{fig:vr-maps-globes:nav-vis-break}(b3)), but no statistical significance with flat map.

\noindent\textbf{\textit{Participant strategies}:}
From the questionnaire we identified three general strategies for direction estimation used for all visualisations:\\
\noindent\textit{ - Mentally following the arrow} -- Those participants ``attempted to `sit behind' the direction of the arrow.'' and ``imagined a marble running down the arrow.''\\
\noindent\textit{ - Moving the mid-point of the two points} -- to the centre of the map or globe.\\
\noindent\textit{ - Moving the start point of the arrow to the centre} -- some participants also tried to vertically/horizontally align the arrow.

\noindent There was one specific strategy for the exocentric globe and the egocentric globe:\\
\noindent\textit{ - Placing the start point of the arrow at the center of the globe}, then rotating the globe following the direction of the arrow.

\subsection{User interactions}
\begin{figure}[t!]
	\centering
	\includegraphics[height=6cm]{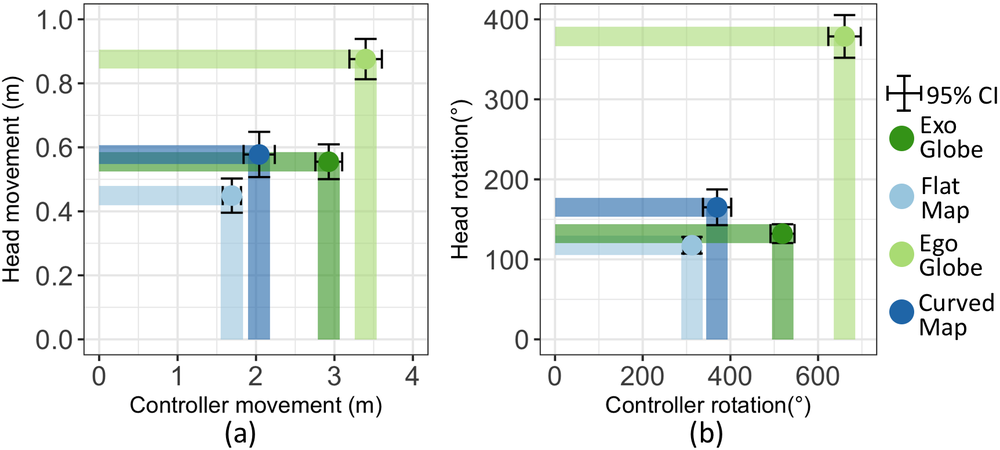}
	\caption{(a) Average accumulated movements of head and controller per task; (b) average accumulated rotations of head and controller per task.}
	\label{fig:vr-maps-globes:interactions-result}
\end{figure}

We analysed aggregate change in positions (euclidean distance) and rotation (in degrees) of the user's head and the controller,
Fig.~\ref{fig:vr-maps-globes:interactions-result}, all results significant by LME ANOVA ($p<.0001$). 

Pairwise post-hoc testing revealed significant differences: egocentric globe $>$ exocentric globe $>$ curved map $>$ flat map; with all $p<.05$.
For head movement, post-hoc test revealed significance: egocentric globe $>$ exocentric globe $\approx$ curved map $>$ flat map; with all $p<.05$, except $\approx$ means no significant differences. For the rotations of the head, the post-hoc test revealed significant differences: egocentric globe $>$ curved map $>$ exocentric globe $\approx$ flat map; with all $p<.05$, except again $\approx$ means no significant differences. 

\subsection{Motion sickness}
A five-point-Likert scale was used for rating participant motion sickness (see Fig.~\ref{fig:vr-maps-globes:ranking} (a)). The Friedman test revealed a significant effect of visualisations on motion sickness rating ($\chi^2(3)=53.84, p < .0001$). From the figure, we can see that the percentage of participants that did not experience motion sickness in the egocentric globe (21.9\%) and the curved map (50\%) is significantly less than the percentage that did not experience motion sickness in the exocentric globe (84.4\%) and the flat map (75\%). The post-hoc test showed that participants experienced significantly more motion sickness in the egocentric globe and the curved map than they did in the exocentric globe and the flat map (all $p<.05$). The egocentric globe also caused {}significantly more motion sickness to participants than the curved map with $p=.0074$.

\subsection{User preference and feedback}
\begin{figure}[t!]
    \captionsetup[subfigure]{justification=centering}
    \centering
    \begin{subfigure}{0.32\textwidth}
    	\label{subfig:vr-maps-globes:dizzy}
        \includegraphics[width=\textwidth]{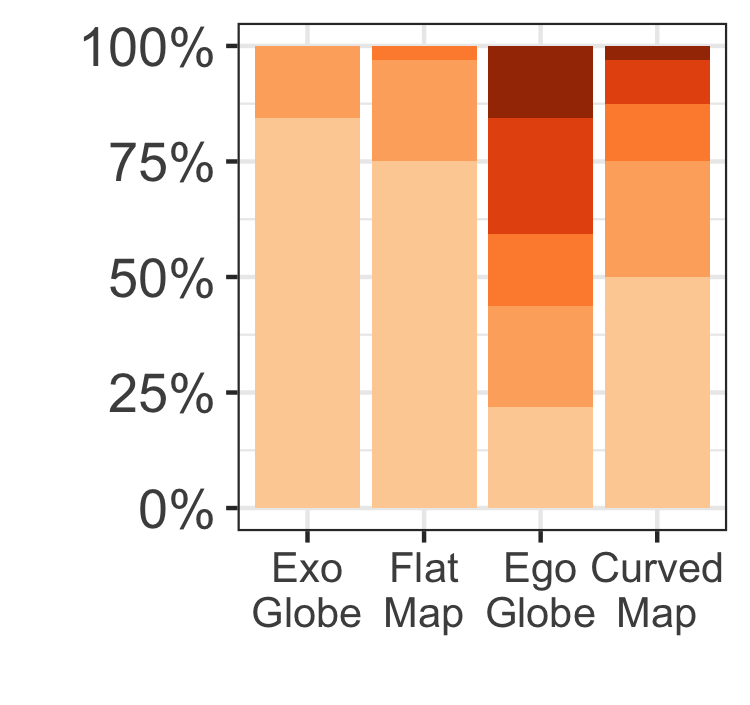}
        \caption{Motion sickness rating}
    \end{subfigure}
    \begin{subfigure}{0.32\textwidth}
    	\label{subfig:vr-maps-globes:visual-ranking}
        \includegraphics[width=\textwidth]{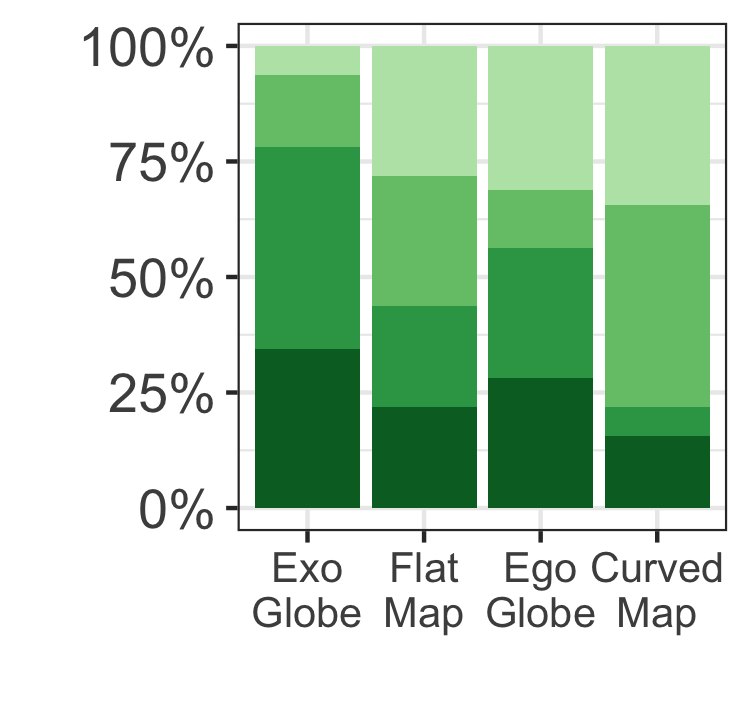}
        \caption{Visual design ranking}
    \end{subfigure}
    \begin{subfigure}{0.32\textwidth}
    	\label{subfig:vr-maps-globes:easy-ranking}
        \includegraphics[width=\textwidth]{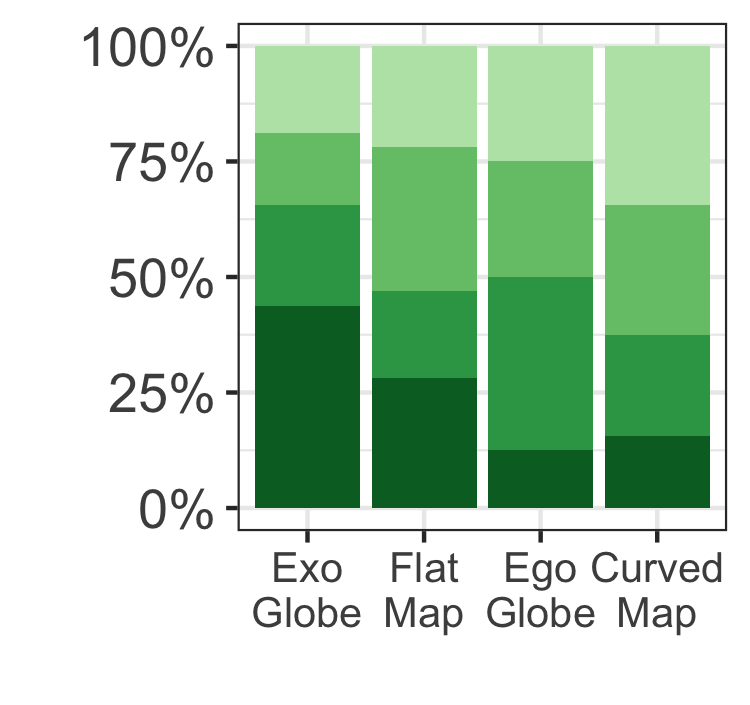}
        \caption{Readability ranking}
    \end{subfigure}
    \caption{Participants motion sickness rating (from \setlength{\fboxsep}{1pt}\colorbox{noDizzy}{\strut no motion sickness} to \setlength{\fboxsep}{1pt}\colorbox{dizzy}{strong motion}\\ \setlength{\fboxsep}{1pt}\colorbox{dizzy}{sickness}) and preference ranking (\setlength{\fboxsep}{1.5pt}\colorbox{first}{\textcolor{white}{$1^{st}$}}, \setlength{\fboxsep}{1pt}\colorbox{second}{\textcolor{white}{$2^{nd}$}}, \setlength{\fboxsep}{1pt}\colorbox{third}{$3^{rd}$} and \setlength{\fboxsep}{1pt}\colorbox{second_fourth}{$4^{th}$})}
    \label{fig:vr-maps-globes:ranking}
\end{figure}

Participant ranking for each of the four visualisations by percentage of respondents is shown by colour: see Fig.~\ref{fig:vr-maps-globes:ranking}~(b) and (c).

For \emph{visual design}, the Friedman test revealed a significant effect of visualisations on preference ($\chi^2(3)=10.612, p = .0140$). 
The strongest preference was for the exocentric globe, with 78.1\% voting it as first or second place. The preference for the flat map and the egocentric globe were similar, each received 43.8\% and 56.3\% votes as first or second place. The curved map was the least preferred, with 21.9\% votes for first or second place. The post-hoc tests only showed a significant difference between the exocentric globe and the curved map with $p=.0080$.

For \emph{readability}, the Friedman test indicated no significant effect of visualisations on preference ($\chi^2(3)=5.363, p = .1471$). The strongest preference is again for the exocentric globe, with 65.6\% of respondents voting it first or second place. The other visualisations have similar preferences, with the flat map, egocentric globe, and curved map receiving 46.9\%, 50\% and 37.5\% votes respectively as first or second place.

The final section of the study allowed participants to give feedback on the pros and cons of each design. Qualitative analysis of these comments reveal (overall):

\noindent\textbf{Exocentric globe} was intuitive, familiar and easy to manipulate: ``It is the same with a physical globe, the way to manipulate it is just nature, you can also walk around it.'', ``I feel like I am most habituated to this kind of representation''. Some also commented on the tasks: ``I think this one is the most suitable for solving the given tasks. It is not distorted at all.'' One noted, ``occlusion is a problem if targets are too far apart.'' Note that even at 180 degrees separation, two points will still be visible at antipodes.  However, perceptual distortion due to curvature of the globe was problematic for our participants in the far condition. Some also suggested a different interaction, which would synchronise the rotations of the exocentric globe with the hand rotations.

\noindent\textbf{Flat map} was again very familiar: ``Similar to what we see on paper''. Some enjoyed the ability to show the whole world at once. Others were excited about the interaction of manipulating the geographic centre: ``it works as fluid and has a feeling of artwork''. However, this interaction was unfamiliar to most participants: ``it took a while to get used to controlling the map in the way that I wanted to.'' Interacting with the map can result in landmasses with unusual orientation, which is also unfamiliar to most participants. Yet, some commented that it is boring to have a flat map in VR.

\noindent\textbf{Egocentric globe} was novel and immersive: ``It is cool and exciting being inside of the world''. However, ``I don't like it when it comes to answering questions''. Some also commented ``it is easier to look around rather than manipulating with the wand'', while some complained about the head movements: ``have to move my head a lot.'' Many participants also reported experiencing motion sickness in it.

\noindent\textbf{Curved map} was generally similar to the flat map. Some commented on their feelings about the curvature: ``it is like facing a curved TV''. Some were impressed by visual design: ``It is the most visually impressive one.'' Yet, others reported it seemed more distorted than the flat map and felt more motion sickness with it than with the flat map.

%% file: content/5-maps-and-globes-in-vr/5-discussion.tex
\section{Discussion}
\label{sec:vr-map-globe:discussion}

\subsection{Exocentric Globe} 
Exocentric globes present geographic information without projection distortion, and this seems to greatly benefit accuracy for our tasks as described below. The overall response time is also comparatively good, except for one task (\emph{small variation} condition in area comparison). Overall, exocentric globe appears to be a good choice for the three fundamental geographic analysis tasks in VR with the following details and caveats:

\noindent\textit{Best for overall accuracy --}
The exocentric globe was the most accurate visualisation in almost all cases. One exception is in \emph{small variation} for area comparison, where it performed the worst. \\
\\
\noindent\textit{Time for distance comparison similar to other visualisations --} The response time for the exocentric globe was similar to the other visualisations when comparing distances. \\
\\
\noindent\textit{Area comparison slower than with maps --} Overall, participants were slower with the exocentric globe than the flat and curved maps. If we break the results down by task difficulties, however, performance for exocentric globe was similar to the flat and curved maps in \emph{easy} and \emph{far distance} conditions. However, it performed much slower than these two in the \emph{small variation} condition. We believe this is due to the relatively small variation of area between the two polygons, and the fact that participants can only view half of the globe. Participants had to rely on their memory to compare areas, and they tended to confirm their choice multiple times by using interactions to switch between viewing each of the two polygons. Interestingly, this did not happen in the \emph{small variation} condition in distance comparison tasks. A likely explanation is that comparing the magnitude of length is easier than area \cite{teghtsoonian1975psychophysics}, thus less interactions were needed in distance comparison tasks. \\
\\
\noindent\textit{Fastest for direction estimation tasks --} Overall, the exocentric globe is the fastest visualisation for finding directions.

\subsection{Flat Map} 
This visualisation is capable of presenting the entire surface of the world within the user's field of view, making it time-efficient for distance and area comparison.  However, projection and perceptual distortion appears to lead to poor accuracy in all direction estimation tasks, and the \emph{small variation} condition of distance and area comparison tasks. 

\noindent\textit{Relatively fast for distance and area comparison,} however, the accuracy is relatively low for distance comparison tasks. \\
\\
\noindent\textit{Drop in accuracy for small variation --} In distance and area comparison tasks, accuracy was relatively high in \emph{easy} and \emph{far distance} conditions, but dropped significantly in the \emph{small variation} conditions. For distance comparison, no map projection can preserve all distances, thus, projection distortion makes comparing distances with small differences more difficult. For area comparison, although the map projection preserves the relative size of areas, the shape of the polygons is distorted (e.g., a polygon is elongated at the edge of a flat map), which affects area perception \cite{Teghtsoonian:1965gf}. \\
\\
\noindent\textit{Poor for direction estimation --} The flat map was the least accurate visualisation and relatively slow in direction estimation.  
We attribute this to distortion effects.

\subsection{Egocentric Globe}
This is the most immersive visualisation. It performed stably in accuracy across different difficulty conditions in the comparison of distances and areas. However, the perceptual distortion introduced by changing view point and the extra effort of body interaction (e.g.\ users needing to turn their head) make it a poor choice in VR for the three tasks tested.

\noindent\textit{Worst performance overall --} The egocentric globe performed significantly worse in almost all cases.  Particularly, for distance comparison, it seems less accurate and slower than other visualisations in \emph{easy} and \emph{far distance} conditions. \\
\\
\noindent\textit{Good for small variation -- } Despite being the worst performer overall, one exception is in \emph{small variation} for distance comparison, in which, it seems to perform the best of all (both faster and more accurate).  One possible explanation is that it was the largest scale visualisation - i.e.\ it maximised the size of the distances relative to the participants' field of view at the default viewing distance.  While it is possible for participants to move closer to the visualisations to achieve a similar relative scale, doing so may be inconvenient or cost them time.\\
\\
\noindent\textit{Stable across difficulties -- }  For distance and area comparison, the accuracy with egocentric globes stayed relatively stable across three difficulty conditions. \\
\\
\noindent\textit{Motion sickness -- } Although we placed two fixed-position references (horizon lines) to help participants perceive their direction, they reported a relatively strong motion-sickness feeling in this visualisation. The egocentric spherical globe covered the full field of view all the time.  Together with the curvature of sphere, this seemed to cause participants to feel more motion-sickness. 

\subsection{Curved Map}
This is generally an improvement on the flat map in VR. The curved map was more accurate than the flat map in direction estimation.  However, motion-sickness is a practical issue, but one that may be mitigated by shrinking the size of the map or by improved hardware.

\noindent\textit{Better than flat map for direction --}
Participants had greater accuracy with curved map than with flat map in direction estimation tasks, and performed similar to flat map in almost all other cases. \\
\\
\noindent\textit{Accurate small variation area comparison --}
In area comparison tasks, curved map seems to be the most accurate visualisation in the \emph{small variation} condition, but was the least accurate one in \emph{far distance} condition. The possible reason might be that, the curved map is not an area-preserving visualisation, and the difference between distortions of areas are larger if two polygons are far apart. \\
\\
\noindent\textit{Direction estimation --} 
In direction estimation tasks, the curved map outperformed the flat map in the \emph{close variation} condition. Less distortions of directions occurs in the curved map than the flat map. \\
\\
\noindent\textit{Second worst for motion-sickness --}
Participants reported more feelings of motion-sickness than for exocentric globe and flat map. The reason might be similar as to the egocentric globe, the curved map covers 108$^\circ$ horizontally of the field of view, and the perception of curvature might produce extra motion-sickness.

\subsection{User interactions}
In the egocentric globe, participants tended to interact significantly more than with other visualisations. This is also reflected in participants' feedback, and could be one reason for slow response times in the egocentric globe condition. 

The exocentric globe needed more controller interactions than curved and flat maps, and a similar degree of head movement. From Figure~\ref{fig:vr-maps-globes:interactions-result} and investigators' observation, it seems participants did not like to move themselves in space for all visualisations, especially in the exocentric globe, as its ratio of $\frac{controller\ interactions}{head\ interactions}$ is significantly larger than with other visualisations. This is possibly due to unfamiliarity with the VR environment, and the ease of manipulating visualisations using the controller.

%% file: content/5-maps-and-globes-in-vr/6-conclusion.tex
\section{Conclusion}
\label{sec:vr-map-globe:conclusion}
We have conducted the first user study evaluating the effectiveness of different visualisations of global geography in VR. 
Of the four conditions and three task types tested, we found that the exocentric globe is generally the best choice of VR visualisation.  This is despite the fact that less of the earth's surface is visible in the exocentric globe than the other representations and that it has the most perceptual distortion, though no distortion due to map projection. We also found that the curved map had benefits over the flat map, but the curved map caused the users greater motion-sickness. In almost all cases the egocentric globe was found to be the least effective visualisation. 

\begin{figure}[t!]
	\centering
	\includegraphics[width=\textwidth]{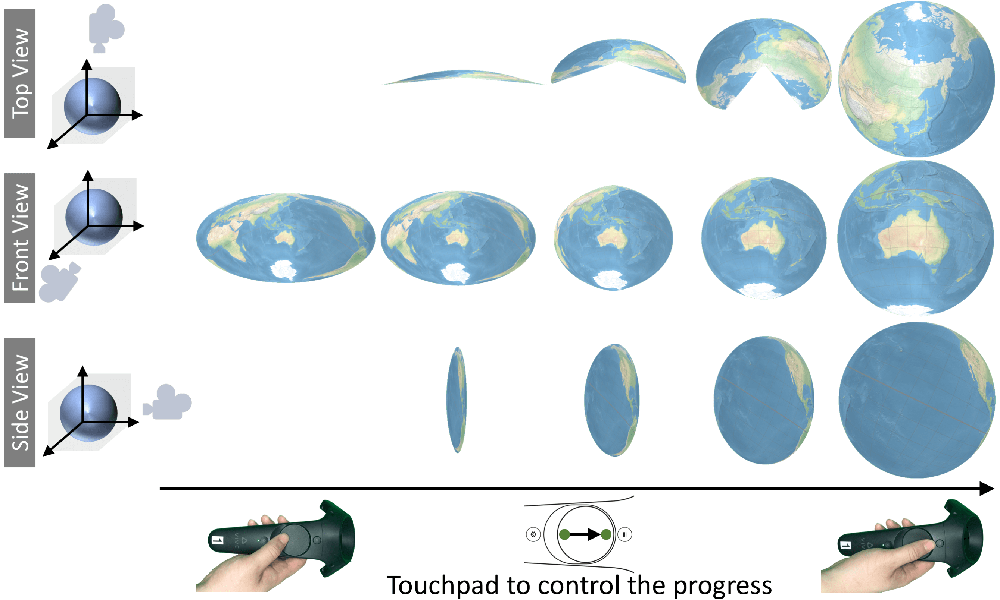}
	\caption{Animated transformation from a flat map to an exocentric globe: top row - top view; second row - front view and bottom row - side view. The touchpad dynamically controls the progress.}
	\label{fig:vr-maps-globes:morphing}
\end{figure}

Our study was performed using VR HMDs as these offer a significantly better field of view compared to the currently available AR devices.  However, it is expected that AR technology will improve in this regard.
Thus, our results have significant implications for the design of geovisualisation applications for VR, AR and MR, providing support for the use of exocentric globes when visualising data with global extent.

While our study found that the exocentric globe had the best overall performance of the four visualisations, the inability to show the entire surface hindered users in some tasks. We therefore think it would be reasonable to combine the exocentric globe with a map representation. As the curved map was more likely to cause motion sickness and the  flat map was more  familiar, we  are currently investigating how to combine the exocentric globe and flat map. As a first possible hybrid we have developed a prototype implementation that allows the viewer to interactively transition between exocentric globe and flat map. 
Due to the complexity of the transition, we allow the user to control the progress of the morphing. Fig.~\ref{fig:vr-maps-globes:morphing} shows the animated transformation from a flat map to an exocentric globe, the reverse transformation  is symmetric. We used linear interpolation to transition between the 3D position of points in the rendered textures of the source visualisation and  the target visualisation. Refer to our video for a demonstration.  Evaluation of this hybrid visualisation remains future work.

Other future work is to design and evaluate other interactions including (non-physical) zooming. We also believe that the curved map has considerable potential and wish to explore how we can reduce motion-sickness and also investigate the use of other map projections.

%% file: content/6-od-flow-maps-in-vr/0-index.tex
\chapter[OD Flow Maps in Immersive Environments]{OD Flow Maps in \\ Immersive Environments}
\label{chapter:flow-maps-vr}
\begin{figure}[ht!]
	\centering
	\includegraphics[width=\textwidth]{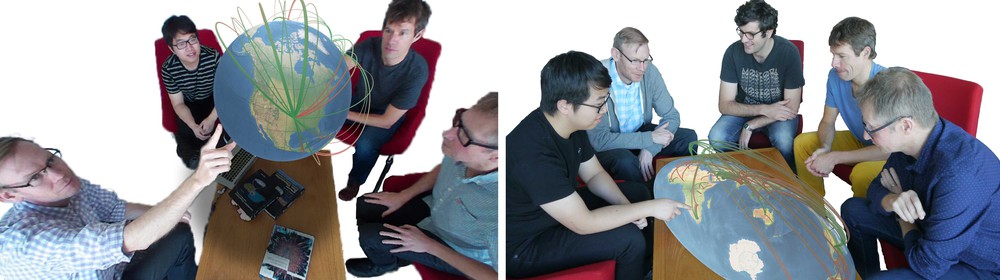}
	\caption{3D Globe (left) was the fastest and most accurate of our tested visualisations. Flat maps with curves of height proportional to distance (right) were more accurate but slower than 2D straight lines. Experiments were conducted individually in virtual reality but our motivation for this work is to support future mixed-reality collaborative scenarios like those envisaged in these figures.}
	\label{fig:vr-flow-maps:teaser}
\end{figure}

In the previous chapter, we explored and evaluated different representations of geographic reference space in immersive environments finding that exocentric globe outperforms other representations in most of the tasks tested in our user study.

In this chapter, we step further into the design space of OD flow maps in immersive environments by investigating different OD flow representations taking advantage of the third dimension, as well as OD flow maps with different representations of the geographic reference space.

We first chart the design space of 3D flow maps into the representation of OD flow and the representation of geographic reference space. We then implement and evaluate five different OD flow representations embedded on a flat map. Next, we implement and evaluate four different OD flow maps varying their representations of the geographic reference space. Finally, we provide guidelines for creating flow maps in immersive environments.

The work presented in this chapter is a collaborative effort with my supervisors and collaborators: Tim Dwyer, Bernhard Jenny, Kim Marriott, Maxime Cordeil and Haohui Chen. This work was accepted and will be presented at IEEE Conference on Information Visualization (InfoVis) 2018. In this work, I collaboratively formulated the hypothesis and research questions, was solely responsible for the implementation and user studies, did the data analysis, and collaboratively wrote the research paper.

\input{content/6-od-flow-maps-in-vr/1-introduction}
\input{content/6-od-flow-maps-in-vr/2-design-space}

\input{content/6-od-flow-maps-in-vr/3-study-1}

\input{content/6-od-flow-maps-in-vr/4-study-2}

\input{content/6-od-flow-maps-in-vr/5-study-3}
\input{content/6-od-flow-maps-in-vr/6-conclusion}

%% file: content/6-od-flow-maps-in-vr/1-introduction.tex
\section{Introduction}
\label{sec:vr-flow-maps:intro}
Visualisation of origin-destination flows is a difficult information visualisation challenge, because both the locations of origins and destinations and the connections between them need to be represented. The most common visualisation is the OD (origin-destination) flow map, where each flow is represented as a line connecting the origin and destination on a map. 

In our previous studies in 2D display space (See Chapter~\ref{chapter:evaluating-2d-od-flow-maps}), we found that
a disadvantage of OD flow maps is that they become cluttered and difficult to read as the number of flows increases. \added{Nonetheless, participants commented on them as intuitive and favoured them for showing a small number of flows.} 
With the arrival of commodity head-mounted displays (HMDs) for virtual-reality (VR), e.g. HTC Vive, and augmented-reality (AR), e.g. Microsoft Hololens, Meta2 and Magic Leap, we would like to see how far the flow maps can scale in these display environments. We can also expect to see more geographic visualisations used in mixed-reality (MR) applications. Such applications include situated analytics~\citep{elsayed2016situated} where visualisations can be made available in almost any environment such as in the field, surgery or factory floor, and collaborative visualisation scenarios, where two or more people wearing HMDs can each see and interact with visualisations while still seeing each other~\citep{Cordeil:2016io}.

The key question we address is whether traditional 2D OD flow maps are the best way to show origin-destination flow in such immersive environments or whether some variant that makes use of a third dimension may be better.  
While current guidelines for information visualisation design caution against the use of 3D spatial encodings of abstract data~\citep[Ch.\ 6]{Munzner:2014wj}, in the case of global flow data visualised in immersive environments the third dimension offers an extended design space that is appealing for a number of reasons:
\begin{itemize}
	\item The height dimension offers the possibility of an additional spatial encoding for data attributes.
	\item Lifting flow curves off the map may reduce clutter and provide better visibility of the underlying map.
	\item In immersive environments, occlusions can be resolved by natural head movements or gesture manipulations to change the view angle.
	\item In 2D flow maps the flows may be perceived as trajectories (highways, shipping routes, etc.), lifting them into the third dimension may resolve this ambiguity.
\end{itemize}

We investigate this design space through three controlled user studies. To the best of our knowledge we are the first to do so. Our paper has three main contributions.

The first contribution is to chart the design space for 3D flow maps (Sec.~\ref{sec:vr-flow-maps:design}). Following D\"ubel \emph{et al.}~\citep{Dubel:2015el} we separate the design space into two orthogonal components: the representation of flow, e.g.\ straight or curved lines in 2 or 3D, and the representation of the geographic reference space, e.g.\ 3D globe or flat map. Furthermore, origins and destinations can either be shown on the same or separate globes or maps. This leads to a rich multi-dimensional design space.

The second contribution is evaluation of different flow maps in VR differing in the representation of flow (Study 1, Sec.~\ref{sec:vr-flow-maps:study-01}). We compared  2D flow representations with (a) straight and (b) curved flow lines, and 3D flow tubes with (c) constant height, height varying with (d) quantity and (e) distance between start and end points. We measured time and accuracy to find and compare the magnitude of flow between two pairs of locations. We found that participants were most accurate using 3D flows on flat maps when flow height was proportional to flow distance and that this was the preferred representation. Participants were less accurate with straight 2D flows than with 3D flows, but faster.

The third contribution is evaluation of different flow map visualisations primarily varying in the representation of the reference space (Studies 2 and 3). For the same task as above, we first compared a flat map with 2D flow lines, a flat map with 3D flow tubes, a 3D globe with 3D tubes, and a novel design called \emph{MapsLink} involving a pair of flat maps linked with 3D tubes. We found (Sec.~\ref{sec:vr-flow-maps:study-02}) that participants were much slower with the linked map pair than all other representations. What surprised us was that participants were more accurate with the 3D globe than with 2D flows on a flat map and linked map pairs. Participants were also faster with the 3D globe than with 3D flows on a flat map. The final experiment (Sec.~\ref{sec:vr-flow-maps:study-03}) was similar but with higher densities of flows. This found the 3D globe to be the fastest and most accurate flow visualisation. It was also the preferred representation.

The performance of the 3D globe is unexpected. While we assumed that 3D flows might reduce the problem of clutter, we did not expect that the 3D globe with its potential shortcomings of occlusion and distortion would be more effective than 2D or 3D flows on a flat map. However, this result accords with \cite{Yang:2018mg}, who found that in VR environments 3D globes were better than maps for a variety of map reading tasks. 

%% file: content/6-od-flow-maps-in-vr/2-design-space.tex
\section{Design Space}
\label{sec:vr-flow-maps:design}

Visualisations of OD flow data can present geographic locations of origins and destinations, the direction of flow and flow weight (magnitude or other quantitative attribute). OD flow maps achieve this by showing each flow as a line or arrow on a map connecting the origin and destination. In this section we explore the design space of 3D OD flow maps.  D\"ubel \emph{et al.}~\citep{Dubel:2015el} categorize geospatial visualizations based on whether the reference space (i.\,e.\ the map or surface) is shown in 2D or 3D and whether the abstract attribute is shown in 2D or 3D. In the case of flow maps,  this categorization implies the design space has two orthogonal components: the representation of flow and the representation of geographic region. 

\subsection{Representation of Flow}
\label{sec:vr-flow-maps:design:flowrep}
Flow on 2D OD flow maps is commonly shown by a straight line from origin to destination with line width encoding magnitude of flow and an arrowhead showing direction. However, with this encoding visual clutter and line crossings are inevitable, even in small datasets. One way to overcome this is to use curved instead of straight lines, such that the paths are carefully chosen to ``fan out'' or maximise the separation between flows at their origins and destinations \citep{Riche:2012ht}. Such curved flow maps have shown to be more effective in ``degree counting'' tasks~\citep{Jenny:2017ci} where the curvature reduced overlaps. Conversely, edge bundling has also been suggested as a way of overcoming clutter. In this approach curved paths are chosen so as to visually combine flows from adjacent regions~\citep{Yang:2017cy}. While bundling can greatly reduce clutter, its disadvantage is that it can make individual flows difficult to follow. Thus, it is probably best suited to overview tasks.

In the case of 3D flow representations, height can be used to
encode flow magnitude or some other quantitative property or to reduce the visual clutter caused by crossings or overlapping. Not only does the use of a third dimension allow flows to be spaced apart, in modern immersive MR environments it also allows the viewer to use motion perspective to better distinguish between flows by either moving their head or by rotating the presentation.
Based on our literature review (see Section~\ref{sec:related:3d}), possible 3D representations include:
\begin{itemize}[leftmargin=1em]
	\item \emph{Constant maximum height:} in which each flow is shown as a curved line connecting the origin and destination and all flows have the same maximum height above the surface of the reference space. This is arguably the simplest way to use height to help address the problem of overlapping flows.
	\item \emph{Height encodes quantity:} height is either proportional or inversely proportional to flow magnitude~\citep{Eick:1996gs}. This allows double-coding the flow magnitude with both thickness and height.
	\item \emph{Height is proportional to distance:} height is proportional to the distance between origin and destination, short flows will be close to the reference space surface while longer flows will be lifted above it~\citep{Cox:1995fp}. This will tend to vertically separate crossings. 	
	\item \emph{Height is inversely proportional to distance:} this was suggested in~\citep{Vrotsou:2017im}. The advantage is that it increases the visual salience of flows between geographically close locations but at the expense of increasing overlap.
\end{itemize} 

\subsection{Representation of Reference Space}
\label{sec:vr-flow-maps:design:refspace}

In 2D flow maps the reference space is always a flat map, which can of course also be used in an immersive environment. 
However, in the case of global flows it also makes sense to use a 3D exocentric globe representation in which the flow is shown on a sphere positioned in front of the viewer. The disadvantage of a globe representation is that the curved surface of the globe causes foreshortening and only half of the globe can be seen at one time. Nonetheless in a previous study (see Chapter~\ref{chapter:maps-globes-vr}~\citep{Yang:2018mg}) we found that this representation had better performance than flat maps for distance comparison and estimation of orientation between two locations.

An alternative is to use a 3D egocentric representation for the globe in which the user is placed inside a large sphere~\citep{Zhang:2016,Yang:2018mg,Zhang:2018jo}. This suffers from similar drawbacks to the exocentric globe: foreshortening and inability to see more than half of the Earth's surface. Furthermore, because of the position of the user it is difficult to see the height of flows. In \citep{Yang:2018mg} we found that the egocentric globe led to worse performance than the exocentric globe for standard map reading tasks and also led to motion sickness. 

Typically the same reference space (map or globe) is used to show both origins and destinations. However, in 2D representations such as \textit{MapTrix} or \textit{Flowstrates} \citep{Boyandin:2011fa,Yang:2017cy} origins and destinations are shown on different  maps. Potential benefits are reduction of clutter in the reference space representations and clearer depiction of  flow direction. In a 3D environment probably the simplest representation using two reference space representations is to show two 2D maps on flat  planes with  flow shown by connecting  tubes. Such a representation is akin to Collins and Carpendale's \emph{VisLink} technique~\citep{Collins:2007ir}, where multiple 2D abstract data representations are viewed in a 3D environment, with lines linking related points across views.  This was the inspiration for our \emph{MapLink} technique evaluated in Study 2, Sec.~\ref{sec:vr-flow-maps:study-02}.

%% file: content/6-od-flow-maps-in-vr/3-study-1.tex
\section{Study 1: 2D and 3D Flows on Flat Maps}
\label{sec:vr-flow-maps:study-01}
The first user study focuses on representation of flow in VR. It compares readability of  five flow representations (two 2D and three 3D) using  the same reference space representation: a flat map.

\subsection{Visualisations and interactions}
\begin{sidewaysfigure}
	\centering
	\includegraphics[width=\textwidth]{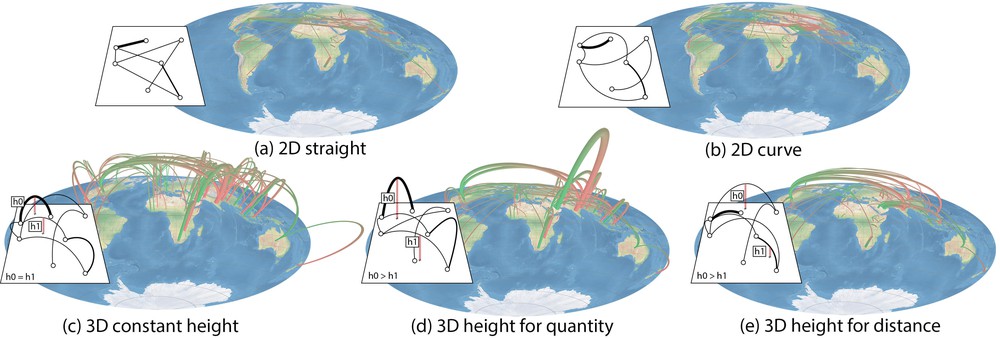}
	\caption{Study 1: Tested 2D and 3D flow maps.}
	\label{fig:vr-flow-maps:first-study-vis}
\end{sidewaysfigure}

The two 2D representations were:\\
\textbf{\textit{2D straight:}} Connecting origins and destinations with straight lines is the most common way to create a 2D flow map (Fig. \ref{fig:vr-flow-maps:first-study-vis}(a)).

\textbf{\textit{2D curve:}} Using curved flow lines that increase separation and acute angle crossings has been shown to increase readability for dense flows in 2D flow maps~\cite{Jenny:2017ci}. The routing technique in \cite{Jenny:2017dy} was used to created 2D curved flows (Fig. \ref{fig:vr-flow-maps:first-study-vis}(b)).

\begin{figure}[b!]
	\centering
	\includegraphics[width=0.4\textwidth]{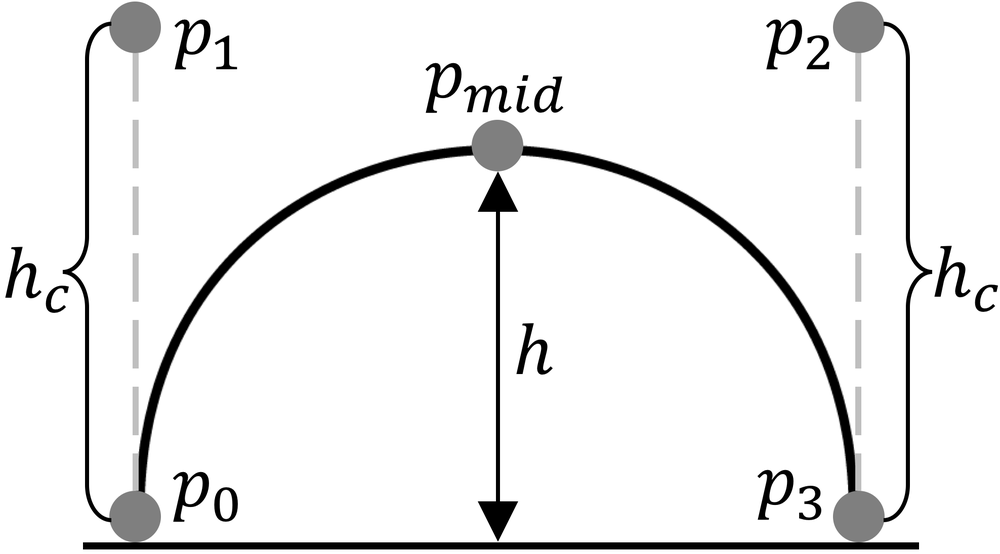} 
	\caption{Control points calculation of cubic B\'ezier curve.}
	\label{fig:vr-flow-maps:bezier-curve}
\end{figure}

The three 3D flow representations used 3D tubes to connect origins and destinations. We used a cubic B\'ezier curve to create the tubes using Equation~(\ref{equ:bezier}), where $P_0$ and $P_3$ are the origin and destination of a flow, $0 \le t \le 1$ is the interpolation factor, and $P_1$ and $P_2$ are the two control points to decide the shape of the tube. 
\begin{equation}
	B(t) = (1-t)^3P_0 + 3(1-t)^2tP_1 + 3(1-t)t^2P_2 + t^3P_3
	\label{equ:bezier}
\end{equation}
For $P_1$ and $P_2$, we make their projected positions on 2D map plane the same as $P_0$ (the origin) and $P_3$ (the destination) respectively, so the projected trajectory on the 2D map plane of the 3D tube is a straight line. This allows users to easily follow the direction of flow lines. $P_1$ and $P_2$ are set to the same height to ensure symmetry
such that the highest point will be at $t=0.5$, the mid-point of the tube. We can use the height of two control points ($h_c$) to precisely control the height of the mid-point ($h$):  $h_c = \frac{h}{6 \times 0.5^3} = \frac{4}{3}h$ (see Fig.~\ref{fig:vr-flow-maps:bezier-curve}). 

Three different height encodings were evaluated:

\textbf{\textit{3D constant:}} All flows have the same height (Fig. \ref{fig:vr-flow-maps:first-study-vis}(c)).

\textbf{\textit{3D quantity:}} Height linearly proportional to flow quantity (Fig.\ \ref{fig:vr-flow-maps:first-study-vis}(d)) such that small quantity flows will be at bottom, while large ones will be on top. In the pilot we also tried the inverse (smaller flows higher) but this was found to be severely cluttered.

\textbf{\textit{3D distance:}}  Height linearly proportional to Euclidean distance between the origin and destination (Fig. \ref{fig:vr-flow-maps:first-study-vis}(e)). Close flows will be lower, while flows further apart will be on top. 
Again the inverse was tried in the pilot but quickly discarded as unhelpful.

\subsection{Encodings Common to All Conditions} 
Quantity was encoded in all conditions using thickness of lines (in 2D) and diameter of tubes (in 3D). Several evaluations informed our choice of  direction encoding.  Holten \emph{et al.}\ evaluated encodings of unweighted edge direction in node-link diagrams~\cite{Holten:2009eq,Holten:2011fp} and found tapering of lines is the only direction encoding to be more effective than colour gradient. In the context of OD flow maps a study by Jenny \emph{et al.}~\cite{Jenny:2017ci} revealed difficulties interpreting tapered connections in geographic context. Furthermore, the  use of  line width/diameter to show weight makes tapered edges impractical (the only part of the line where width could be reliably compared would be at the origin).  We therefore chose to use colour gradient in both 2D and 3D conditions, using the same colour gradient (red-green) from~\cite{Holten:2011fp} to present direction information in our study.
To reduce the distortion of shapes and also for aesthetic reasons, we chose to use the Hammer map projection, an equal-area projection with an elliptical boundary. See \cite{Yang:2018mg} for additional details of our use of this projection.

The Natural Earth raster map from \url{naturalearthdata.com} was used as the base texture. Originally, we had concerns about the texture colour interfering with flow readability but in our pilot tests participants had no problem with this. 
A legend was presented with all flow maps, indicating direction, quantity and  other encodings (e.g.\ height for distance).

\subsection{Rendering}  
Geometry computation was accellerated with the GPU, tessellation was used for curve and tube interpolation, and a geometry shader was used to build the structure of line or tube segments.

\subsection{Interactions} 
We provided the same interaction across the five visualisations. First, viewers can move in space to change their viewpoint. Second, we allowed viewers to change the 3D position and rotation of the map. They could pick up the map using a standard handheld VR controller, and reposition or rotate it in 3D space. We did not provide explicit interactive widgets or dedicated manipulations for adjusting the scale of the different maps. However, viewers could either move closer to the maps, or pick maps with a VR controller and bring them closer to their HMD. We did not allow other interaction such as filtering as we wished to focus on base-line readability of the representations.

\subsection{Experiment}
\noindent\textbf{Stimuli and Task Data}\\ 
We used  datasets based on real international migration flows between countries~\cite{Abel:2014iz} for the study. We show only a single net flow between each pair of countries. To control the number of flows for our different difficulty conditions we symmetrically filtered the data by dropping the same percentage of small and large flows. We randomised the origin and destination of the original dataset to ensure different data for each question.

Task: To keep the study duration for each participant to around one hour, we could only test a single task. In Chapter~\ref{chapter:interviews}, we realised that \textit{Search} is a required primitive action for analysing flow data. Meanwhile, the most common types of queries are: \textit{Identify}, \textit{Summarise} and \textit{Compare}, among which \textit{Compare} is preliminary to others. That is, in order to identify or summarise flow map features, you first need to be able to compare individual flows. Thus, we choose the task of finding and comparing the flow between two given pairs of locations:

\centerline{\textit{For the two flows from A to B and X to Y, which is greater?}} 

This task was chosen because it combines two fundamental sub-tasks: \emph{searching} for the flow line between two given locations and \emph{comparison} of magnitude of two flows. We would expect visual clutter to negatively impact on both of these sub-tasks while dual encoding of magnitude might help with comparison.
Besides the choice of flow representions, two factors may affect user performance: the number of flows, and the relative difference of quantity between two given flows.

Following Feiner \textit{et al.}~\cite{Feiner:1993ip}, leader lines were used to link  labels ``A'', ``B'' with the origin and destination of one flow and ``X'', ``Y'' with the origin and destination of the other flow. Labels were horizontal and rotated in real-time so as to remain oriented towards the viewer.

Number of flows: In the study by Jenny \textit{et al.}~\cite{Jenny:2017ci}, the largest number of flows tested was around 40.  To better understand scalability, we decided to test three different difficulty levels in this study: (1) 40 flows with 20\% difference, (2) 40 flows with 10\% difference and (3) 80 flows with 20\% difference. We required the two flows in question to be separated by at least 15\textdegree~on the great circle connecting them so as to avoid situations where the origin and destination of a flow were too close to be clearly distinguished. To balance the difficulty of searching for a flow, all  origins and destinations of flows under comparison were required to have more than three flows.

Quantity encoding: The smallest (largest) flow magnitude was mapped to the thinnest (widest) flow width, and intermediate values were linearly encoded.

\noindent\textbf{Experimental Set-up}\\
We used an HTC Vive with $110^{\circ}$ field of view and 90Hz refresh rate as the VR headset for the experiment. The PC was equipped with an Intel i7-6700K 4.0GHz processor and NVIDIA GeForce GTX 1080 graphics card. Only one handheld VR controller was needed in the experiment: participants could use this to reposition and rotate the map in 3D space. The frame rate was around 110FPS, \textit{i.e.} computation was faster than the display refresh rate.

Visuals were positioned comfortably within the users' reach and sized by default to occupy approximately 60\% of the viewers' horizontal field of view. The map was texture-mapped onto a quad measuring 1$\times$0.5 metre and placed at 0.55 metre in front and 0.3 metre under participants' eye position, and tilted to 45\textdegree. The map was centred on 0\textdegree~longitude and 0\textdegree~latitude. We repositioned the map at the beginning of every question. The thickness of lines in 2D and diameter of tubes in 3D were in the range of 2mm to 16mm. The height of \textit{3D quantity} and \textit{3D distance} was linearly mapped to the range of 5cm to 25cm. The constant height for \textit{3D constant} was 15cm.

\noindent\textbf{Participants}\\
We recruited 20 participants (8 female) from our university.  All had normal or corrected-to-normal vision and included university students and researchers. 1 participant was under 20, 15 participants were within the age group 20$-$30, 1 participant was between 30$-$40, and 3 participants were over 40. VR experience varied: 13 participants had less than 5 hours of prior VR experience, 5 participants had 6$-$20 hours, and 3 participant had more than 20 hours. While our encoding used colour to indicate direction, the tasks used did not involve ambiguity regarding direction.  Therefore, we did not test participants for colour blindness.

\noindent\textbf{Design and Procedure}\\
The experiment was within-subjects: 20 participants $\times$ 5 visualisations $\times$ 1 task $\times$ 3 difficulty levels $\times$ 5 repetitions = 1,500 responses (75 responses per participant) with performance measures and lasted one hour on average. Latin square design was used to balance the order of visualisations. 

Participants were first given a brief introduction to the experiment. Before they put on the VR headset, we measured the pupil distance (PD) of the participants, and adjusted the PD on the VR headset. Two types of training were included in this experiment: interaction training and task training. Both were  conducted when each flow map representation was presented to the participants for the first time.

During \textit{interaction training} participants were introduced to the flow map with details of the encodings and given sufficient time to familiarise themselves with  interaction. They were then asked to pick up the map and put it on a virtual table in VR. This activity familiarised participants with each flow map representation as well as the VR headset and controller. This was followed by \textit{task training}. Two sample tasks were given to participants with unlimited time. We asked the participants to check their strategies both when they were doing the training tasks and when the correct answers for those tasks were shown.

Participants were presented with the five flow map representations in counterbalanced order. A posthoc questionnaire recorded feedback on: 
\begin{itemize}
	\item preference ranking of visualisations in terms of visual design and ease of use for the tasks;
	\item advantages and disadvantages of each visualisation;
	\item strategies for different flow maps;
	\item background information about the participant.
\end{itemize}

In the questionaire the visualisations were listed in the same order that they were presented to participants during the experiment. All experimental materials are available for download from \url{https://vis.yalongyang.com/VR-Flow-Maps.html}.

\noindent\textbf{Measures}\\
We measured the time between the first rendering of the visualisation and the double-click on the controller trigger button. After participants double-clicked, the visualisation was replaced by two buttons to answer the question. Collected answers were binary (i.e. participants chose between two options) and we therefore used the  accuracy score from \cite{Willett:2015fv} to indicate perfect performance with 1, and a result equal to pure chance (i.e. randomly guessing) with 0: $(\frac{number\ of\ correct\ responses}{number\ of\ total\ responses} - 0.5) \times2$.

We also recorded the number of clicks, head position, head rotation, controller position, and map position every 0.1s. Recording these parameters is important, as users can move in a relatively large open space with the HTC Vive HMD. 

\subsection{Results}
Accuracy scores were not normally distributed (checked with histograms and Q$-$Q plots). Significance was tested with the Friedman test because we have more than two conditions; the Wilcoxon-Nemenyi-McDonald-Thompson post-hoc test was used to compare pairwise \cite{Hollander:1999ns}.

\begin{figure}[b!]
	\centering
	\includegraphics[height=20cm]{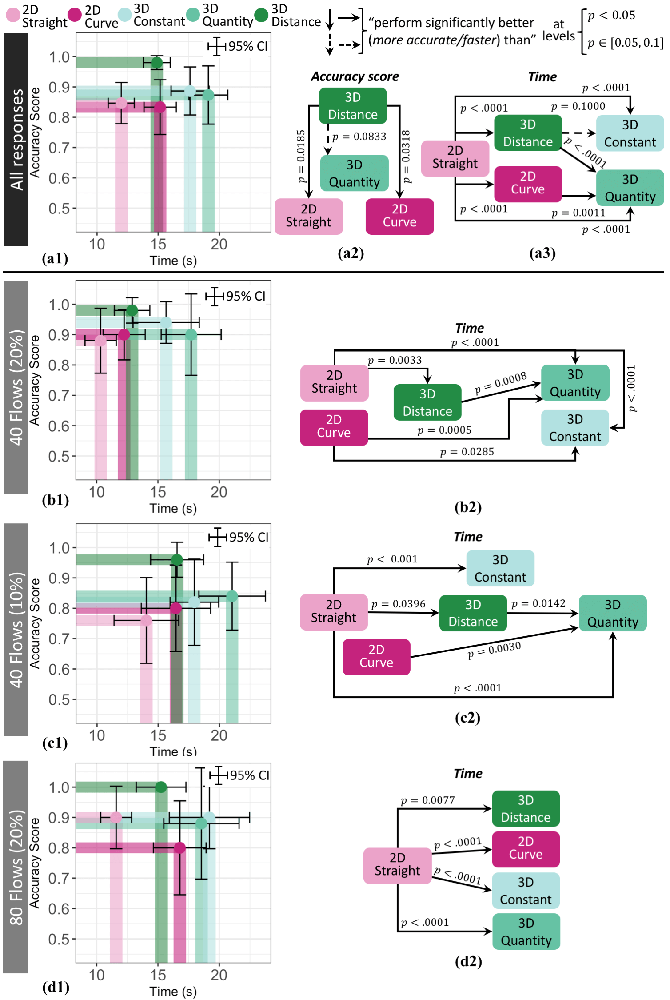}
	\caption{Study 1: Accuracy score and response time for different flow map representations in first study: (a1, b1, c1, d1) Average performance with 95\% confidence interval, (a2, a3, b2, c2, d2) graphical depiction of results of pairwise post-hoc test.}
	\label{fig:vr-flow-maps:first-study-result-all}
\end{figure}

Response times were log-normal distributed (checked with histograms and Q$-$Q plots), so a log-transform was used for statistical analysis~\citep{howell2012statistical}.
We chose one-way repeated measures ANOVA with linear mixed-effects model to check for significance and applied Tukey's HSD post-hoc tests to conduct pairwise comparisons~\citep{Andy:2012ds}. For user preferences we again used the Friedman test and the Wilcoxon-Nemenyi-McDonald-Thompson post-hoc test to test for significance. 

The Friedman test revealed a statistically significant effect of visualisations on accuracy ($\chi^2(4) = 12.29, p = .015$). Fig.~\ref{fig:vr-flow-maps:first-study-result-all}(a1) shows the average accuracy score of \textit{3D distance} (0.98) was higher than that of \textit{3D quantity} (0.87) and of 2D flow maps (\textit{straight} with 0.85 and \textit{curve} with 0.83). While \textit{3D distance} also outperformed \textit{3D constant} (0.89), this was not found to be statistically significant. A post-hoc test showed statistical significances as per Fig.~\ref{fig:vr-flow-maps:first-study-result-all}(a2).

The ANOVA analysis showed significant effect of visualisations on time ($\chi^2(4) = 50.63, p < .0001$). \textit{2D straight} (avg.\ 12.0s) was significantly faster than other flow maps. \textit{3D distance} (avg.\ 14.9s) and \textit{2D curve} (avg.\ 15.2s) were significantly faster than \textit{3D constant} (avg.\ 17.6s) and \textit{3D quantity} (avg.\ 19.1s) (see Fig.~\ref{fig:vr-flow-maps:first-study-result-all}(a3)).

By difficulty condition the Friedman test did not reveal significant effect for accuracy. The ANOVA analysis revealed:\\
\noindent\textbf{\textit{40 flows (20\%)}:} $\chi^2(4) = 36.39, p < .0001$. \textit{2D straight} (avg.\ 10.3s) was significantly faster than other flow maps except \textit{2D curve} (avg.\ 12.2s). \textit{3D distance} (avg.\ 12.9s) was only slower than \textit{2D straight}. \textit{3D quantity} (avg. 17.7s) was slower than other flow maps, except \textit{3D constant} (avg.\ 15.7s).\\
\noindent\textbf{\textit{40 flows (10\%)}:} $\chi^2(4) = 31.30, p < .0001$. \textit{2D straight} (avg.\ 14.0s) was significantly faster than other flow maps except \textit{2D curve} (avg.\ 16.4s). \textit{3D quantity} (avg.\ 21.0s) was significantly slower than \textit{3D distance} (avg. 16.5s). It also seemed to be slower than \textit{3D constant} (avg. 18.0s), but no statistical significance was found.\\
\noindent\textbf{\textit{80 flows (20\%)}:} $\chi^2(4) = 34.39, p < .0001$. \textit{2D straight} (avg.\ 11.6s) was significantly faster than other flow maps: \textit{2D curve} (avg.\ 16.8s), \textit{3D distance} (avg.\ 15.3s), \textit{3D constant} (avg.\ 19.2s) and \textit{3D quantity} (avg.\ 18.5s).

\begin{figure}[t!]
	\centering
	\includegraphics[width=0.44\textwidth]{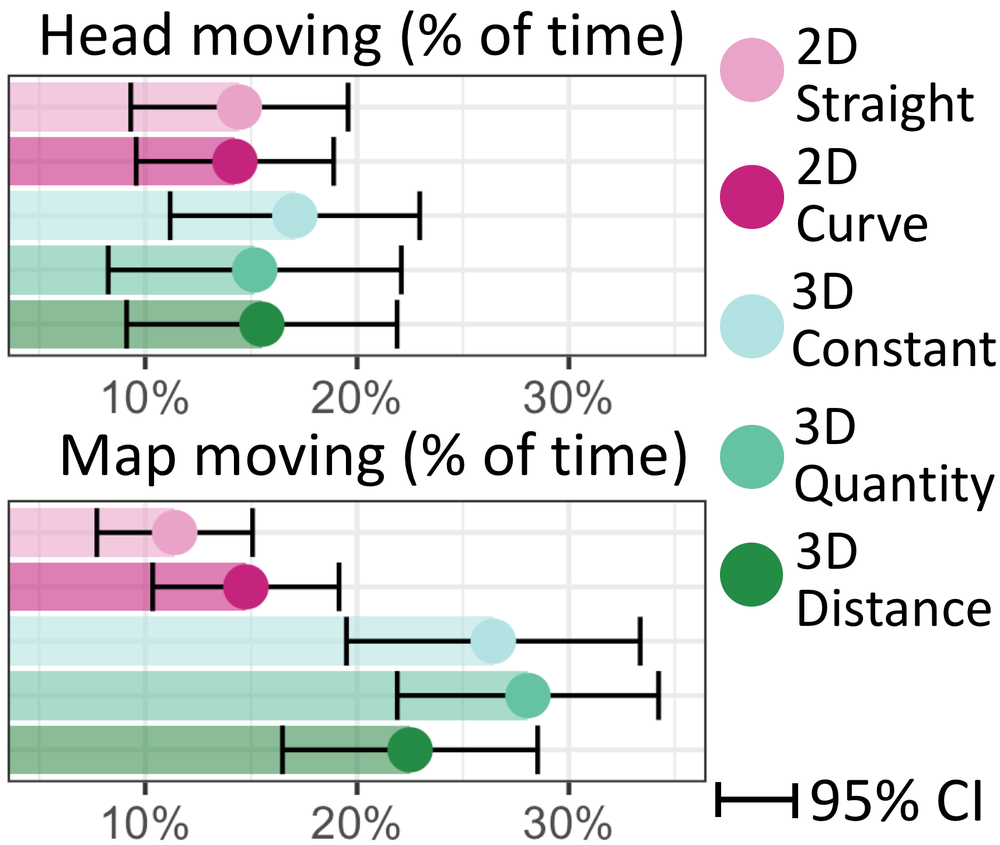}
	\caption{Study 1: Interaction time percentage with 95\% confidence interval}
	\label{fig:vr-flow-maps:first-interaction}
\end{figure}
\begin{figure}[t!]
	\centering
	\includegraphics[width=0.8\textwidth]{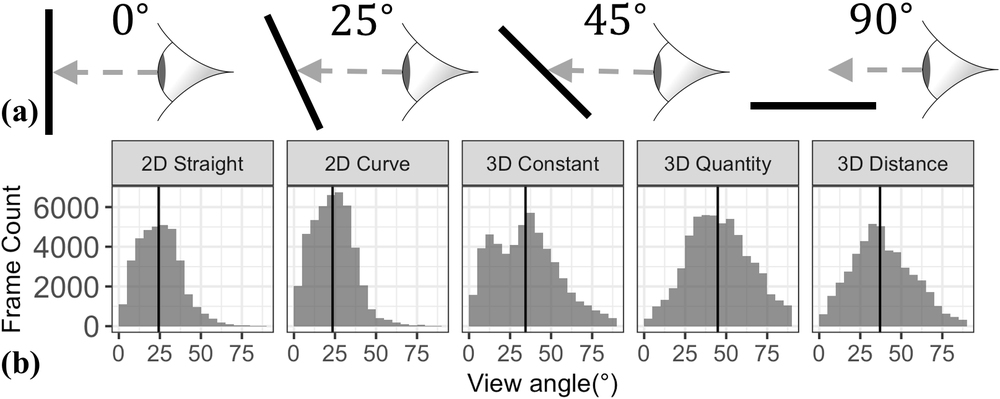}
	\caption{(a)~Demonstration of different view angles, (b)~view angle distribution among different flow maps with median line.}
	\label{fig:vr-flow-maps:first-angle}
\end{figure}

When analysing the details of interaction, we sampled every second frame. If the head or the map moved more than 1cm or rotated more than 5\textdegree, we considered it as an interaction, and accumulated the interaction time for every user and then normalised the time related to the percentage of time spent on that question (see Fig.~\ref{fig:vr-flow-maps:first-interaction}). Friedman test revealed a statistically significant effect of visualisations on map movements ($\chi^2(4) = 30.4, p < .0001$). Participants tended to move the map significantly more in all 3D conditions than 2D conditions (all $p < .05$). Wilcoxon signed rank test also revealed participants spent statistically significant more percentage of time moving the map than their head in \textit{3D distance} (at level $p=.10$) and \textit{3D quantity} (at level $p=.06$).

We also analyzed the view angle, i.e. the angle between viewers' heads forward vector and the normal vector of the map plane.  The Friedman test revealed that the effect of visualisations on the percentage of time spent with a view angle larger than 45\textdegree~per user  was statistically significant  ($\chi^2(4) = 64.88, p < .0001$). As one might expect, participants spent significantly more percentage of time with large view angles in all 3D conditions than all 2D conditions (see Fig.~\ref{fig:vr-flow-maps:first-angle}). 

\begin{figure}[b!]
	\captionsetup[subfigure]{justification=centering}
    \centering 
    \begin{subfigure}{0.48\textwidth}
        \includegraphics[width=\textwidth]{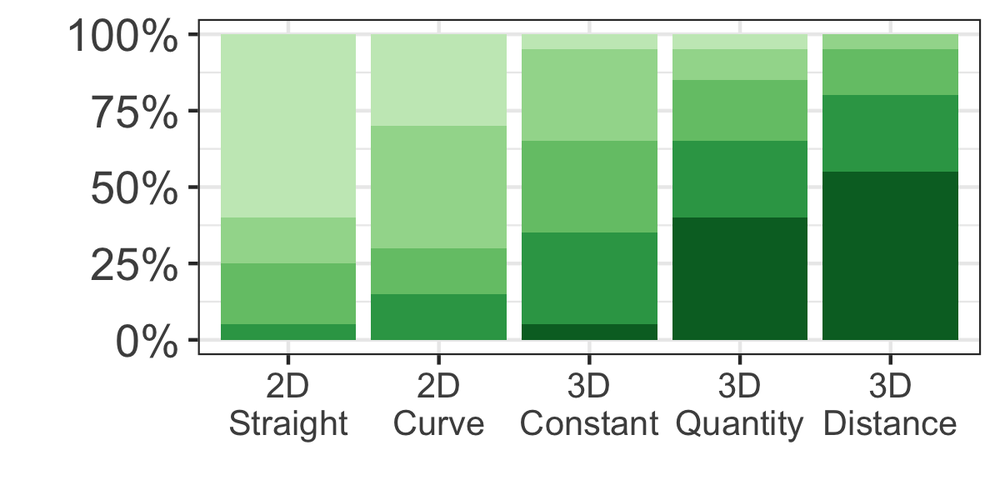}
        \caption{Visual Design Ranking}
    \end{subfigure}
    \begin{subfigure}{0.48\textwidth}
        \includegraphics[width=\textwidth]{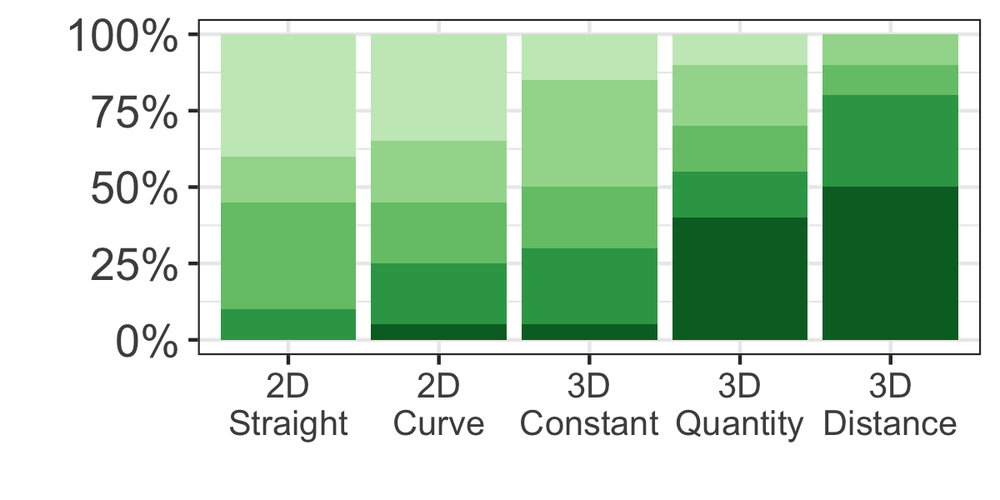}
        \caption{Readability Ranking}
    \end{subfigure}
    \caption{User preference ranking (\setlength{\fboxsep}{1.5pt}\colorbox{first}{\textcolor{white}{$1^{st}$}}, \setlength{\fboxsep}{1pt}\colorbox{second}{\textcolor{white}{$2^{nd}$}}, \setlength{\fboxsep}{1pt}\colorbox{third}{$3^{rd}$}, \setlength{\fboxsep}{1pt}\colorbox{fourth}{$4^{th}$} and \setlength{\fboxsep}{1pt}\colorbox{fifth}{$5^{th}$}).}
    \label{fig:vr-flow-maps:study-01-preference}
\end{figure} 

\noindent\textbf{User preference and feedback}\\
Participant ranking for each of the four visualisations by percentage of respondents is shown by colour (see Fig.~\ref{fig:vr-flow-maps:study-01-preference}). 

For \emph{visual design}, the Friedman test revealed a significant effect of visualisations on preference ($\chi^2(4)=38.6, p < .0001$). The strongest preference was for \textit{3D distance}, with 95\% voting it as top three. The post-hoc tests also found a stronger preference for \textit{3D distance} than \textit{3D constant} (65\% voting it as top three), \textit{2D curve} (40\% voting it as top three) and \textit{2D straight} (30\% voting it as top three). Participants also seemed to prefer \textit{3D quantity} (85\% voting it as top three), however, the post-hoc tests only suggested it was preferred to \textit{2D straight}. 

For \emph{readability}, the Friedman test indicated significant effect of visualisations on preference ($\chi^2(4)=23.32, p = .0001$). The strongest preference is again for the \textit{3D distance}, with 90\% of respondents voting it top three. The post-hoc tests again showed stronger preference for \textit{3D distance} than \textit{3D constant} (50\% voting it as top three), \textit{2D curve} (45\% voting it as top three) and \textit{2D straight} (45\% voting it as top three). Participants also seemed to prefer \textit{3D quantity} (70\% voting it as top three), however, the post-hoc tests again only revealed it was preferred over \textit{2D straight}. 

The final section of the study allowed participants to give feedback on the pros and cons of each design. Qualitative analysis of these comments reveal (overall):

\noindent\textbf{\textit{2D straight}} was found to be easy for small data sets, however, lines were found to be hard to distinguish due to increasing overlaps in large data sets. Several participants reported: \emph{``I answered sometimes with a very low confidence, close to luck.''}

\noindent\textbf{\textit{2D curve}} was found to have fewer overlaps than \textit{2D straight}. However, many participants reported the curvature made it difficult to follow lines, and sometimes, the curvature was found to be unexpected, which apparently increased difficulty.

\noindent\textbf{\textit{3D constant}} was found to have considerable numbers of overlaps by most participants. However, some participants also found it efficient with interaction: \emph{``I could look at the line from a straight angle, plus wiggle the map around a little to confirm the line.''} 

\noindent\textbf{\textit{3D quantity}} was more trusted. Many participants reported: \emph{``If I couldn't work out from thickness, I could move the visualisation to compare heights from the side to confirm my answer. I felt more confident.''} However, they also commented about the extra time they spent for this confirmation.

\noindent\textbf{\textit{3D distance}} was easy to distinguish flows. \emph{``This one left enough gaps between the curves to clearly distinguish the curves''} and \emph{``it was visually appealing.''} However, a few participants commented they felt more confident with \textit{3D quantity}, and a few commented that this might be due to the double encoding used by \textit{3D quantity} (\textit{3D quantity} encodes quantity with height and width).

\subsection{Key Findings}
The main finding of this study was that the \textit{3D distance} was more accurate than the other 3D conditions and both 2D visualisations. It was also the preferred visualisation. We also found that:
\begin{itemize}
	\item The \textit{2D straight-line} flow map was the fastest in almost all conditions but least preferred.	
	 \item Participants tended to look more often from the side in 3D conditions than in 2D conditions.
	\item Participants tended to interact with the map more in 3D conditions than 2D conditions.
	\item Participants tended to move the map more than their heads in \textit{3D distance} and \textit{3D quantity}.
\end{itemize}

%% file: content/6-od-flow-maps-in-vr/4-study-2.tex
\section{Study 2: Flows on Flat Maps, Globes and Map Pairs}
\label{sec:vr-flow-maps:study-02}
In the second study, we focused on exploring different 2D and 3D representations of the reference space. 

\subsection{Visualisations and Interactions}
\label{sec:vr-flow-maps:globes-mapslink}
We evaluated 4 different representations. 

\noindent\textbf{\textit{2D straight}} and \textbf{\textit{3D distance}}\\ 
The first two used a flat map to represent the reerence space. These were the best performing representations from our first study (fastest with  \textit{\textit{2D straight}}, most accurate with \textit{\textit{3D distance}}). 

\begin{figure}[t!]
	\centering
	\includegraphics[width=\textwidth]{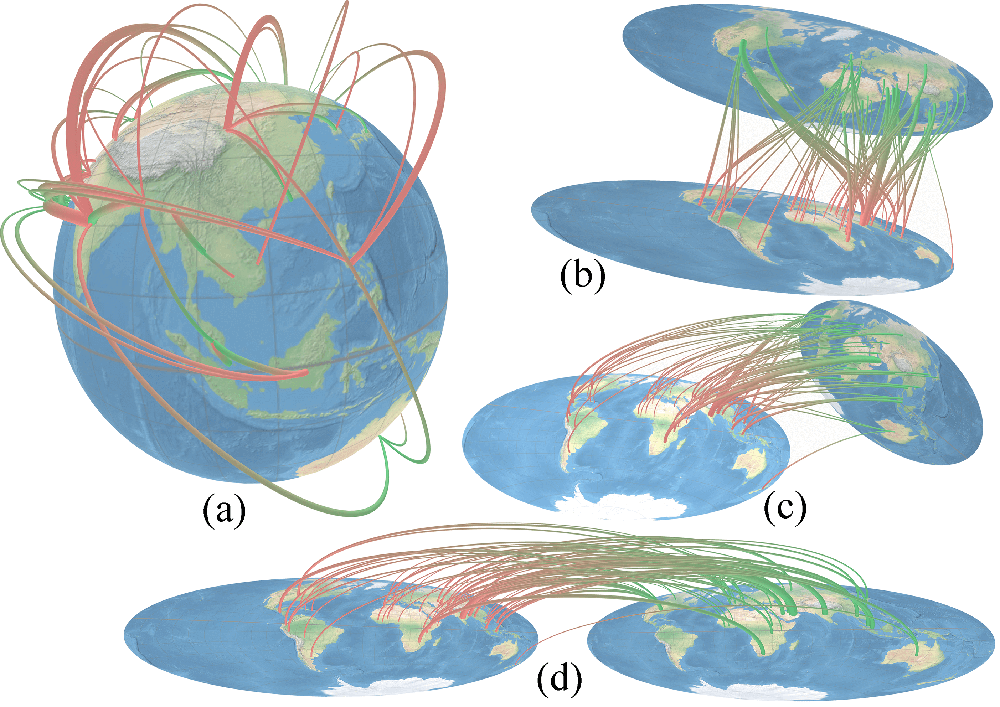}
	\caption{Study 2: (a) 3D globe flow map, (b, c, d) MapsLink: flow tubes linking a pair of flat maps.}
	\label{fig:vr-flow-maps:second-study-vis}
\end{figure}

\noindent\textbf{\textit{Globe}}\\
A 3D globe has proven to be an effective way to present global geometry \cite{Yang:2018mg} but the effectiveness for showing OD flows has not been previously evaluated. 
We represented flow in the globe using 3D tubes, i.e. we linked two locations on the globe with their great circle trajectory (Fig. \ref{fig:vr-flow-maps:second-study-vis}(a)).
Based on the result of our first study, we chose to use curve height to encode the great circle distance between two points where height here refers to the distance between the curve's centre point and the centre of the globe. 
We used a cubic transformation with interpolation factor $0 \le t \le 1$, $h_t = ((-|t-0.5| / 0.5)^3 + 1) \times h + radius$.

\noindent\textbf{\textit{MapsLink}}\\
We also evaluated a novel flow map representation which used a separate reference space for the origin and destination. 
This used two flat maps in 3D space: the \textit{origin map} showing  origins and the \textit{destination map} showing  destinations (Fig. \ref{fig:vr-flow-maps:second-study-vis}~(b)). Flows from origin to destination were rendered with curved tubes linking origins in the origin map and destinations in the destination map. The 3D tubes were cubic B\'ezier curves (see Equ. \ref{equ:bezier}) with orgin and destination as the first and last control point. As the two maps might not be in the same plane, we could not control the height in the same way as the first experiment, instead, the two control points were raised from the orign and destination maps to the same height, which was proportional to the Euclidean distance in 3D space between the origin and destination points. This meant when the origin and destination map were  facing each other, origins and destinations were linked by straight lines and by smooth curves at other orientations. 

\noindent{\textbf{Interactions}}\\
As in the first study participants could rotate and reposition the visualisation. In addition we allowed participants to adjust the centre of the geographic area in the visualisation in VR. Viewers could pick any location and drag it to a new position using the VR controller. This interaction (called \textit{geo-rotation}) was presented in~\cite{Yang:2018mg}
and allows the viewer to bring the geographic area of interest to the centre of the visualisation. This changes the relative position of points and straight line flows on the flat map, thus providing some of the benefits that changing viewpoint provided for the 3D representations of flow. We were interested to see if it improved accuracy of the flat map with straight line 2D flows. 

\subsection{Experiment}
\noindent\textbf{Stimuli and tasks}\\
The same task of finding and comparing flows between two origin-destination pairs was used in this study. The same raw data was used as well. With the addition of the \textit{geo-rotation} interaction, we assumed participants could complete the task more easily. We therefore increased the difficulty of the three conditions: (a) 80 flows with 20\% difference, (b) 80 flows with 10\% difference and (c) 120 flows with 20\% difference. Pilots demonstrated participants could handle these difficulty conditions.

\noindent\textbf{Set-up}\\
The headset and PC setup were the same as used in Study 1, except that two controllers were given to the participants so that they could use one controller to position and rotate the map/globe while the other is used for \textit{geo-rotation}.
In \textit{MapsLink} two controllers also affords bimanual gestures to manipulate both maps simultaneously.

The \textit{2D straight} and \textit{3D distance} setup was the same as for Study 1. As for was the case for 3D maps in Study 1, the thickness of tubes in \textit{globe} and \textit{MapsLink} was linearly mapped to the range of 0.1cm to 0.8cm.

The globe had a radius of 0.4 metre. The starting position for the centre of the globe was 1 metre in front and 0.3 metre under the participant's eye position. The geographic centre of the globe was set at 0\textdegree longitude and 0\textdegree latitude, facing towards the viewer. As for \textit{3D distance}, the height was linearly mapped to the range of 5cm to 25cm.

The two maps of \textit{MapsLink} measure 75\% of the size of the flat map in the first study (0.75 $\times$ 0.375 metre). We reduced size to reduce the chance of the two maps intersecting. The two maps were first placed 0.55 metre in front and 0.3 metre under participant's eye position, then the origin map was moved left 0.4 metre, and the destination map was moved right 0.4 metre. Finally, both maps were tilted towards participants around the y-axis by 30\textdegree and around the x-axis by 45\textdegree. Flows were modeled with cubic B\'ezier curves: The two control points were placed on a line orthogonal to map planes; distances between control points and planes were between 5 and 50cm, and were proportional to the distances between origins and detinations (which were assumed to be between 0 and 2m).

\noindent\textbf{Participants}\\
We recruited 20 participants (6 female) from our university campus, all with normal or corrected-to-normal vision. Participants included university students and researchers. 14 participants were within the age group 20$-$30, 5 participants were between 30$-$40, and 1 participant was over 40. VR experience varied: 14 participants had less than 5 hours of prior VR experience, 4 participants had 6$-$20 hours, and 2 participants had more than 20 hours.

\noindent\textbf{Design and Procedure}\\
A similar design to the first user study was used, within-subjects: 20 participants $\times$ 4 visualisations $\times$ 1 task $\times$ 3 difficulty levels $\times$ 5 repetitions = 1,200 responses (60 responses per participant) with performance measures and lasted one hour on average. Latin square design was used to balance the order of visualisations, and 4 data sets were ordered  to balance the effect of tasks across participants (i.e. every flow map was tested on all data sets).

The procedure was similar to Study 1 but with two modifications.
In \textit{interaction training}, in addition to asking participants to place the flow maps on top of a table, we also asked them to use \textit{geo-rotation} to rotate Melbourne to the centre of the map or to the centre of participant's view. In the \textit{posthoc questionnaire}, we added a question to rate their confidence with each flow map with a five-point-Likert scale.

\noindent\textbf{Measures}\\
In addition to real-time recording of participant's head, controller and map  position and rotation information, we also recorded the time duration whenever a participant used \textit{geo-rotation}.

\subsection{Results}
\begin{figure}[b!]
	\centering
	\includegraphics[height=20cm]{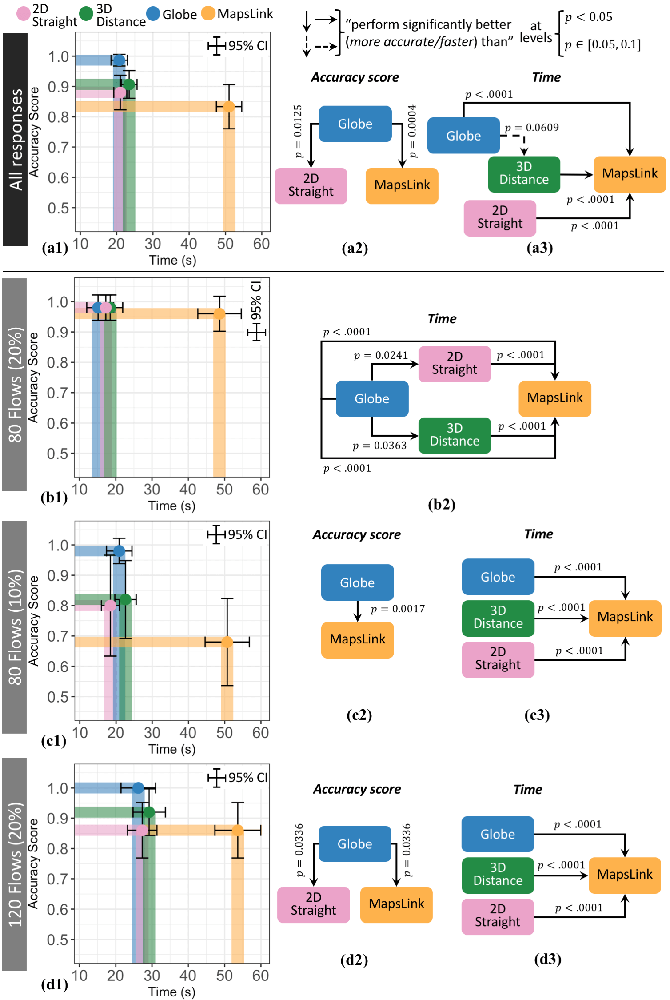}
	\caption{Study 2: (a1, b1, c1, d1) Average performance with 95\% confidence interval, (a2, a3, b2, c2, c3, d2, d3) graphical depiction of results of pairwise post-hoc test.}
	\label{fig:vr-flow-maps:second-study-result-all}
\end{figure}
As in the first study, after checking normality of the data with histograms and Q$-$Q, we used the Friedman test to check for significance of accuracy score and applied the Wilcoxon-Nemenyi-McDonald-Thompson post-hoc test to conduct pairwise comparisons. For response time, we chose one-way repeated measures ANOVA with linear mixed-effects model to check for significance of its $log$ transformed values and applied Tukey's HSD post-hoc to conduct pairwise comparisons. 

The Friedman test revealed a statistically significant effect of visualisations on accuracy ($\chi^2(3) = 18.06, p = 0.0004$). Fig.~\ref{fig:vr-flow-maps:second-study-result-all}~(a1) shows the average accuracy score of \textit{globe} (avg. 0.99) was higher than that of \textit{2D straight} (avg. 0.88) and of \textit{MapsLink} (avg. 0.83). While \textit{globe} also outperformed \textit{3D distance} (avg. 0.91), this was not found to be statistically significant. A post-hoc test showed statistical significances as per Fig.~\ref{fig:vr-flow-maps:second-study-result-all}~(a2).

The ANOVA analysis showed significant effect of visualisations on time ($\chi^2(3) = 107.87, p < .0001$). \textit{MapsLink} (avg. 50.9s) was significantly slower than other visualisations. \textit{Globe} (avg. 20.7s) was significantly faster than \textit{3D distance} (avg. 23.4s). \textit{2D straight} (21.0s) had no significances between \textit{globe} and \textit{3D distance}.

By difficulty condition the Friedman test revealed a significant effect on accuracy score:\\
\noindent\textbf{\textit{80 flows (20\%)}:} $\chi^2(3) = 0.69, p = .8750$. All visualisations had similar performance in this condition.\\
\noindent\textbf{\textit{80 flows (10\%)}:} $\chi^2(3) = 13.28, p = .0041$. \textit{Globe} (avg. 0.98) was significantly more accurate than \textit{MapsLink} (avg. 0.68). \textit{2D straight} (avg. 0.80) and \textit{3D distance} (avg. 0.82) had no statistical significance with other visualisations.\\
\noindent\textbf{\textit{120 flows (20\%)}:} $\chi^2(3) = 9.9, p = .0194$. Responses with \textit{globe} were perfect (with an accuracy score 1), and it was significantly more accurate than both \textit{2D straight} (avg. 0.86) and \textit{MapsLink} (avg. 0.86). \textit{3D distance} (avg. 0.92) had no statistical significance with other visualisations.

By difficulty condition the ANOVA analysis revealed significant effect on time:\\
\noindent\textbf{\textit{80 flows (20\%)}:} $\chi^2(3) = 97.62, p < .0001$. \textit{MapsLink} (avg. 48.6s) was significantly slower than other visualisations. \textit{Globe} (avg. 15.1s) was also significantly faster than \textit{2D straight} (avg. 17.3s) and \textit{3D distance} (avg. 18.5s).\\
\noindent\textbf{\textit{80 flows (10\%)}:} $\chi^2(3) = 71.6, p < .0001$. Again, \textit{MapsLink} (avg. 50.7s) was significantly slower than other visualisations: \textit{2D straight} (avg. 18.5s), \textit{3D distance} (avg. 22.6s) and \textit{globe} (avg. 20.9s).\\
\noindent\textbf{\textit{120 flows (20\%)}:} $\chi^2(3) = 58.72, p < .0001$. Again, \textit{MapsLink} (avg. 53.6s) was found to be significantly slower than other visualisations: \textit{2D straight} (avg. 27.3s), \textit{3D distance} (avg. 29.2s) and \textit{globe} (avg. 26.2s).

\begin{figure}[b!]
	\centering
	\includegraphics[width=\textwidth]{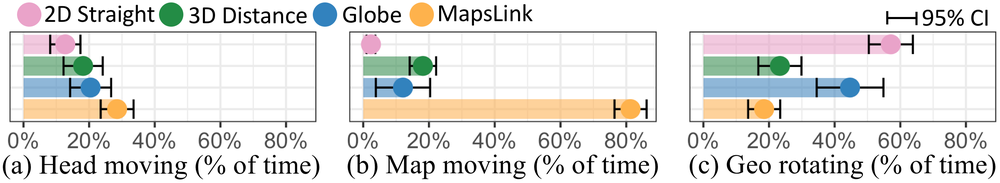}
	\caption{Study 2: Interaction time percentage with 95\% confidence interval}
	\label{fig:vr-flow-maps:second-interaction}
\end{figure}

\noindent\textbf{Interactions}\\
The percentage of time spent in different interactions per user was investigated (see Fig.~\ref{fig:vr-flow-maps:second-interaction}). The Friedman test was used to determine statistical significance between different visualisations and between different interactions.

For \textit{head movement}, participants tended to spend a smaller percentage of time moving their heads in \textit{2D straight} than \textit{3D distance} ($p=.09$), \textit{globe} ($p=.03$) and \textit{MapsLink} ($p<.0001$). In \textit{map movement}, participants tended to move \textit{MapsLink} significantly more than other visualisations and there was more head movement for \textit{3D distance}  than for \textit{2D straight} (all $p<.05$). In \textit{geo-rotation}, participants used \textit{geo-rotation} significantly more in \textit{globe} and in \textit{2D straight} than with \textit{3D distance} and \textit{MapsLink} (all $p<.05$).

In \textit{2D straight}, the percentage of time was significantly different across interaction types: \textit{geo-rotation} $>$ head movement $>$ map movement (all $p<.05$). In \textit{3D distance}, no significant difference was found among different interactions. In \textit{globe}, \textit{geo-rotation} $>$ map movement ($p=.0025$). In \textit{MapsLink}, map movement $>$ head movement $>$ \textit{geo-rotation} (all $p<.05$).

We also investigated the benefits of adding \textit{geo-rotation} to \textit{2D straight} and \textit{3D distance}. We compared the responses of \textit{80 flows, 20\%} in the first (without \textit{geo-rotation}) and second (with \textit{geo-rotation}) studies.

For \textit{2D straight}, Exact Wilcoxon-Mann-Whitney test revealed \cite{Hothorn:2008fy} an increase of accuracy with \textit{geo-rotation} at level $p = .0530$ with $Z=-1.5072$. Log-transformed time values have been analysed with mixed ANOVA, the result demonstrated a significant increase in response time with \textit{geo-rotation} ($\chi^2(1) = 13.07, p < .0001$). For \textit{3D distance}, Exact Wilcoxon-Mann-Whitney test and mixed ANOVA did not show a significant difference between with and without \textit{geo-rotation}. 

\begin{figure}[b!]
	\centering
	\includegraphics[width=\textwidth]{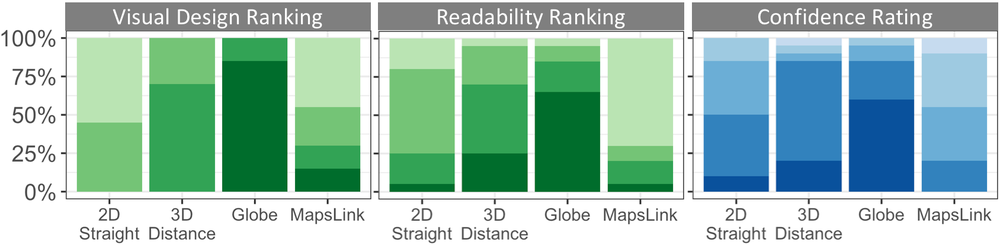}
	\caption{Study 2: Participants preference ranking (\setlength{\fboxsep}{1.5pt}\colorbox{first}{\textcolor{white}{$1^{st}$}}, \setlength{\fboxsep}{1pt}\colorbox{second}{\textcolor{white}{$2^{nd}$}}, \setlength{\fboxsep}{1pt}\colorbox{third}{$3^{rd}$} and \setlength{\fboxsep}{1pt}\colorbox{second_fourth}{$4^{th}$}) and confidence rating (from \setlength{\fboxsep}{1pt}\colorbox{first_conf}{\strut \textcolor{white}{fully confident}} to \setlength{\fboxsep}{2pt}\colorbox{last_conf}{not confident at all}).}
	\label{fig:vr-flow-maps:second-rankings}
\end{figure}

\noindent\textbf{User preference and feedback}\\
Participant ranking for each of the four visualisations by percentage of respondents is shown by colour (see Fig. \ref{fig:vr-flow-maps:second-rankings}). 

For \emph{visual design}, the Friedman test revealed a significant effect of visualisations on preference ($\chi^2(3)=38.58, p < .0001$). The strongest preference was for the \textit{globe}, with 100\% voting it in the top two. The post-hoc tests also proved the strongest preference for \textit{globe} compared to other flow maps with all $p < .05$. \textit{3D distance} (70\% voting it top two) was also statistically preferred to \textit{2D straight} (0\% voting it top two). \textit{MapsLink}, with 30\% voting it top two, did not show statistical difference between \textit{2D straight} or \textit{3D distance}. 

For \emph{readability}, the Friedman test indicated significant effect of visualisations on preference ($\chi^2(3)=25.5, p < .0001$). The strongest preference is again for the \textit{globe}, with 85\% of respondents voting it top two. The post-hoc revealed stronger preference of \textit{globe} than \textit{2D straight} (25\% voting it top two) and \textit{MapsLink} (20\% voting it as top two). \textit{3D distance} (70\% voting it top two) was also statistically preferred to \textit{MapsLink}.

A five-point-Likert scale was used for rating participants' confidence (see Fig. \ref{fig:vr-flow-maps:second-rankings}). The Friedman test revealed a significant effect of visualisations on confidence ($\chi^2(3)=24.82, p < .0001$). Participants felt significantly more confident in \textit{globe} and \textit{3D distance} than \textit{MapsLink}. \textit{Globe} was also found more confident than \textit{2D straight} (all $p<.05$).

The final section of the study allowed participants to give feedback on the pros and cons of each design. Qualitative analysis of these comments reveal (overall):

\noindent\textbf{\textit{2D straight}} was found to be very difficult at the beginning. However, many participants commented: \emph{``With the map rotating, it is usually possible to keep track of the pair of points.''}

\noindent\textbf{\textit{3D distance}} was found to be more visually appealing than \textit{2D straight}, and more efficient than \textit{2D straight} for small data. However, \emph{``things become very difficult when data size increases''}, and \emph{``sometimes, it felt more difficult than 2D (straight) map''}.

\noindent\textbf{\textit{Globe}} was found to be the most intuitive. Many participants also commented that the visualisation \emph{``felt very sparse so it was easy to tell which lines were connected to the points''}. Some participants also suggested to have a snapshot functionality to store the current globe rotation or two globes  positioned side by side.

\noindent\textbf{\textit{MapsLink}}: was found to be \emph{``very interesting to play with, but very difficult to use when it comes to the questions''}. However, some participants liked the freedom of manipulating it: \emph{``You can almost find the certain answer for each question by patiently manipulating map positions and rotating the maps.''} Meanwhile, many participants reported it took a long time to answer questions.

\subsection{Key Findings}
The main finding of this study was that user performance with the \textit{globe} was significantly more accurate than with \textit{2D straight} and \textit{MapsLink}. There was also some evidence that the \textit{globe} was more accurate than \textit{3D distance}, but this was not statistically significant. There was also some evidence that the \textit{globe} is resistant to increased clutter density (performance was stable with increasing flows, while other visualisations degraded with the number of flows). Additionally, we found:
\begin{itemize}
	\item \textit{MapsLink} was significantly slower than other representations and that participants spent most of their time moving the maps in \textit{MapsLink} (more than 80\% in average). 
	\item \textit{Geo-rotation} increased accuracy and slowed response time for \textit{2D straight}. With \textit{geo-rotation} there was no longer a significant difference in accuracy or speed between \textit{2D straight} and \textit{3D distance}.
	\item \textit{Globe} had the strongest preference in terms of visual design, while participants were more confident with both \textit{globe} and \textit{3D distance} and preferred them for readability.
	\item Participants chose to use different interactions in different representations. Compared to other visualisations, participants do not like to move their heads in \textit{2D straight} and participants liked to use \textit{geo-rotation} in \textit{2D straight} and \textit{globe}.
\end{itemize}

%% file: content/6-od-flow-maps-in-vr/5-study-3.tex
\section{Study 3: Dense Flow Data Sets}
\label{sec:vr-flow-maps:study-03}
The third study was designed to investigate the scalability of the different flow maps. As participants spent significantly more time on \textit{MapsLink} and qualitative feedback indicated limited scalability for this design, we decide to test only the other three visualisations: \textit{2D straight}, \textit{3D distance} and \textit{globe}. We tested them with 200 and 300 flows, both with 10\% difference. 

We recruited 12 participants (6 female) from our university campus, all with normal or corrected-to-normal vision. Participants included university students and researchers. 9 participants were within the age group 20--30, 2 participant was between 30--40, and 1 participant was over 40. VR experience varied: 10 participants had less than 5 hours of prior VR experience, 2 participants had 6--20 hours.

Otherwise, experimental design and setup was identical to Study 2, within-subjects: 12 participants $\times$ 3 visualisations $\times$ 1 task $\times$ 2 difficulty levels $\times$ 8 repetitions = 576 responses (48 responses per participant) with performance measures and duration of one hour on average. 

\begin{figure}[t!]
	\centering
	\includegraphics[height=16cm]{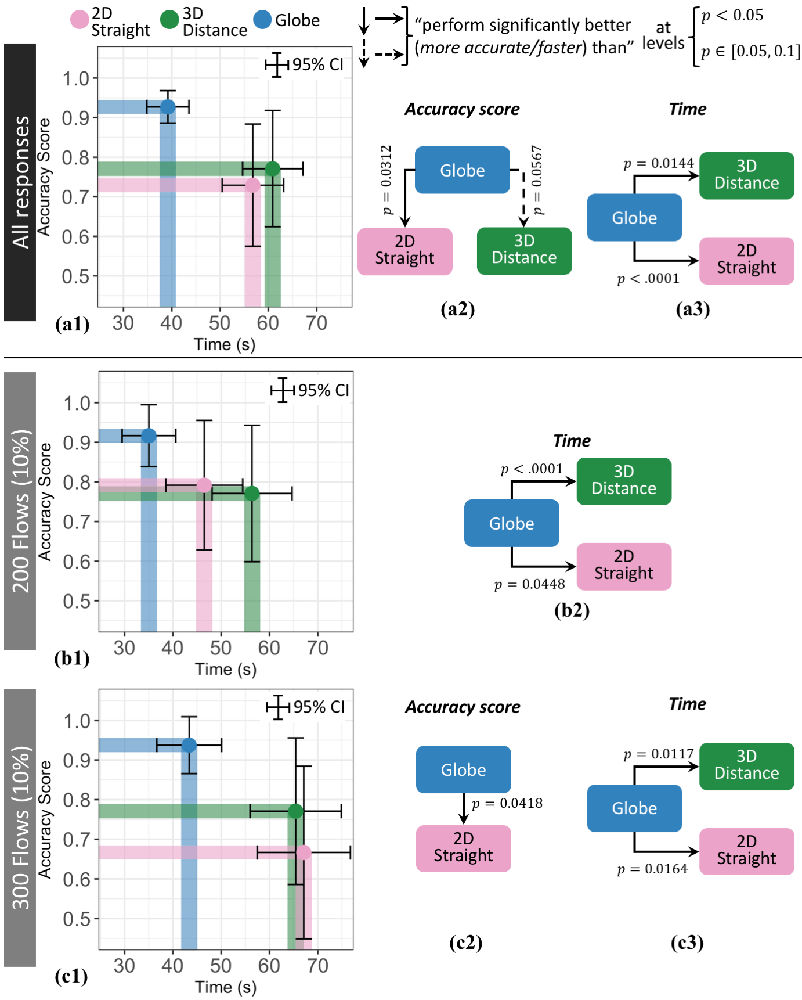}
	\caption{Study 3: (a1, b1, c1) Average performance with 95\% confidence interval, (a2, a3, b2, c2, c3) graphical depiction of results of pairwise post-hoc test.}
	\label{fig:vr-flow-maps:third-study-result-all}
\end{figure}
\subsection{Results} 
The same statistical analysis methods were used for accuracy score and $log$ transformed responding time. The Friedman test revealed a statistically significant effect of visualisations on accuracy ($\chi^2(2) = 7.79, p = .0203$). Fig.~\ref{fig:vr-flow-maps:third-study-result-all}~(a1) shows \textit{Globe} (avg. 0.93) was significantly more accurate than \textit{3D distance} (avg. 0.77) and \textit{2D straight} (avg. 0.73). The ANOVA analysis also showed significant effect of visualisations on time ($\chi^2(2) = 11.82, p = .0027$). \textit{Globe} (avg. 39.2s) again was significantly faster than \textit{3D distance} (avg. 60.9s) and \textit{2D straight} (avg. 56.8s). While \textit{3D distance} was slightly more accurate than \textit{2D straight} this was not statistically significant.

By difficulty condition the Friedman test revealed significant effect for accuracy score:\\
\noindent\textbf{\textit{200 flows (10\%)}:} $\chi^2(2) = 2.47, p = .2910$. No statistical significance effect was found in this condition of visualisations.\\
\noindent\textbf{\textit{300 flows (10\%)}:} $\chi^2(2) = 6.26, p = .0437$. \textit{Globe} (avg. 0.94) was significantly more accurate than \textit{2D straight} (avg. 0.67). No significant difference between \textit{3D distance} (avg. 0.77) and other flow maps.

By difficulty condition the ANOVA analysis revealed significant effect for time:\\
\noindent\textbf{\textit{200 flows (10\%)}:} $\chi^2(2) = 13.73, p = .0010$. \textit{Globe} (avg. 35.0s) was significantly faster than both \textit{2D straight} (avg. 46.5s) and \textit{3D distance} (avg. 56.4s).\\ 
\noindent\textbf{\textit{300 flows (10\%)}:} $\chi^2(2) = 8.72, p = .0128$. \textit{Globe} (avg. 43.4s) again was significantly faster than both \textit{2D straight} (avg. 67.1s) and \textit{3D distance} (avg. 65.4s).

\begin{figure}[t!]
	\centering
	\includegraphics[width=\textwidth]{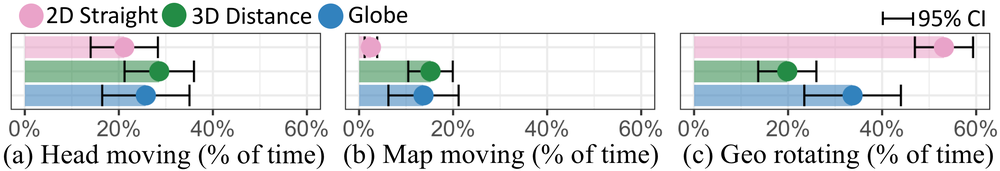}
	\caption{Study 3: Interaction time percentage with 95\% confidence interval}
	\label{fig:vr-flow-maps:third-interaction}
\end{figure}

\noindent\textbf{Interactions}\\
The percentage time difference between interactions per user is demonstrated in Fig.~\ref{fig:vr-flow-maps:third-interaction}. The Friedman test was used to analyse the relationship between interactions and visualisations.
In \textit{head movement}, there is no significant difference among the visualisations. In \textit{map movement}, \textit{3D distance} $\approx$ \textit{globe} $>$ \textit{2D straight} ($\approx$ means no statistical significance found between two visualisations). In \textit{geo-rotation}, \textit{2D straight} $>$ \textit{3D distance} ($p<.0001$), \textit{2D straight} $>$ \textit{globe} ($p=.0637$) and \textit{globe} $>$ \textit{3D distance} ($p=.0637$).
In \textit{\textit{2D straight}}, \textit{geo-rotation} $>$ head movement $>$ map movement (all $p<.05$). In \textit{\textit{3D distance}}, no statistical significance found among different interactions. In \textit{globe}, \textit{geo-rotation} $>$ map movement ($p=.0216$).

\begin{figure}[b!]
	\centering
	\includegraphics[width=\textwidth]{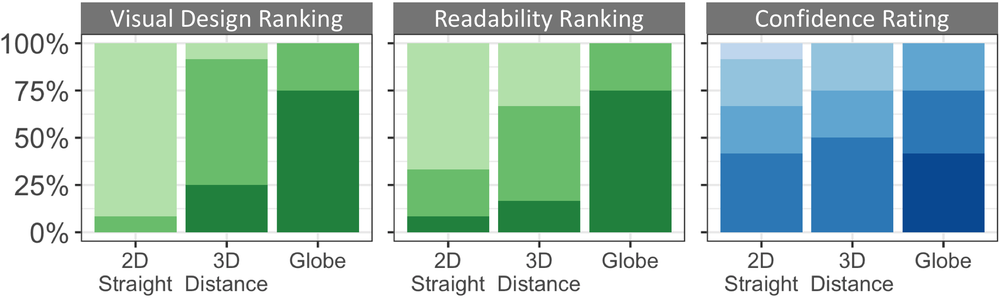}
	\caption{Study 3: Participants preference ranking (\setlength{\fboxsep}{1.5pt}\colorbox{third_first}{\textcolor{white}{$1^{st}$}}, \setlength{\fboxsep}{1pt}\colorbox{third}{\textcolor{black}{$2^{nd}$}} and \setlength{\fboxsep}{1pt}\colorbox{second_fourth}{$3^{rd}$}) and confidence rating (from \setlength{\fboxsep}{1pt}\colorbox{first_conf}{\strut \textcolor{white}{fully confident}} to \setlength{\fboxsep}{2pt}\colorbox{last_conf}{not confident at all}).}
	\label{fig:vr-flow-maps:third-rankings}
\end{figure}

\noindent\textbf{User preference}\\
Participant ranking for each of the three visualisations by percentage of respondents is shown by colour (see Fig.~\ref{fig:vr-flow-maps:third-rankings}). 

For \emph{visual design}, the Friedman test revealed a significant effect of visualisations on preference ($\chi^2(2)=17.17, p = .0001$). Both \textit{globe} (75\% voting it the best) and \textit{3D distance} (25\% voting it the best) were preferred to \textit{2D straight} (0\% voting it as the best) with all $p < .05$. 

For \emph{readability}, the Friedman test revealed a significant effect of visualisation on preference ($\chi^2(2)=11.17, p = .0038$). \textit{Globe} (75\% voting it the best) was preferred to \textit{2D straight} (8.33\% voting it the best) at significant level $p=.0030$ and \textit{3D distance} (16.67\% voting it the best) at significance level $p=.0638$. 

As in Study 2, a five-point-Likert scale was used for rating participants' confidence (see Fig.~\ref{fig:vr-flow-maps:third-rankings}). The Friedman test revealed a significant effect of visualisations on confidence ($\chi^2(2)=5.19, p = .07463$). Participants felt more confident with \textit{globe} than \textit{2D straight} at significance level $p=.0954$.

\subsection{Key Findings}
The main finding of Study 3 was confirmation that the findings of Study 2 extend to larger data sets. We found that \textit{globe} was more accurate and faster than \textit{2D straight} and \textit{3D distance} for larger datasets. Overall, \textit{globe} was the preferred visualisation and again participants tended to use \textit{geo-rotation} more than other interactions with both \textit{2D straight} and \textit{globe}. There was some evidence that even with \textit{geo-rotation} \textit{3D distance} scaled better to larger data sets than \textit{2D straight} but this was not statistically significant.

We were surprised by the performance of \textit{globe}, as viewers can only see half of the globe at a time. We therefore investigated performance on items where the OD flows were more than 120\textdegree \ apart. Again we found that in both Studies 2 and 3, performance was better with the \textit{globe} than the other two representations. 

%% file: content/6-od-flow-maps-in-vr/6-conclusion.tex
\section{Conclusion}
\label{sec:vr-flow-maps:conclusion}
This chapter significantly extends the understanding of how to visualise spatially embedded data in modern immersive environments by systematically investigating and evaluating different 2D and 3D representations for OD flow maps.
We have conducted the first investigation and empirical evaluation of OD flow map visualisation with a modern head-tracked binocular VR HMD. We have found strong evidence that 2D OD flow maps are \textit{not} the best way to show origin-destination flow in such  an environment, and that the use of 3D flow maps can allow viewers to resolve overlapping flows by changing the relative position of the head and object.  However, the particular 3D design choices of the visualisation have a significant effect, for example, encoding flow height to distance was clearly better than to quantity while our most novel use of 3D space, \emph{MapLink}, had the worst performance.

We found that for global flows, the most accurate and preferred representation was a 3D globe with raised flows whose height is proportional to the flow distance, while for regional flow data the best view would be a flat map with distance-proportional raised 3D flows. We found that accuracy of a standard flat flow map with straight lines could be significantly improved by the use of \emph{geo-rotation}, in which the user can interactively reposition the centre of the map. Nonetheless, the 3D representations were still more accurate and preferred.

Our findings suggest that globes are preferable for visualising global OD flow data in immersive environments. For regional OD data, flow should be shown using 3D flows with heights proportional to flow distance though it should be further tested. Further work could include testing additional encodings and additional tasks including collaborative tasks in multi-user immersive environments. We only investigated net (or one directional) OD flows. Bi-directional OD flows are another type of OD flow data commonly used in many analysis tasks. More complex layout are required to clearly present such bi-directional flows between one pair of locations and how to effectively lay them out in immersive environments is an interesting research topic. Another future direction could be exploring the design space of multivariate OD flow visualisation (e.g. how to represent volume and speed of road traffic) in immersive environments. This study used VR HMDs, as these currently offer the best field-of-view.  The results should be applicable to improved AR headsets as they become available, but this should also be tested in future work. 

%% file: content/7-conclusion/conclusion.tex
\chapter{Conclusions}
\label{chapter:conclusion}
Visualising origin-destination (OD) flow data has been an active area of research for more than one century. But, while researchers have explored many aspects of the design space, many other aspects are yet to be studied. We have contributed by creating novel visualisations that extend the design space for both 2D and immersive environments and conducting some of the first user studies evaluating OD flow data visualisations.

In this chapter, we take some time to summarise this research project. 
When examining the literature, we realised that some 2D OD flow visualisations have better scalability, while some other visualisations are better at presenting geographic context.
We discuss this trade-off in Section~\ref{sec:conclusion:scalability}. Meanwhile, to better justify the general design choices in immersive environments, the dimensionality of visualisations in immersive environments is discussed in Section~\ref{sec:conclusion:2d}. The summarised contributions are presented in Section~\ref{sec:conclusion:contribution}. Thoughts for future work are discussed in Section~\ref{sec:conclusion:future}.

\section{Scalability and Presenting Geographic Context}
\label{sec:conclusion:scalability}
In Section~\ref{sec:related:2d} and Chapter~\ref{chapter:od-flow-maps-2d}, we explore the design space of 2D OD flow visualisations. Two main difficulties were identified for visualising OD flow data:
\begin{itemize}
	\item Scalability; i.e. how to show large OD flow data with many flows while still retaining the ability to read single OD flows effectively.
	\item Presenting geographic context; i.e. how to present geographic information so as to reveal geographic patterns effectively.
\end{itemize}

Flow map approaches present geographic context in a straightforward way by placing the OD flows representations over a geographic map. However, such flow maps do not readily scale to large data because of crossings and overlaps of the comprised OD flows. Bundled flow maps attempt to improve scalability by using routing algorithms to reduce crossings and overlaps. However, the gained scalability is limited and rerouting the OD flows makes it difficult for people to follow individual flows (see Section~\ref{sec:maptrix:flow-map}).

\begin{figure}[b!] 
	\centering
	\includegraphics[width=0.5\textwidth]{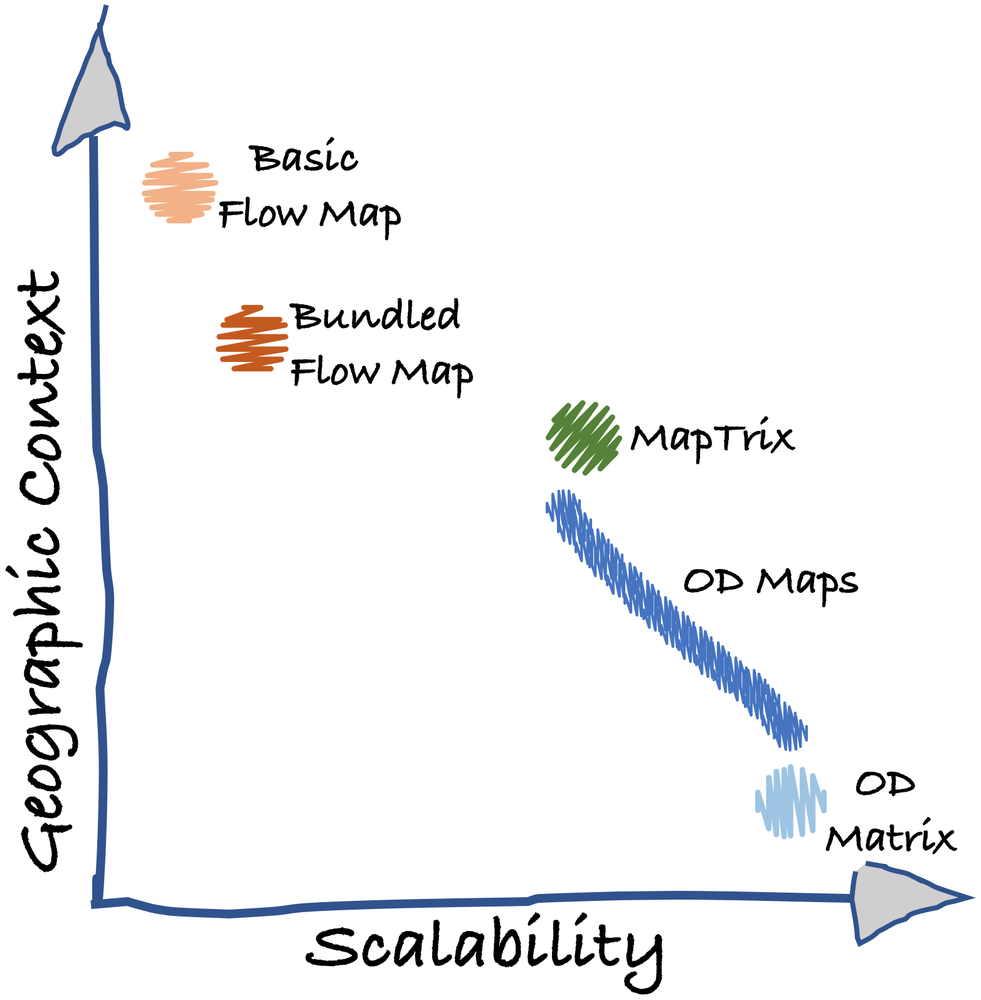}
	\caption{Graphical depiction of the trade-off between scalability and presenting geographic context for different visualisations. Positions of points in the plot are relative and qualitative, \textbf{NOT} based on precise metrics. The chart is therefore presented in a ``sketchy'' style.}
	\label{fig:conclusion:trade-off}
\end{figure}

OD matrices present OD flow data in a compact matrix which can scale to large data. However, geographic context is almost totally missing in this visual representation. 
\added{
Theoretically, any OD Maps can be rearranged and compacted into an OD matrix, thus in the limit OD Maps scale the same as OD matrix. However, such rearrangement may lose geographic fidelity. In this case, blank cells will need to be introduced to preserve fidelity and so the layout will be less compact. In Fig.~\ref{fig:conclusion:trade-off}, we use a sketchy line to present the flexibility of the OD Maps’ rearrangement process.
}
Compared to \emph{flow map} approaches, \emph{matrix-based} approaches provide better scalability. Furthermore, OD Maps provide a way to partially encode the underlying geographic information in the layout. But the grid-based layout can introduce distortions that can hinder the perception of geographic patterns.

Our method, MapTrix, which visualises OD flows by connecting two undistorted geographic maps to a compact OD matrix, is aimed at providing a better balance between scalability and presenting geographic context (see Section~\ref{sec:maptrix:maptrix}). Compared  to \emph{flow map} approaches, MapTrix achieves a better scalability by presenting OD flows using a standard OD matrix. Compared to \emph{matrix-based} approaches, MapTrix provides two undistorted geographic maps to help viewers find geographic patterns. However, although there are leader lines connecting the maps and the matrix, OD flows in MapTrix are more isolated from the geographic reference space (maps) than in \emph{flow map} approaches. Meanwhile, with two maps occupying the screen space, less scalability can be achieved compared to OD matrices. A similar situation applies to OD Maps, which were commonly presented with one origin-destination (OD) map and one destination-origin (DO) map~\citep{Kelly:2013wm}. Thus either map can take roughly half of the screen space.

In our study (see Chapter~\ref{chapter:evaluating-2d-od-flow-maps}), MapTrix and OD Maps had a similar performance, with each demonstrating a better scalability than bundled flow maps. All tasks tested in the user studies were geographically related, but we realised a possible limitation of the way we described the tasks in the studies. We labelled geographic locations in both MapTrix and OD maps, and randomly chose one or two labels to construct tasks. Such tasks could be focusing more on locating labels in visualisations rather than querying geographic information. If we describe the tasks in a more geographic way, participants may find MapTrix easier to locate geographic locations in its undistorted maps than OD maps with its grid-like layout. For example, instead of just using text description of tasks, an additional geographic map with highlighted labels can be used. This hypothesis needs to be confirmed with evaluations. 

In summary, a graphical depiction of the trade-off between scalability and presenting geographic context for different visualisations discussed is presented in Fig.~\ref{fig:conclusion:trade-off}. However, positions of points in the plot are relative and qualitative, \textbf{NOT} based on precise metrics. More precise tasks need to be designed to properly evaluate these visualisations for comparing scalability and presenting geographic context.

\section{2D and 3D in Immersive Environments}
\label{sec:conclusion:2d}

The most straightforward way to visualize data in immersive environments is by creating virtual 2D canvases and drawing traditional 2D visualisations on top of them. In this case the visual representations on such are identical to those displayed on the 2D screens. Mouse based interactions on 2D screens can be easily simulated with controllers or hand gestures in immersive environments. In addition, taking advantage of the space-tracking in immersive environments, physical interactions can be achieved in immersive environments, e.g. tilting or moving a 2D physical screen. In summary,  visualisation and interaction designs developed for 2D screens can be  adapted directly to immersive environments.  Such a simple use of visualisations in immersive environments is already available in some application scenarios. These scenarios include situated analytics~\citep{elsayed2016situated} where visualisations can be made available in almost any environmentsuch as in the field, surgery or factory floor.

One thing to be noted is that most current immersive applications are developed through game engines (e.g. Unity3D, Unreal etc). In these systems, all visual elements were rendered as meshes with triangles. Such a rendering technique is different from some of the commonly used 2D techniques (e.g. SVG, HTML5 Canvas Drawing). From our development experience (see Chapter~\ref{chapter:maps-globes-vr} and Chapter~\ref{chapter:flow-maps-vr}), with current developer tools, extra efforts were needed to port a 2D interactive visualisation to an immersive environment. There are emerging tools~\citep{cordeil:2018iatk,Cordeil:2017hy,Sicat:2018bo} focusing on building infrastructures in immersive environments to simplify this process. We can expect techniques to be more available and mature in the future.

In immersive environments, instead of porting 2D visualisations, we can also consider reimagining data visualisations to take advantage of the additional dimension in the 3D space surrounding the user. We demonstrated our attempts in Chapter~\ref{chapter:maps-globes-vr} and Chapter~\ref{chapter:flow-maps-vr}. Some of these redesigned visualisations outperformed some ported 2D visualisations in immersive environments in our controlled user studies. However, there is \textbf{NO} promise that redesigned 3D visualisations will always be more effective than 2D ones. Certain issues~\citep[Chap. 6]{Munzner:2014wj} need to be considered in the redesign process and the final design needs to be well justified with evaluations.

3D rendered visualisations with depth cue on 2D screens (see examples in Section~\ref{sec:intro:immersive}), can usually be directly rendered with a better depth perception in immersive environments. Furthermore, there is a potential to interact with those visualisations more intuitively in an immersive environment than on a 2D screen, especially for interactions like rotation and selection~\citep{Cordeil:2017dz,Cordeil:2013up}.

Immersive environments expand the design space of visualisations. However, there are limitations we need to notice. One concern raised from the literature is the limited resolution of immersive displays~\citep[Chap. 6]{Munzner:2014wj}. One of the most recent VR system, HTC Vive Pro, provides a resolution of 1440$\times$1600 pixels per eye. A recent AR system, Meta2, can achieve a similar but slightly less resolution of 1280$\times$1440 per eye. It seems the resolutions of these immersive systems can be comparable with 2D screens. However, please note that these are resolutions for the full Field of View (FOV) in immersive environments. The visualisations are usually placed at a distance to the viewer and only occupy a small portion of the FOV. As in the physical world, a 2D screen on top of a desk only take roughly 20\% of the user's FOV. The immersive technology is rapidly developing. For example, a prototype of 8K VR HMD (3840$\times$2160 per eye), Pimax, was demonstrated in late 2017. We can expect the resolution of the focus area in immersive environments to match that of a physical 2D screen in the near future.

\section{Contributions}
\label{sec:conclusion:contribution}
The main objective of our research was to design OD flow visualisations that better support analytic tasks. We summarise the contribution of this thesis in fulfilling this aim as follows:

\textbf{A novel 2D Many-to-Many OD Flow Visualisation --- MapTrix.}

	\vspace{-1em}
	\hangindent=2em \qquad For the two main 2D visualisation approaches for OD flow data (see Section~\ref{sec:intro:motivation}): \emph{flow map} approaches are intuitive, but become cluttered with a large number of OD flows. \emph{Matrix-based} approaches are more scalable, but lose all or part geographic contexts. We have introduced a new method, \emph{MapTrix}, for visualising many-to-many flows by connecting an OD matrix with origin and destination maps. We have provided a detailed analysis of the design alternatives and have given an algorithm for computing a layout  with crossing-free leader lines.

\textbf{The first experimental results regarding the effectiveness of different visual representations of many-to-many flow.} We find that OD Maps and MapTrix had similar performance while bundled flow maps did not scale at all well.

	\vspace{-1em}
	\hangindent=2em \qquad To better justify the design space of visualising OD flow data in 2D, we compared our novel visualisation MapTrix with other two state-of-the-art visualisation techniques. In our first study we compared MapTrix with a flow map with bundled edges and with OD Maps for different country maps. All three visualisations performed well for the smallest data set (Australia - 8 locations), but MapTrix and OD Maps were far better for Germany and New Zealand (16 locations). There was no statistically significant difference between MapTrix and OD Maps on data sets of this size.  Surprisingly, we did not find that country shape affected performance: in particular we had expected this to affect OD Maps. In our second study we compared MapTrix and OD Maps on two larger data sets (China - 34 locations and United States - 51 locations). Both performed relatively well for all tasks and we did not find that one method outperformed the other even for individual tasks. We did find in the pilot that analysing flow between or within regions for data sets of this size was extremely difficult with both methods, though slightly easier with OD Maps. Thus, in the study we used highlighting to help with analysis of regional flow. In the first study users ranked MapTrix highest in terms of design and readability while in the second study  MapTrix was preferred for design but OD Maps for readability.

\textbf{Four interactive visual representations of world-wide geographic maps in immersive environments.}

	\vspace{-1em}
	\hangindent=2em \qquad We designed: a 3D exocentric globe, where the user’s viewpoint is outside the globe; a flat map (rendered to a plane in VR); an egocentric 3Dglobe, with the viewpoint inside the globe; and curved map, created by projecting the map onto a section of a sphere which curves around the user. In all four visualisations the geographic centre could be smoothly adjusted with a standard hand held VR controller. The user, through a head-tracked headset, could physically move around the visualisation.

\textbf{The first experimental results regarding the effectiveness of different visualisations of global geography in VR.} Overall, our results provide support for the use of exocentric globes for geographic visualisation in immersive environments.

	\vspace{-1em}
	\hangindent=2em \qquad We tested these four different interactive visual representations in VR for three analytic tasks: distance comparison, area comparison and direction estimation. Of the four conditions and three task types tested, we found that the exocentric globe was generally the best choice of VR visualisation. This was despite the fact that less of the earth’s surface was visible in the exocentric globe than the other representations and that it had the most perceptual distortion, though no distortion due to map projection. We also found that the curved map had benefits over the flat map, but the curved map caused the users greater motion-sickness. In almost all cases the egocentric globe was found to be the least effective visualisation.

\textbf{The first experimental results regarding the effectiveness of different 2D and 3D representations for OD flows on flat maps in VR.} We found that participants were significantly more accurate with raised flow paths whose heights were proportional to flow distance but fastest with traditional straight line 2D flows.

	\vspace{-1em}
	\hangindent=2em \qquad We have found strong evidence that 2D representations are \textbf{not} the best way to show OD flow data in an such environment, and the use of 3D flow maps can allow viewers to resolve overlapping flows by changing the relative position of the head and object. However, the particular 3D design choices of the visualisation had a significant effect. For example, encoding flow height to distance was clearly better than to quantity.

\textbf{Four interactive flow maps for global flows in immersive environments.}

	\vspace{-1em}
	\hangindent=2em \qquad We designed: a flat map with 2D flat lines and 3D tubes, 3D tubes on a globe and a novel interactive design we called MapsLink, involving a pair of linked flat maps with 3D tubes. Again, in all four visualisations, the geographic centre could be smoothly adjusted and the user could physically move around the visualisation. In addition, user could manipulate the rotation and position of the visualisation by using a standard hand held VR controller to grab the visualisation.

\textbf{The first experimental results regarding the effectiveness of different OD flow maps in VR.} We found that participants took significantly more time with MapsLink than other flow maps while the 3D globe with raised flows was the fastest, most accurate, and most preferred method.

	\vspace{-1em}
	\hangindent=2em \qquad We found that for global flows, the most accurate and preferred representation was a 3D globe with raised flows whose height is proportional to the flow distance, while for regional flow data the best view would be a flat map with distance-proportional raised 3D flows. Nonetheless, the 3D representations were still more accurate and preferred. Our findings suggest that globes are preferable for visualising global OD flow data in  immersive environments. For regional OD data, flow should be shown using 3D flows with heights proportional to flow distance.

	\vspace{-1em}
	\hangindent=2em \qquad Addtionally, we also compared a standard flat flow map with straight lines with or without \emph{geo-rotation} interation, in which the user could interactively reposition the centre of the map. We found that accuracy could be significantly improved by the use of \emph{geo-rotation}.

\section{Future Work}
\label{sec:conclusion:future}
The design space of visualising OD flow data is huge in both 2D and immersive environments. We highlight in this section some possible directions for future research:

\textbf{Hybrid OD flow visualisation in immersive environments.}

	\vspace{-1em}
	\hangindent=2em \qquad In Chapter~\ref{chapter:od-flow-maps-2d}, we explored three different methods of visualising OD flow data in 2D displays, namely bundled flow maps (a node-link based approach), OD maps (a matrix-based approach) and MapTrix (a hybrid approach). In the user studies (see Chapter~\ref{chapter:evaluating-2d-od-flow-maps}), we found that matrix-based and hybrid approaches scale better than node-link based approaches. In Chapter~\ref{chapter:flow-maps-vr}, we explored the design space of flow maps (node-link based approaches) in immersive environments mainly because such approaches are intuitive and familiar to potential users. 

	\begin{figure}[t!]
		\centering
		\includegraphics[width=\textwidth]{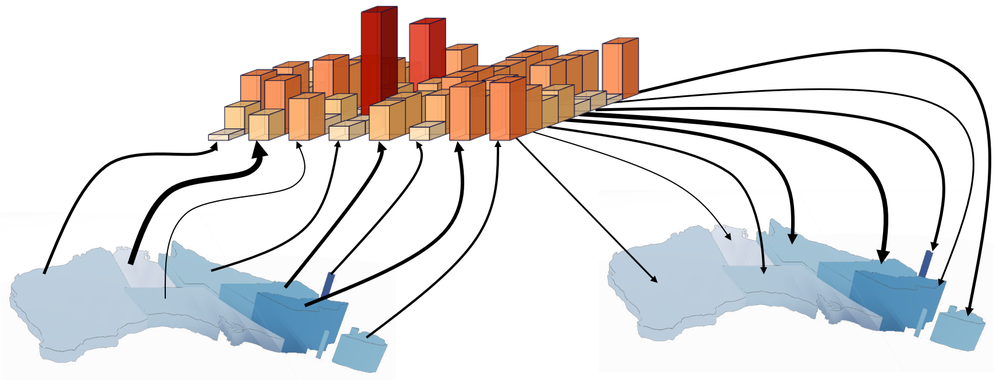} 
		\caption{
		\label{fig:conclusion:maptrix-vr}
		A mock-up of one possible hybrid OD flow visualisation design in immersive environments.
		}
	\end{figure}

	\vspace{-1em}
	\hangindent=2em \qquad A further research question worth pursuing is “How can we visualise OD flow data effectively in a MapTrix-based way within immersive environments?” There are many components that need to be considered to explore such a design space: (a) the matrix, (b) the maps and (c) the leader lines. The additional third dimension (height) in immersive environments may allow us to encode additional information and/or encode the existing information in a different way. Fig.~\ref{fig:conclusion:maptrix-vr} demonstrates one possible hybrid OD flow visualisation design in immersive environments, in which (a) a 3D bar chart is used to present OD matrix (inspired by~\citep{Jansen:2013kzb}), (b) prism maps are used as origin and destination maps, the height can be used to present an additional property like population density, (c) as for MapTrix, the leader lines are connecting the maps with the OD matrix. Due to the additional dimension, 3D leader line layout algorithms should be considered to fully take the advantage of the additional space. Furthermore, the algorithm also needs to consider the high interactivity (e.g. manipulation of objects in 3D space) in immersive environments.
 
\textbf{Present OD flow data with temporal information.}

	\vspace{-1em}
	\hangindent=2em \qquad In the interviews with the domain experts (see Chapter~\ref{chapter:interviews}), temporal information was available in almost all cases, and in some cases was crucial for their analysis. Thus, how to present temporal information with OD flow data in an effective way is a practical problem. Scalability is already a challenging issue for visualising OD flow data. With the requirement of presenting an additional temporal dimension, understanding how to visualise such large data in an effective way became an even more serious problem.

	\vspace{-0.5em}
	\hangindent=2em \qquad Two general approaches have been evaluated qualitatively to present temporal information in flow maps: animation and small-multiples. Animation representation the changes of OD flow data with interpolated transitions between the time periods.  Small-multiples arranged static visualisations in a grid format   with each static visualization showing a snapshot of the data at a specific time stamp. In the qualitative study with animation, participants tended to make more findings concerning geographically local events and changes between subsequent years. While with small-multiples, more findings concerning longer time periods were made by participants ~\citep{Boyandin:2012ip}. 

	\vspace{-0.5em}
	\hangindent=2em \qquad Boyandin \textit{et al.} explored encoding the temporal information of OD flow data in another way. Their system, Flowstrates, displays origins and destinations of the flows in two separate maps, and presents changes over time of the flow magnitudes in a separate heatmap view in the middle~\citep{Boyandin:2011fa}. However, scalability still remains to be an issue with their design as the number of origins and destinations increases.

	\begin{figure}[t!]
		\centering
		\includegraphics[width=\textwidth]{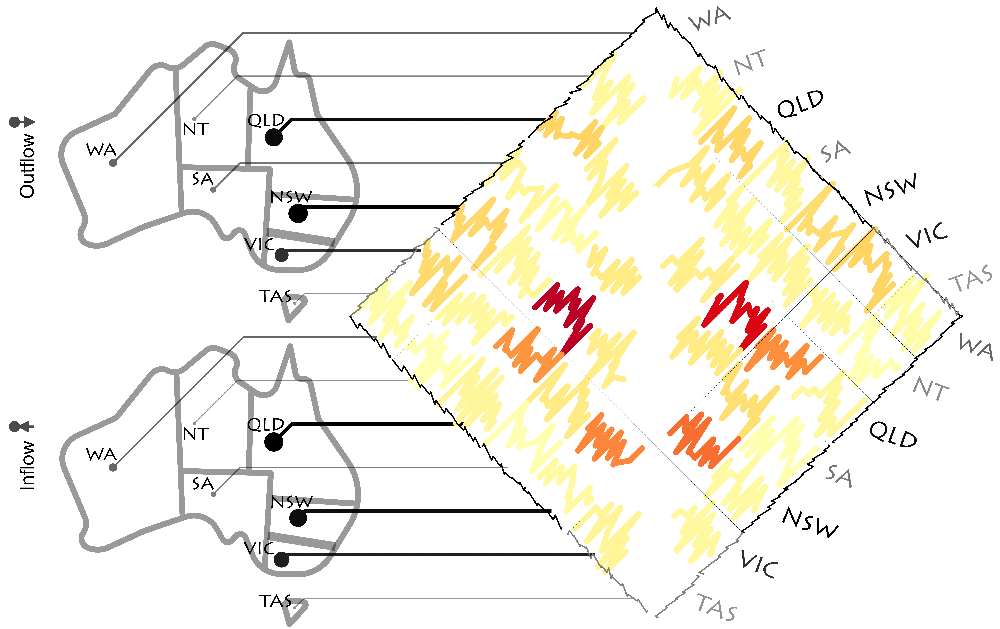} 
		\caption{
		\label{fig:conclusion:maptrix-time}
		A hand-drawn sketch of embedding line charts into the cells of MapTrix to present temporal information (random data is presented in this sketch).
		}
	\end{figure}

	\vspace{-0.5em}
	\hangindent=2em \qquad An interesting direction for us might be looking at modifying the design of MapTrix to embed temporal information. Inspired by Wood \textit{et al.}\citep{Wood:2011ez}, we can put small static charts in the cells of MapTrix as demonstrated in Fig.~\ref{fig:conclusion:maptrix-time}.

	\vspace{-0.5em}
	\hangindent=2em \qquad An alternative direction can be investigating using these similar static chart on traditional flow maps and adding sophisticated interactions to support iterative exploration. Part of the design space has been explored recently by Andrienko~\textit{et al.} in~\citep{Andrienko:2017fe}. There is still remaining design space we can explore.

\textbf{Visualise trajectory data in immersive environments.}

	\vspace{-1em} 
	\hangindent=2em \qquad As mentioned in Section~\ref{sec:intro:motivation}, trajectory data is another type of geographically-embedded flow data. Trajectory data contains continuous information about the physical routes between origins and destinations. Such trajectories are collected in many application domains, e.g. air traffic (more details discussed with our interviewee in Section~\ref{sec:interviews:airline}), vehicle GPS, vessel traffic, etc. In these cases, the continuous physical routes are important, so we cannot re-route these trajectories for aesthetic or clutter reduction reasons as is possible for the more abstract flow maps.

	\vspace{-0.5em}
	\hangindent=2em \qquad Due to detailed information of trajectory data, aggregation and clustering are widely used and different techniques have been investigated to reduce the size of data or show only the most important features~\citep{Andrienko:2013dx,Andrienko:2017fp,Andrienko:2012fu,Andrienko:2010gx,Andrienko:2013kc,Rinzivillo:2008bc}. Several conceptual frameworks have been proposed by Andrienko~\textit{et al.} to support systematically analysing such trajectory data with visualisations equipped with interactions including these aggregation and clustering techniques~\citep{Andrienko:2013hs,Andrienko:2006up,Andrienko:2018iv}. 

	\vspace{-0.5em}
	\hangindent=2em \qquad However, these techniques have not been well explored in immersive environments. To the best of our knowledge, the only preliminary exploration has been done by Hurter~\textit{et al.} with interactive ways of selecting trajectories in immersive environments~\citep{Hurter:2019hz}. 
	There is still a large unexplored design space of interaction and visualisation with respect to analysing trajectory data in immersive environments.

\textbf{Further explore the design space of interactions in immersive environments.}
	
	\vspace{-1em} 
	\hangindent=2em \qquad In our VR user studies (see Chapter~\ref{chapter:maps-globes-vr} and Chapter~\ref{chapter:flow-maps-vr}), we provided some standard interactions, such as: users moving physically and users rotating or repositioning virtual objects. We also introduced geo-rotation into the immersive environments allowing a user to smoothly change the geographic centre of a map. However, there are possibly more interactions we should consider. For example, in Section ~\ref{sec:maptrix:interaction}, we implemented several standard interactions for MapTrix based on participants' feedback, including:
	\begin{itemize}[leftmargin=5em]
	 	\item Selecting a single OD flow to highlight, or a location to highlight all flows associated with it; 
	 	\item Filtering flows with a certain range of magnitudes, or with the total flow magnitudes of locations.
	 	\item Grouping regional flows.
	\end{itemize}

	\vspace{-0.5em} 
	\hangindent=2em \qquad There are also other standard interactions in geographic visualisation, like zooming. Those interaction techniques should be considered in immersive flow maps. Among those techniques, selection has been detailed and discussed in recent research with the respect of trajectory visualisation in immersive environments~\citep{Hurter:2019hz}. However, other interaction techniques in immersive environments are understudied.

\textbf{Evaluate and improve our tools with domain specialists.}

	\vspace{-1em} 
	\hangindent=2em \qquad It has been widely accepted by the research community that evaluation is an important part in human-centred visualisation design and that it is necessary to provide empirical evidence of the effectiveness and efficiency as well as usability of new visualisation methods~\citep{Aigner:2013,Plaisant:2004eh}. We evaluated our tools with controlled user studies focusing on perception and low-level analytic tasks (see Chapter~\ref{chapter:evaluating-2d-od-flow-maps}, Chapter~\ref{chapter:maps-globes-vr} and Chapter~\ref{chapter:flow-maps-vr}). However, we have not systematically evaluated the performance of our tools for high-level analytic tasks, like thinking, discovering, deciding and exploring (examples discussed in Chapter~\ref{chapter:interviews}). 

	\vspace{-0.5em}
	\hangindent=2em \qquad Asking domain specialists to test a visualisation technique in their daily  tasks is another way of evaluation~\citep{Tory:2005df}. 
	Compared to a few minutes training time in a controlled user study, participants can have much more time to get familiar with the visualisation technique. High-level analytic tasks are performed by participants with the tools, and usually qualitative feedback data are collected. Lloyd \textit{et al.} conducted a three year study on a specific geovisualisation design project with domain specialists, and they reported that a long term cooperation with domain specialists can elicit requirements from the end users~\citep{Lloyd:2011ga}. The expert interviews described in Chapter~\ref{chapter:interviews} is our initial step. We believe that by working closely with domain specialists, we can identify the existing problems of our tools and potentially discover novel ideas of visualising the OD flow data. 

\section{Closing Remarks}
\label{sec:conclusion:closing}
This thesis has presented paradigms of OD flow visualisations in 2D and immersive environments, as well as guidelines of some design choices in these implementations. These are intended to assist analysing OD flow data. The domain specialists we spoke to are all keen to gain deep insights from geographically-embedded flow data (see Chapter~\ref{chapter:interviews}), however, there is a clear gap between the information they want to extract and the capability of existing commonly used analytical tools. While geographically-embedded flow data is widely collected in different industrials, compared to other geographic data (e.g. demography data), only a limited amount can be easily accessed. It is hoped that geographically-embedded flow data will become more available, and knowledge of how to analyse them effectively will eventually lead to a variety of different visualisation designs to meet the needs in different domains.

The emerging immersive technologies open up new possibilities for exploring the design space of visualisation in such immersive environments~\citep{Marriott:2018ig}. We demonstrated our initial attempts of immersive OD flow maps in this thesis. We can expect that as these techniques become more sophisticated and mature, the physical screens will be no longer needed, and powerful immersive applications will be available in our daily life and make our work flows and communication more efficient.